\documentclass[a4paper,11pt]{article}

\usepackage{amsmath,amsthm,amsbsy,amssymb,amsfonts,graphicx,mathrsfs,bm,cite,color,accents}
\usepackage{hyperref,bookmark}
\usepackage[scr=boondoxo]{mathalfa}

\DeclareRobustCommand{\vect}[1]{\bm{#1}}
\makeatletter
\catcode`\@=11
\@addtoreset{equation}{section}

\makeatother

\usepackage[utf8]{inputenc}

\renewcommand{\thefootnote}{\arabic{footnote}}

\newcommand{\Exp}[1]{\operatorname{e}^{#1}}
\newcommand{\diag}{\operatorname{diag}}
\newcommand{\tr}{\operatorname{tr}}
\newcommand{\abs}[1]{\lvert {#1} \rvert}
\newcommand{\rmd}{{\mathrm{d}}}
\newcommand{\rmT}{{\tt T}}
\newcommand{\nn}{\nonumber}
\newcommand{\Lie}{\pounds}
\newcommand{\gLie}{\hat{\pounds}}

\newcommand{\cA}{\mathcal A}\newcommand{\cB}{\mathcal B}
\newcommand{\cC}{\mathcal C}\newcommand{\cD}{\mathcal D}
\newcommand{\cE}{\mathcal E}\newcommand{\cF}{\mathcal F}
\newcommand{\cG}{\mathcal G}\newcommand{\cH}{\mathcal H}
\newcommand{\cI}{\mathcal I}\newcommand{\cJ}{\mathcal J}
\newcommand{\cK}{\mathcal K}\newcommand{\cL}{\mathcal L}
\newcommand{\cM}{\mathcal M}\newcommand{\cN}{\mathcal N}
\newcommand{\cO}{\mathcal O}\newcommand{\cP}{\mathcal P}
\newcommand{\cR}{\mathcal R}
\newcommand{\cS}{\mathcal S}

\newcommand{\cW}{\mathcal W}\newcommand{\cX}{\mathcal X}
\newcommand{\cY}{\mathcal Y}\newcommand{\cZ}{\mathcal Z}

\newcommand{\SL}{\text{SL}}
\newcommand{\SU}{\text{SU}}
\newcommand{\SO}{\text{SO}}
\newcommand{\UU}{\text{U}}
\newcommand{\GL}{\text{GL}}

\newcommand{\OO}{\text{O}}

\newcommand{\sfi}{\mathsf{i}}
\newcommand{\sfj}{\mathsf{j}}
\newcommand{\sfk}{\mathsf{k}}
\newcommand{\sfl}{\mathsf{l}}

\newcommand{\sfI}{\mathsf{I}}
\newcommand{\sfJ}{\mathsf{J}}
\newcommand{\sfK}{\mathsf{K}}

\newcommand{\MM}{\check{\cE}}

\newcommand{\bra}[1]{\langle #1 \rvert}
\newcommand{\ket}[1]{\lvert #1 \rangle}
\newcommand{\sla}[1]{\setbox0=\hbox{$#1$} 
\dimen0=\wd0 \setbox1=\hbox{/} \dimen1=\wd1 
\ifdim\dimen0>\dimen1 \rlap{\hbox to \dimen0{\hfil/\hfil}} #1 
\else\rlap{\hbox to \dimen1{\hfil$#1$\hfil}} / \fi}
\DeclareMathOperator{\arcsinh}{arcsinh}

\makeatletter
\newcommand{\ubar}[1]{\underaccent{\bar}{#1}}
\makeatother

\newcommand{\bcheck}[1]{\,\check{\bm{\!#1}}}

\renewcommand{\div}{\operatorname{div}}

\newcommand{\CE}{\cE}

\newcommand{\VV}{K}
\newcommand{\KK}{K}
\newcommand{\kk}{k}
\newcommand{\hh}{m}

\allowdisplaybreaks[3]

\begin{document}

\begin{titlepage}
\renewcommand{\thefootnote}{\fnsymbol{footnote}}

\vspace*{1cm}

\begin{center}
\LARGE\textbf{Poisson--Lie $T$-plurality for dressing cosets}
\end{center}

\vspace{1.2cm}

\centerline{
Yuho Sakatani%
\footnote{E-mail address: \texttt{yuho@koto.kpu-m.ac.jp}}
}

\begin{center}
{\it Department of Physics, Kyoto Prefectural University of Medicine,}\\
{\it 1-5 Shimogamohangi-cho, Sakyo-ku, Kyoto, Japan}
\end{center}

\vspace*{2mm}

\begin{abstract}
The Poisson--Lie $T$-plurality is an equivalence of string theories on various cosets $\mathcal{D}/\tilde{G}$, $\mathcal{D}/\tilde{G}'$, $\cdots$, where $\mathcal{D}$ is a Drinfel'd double and $\tilde{G}$, $\tilde{G}'$, $\cdots$ are maximal isotropic subgroups. This can be extended to the equivalence for dressing cosets, i.e., $F\backslash\mathcal{D}/\tilde{G}$, $F\backslash\mathcal{D}/\tilde{G}'$, $\cdots$, where $F$ is an isotropic subgroup of $\mathcal{D}$. We explore this extended Poisson--Lie $T$-plurality, clarifying the relation between several previous approaches. We propose a gauged sigma model for a general gauge group $F$ and obtain the formula for the metric and the $B$-field on the dressing coset. Using this formula and an ansatz for the dilaton, we show that the Poisson--Lie $T$-plurality for dressing cosets (with spectator fields) is a symmetry of double field theory. The formula for the R--R field strength is also proposed such that the equations of motion for the NS--NS fields are transformed covariantly. In addition, we provide specific examples of the PL $T$-plurality for dressing cosets. 
\end{abstract}

\thispagestyle{empty}
\end{titlepage}

\setcounter{footnote}{0}

\newpage

\renewcommand{\baselinestretch}{0.7}\normalsize
\tableofcontents
\thispagestyle{empty}
\setcounter{page}{0}
\renewcommand{\baselinestretch}{1.35}\normalsize

\newpage

\section{Introduction}
\label{sec:intro}

\subsection{Poisson--Lie $T$-duality}

$T$-duality is an equivalence of string theories on several different spaces. 
Some extensions of $T$-duality have now been proposed, but the original one requires the presence of $D$ mutually commuting (i.e., Abelian) Killing vector fields and is usually called Abelian $T$-duality \cite{Kikkawa:1984cp,Sakai:1985cs,Buscher:1987sk,Buscher:1987qj}. 
The supergravity background fields of two mutually dual models are related to each other through an $\OO(D,D)$ transformation.
This transformation is also a symmetry of supergravity equations of motion, and if we find a supergravity solution admitting $D$ Abelian Killing vector fields, we can generate new supergravity solutions by acting various $\OO(D,D)$ transformations. 
However, the assumption that $D$ Abelian Killing vector fields exist is rather restrictive, and in a general space, we cannot perform such $\OO(D,D)$ transformations. 

It was later realized that the $T$-duality, as a symmetry of supergravity equations of motion, can be extended to the case where the Killing vector fields form a non-Abelian algebra \cite{hep-th:9210021,hep-th:9309039,hep-th:9403155}. 
This symmetry is called non-Abelian $T$-duality, and has provided various new supergravity solutions. 
An issue in this generalized $T$-duality is that, after the dualization, we may get a space where the Killing vector field does not exist. 
Then, we cannot go back to the original background, and in this sense, this is not a usual duality. 
This issue can be resolved if we consider a further generalization, called the Poisson--Lie (PL) $T$-duality \cite{hep-th:9502122,hep-th:9509095}. 

To perform the PL $T$-duality, again the target space $M$ should satisfy some dualizability condition. 
If we denote a combination of the metric $g_{mn}$ and the Kalb--Ramond $B$-field $B_{mn}$ as $E_{mn}\equiv g_{mn}+B_{mn}$ ($m,n=1,\dotsc,D$), the condition can be expressed as \cite{hep-th:9502122}
\begin{align}
 \Lie_{v_a}E_{mn} = - f_a{}^{bc}\,E_{mp}\,v^p_b\,v_c^q\,E_{qn}\,,\qquad [v_a,\,v_b]=f_{ab}{}^c\,v_c\qquad (a,b=1,\dotsc,D)\,. 
\label{eq:PL-dualizability}
\end{align}
The conditions to perform Abelian and non-Abelian $T$-duality are reproduced as specific cases, $f_{ab}{}^{c}=f_a{}^{bc}=0$ and $f_a{}^{bc}=0$\,, respectively. 
When $f_a{}^{bc}\neq 0$\,, even though the background fields $E_{mn}$ do not satisfy the usual Killing equation, we can still perform the PL $T$-duality. 
Some consistency conditions suggest that $f_{ab}{}^c$ and $f_a{}^{bc}$ can be identified as the structure constants of a $2D$-dimensional Lie algebra $\mathfrak{d}$, called the Drinfel'd double \cite{hep-th:9502122}
\begin{align}
 [T_a,\,T_b]=f_{ab}{}^c\,T_c\,,\qquad
 [T^a,\,T^b]=f_c{}^{ab}\,T^c\,,\qquad
 [T^a,\,T_b]= - f_b{}^{ac}\,T_c + f_{bc}{}^a\,T^c\,,
\label{eq:DD-alg}
\end{align}
which admits an adjoint-invariant bilinear form
\begin{align}
 \langle T_A,\,T_B\rangle \equiv \begin{pmatrix} 0 & \delta_a^b \\ \delta^a_b & 0 \end{pmatrix} \equiv \eta_{AB}\,, \qquad (T_A)\equiv (T_a,\,T^a) \,.
\label{eq:eta-AB}
\end{align}
Under this identification, we can find a solution of the differential equation \eqref{eq:PL-dualizability} as
\begin{align}
 E_{mn} = r_m^a\,[(\hat{E}+\pi)^{-1}]_{ab}\,r^b_n\,,
\label{eq:PL-symmetric}
\end{align}
where the matrix $\hat{E}^{ab}$ is the integration ``constant''\footnote{Here `` '' indicate that $\hat{E}^{ab}$ may depend on the coordinates on a certain space orthogonal to $M$. These coordinates are invariant under $T$-duality, and are sometimes called spectator fields or external coordinates.} and $r_m^a$ and $\pi^{ab}$ are some geometric quantities associated with the Drinfel'd double. 
Backgrounds of the form \eqref{eq:PL-symmetric} are called PL symmetric, and they serve as a starting point to perform the PL $T$-duality. 

The Lie algebra of the Drinfel'd double \eqref{eq:DD-alg} contains two $D$-dimensional subalgebras: $\mathfrak{g}$ defined by $[T_a,\,T_b]=f_{ab}{}^c\,T_c$ and $\tilde{\mathfrak{g}}$ defined by $[T^a,\,T^b]=f_c{}^{ab}\,T^c$ (whose Lie groups are denoted as $G$ and $\tilde{G}$).
Accordingly, the Lie algebra $\mathfrak{d}$ of the Drinfel'd double is sometimes denoted as the pair $(\mathfrak{g}|\tilde{\mathfrak{g}})$. 
The two subalgebras appear symmetrically in the algebra \eqref{eq:DD-alg}, and if we consider a redefinition, $(\tilde{T}_a,\,\tilde{T}^a)\equiv (T^a,\,T_a)$, they also satisfy the Lie algebra of the Drinfel'd double with the new structure constants $\tilde{f}_{ab}{}^c=f_c{}^{ab}$ and $\tilde{f}_a{}^{bc}=f_{bc}{}^a$. 
Since the role of $\mathfrak{g}$ and $\tilde{\mathfrak{g}}$ are interchanged, the algebra after the redefinition can be denoted as $(\tilde{\mathfrak{g}}|\mathfrak{g})$.
From the point of view of the Drinfel'd double $\mathfrak{d}$, this is merely a change of basis, but the change in decomposition has an important consequence. 
The decomposed pair $(\mathfrak{g}|\tilde{\mathfrak{g}})$ or the triple $(\mathfrak{d},\,\mathfrak{g},\,\tilde{\mathfrak{g}})$ is called the Manin triple, and the new Manin triple $(\tilde{\mathfrak{g}}|\mathfrak{g})$ gives new geometric objects $\tilde{r}_{am}$ and $\tilde{\pi}_{ab}$\,.\footnote{The position of the indices $a,b$ is changed due to the exchange $T_a\leftrightarrow T^a$.} 
Using these, we obtain a new PL-symmetric background
\begin{align}
 \tilde{E}_{mn} = \tilde{r}_{am}\,[(\tilde{\hat{E}}+\tilde{\pi})^{-1}]^{ab}\,\tilde{r}_{bn}\,,
\label{eq:PL-symmetric-d}
\end{align}
where $\tilde{\hat{E}}_{ab}$ is related to $\hat{E}^{ab}$ through the Buscher rule: $\tilde{\hat{E}} = \hat{E}^{-1}$. 
Then, the PL $T$-duality can be stated as the equivalence of string theories on the two backgrounds $E_{mn}$ and $\tilde{E}_{mn}$\,. 

In general, a single Drinfel'd double has several Manin triples $(\mathfrak{g}|\tilde{\mathfrak{g}})\cong (\mathfrak{g}'|\tilde{\mathfrak{g}}')\cong \cdots$, and we can construct a PL-symmetric background for each Manin triple. 
The equivalence of string theories on these multiple backgrounds is called the PL $T$-plurality \cite{hep-th:0205245} (though it is often referred to simply as the PL $T$-duality). 

\subsection{Dressing cosets}

Various physically interesting backgrounds can be constructed as PL-symmetric backgrounds, but the expression \eqref{eq:PL-symmetric} is still restrictive. 
To illustrate this point, let us consider the setup of the traditional non-Abelian $T$-duality, i.e., $f_a{}^{bc}=0$\,. 
Here, the background admits Killing vector fields $v_a^m$ satisfying the algebra $[v_a,\,v_b]=f_{ab}{}^c\,v_c$\,. 
They can be obtained as the left-invariant vector fields associated with the Lie group $G$, and the flow of the Killing vector fields $v_a^m$ generates a group action of $G$ on the background. 
The key point is that the left-invariant vector fields $v_a^m$ do not vanish anywhere and the isometry group $G$ acts on the background without any fixed point. 
In the context of non-Abelian $T$-duality, this is known to be a little specific setup (see e.g., section 3.2 of \cite{1104.5196}). 
For example, if we consider a coset space $H\backslash G$ such as the 2-sphere $\text{S}^2$, group actions of $G$ can have certain fixed points on the coset space $H\backslash G$. 
To perform the PL $T$-duality in such a space, a more general framework needs to be developed.\footnote{This does not mean that the standard PL $T$-duality cannot be applied to a coset space. For example, $\text{S}^3$ is a coset space $\OO(3)\backslash \OO(4)$. To perform a PL $T$-duality with respect to the isometry group $\OO(4)$, we need the general framework (see e.g., \cite{1104.5196}). However, $\text{S}^3$ is isomorphic to $\SU(2)$, and we can perform the standard PL $T$-duality with respect to this $\SU(2)$. Other examples are the PL $T$-duality of coset spaces, AdS$_2$ and $\text{S}^2$. Their PL $T$-dualities are discussed in this paper by using the general framework, but a recent paper \cite{2111.07700} found that some two-dimensional group acts freely and transitively on a certain coordinate patch of AdS$_2$ or $\text{S}^2$. Then, we can perform the standard PL $T$-duality with respect to the two-dimensional isometry group.}
Such a framework has been studied in \cite{hep-th:9602162,hep-th:9904188}, and novel types of geometries, called dressing cosets, have been proposed.

To explain the notion of dressing cosets, let us comment on the PL-symmetric backgrounds \eqref{eq:PL-symmetric} in more detail. 
The string sigma model on the background \eqref{eq:PL-symmetric} can be reproduced from a gauged sigma model, where the target space is the group manifold of a Drinfel'd double $\cD=\exp\mathfrak{d}$ and multiplications of group elements $\tilde{h}\in \tilde{G}$ from the right are gauged (see section \ref{sec:Hull-Reid-Edwards} for a review). 
Then, the background \eqref{eq:PL-symmetric} can be regarded as a coset space $M=\cD/\tilde{G}$ \cite{hep-th:9605212}. 
For simplicity, we introduce an embedding function of a string $l(\sigma)\in \cD$ and assume that this can be decomposed as $l(\sigma)=g(\sigma)\,\tilde{g}(\sigma)$, where $g(\sigma)\in G$ and $\tilde{g}(\sigma)\in \tilde{G}$.
Then we can choose $\tilde{g}(\sigma)=1$ by using the gauge symmetry, and we can consider that the string is propagating on the group manifold $G$ with the embedding function $g(\sigma)$. 
Since $g(\sigma)$ is an element of the Drinfel'd double, we can consider an action of $\cD$ on $g(\sigma)$ from the left. 
If we multiply an element $h\in G$ from the left, $g(\sigma)$ is transformed as $g'(\sigma)=h\,g(\sigma)$ but $\tilde{g}(\sigma)=1$ is unchanged. 
Instead, if we multiply an element $\tilde{h}\in \tilde{G}$ from the left, we obtain
\begin{align}
 g(\sigma)\to \tilde{h}\,g(\sigma)= g'(\sigma)\,\tilde{h}'(\sigma)\,,\qquad \bigl[ g'(\sigma)\in G,\quad \tilde{h}'(\sigma)\in\tilde{G}\bigr]\,,
\end{align}
and $\tilde{h}'(\sigma)$ can be removed by using the gauge symmetry. 
In this way, a general element of $\cD$ acts on $G$ and produces an element of $G$ in a non-trivial manner using the gauge symmetry. 
This type of the action of $\cD$ on $G$ is called the dressing action. 

The dressing coset \cite{hep-th:9602162} is defined as a double coset $\check{M}=F\backslash \cD/\tilde{G}$ where $F$ is a subgroup of $\cD$ satisfying some properties. 
This type of background can be obtained by gauging the left multiplication by $F$ (i.e., the dressing action is gauged). 
When $F$ is a subgroup of $G$, the action of $F$ on $G$ is the usual one and we obtain a standard coset, but for a general $F$ that is not a subgroup of $G$, we obtain a dressing coset. 
The important finding of \cite{hep-th:9602162,hep-th:9904188} is that the PL $T$-duality, which exchanges $G$ and $\tilde{G}$, is a symmetry of string theory even for the dressing cosets $\check{M}=F\backslash \cD/\tilde{G}$. 
Namely, it has been shown that string theory on $\check{M}=F\backslash \cD/\tilde{G}$ is equivalent to that on $\tilde{\check{M}}=F\backslash \cD/G$, where the gauge group $F$ is not changed. 
The main motivation of this paper is to clarify the structure of this extended PL $T$-duality by using $T$-duality-covariant formulations, which are developed recently. 

\subsection{Previous studies}

Here, we briefly summarize the history of the PL $T$-duality for dressing cosets and comment on possible challenges that need to be addressed.

\subsubsection{Original proposal}
\label{sec:K-S-review}

The PL $T$-duality of dressing cosets has been proposed initially by Klim\v{c}\'\i{}k and \v{S}evera \cite{hep-th:9602162}. 
They studied the string sigma model on a PL-symmetric background $M=\cD/\tilde{G}$
\begin{align}
 \check{E}_{mn} = r_m^a\, \bigl[(\cR+\pi)^{-1}\bigr]_{ab}\,r^b_n\,.
\label{eq:orig-model}
\end{align}
The difference from the usual PL $T$-duality is that the ``constant'' matrix $\cR^{ab}$ satisfies some special properties. 
Let us introduce two linear subspaces of $\mathfrak{d}$
\begin{align}
 \cR\equiv\text{span}\bigl\{T^a+T_b\,\cR^{ba}\bigr\}\,,\qquad \cR^\perp\equiv\text{span}\bigl\{T^a - \cR^{ab}\,T_b\bigr\}\,,
\label{eq:KS-gauge}
\end{align}
and suppose that their intersection $\mathfrak{f}=\cR\cap\cR^\perp$ forms a Lie subalgebra. 
Here, in order to obtain a non-trivial intersection $\cR\cap\cR^\perp$, the symmetric part of $\cR^{ab}$ should be degenerate. 
Moreover, we suppose that the matrix $\cR^{ab}$ is chosen such that the subspace $\cR$ (or $\cR^\perp$) is invariant under the adjoint action of the Lie group $F$ associated with $\mathfrak{f}$\,,
\begin{align}
 f\,(T^a+T_b\,\cR^{ba})\,f^{-1} \in \cR \qquad (\forall f\in F)\,.
\end{align}
Under such a situation, it was found that the sigma model on $M=\cD/\tilde{G}$ develops a gauge symmetry that corresponds to the local left multiplication by $F$, provided that the right-invariant 1-form $\rmd l(\sigma)\,l^{-1}(\sigma)$ is constrained to be orthogonal to the linear subspace $\mathfrak{f}$\,:
\begin{align}
 \langle \rmd l\,l^{-1},\, \mathfrak{f} \rangle = 0\,.
\label{eq:r-f-orthogonal}
\end{align}
Due to the appearance of this gauge symmetry, the sigma model can be found to be defined on the dressing coset $\check{M}= F\backslash \cD/\tilde{G}$. 
Apparently, the background fields $\check{E}_{mn}$ are defined on a $D$-dimensional space, but by imposing certain gauge-fixing conditions (e.g., several coordinates are fixed to constants), the matrix $\check{E}_{mn}$ is reduced to a $d\times d$ matrix $E_{\check{m}\check{n}}$ ($\check{m},\check{n}=1,\dotsc,d$, $d\equiv D-n$, $n\equiv \dim F$), which plays the role of the supergravity background fields on the dressing coset $\check{M}=F\backslash \cD/\tilde{G}$. 
In \cite{hep-th:9602162}, the PL $T$-duality was extended to this general situation, and it was found that the sigma model on the dressing coset $\check{M}= F\backslash \cD/\tilde{G}$ is equivalent to that on the dual dressing coset $\tilde{\check{M}}=F\backslash \cD/G$, whose background fields are\footnote{In general, the matrix $\cR^{ab}$ can be degenerate. This is indeed the case in the examples discussed in \cite{hep-th:9602162}. Therefore, it is more reasonable to rewrite Eq.~\eqref{eq:dual-model} as $\tilde{\check{E}}_{mn} = \tilde{r}_{am}\,\, \bigl[(\bm{1}+ \cR\,\tilde{\pi})^{-1}\,\cR\bigr]^{ab}\,\tilde{r}_{bn}$\,.}
\begin{align}
 \tilde{\check{E}}_{mn} = \tilde{r}_{am}\, \bigl[(\cR^{-1} + \tilde{\pi})^{-1}\bigr]^{ab}\,\tilde{r}_{bn}\,.
\label{eq:dual-model}
\end{align}
This extended duality is called the PL $T$-duality of dressing cosets. 

\subsubsection{Singular limit}

The PL $T$-duality of coset spaces was also studied in \cite{hep-th:9903170,hep-th:9904188}. 
In particular, in the paper by Sfetsos \cite{hep-th:9904188}, a singular limit that constructs $\check{E}_{mn}$ from the usual $D$-dimensional background fields $E_{mn}$ was proposed. 
There, the gauge group $F$ has been chosen to be a subgroup of $G$, and thus the original background $\check{M}=F\backslash \cD/\tilde{G}$ is a usual cost. 
The dual geometry $\tilde{\check{M}}=F\backslash \cD/G$ obtained after performing the PL $T$-duality is a non-trivial dressing coset.

To explain the singular limit, let us decompose the generators of $\mathfrak{d}$ as $\{T_a\}=\{T_{\check{a}},\,T_{\sfi}\}$ and $\{T^a\}=\{T^{\check{a}},\,T^{\sfi}\}$ ($\check{a}=1,\dotsc,d$), and suppose that the gauge algebra $\mathfrak{f}$ is generated by $T_{\sfi}$\,. 
Under this decomposition, the matrix $\hat{E}^{ab}$ and its inverse $\hat{E}_{ab}$ is decomposed as
\begin{align}
 \hat{E}=(\hat{E}^{ab}) = \begin{pmatrix} \hat{E}^{\check{a}\check{b}} & \hat{E}^{\check{a}\sfj} \\ \hat{E}^{\sfi\check{b}} & \hat{E}^{\sfi\sfj}\end{pmatrix}, \qquad
 \hat{E}^{-1}=(\hat{E}_{ab}) = \begin{pmatrix} \hat{E}_{\check{a}\check{b}} & \hat{E}_{\check{a}\sfj} \\ \hat{E}_{\sfi\check{b}} & \hat{E}_{\sfi\sfj} \end{pmatrix} .
\end{align}
Then, the singular limit is defined as a rescaling $\hat{E}^{\sfi\sfj}\to \lambda^{-1}\,\hat{E}^{\sfi\sfj}$ and a limit $\lambda\to 0$\,. 
Under this limit, we obtain
\begin{align}
 \hat{E}^{-1} \to \cR^{-1}\equiv \begin{pmatrix} \hat{E}_{\check{a}\check{b}} & 0 \\ 0 & 0\end{pmatrix}, \qquad
 \bigl(\hat{E}+\pi\bigr)^{-1} \to \bigl(\cR+\pi\bigr)^{-1}\equiv \begin{pmatrix} [(\hat{E}^{\check{c}\check{d}} + \pi^{\check{c}\check{d}})^{-1}]_{\check{a}\check{b}} & 0 \\ 0 & 0 \end{pmatrix} ,
\label{eq:E-limit}
\end{align}
where $\hat{E}_{\check{a}\check{b}}$ coincides with the inverse matrix of $\hat{E}^{\check{a}\check{b}}$. 
This also gives
\begin{align}
 \bigl(\hat{E}^{-1}+\tilde{\pi}\bigr)^{-1} \to \bigl(\cR^{-1}+\tilde{\pi}\bigr)^{-1} = \begin{pmatrix} \hat{E}_{\check{a}\check{b}} + \tilde{\pi}_{\check{a}\check{b}} & \tilde{\pi}_{\check{a}\sfj} \\ \tilde{\pi}_{\sfi\check{b}} & \tilde{\pi}_{\sfi\sfj} \end{pmatrix}^{-1},
\label{eq:E-limit-d}
\end{align}
and the backgrounds \eqref{eq:orig-model} and \eqref{eq:dual-model} are reproduced through the limit
\begin{align}
 E_{mn}\to \check{E}_{mn}=r_m^a\,\bigl[\bigl(\cR+\pi\bigr)^{-1}\bigr]_{ab}\,r_n^b \,,\qquad
 \tilde{E}_{mn}\to \tilde{\check{E}}_{mn}=\tilde{r}_{am}\,\bigl[\bigl(\cR^{-1}+\tilde{\pi}\bigr)^{-1}\bigr]^{ab}\,\tilde{r}_{bn} \,.
\end{align}
By considering the case of $F\subset G$ concretely, it has become easier to understand the meaning of the degenerate matrix $\cR^{ab}$. 
Additionally, the canonical equivalence of string theories on the two backgrounds was shown, and some quantum aspects were also discussed in \cite{hep-th:9904188}. 

An unsatisfying point is that the meaning of the singular limit is not very clear. 
The important quantities such as $(\cR+\pi)^{-1}$ and $(\cR^{-1}+\tilde{\pi})^{-1}$ are finite, but the matrix $\hat{E}$ is divergent. 
In the $T$-duality covariant approach, $E_{mn}$ is usually redefined as the generalized metric $\cH_{MN}$, but this object also diverges under the singular limit. 
This makes it difficult to describe the singular limit in the $T$-duality covariant approaches. 
In addition, when $F$ is a subgroup of $G$ we can introduce the parameter $\lambda$ as $\hat{E}^{\sfi\sfj}\to \lambda^{-1}\,\hat{E}^{\sfi\sfj}$, but when the gauge group $F$ is chosen more generally, it is not obvious how to introduce such a parameter $\lambda$ into $\hat{E}^{ab}$. 

\subsubsection{Dressing cosets revisited}
In \cite{1105.0162}, the proposals of \cite{hep-th:9602162} and \cite{hep-th:9904188} were discussed from a slightly different perspective. 
By using a first-order action, a general expression for the reduced background fields $\check{E}_{mn}$ was found after some technical computation. 
The gauge invariance of the string action was also shown in a general setup. 
A drawback is that two operators $R$ and $\hat{R}$ have been introduced, but it is not clear how to choose these when actually studying concrete examples. 

\subsubsection{Gauged action}

In the context of the standard PL $T$-duality, a first-order action that relates the mutually dual sigma models has been found in \cite{hep-th:9512040}. 
Similar to Tseytlin's double sigma model \cite{Tseytlin:1990nb,Tseytlin:1990va}, this action is not covariant under worldsheet diffeomorphisms. 
A gauged version of this action was recently proposed by Klim\v{c}\'\i{}k \cite{1903.00439}. 
This gauged action is invariant under left multiplications by the gauge group $F$ and is suited for studying the PL $T$-duality of dressing cosets. 
In a special case where $\mathfrak{f}$ is a subalgebra of $\mathfrak{g}$ or $\tilde{\mathfrak{g}}$, this gauged action was shown to reproduce the results by Sfetsos \cite{hep-th:9904188}. 
Moreover, for a particular choice of the pair $\{F,\,\cD,\,\tilde{G}\}$, a certain class of integrable sigma models was reproduced from the first-order action. 
This approach has a great benefit that, for a general choice of $\{F,\,\cD,\,\tilde{G}\}$, one can in principle compute the reduced background fields $\check{E}_{mn}$ by eliminating the gauge field from the action. 
In addition, we do not need to take any singular limit in this approach. 

\subsubsection{Several examples}

Several examples of the PL $T$-duality for dressing cosets have been studied in \cite{hep-th:9602162,hep-th:9904188,hep-th:9906163,1104.5196,1312.4560,1506.05784,1709.01448,1807.04608,1810.07763,1903.00439}. 
However, to the best of the author's knowledge, in all of these examples, the gauge group $F$ is a subgroup of $G$ or $\tilde{G}$, and Sfetsos's limit can be applied. 

\subsubsection{Supergravity}

The PL $T$-duality of dressing cosets at the level of supergravity equations of motion was studied in \cite{1810.07763} by using the language of Courant algebroids.
The PL $T$-duality of dressing cosets was shown to be a symmetry of the (generalized) supergravity equations of motion. 
However, their discussion is not relying on the explicit expression for the background fields $\check{E}_{mn}$ studied in earlier literature, and the relation to the earlier proposals is less clear. 

We also note that, in more recent papers \cite{1912.11036,2002.11144}, geometry of the dressing cosets have been developed by using a $T$-duality-covariant approach. 

\subsection{Outline and results}
\label{sec:results}

Here we describe the structure of this paper and summarize the main results.

\subsubsection{Double Field Theory}

The purpose of this paper is to clarify the PL $T$-plurality of dressing cosets by using $T$-duality-covariant formulations. 
For this, in section \ref{sec:DFT}, we review the basics of double field theory (DFT) \cite{Siegel:1993th,Siegel:1993xq,0904.4664,1006.4823}, which is a manifestly $T$-duality-covariant formulation of supergravity. 
In addition to the basic review, we also explain another approach of supergravity \cite{1810.07763} and show its equivalence to DFT. 

\subsubsection{Standard PL $T$-plurality}

In section \ref{sec:PL-T-review}, we review the standard PL $T$-plurality. 
In section \ref{sec:Hull-Reid-Edwards}, using the gauged sigma model proposed by Hull and Reid-Edwards \cite{0902.4032,1001.2479}, we review how to reproduce the string sigma model on a PL-symmetric background from a duality-invariant action. 
In section \ref{sec:PL-T-DFT}, we review recent studies \cite{1707.08624,1810.11446,1903.12175} on how to express the DFT equations of motion in a covariant form under the PL $T$-plurality. 

The contents of section \ref{sec:left-mult} may not be standard. 
We show that an infinitesimal left multiplication by $(1+\epsilon^A\,T_A)$ is generated by the generalized Lie derivative along a generalized vector field $\epsilon^A E_A$\,. 
We then find the condition that $\hat{\cH}_{AB}$ must obey for the gauged action to be invariant under the left multiplication. 
The same condition in terms of $\hat{E}^{ab}$ is given in Appendix \ref{app:left-mult} and the known conditions \cite{hep-th:9904188,1105.0162} are reproduced as specific cases. 

\subsubsection{Gauged sigma model}

In section \ref{sec:Dressing-cosets}, to study the dressing cosets, we propose a gauged sigma model
\begin{align}
 S = \frac{1}{4\pi\alpha'}\int_\Sigma \Bigl(\frac{1}{2}\,\hat{\cH}_{AB}\,\mathbb{P}^A \wedge * \mathbb{P}^B + \langle \cP \overset{\wedge}{,}\, l\,\cC\,l^{-1}\rangle 
 - \langle \hat{\cP} \overset{\wedge}{,}\, \cA\rangle \Bigr)
 - \frac{1}{4\pi\alpha'}\int_{\cB} \frac{1}{3!}\, \langle \cP\overset{\wedge}{,}\, [\cP,\, \cP]\rangle \,,
\label{eq:our-action}
\end{align}
which is invariant under local left multiplications by $F$. 
This will be equivalent to the gauged action proposed in \cite{1903.00439}, but unlike the first-order action of \cite{1903.00439}, our gauged action is manifestly covariant on the worldsheet. 
For some specific choices of $F$, we check the consistency of the two gauged actions by eliminating the gauge field $\cC\in\tilde{\mathfrak{g}}$\,.

In section \ref{sec:Hull-Spence}, we find that our gauged action reproduces the gauged action by Hull and Spence \cite{Hull:1989jk} by eliminating the gauge fields $\cC$\,. 
By further eliminating the gauge fields $\cA\in \mathfrak{f}$\,, we obtain the sigma model on the dressing cosets $\check{M}=F\backslash \cD/\tilde{G}$. 

\subsubsection{Reduced background fields}

Using the above gauged action, we find a general formula for the background fields $\check{E}_{mn}$ on the dressing coset $\check{M}=F\backslash \cD/\tilde{G}$. 
The formula is simple. 
Suppose that a $D$-dimensional PL symmetric space $E_{mn}=g_{mn}+B_{mn}$ admits some set of (generalized) Killing vector fields
\begin{align}
 \Lie_{\kk_{\sfi}} g_{mn}=0\,,\qquad \Lie_{\kk_{\sfi}} H_3=0\,,\qquad \iota_{\kk_{\sfi}}H_3 + \rmd \hat{\kk}_{\sfi} =0\qquad (H_3\equiv \rmd B_2)\,,
\end{align}
satisfying $\Lie_{\kk_{\sfi}}\kk_{\sfj} =-f_{\sfi\sfj}{}^{\sfk}\,\kk_{\sfk}$\,, $\Lie_{\kk_{\sfi}}\hat{\kk}_{\sfj} =-f_{\sfi\sfj}{}^{\sfk}\,\hat{\kk}_{\sfk}$\,, and $\kk_{(\sfi}^m\,\hat{\kk}_{\sfj) m}=0$\,, where $f_{\sfi\sfj}{}^{\sfk}$ denote the structure constants of the gauge algebra $\mathfrak{f}\subset \mathfrak{d}$\,. 
Then the obtained formula is
\begin{align}
 \check{E}_{mn} = E_{mn} - \bigl(\kk_{\sfi m}-\hat{\kk}_{\sfi m}\bigr)\,N^{\sfi\sfj}\,\bigl(\kk_{\sfj n}+\hat{\kk}_{\sfj n}\bigr)\qquad \bigl(\kk_{\sfi m}\equiv g_{mn}\,\kk_{\sfi}^n\bigr)\,,
\label{eq:formula1}
\end{align}
where $N^{\sfi\sfj}$ is the inverse matrix of
\begin{align}
 N_{\sfi\sfj} \equiv \kk_{\sfi}^m\,g_{mn}\,\kk_{\sfj}^n +\hat{\kk}_{\sfi m}\,\kk_{\sfj}^{m} \,.
\end{align}
Similar to the approach of \cite{1903.00439}, the singular limit is not necessary. 
When $F\subset G$, we find that our $\check{E}_{mn}$ reproduces the background fields obtained by the singular limit: $\check{E}_{mn}={\displaystyle\lim_{\lambda\to 0}}E_{mn}$\,. 
By making a gauge fixing, $\check{E}_{mn}$ reduces to the reduced background fields $\check{E}_{\check{m}\check{n}}$\,. 

We find that all of the examples studied in this paper can be reproduced from this formula.
However, in general, $N_{\sfi\sfj}$ may be degenerate, in which case special care should be taken.
As a specific one of such cases, we consider the case $N_{\sfi\sfj}=0$\,. 
In this case, we find that the original $D$-dimensional background is not modified $\check{E}_{mn} = E_{mn}$\,. 

\subsubsection{A simple rewriting of the reduced background fields}

In the derivation of the formula \eqref{eq:formula1}, to solve the equations of motion for $\cA$ as $\cA=\cdots$\,, we have assumed that the matrix $N_{\sfi\sfj}$ is non-degenerate. 
However, by using some technical trick (which is inspired by the computation given in \cite{1105.0162}), we can eliminate the gauge fields $\cA$ by using the equations of motion, without using the explicit expression for $\cA$\,. 
In section \ref{sec:revisited}, by taking this approach, we obtain a simple rewriting of the formula \eqref{eq:formula1}, i.e.,
\begin{align}
 \check{E}_{mn} = r_m^a\,[(\cR + \pi)^{-1}]_{ab}\,r_n^b\,.
\label{eq:gauge-to-original}
\end{align}
This rewriting makes the consistency with the original proposal by Klim\v{c}\'\i{}k and \v{S}evera clear. 
Moreover, the non-degeneracy of $N_{\sfi\sfj}$ is no longer necessary. 
We also relax a mild assumption made in the original proposal and find a formula that slightly extends Eq.~\eqref{eq:gauge-to-original}. 
Additionally, we find an explicit expression for the matrix $\cR^{ab}$ for a given matrix $\hat{E}^{ab}$ and a general gauge group $F$. 
This formula is obtained in section \ref{sec:non-deg} by assuming the non-degeneracy of a certain metric $\langle T_{\sfi},\,\hat{\cH}(T_{\sfj})\rangle$\,, but this assumption is lifted in Appendix \ref{app:general-case}. 

\subsubsection{Sfetsos's limit in $T$-duality-covariant approaches}

In $T$-duality-covariant approaches, the matrix $\hat{E}^{ab}$ is embedded into an $\OO(D,D)$ matrix $\hat{\cH}_{AB}$\,, but this matrix diverges under Sfetsos's limit. 
A prescription to avoid the divergence has been proposed in \cite{hep-th:9904188}: the divergent matrix $R$ (i.e., our $\hat{\cH}_{AB}$) is replaced by a finite matrix $R_{g/h}$. 
In section \ref{sec:Sfetsos-limit}, we discuss this prescription in more detail by using our gauged sigma model. 
We find that the divergent part of $\hat{\cH}_{AB}$ does not contribute to the string equations of motion, and can be subtracted. 
After the subtraction, $\hat{\cH}_{AB}$ is reduced to a degenerate matrix $\hat{\mathbb{H}}_{AB}$ that corresponds to the operator $R_{g/h}$\,. 
The finite matrix $\hat{\mathbb{H}}_{AB}$ contains information of $\check{E}_{mn}$\,. 
In section \ref{sec:Sfetsos-revisted}, for the case $F\subset G$, we explain how to compute $\check{E}_{mn}$ from $\hat{\mathbb{H}}_{AB}$\,. 

Sfetsos's limit has been defined only when $F\subset G$ (or $F\subset \tilde{G}$). 
In section \ref{sec:g-Sfetsos}, to consider a general $F$, we propose a generalization of the Sfetsos limit. 
In section \ref{sec:dGM}, using this limit, the degenerate matrix $\hat{\mathbb{H}}_{AB}$ and the degenerate generalized metric $\hat{\mathbb{H}}_{MN}$ are constructed in the general case. 
To reproduce the reduced background fields $\check{E}_{mn}$\,, we rewrite $\hat{\mathbb{H}}_{MN}$ by using certain gauge-invariant generalized vector fields $H_{\bcheck{A}}$ ($\bcheck{A}=1,\dotsc,2d$). 
In section \ref{sec:reduced-GM}, using $H_{\bcheck{A}}$\,, we construct some generalized vector fields $\CE_{\bcheck{A}}$ on $M$ and truncate these into $\check{\CE}_{\bcheck{A}}$ defined on $\check{M}$. 
We find that $\check{\CE}_{\bcheck{A}}$ play the role of the vielbeins associated with the reduced generalized metric $\check{\cH}_{\check{M}\check{N}}$ on $\check{M}$, which corresponds to the reduced background fields $\check{E}_{\check{m}\check{n}}$\,.

The construction of $\CE_{\bcheck{A}}$ is purely geometric and is independent of the gauged sigma model. 
Once a PL-symmetric background $E_{mn}$ and a gauge group $F$ is given, we can in principle compute $\CE_{\bcheck{A}}$\,, and then we obtain the reduced background fields $\check{E}_{\check{m}\check{n}}$\,. 
Thus, this gives another independent approach to compute $\check{E}_{\check{m}\check{n}}$\,. 

\subsubsection{PL $T$-plurality in supergravity}

In section \ref{sec:PL-DFT}, using the idea of \cite{1810.07763}, we discuss the symmetry of the supergravity equation of motion under the PL $T$-plurality. 
Using the properties of $\CE_{\bcheck{A}}$ and the standard formula for the dilaton, we show that the PL $T$-plurality is a symmetry of the DFT equations of motion. 
Unlike the discussion of \cite{1810.07763}, our discussion is based on the Sfetsos limit. 
The result of this section shows that the PL $T$-plurality of dressing cosets is a solution-generating transformation (i.e., a transformation which maps a solution to another solution). 
However, our discussion is not complete when the Ramond--Ramond (R--R) fields exist. 
We determine the transformation rule of the R--R field strength such that the self-duality relation and the equations of motion for the NS--NS fields are satisfied in each duality frame. 
However, we have not succeeded in showing that the Bianchi identities (or equivalently, the equations of motion for the R--R fields) are satisfied in each duality frame. 
Of course, the Bianchi identities are always satisfied in our specific examples, and this would be merely a technical difficulty.

\subsubsection{Rlations between three approaches}
At least three approaches have been proposed for the construction of dressing cosets. 
The original one \cite{hep-th:9602162} plays a different role than the others and gives a minimal construction of the $d$-dimensional dressing cosets by introducing a (possibly) degenerate matrix $\cR^{ab}$. 
This does not consider an embedding of the dressing cosets into a $D$-dimensional space, but the other two approaches reproduce the dressing cosets from a non-degenerate $D$-dimensional space. 
In the approach by Sfetsos, the reduction is achieved by the singular limit. 
In the approach that is based on the gauged action (which may be the one by Klim\v{c}\'\i{}k \cite{1903.00439} or ours \eqref{eq:our-action}), the reduction is caused by the gauging procedure. 
The presence of the $D$-dimensional uplift makes it easy to study the symmetry of the DFT equations of motion under the PL $T$-plurality. 

In the approach based on the gauged action, we find that the reduced background fields $\check{E}_{mn}$ have the same form as the original proposal \eqref{eq:orig-model}. 
By using our extension of Sfetsos's approach, we can construct a set of vielbeins $\check{\CE}_{\bcheck{A}}$ for the dressing coset.
We then find that these vielbeins construct the same reduced background fields $\check{E}_{mn}$ as those obtained in the approach based on the gauged action. 
Combining these results, we show that all of the three approaches are equivalent for any choice of the gauge group $F$. 

\subsubsection{Examples}

In section \ref{sec:examples}, we study specific examples. 
We begin by providing useful formulas to perform the PL $T$-plurality. 
In section \ref{sec:S2}, we consider the PL $T$-plurality of S$^2$, which has been already discussed in \cite{hep-th:9210021,hep-th:9602162,hep-th:9904188}. 
Using this well-known example, we explain how the obtained formulas can be applied to concrete examples. 
In section \ref{sec:hYB}, we study a class of PL $T$-pluralities, called the homogeneous Yang--Baxter deformations of the standard cosets. 
For this type of PL $T$-plurality, using a solution of the homogeneous classical Yang--Baxter equation, we can easily write down the deformed background. 
In section \ref{sec:PL-AdS2}, we study the PL $T$-plurality of AdS$_2$\,. 
More specifically, we consider the $T$-plurality of an $\text{AdS}_2\times \text{S}^2\times T^6$ solution with $H$-flux. 
The Drinfel'd double employed here has ten Manin triples, and we find ten DFT solutions. 
To the best of the author's knowledge, the PL $T$-dualities studied in the literature is a duality between a standard coset and a dressing coset. 
However, we find more general examples, where both the original and its dual geometries are dressing cosets. 
Such examples can be found in the two dual pairs of Manin triples $(\vect{7_0}|\vect{5.i})\leftrightarrow (\vect{5.i}|\vect{7_0})$ and $(\vect{5}|\vect{2.ii})\leftrightarrow (\vect{2.ii}|\vect{5})$. 
There, both $\mathfrak{g}$ and $\tilde{\mathfrak{g}}$ are non-Abelian algebras, and the gauge algebra $\mathfrak{f}$ is generated by a linear combination of these two subalgebras $T_{\sfi}=\mathsf{\kk}_{\sfi}^a\,T_a+\tilde{\kk}_{\sfi a}\,T^a$. 
In section \ref{sec:PL-AdS2-RR}, using the results of section \ref{sec:PL-AdS2}, we study the $T$-plurality of an $\text{AdS}_2\times \text{S}^2\times T^6$ solution with R--R fields. 

\section{Double Field Theory}
\label{sec:DFT}

Before considering the Poisson-Lie $T$-plurality, we provide a short review of DFT \cite{Siegel:1993th,Siegel:1993xq,0904.4664,1006.4823}. 
This allows us to simplify various expressions in later sections. 
We also explain the relation between DFT and the approach \cite{1810.07763} that is based on Courant algebroids. 

\subsection{Generalized Lie derivative}
In DFT, we consider a $2D$-dimensional doubled space which admits the $\OO(D,D)$ metric $\eta_{MN}$. 
We introduce the coordinates $(x^M)=(x^m,\,\tilde{x}_m)$ $(M=1,\dotsc,2D)$ such that the $\OO(D,D)$ metric and its inverse take the form
\begin{align}
 \eta_{MN} = \begin{pmatrix} 0 & \delta_m^n \\ \delta^m_n & 0 \end{pmatrix},\qquad
 \eta^{MN} = \begin{pmatrix} 0 & \delta^m_n \\ \delta_m^n & 0 \end{pmatrix},
\end{align}
and do not consider coordinate transformations that change the form. 
Throughout this paper, the indices $M,N$ are raised or lowered with the matrices $\eta^{MN}$ and $\eta_{MN}$. 
In the doubled space, infinitesimal diffeomorphisms that preserve the metric $\eta_{MN}$ are generated by the generalized Lie derivative $\gLie_V$, which acts on a generalized vector field $W^M$ as
\begin{align}
 \gLie_{V} W^M = V^N\,\partial_N W^M - \bigl(\partial_N V^M - \partial^M V_N\bigr)\,W^N \,,
\end{align}
where $(\partial_M)\equiv(\partial_m,\,\tilde{\partial}^m)\equiv (\frac{\partial}{\partial x^m},\,\frac{\partial}{\partial \tilde{x}_m})$. 
In this paper, $\gLie_{V} W$ may also be denoted as $[V,\,W]_{\text{D}}$\,, which is called the D-bracket, and we use both expressions interchangeably. 
From its definition, the D-bracket is not anti-symmetric. 
However, when the inner product $\langle V,\,W\rangle\equiv \eta_{MN}\,V^M\,W^N$ is constant, one can easily show $[V,\,W]_{\text{D}}=-[W,\,V]_{\text{D}}$\,. 

For arbitrary two generalized vector fields $V^M$ and $W^M$, $\gLie_{V} W^M$ also should transform as a generalized vector field.
However, this property is not automatically satisfied in DFT. 
To ensure this, we usually assume that any fields, say $A$ and $B$, defined in the doubled space satisfy the so-called section condition
\begin{align}
 \eta^{MN}\,\partial_M \partial_N A = 0\,,\qquad
 \eta^{MN}\,\partial_M A\,\partial_N B = 0\,.
\end{align}
This condition removes the coordinate-dependence on half of the generalized coordinates, and in this paper, we choose a section where all of the fields are independent of $\tilde{x}_m$\,: $\tilde{\partial}^m =0$\,. 
Under this section, the components of the generalized Lie derivative become
\begin{align}
 \gLie_{V} W^M = \begin{pmatrix} (\gLie_{V} W)^m \\ (\gLie_{V} W)_m \end{pmatrix} = \begin{pmatrix} \Lie_v w^m \\ \Lie_v\tilde{w}_m - 2\,w^n\,\partial_{[n} \tilde{v}_{m]} \end{pmatrix},
\end{align}
where we have used the parameterizations, $(V^M)=(v^m,\,\tilde{v}_m)$ and $(W^M)=(w^m,\,\tilde{w}_m)$, and $\Lie_v$ denotes the usual Lie derivative along the $D$-dimensional vector field $v^m$. 

\subsection{Supergravity fields}
In DFT, the supergravity fields in the NS--NS sector are the generalized metric $\cH_{MN}$ satisfying $\cH_{MN} = \cH_{NM}$ and $\cH_{M}{}^P\,\cH_P{}^N = \delta_M^N$\,, and the DFT dilaton $d$ where $\Exp{-2d}$ behaves as a scalar density: $\gLie_V \Exp{-2d} = V\cdot{} \Exp{-2d} + \partial_M V^M\Exp{-2d}$\,. 
Here and hereafter, we may use the notation
\begin{align}
 V\,\cdot \,\equiv V^M\,\partial_M = v^m\,\partial_m \,.
\end{align}
In terms of the standard supergravity fields $\{g,\,B,\,\Phi\}$\,, these DFT fields are parameterized as
\begin{align}
 \cH_{MN} = \begin{pmatrix} (g-B\,g^{-1}\,B)_{mn} & (B\,g^{-1})_m{}^n \\ -(g^{-1}\,B)^m{}_n & g^{mn} \end{pmatrix},\qquad 
 \Exp{-2d} = \Exp{-2\Phi}\sqrt{\abs{\det(g_{mn})}}\,.
\label{eq:SUGRA-param}
\end{align}
We can also introduce the R--R fields and fermions, but for simplicity, we here concentrate only on the NS--NS sector (see section \ref{sec:R-R} for the R--R fields). 
Due to the section condition, all of these supergravity fields are functions of $x^m$. 

\subsection{Equations of motion}
Using the NS--NS sector fields $\cH_{MN}$ and $d$\,, we can define a generalized Ricci scalar $\cS$ and the Ricci tensor $\cS_{MN}$ \cite{1011.1324,1105.6294,1112.5296}. 
Using these curvature tensors, the equations of motion of DFT can be summarized as
\begin{align}
 \cS = 0\,,\qquad \cS_{MN} = 0\,.
\label{eq:EOM-gmDFT}
\end{align}
The former corresponds to the dilaton equations of motion while the latter corresponds to the equations of motion for $\cH_{MN}$\,. 
Here we do not review the construction of the curvature tensors, but the results can be summarized as follows:
\begin{align}
 \cS &= \tfrac{1}{8}\,\cH^{MN}\,\partial_M \cH^{PQ}\partial_N \cH_{PQ} -\tfrac{1}{2}\, \cH^{PQ}\,\partial_Q\cH^{MN}\,\partial_N \cH_{PM} 
 + 4\,\partial_M d\,\partial_N\cH^{MN} 
\nn\\
 &\quad -4\,\cH^{MN}\,\partial_M d\,\partial_N d
 - \partial_M\partial_N\cH^{MN} 
 +4\,\cH^{MN}\,\partial_M\partial_N d\,,
\\
 \cS_{MN}&= -2\,\bigl(P_M{}^P\,\bar{P}_N{}^Q + \bar{P}_M{}^P\,P_N{}^Q\bigr)\, \cK_{PQ} \,,
\\
 \cK_{MN} &\equiv \tfrac{1}{8}\,\partial_M \cH^{PQ}\,\partial_N \cH_{PQ} 
 - \tfrac{1}{2}\,\partial_{(M|} \cH^{PQ}\,\partial_P \cH_{|N)Q}
 + 2\,\partial_M \partial_N d 
\nn\\
 &\quad + \bigl(\partial_P - 2\,\partial_P d\bigr)\,\bigl( 
        \tfrac{1}{2}\,\cH^{PQ}\,\partial_{(M} \cH_{N)Q} 
        + \tfrac{1}{2}\,\cH^Q{}_{(M|}\,\partial_Q \cH^P{}_{|N)} 
        -\tfrac{1}{4}\,\cH^{PQ}\,\partial_Q \cH_{MN} \bigr) \,.
\end{align}
Here, have defined $P_{MN}$ and $\bar{P}_{MN}$ as
\begin{align}
 P_{MN} \equiv \frac{1}{2}\,\bigl(\eta_{MN}+\cH_{MN}\bigr)\,,\qquad
 \bar{P}_{MN} \equiv \frac{1}{2}\,\bigl(\eta_{MN}-\cH_{MN}\bigr)\,,
\end{align}
which behave as projectors,
\begin{align}
 P_M{}^P\,P_P{}^N=P_M{}^N\,,\qquad \bar{P}_M{}^P\,\bar{P}_P{}^N=\bar{P}_M{}^N\,,\qquad
 P_M{}^P\,\bar{P}_P{}^N=0\,.
\end{align}
Using the parameterization \eqref{eq:SUGRA-param} and $\tilde{\partial}^m=0$\,, $\cS$ and $\cS_{MN}$ reduce to
\begin{align}
\begin{split}
 \cS &=R + 4\,\nabla^m \partial_m \Phi - 4\,\nabla^m \Phi\,\nabla_m\Phi - \frac{1}{12}\,H_{mnp}\,H^{mnp} \,,
\\
 (\cS_{MN})&= \begin{pmatrix}
 2\,g_{(m|k}\,s^{[kl]} \,B_{l|n)} - s_{(mn)} 
 - B_{mk}\,s^{(kl)}\,B_{ln}\quad & B_{mk}\,s^{(kn)} - g_{mk}\,s^{[kn]} \\
 s^{[mk]}\,g_{kn} -s^{(mk)}\,B_{kn}\quad & s^{(mn)}
 \end{pmatrix} \,,
\\
 s_{mn} &\equiv R_{mn}-\frac{1}{4}\,H_{mpq}\,H_n{}^{pq} + 2\,\nabla_m \partial_n \Phi 
 - \frac{1}{2}\,\nabla^p H_{pmn} + \partial_p\Phi\,H^p{}_{mn} \,,
\end{split}
\label{eq:Ss-explicit}
\end{align}
where the indices $m,n$ are raised by using $g^{mn}$, $H_{mnp}\equiv 3\,\partial_{[m}B_{np]}$\,, and $\nabla_m$\,, $R_{mn}$\,, $R$ are the covariant derivative, the Ricci tensor, and the Ricci scalar associated with the metric $g_{mn}$\,, respectively.
Eq.~\eqref{eq:Ss-explicit} shows that the DFT equations of motion \eqref{eq:EOM-gmDFT} are equivalent to the usual supergravity equations of motion. 

\subsection{Flux formulation}
The above formulation of DFT is called the generalized metric formulation because it is based on the generalized metric $\cH_{MN}$. 
Another formulation of DFT, which is based on the generalized frame fields $E_A{}^M$, has been proposed in \cite{1109.0290,1109.4280,1201.2924,1304.1472} and is called the flux formulation.
The generalized frame fields $E_A{}^M\in \OO(D,D)$ are defined to satisfy
\begin{align}
 \cH_{MN}= E_M{}^A\,E_N{}^B\,\hat{\cH}_{AB}\,,
\end{align}
where $\hat{\cH}_{AB}$ is a constant $\OO(D,D)$ matrix and $(E_M{}^A)$ denotes the inverse matrix of $(E_A{}^M)$. 
Using the generalized frame fields, we define two types of generalized fluxes
\begin{align}
 \cF_{ABC} \equiv 3\,\cW_{[ABC]}\,, \qquad 
 \cF_A \equiv \cW^B{}_{AB} + 2\, D_A d \,,
\label{eq:gen-fluxes-def}
\end{align}
where the indices are raised or lowered with the metric $\eta_{AB}$ given in Eq.~\eqref{eq:eta-AB} and
\begin{align}
 \cW_{ABC} \equiv - D_A E_B{}^M\, E_{MC} \,,\qquad D_A \equiv E_A{}^M\,\partial_M \,.
\end{align}
Using the former flux, the generalized frame fields satisfy the relation
\begin{align}
 [E_A,\, E_B]_{\text{D}} = - \cF_{AB}{}^C\,E_C \,.
\label{eq:frame-alg}
\end{align}
The equations of motion in this formulation can be expressed as
\begin{align}
 \cR = 0 \,,\qquad \cG^{AB} = 0\,,
\end{align}
where
\begin{align}
 \cR &\equiv \hat{\cH}^{AB}\, \bigl(2\,D_{A} \cF_{B} - \cF_{A}\,\cF_{B}\bigr) 
 + \tfrac{1}{12}\, \hat{\cH}^{AD}\,\bigl(3\,\eta^{BE}\,\eta^{CF} - \hat{\cH}^{BE}\, \hat{\cH}^{CF}\bigr) \,\cF_{ABC}\,\cF_{DEF} \,,
\\
 \cG^{AB} &\equiv 2\,\hat{\cH}^{C[A}\, D^{B]} \cF_C 
 -\tfrac{1}{2}\,\hat{\cH}^{CD}\,(\eta^{AE}\,\eta^{BF}-\hat{\cH}^{AE}\,\hat{\cH}^{BF})\,\bigl(\cF_C-D_C\bigr)\, \cF_{DEF} 
\nn\\
 &\quad -\hat{\cH}_D{}^{[A}\,\bigl(\cF_C-D_C\bigr)\, \cF^{B]CD} 
  +\tfrac{1}{2}\,\bigl(\eta^{CE}\,\eta^{DF} - \hat{\cH}^{CE}\, \hat{\cH}^{DF}\bigr)\, \hat{\cH}^{G[A}\,\cF_{CD}{}^{B]}\,\cF_{EFG} \,. 
\end{align}

\subsection{An approach based on Courant algebroids}
Here we translate the approach of \cite{1810.07763} (based on Courant algebroids) into the DFT language. 
We introduce the same generalized frame fields $E_A{}^M$ as those used the flux formulation, and define the divergence operator as
\begin{align}
 \div V \equiv \Exp{2d} \partial_M\bigl(\Exp{-2d} V^M\bigr)\,.
\end{align}
In \cite{1810.07763}, this operator has been defined in a more general sense, but here we restrict ourselves to this one. 
Then, introducing the projectors $P_\pm$ as
\begin{align}
 P_+(V)^M \equiv P^M{}_N\,V^N\,,\qquad P_-(V)^M \equiv \bar{P}^M{}_N\,V^N\,,
\end{align}
we define the generalized Ricci tensor $\text{GRic}_+$ or $\text{GRic}_-$ as
\begin{align}
\begin{split}
 \text{GRic}_{\pm}(X,\,Y)
 &\equiv \div \bigl[P_\pm([P_\mp(Y),\,P_\pm(X)]_{\text{D}})\bigr] - P_\mp(Y)\cdot \div P_\pm(X)
\\
 &\quad - \langle [P_\mp([P_\pm(E_A),\,P_\mp(Y)]_{\text{D}}),\,P_\pm(X)]_{\text{D}},\,P_\pm(E^A) \rangle\,.
\end{split}
\label{eq:GRic}
\end{align}
We can easily show that the generalized Ricci tensor is invariant under a $x^m$-dependent $\OO(D,D)$ redefinition of $E_A$ (and $E^A\equiv \eta^{AB}\,E_A$),
\begin{align}
 E_A(x) \to E'_A(x) \equiv \Lambda_A{}^B(x)\,E_B(x)\,.
\end{align}
Namely, $E_A{}^M$ appearing in Eq.~\eqref{eq:GRic} can be an arbitrary $\OO(D,D)$ matrix, and $\text{GRic}_{\pm}$ depends only on $\cH_{MN}$ and $d$\,. 
We also define the Laplacian associated with $\{\hat{\cH}_{AB},\,E_A{}^M\}$ as
\begin{align}
\begin{split}
 \Delta_+ &\equiv 4\,\gLie_{E_{\ubar{A}}}\gLie_{E^{\ubar{A}}} -\tfrac{1}{6}\,\cF_{\ubar{A}\ubar{B}\ubar{C}}\,\cF^{\ubar{A}\ubar{B}\ubar{C}}
 -\tfrac{1}{2}\,\cF_{\ubar{A}\ubar{B}\bar{C}}\,\cF^{\ubar{A}\ubar{B}\bar{C}} \,,
\\
 \Delta_- &\equiv -4\,\gLie_{E_{\bar{A}}}\gLie_{E^{\bar{A}}} + \tfrac{1}{6}\,\cF_{\bar{A}\bar{B}\bar{C}}\,\cF^{\bar{A}\bar{B}\bar{C}}
 +\tfrac{1}{2}\,\cF_{\bar{A}\bar{B}\ubar{C}}\,\cF^{\bar{A}\bar{B}\ubar{C}} \,.
\end{split}
\end{align}
Here, the bar and the underbar denote the projections, e.g.,
\begin{align}
 E_{\ubar{A}} \equiv \hat{P}_A{}^B\,E_B\,,\quad 
 E_{\bar{A}} \equiv \bar{\hat{P}}_A{}^B\,E_B\,,\quad 
 \cF^{\bar{A}\bar{B}\bar{C}} \equiv \bar{\hat{P}}^A{}_D\,\bar{\hat{P}}^B{}_E\,\bar{\hat{P}}^C{}_F\,\cF^{DEF}\,,
\end{align}
where the projectors have been defined as $\hat{P}_{AB} \equiv \frac{1}{2}\,\bigl(\eta_{AB}+\hat{\cH}_{AB}\bigr)$ and $\bar{\hat{P}}_{AB} \equiv \frac{1}{2}\,\bigl(\eta_{AB}-\hat{\cH}_{AB}\bigr)$.
The generalized Ricci scalars can be defined as
\begin{align}
 \text{GR}_\pm \equiv -2\Exp{d}\Delta_\pm \Exp{-d} .
\end{align}
Using a formula
\begin{align}
 2\Exp{d}\gLie_{E_A}\Exp{-d} = -2\,E_A\cdot d + (\partial_M E_A{}^M) 
 = \div E_A = -\cF_A\,,
\label{eq:dilaton-flux}
\end{align}
we can express these as
\begin{align}
\begin{split}
 \text{GR}_+ &= 2\,P^{AB}\, \bigl(2\,D_{A} \cF_{B} - \cF_{A}\,\cF_{B}\bigr) + \tfrac{1}{3}\,\cF_{\ubar{A}\ubar{B}\ubar{C}}\,\cF^{\ubar{A}\ubar{B}\ubar{C}}
 + \cF_{\ubar{A}\ubar{B}\bar{C}}\,\cF^{\ubar{A}\ubar{B}\bar{C}}\,,
\\
 \text{GR}_- &= -2\,\bar{P}^{AB}\, \bigl(2\,D_{A} \cF_{B} - \cF_{A}\,\cF_{B}\bigr) - \tfrac{1}{3}\,\cF_{\bar{A}\bar{B}\bar{C}}\,\cF^{\bar{A}\bar{B}\bar{C}}
 - \cF_{\bar{A}\bar{B}\ubar{C}}\,\cF^{\bar{A}\bar{B}\ubar{C}}\,.
\end{split}
\label{eq:gR}
\end{align}
The supergravity equations of motion in the approach of \cite{1810.07763} can be expressed as
\begin{align}
 \text{GR}_+=0\,,\qquad \text{GRic}_+ =0\,.
\end{align}
As we show below, they are equivalent to $\text{GR}_-=0$ and $\text{GRic}_-=0$ under the section condition. 

\subsection{Equivalence of three formulations}
Now, we show the equivalence of the equations of motion in the three formulations. 
To this end, we note that the following quantities vanish under the section condition \cite{1304.1472}
\begin{align}
\begin{split}
 \cZ &\equiv D^A \cF_A - \frac{1}{2}\,\cF_A\,\cF^A + \frac{1}{12}\,\cF_{ABC}\,\cF^{ABC}\,,
\\
 \cZ_{AB} &\equiv D_A \cF_B - D_B\cF_A + \bigl(\cF_C - D_C\bigr)\,\cF^C{}_{AB} \,.
\end{split}
\end{align}

From \eqref{eq:gR}, we can easily show
\begin{align}
 \text{GR}_+ - \text{GR}_- = 4\,\cZ\,,\qquad 
 \text{GR}_+ + \text{GR}_- = 2\,\cR\,,
\end{align}
and then we obtain $\text{GR}_+ = \text{GR}_- = \cR$ under the section condition. 
Under the section condition, we can also show that $\cR=\cS$ \cite{1304.1472}. 
Therefore, the dilaton equations of motion are equivalent in all the three formulations. 

Using the relations \eqref{eq:frame-alg} and \eqref{eq:dilaton-flux}, we find
\begin{align}
\begin{split}
 \text{GRic}_+(E_A,\,E_B)
 &= - \bigl(\cF_{\ubar{C}} - D_{\ubar{C}}\bigr) \cF_{\ubar{A}\bar{B}}{}^{\ubar{C}}
 + D_{\bar{B}}\cF_{\ubar{A}}
 + \cF_{\ubar{A}\ubar{C}\bar{D}}\,\cF_{\bar{B}}{}^{\ubar{C}\bar{D}}\,,
\\
 \text{GRic}_-(E_A,\,E_B)
 &= - \bigl(\cF_{\bar{C}} - D_{\bar{C}}\bigr) \cF_{\bar{A}\ubar{B}}{}^{\bar{C}}
 + D_{\ubar{B}}\cF_{\bar{A}}
 + \cF_{\bar{A}\ubar{C}\bar{D}}\,\cF_{\ubar{B}}{}^{\ubar{C}\bar{D}}\,.
\end{split}
\end{align}
From these expressions, we can show
\begin{align}
\begin{split}
 \text{GRic}_+(E_A,\,E_B) - \text{GRic}_-(E_B,\,E_A) &= -\cZ_{\ubar{A}\bar{B}}\,,
\\
 \text{GRic}_+(E_A,\,E_B) + \text{GRic}_-(E_B,\,E_A) &= \cG_{\ubar{A}\bar{B}}\,,
\end{split}
\end{align}
and, under the section condition, we get
\begin{align}
 \text{GRic}_+(E_A,\,E_B)=\text{GRic}_-(E_B,\,E_A)=\tfrac{1}{2}\,\cG_{\ubar{A}\bar{B}}\,.
\end{align}
Moreover, under the section condition, we find
\begin{align}
 \text{GRic}_\pm(X,\,Y) = P_{\mp}(X)^M\, P_{\pm}(Y)^N\,\cK_{MN}\,.
\end{align}
If we denote $\cS(X,\,Y)\equiv \cS_{MN}\,X^M\,Y^N$ and $\cG(X,\,Y)\equiv \cG^{AB}\,E_{AM}\,E_{BN}\,X^M\,Y^N$\,, we have
\begin{align}
\begin{split}
 -\tfrac{1}{2}\,\cS(X,\,Y) &= \text{GRic}_+(X,\,Y) + \text{GRic}_-(X,\,Y)=\text{GRic}_+(X,\,Y) + \text{GRic}_+(Y,\,X)\,,
\\
 \tfrac{1}{2}\,\cG(X,\,Y) &= \text{GRic}_+(X,\,Y) - \text{GRic}_-(X,\,Y) =\text{GRic}_+(X,\,Y) - \text{GRic}_+(Y,\,X)\,,
\end{split}
\end{align}
under the section condition. 
Using these expressions, we can easily see the equivalence of the equations of motion, $\cS_{MN}=0$\,, $\cG^{AB}=0$\,, and $\text{GRic}_+=0$\,. 

\section{Poisson--Lie $T$-duality}
\label{sec:PL-T-review}

\subsection{Conventions}

\subsubsection{Drinfel'd double}

As was mentioned in the introduction, the PL $T$-duality is based on a Lie algebra $\mathfrak{d}$
\begin{align}
 [T_A,\,T_B] = F_{AB}{}^C\,T_C\qquad (A,B=1,\dotsc,2D)\,,
\label{eq:2D-alg}
\end{align}
which admits an adjoint-invariant metric
\begin{align}
 \langle T_A,\,T_B\rangle = \eta_{AB}\,.
\end{align}
Since the indices $A,B$ have been raised or lowered with $\eta_{AB}$ and its inverse $\eta^{AB}$, the adjoint invariance means that the structure constants $F_{ABC}$ are totally antisymmetric. 
We suppose that the generators $(T_A)=(T_a,\,T^a)$ can be decomposed such that $\eta_{AB}$ takes the form
\begin{align}
 \eta_{AB}=\begin{pmatrix} 0 & \delta_a^b \\ \delta^a_b & 0 \end{pmatrix}.
\label{eq:etaAB}
\end{align}
Under this decomposition, we can decompose the algebra \eqref{eq:2D-alg} as
\begin{align}
\begin{split}
 [T_a,\,T_b] &= f_{ab}{}^c\,T_c + H_{abc}\,T^c\,,
\\
 [T_a,\,T^b] &= f_a{}^{bc}\,T_c - f_{ac}{}^b\,T^c\,,
\\
 [T^a,\,T^b] &= R^{abc}\,T_c + f_c{}^{ab}\,T^c \,,
\end{split}
\end{align}
where $f_{ab}{}^c=f_{[ab]}{}^c$, $f_a{}^{bc}=f_a{}^{[bc]}$, $H_{abc}=H_{[abc]}$, and $R^{abc}=R^{[abc]}$ follow from the adjoint invariance. 
In this paper, we restrict our consideration to the algebra with $H_{abc}=0$ and $R^{abc}=0$\,, and then the Lie algebra $\mathfrak{d}$ is called the Drinfel'd double.
The Drinfel'd double $\mathfrak{d}$ contains two maximal isotropic\footnote{The isotropy means the null property $\langle T_a,\,T_b\rangle=0$ and the maximal means that we cannot extend the linear subspace $\text{span}(T_a)$ without breaking the isotropy.} algebras $\mathfrak{g}$ and $\tilde{\mathfrak{g}}$ generated by $\{T_a\}$ and $\{T^a\}$, respectively. 
The Lie groups associated with $\{\mathfrak{d},\, \mathfrak{g},\, \tilde{\mathfrak{g}}\}$ are denoted as $\{\cD,\,G,\,\tilde{G}\}$. 

\subsubsection{Our convention}

The PL $T$-duality is the equivalence of string sigma models on the two cosets $\cD/\tilde{G}$ and $\cD/G$. 
Under the PL $T$-duality, the Drinfel'd double is not changed and the structure constants $F_{AB}{}^C$ are invariant. 
However, in this paper, we adopt a convention that the group to be modded out from the right is generated by $\{T^a\}$, and then the structure constants $F_{AB}{}^C$ are rather transformed covariantly. 
Namely, when we consider the coset $\cD/\tilde{G}$, $\tilde{G}$ is generated by $\{T^a\}$ and we need not do anything. 
We simply take the coset.
Instead, when we consider the coset $\cD/G$, we perform a redefinition of generators $\{T_A\}\to \{T'_A\}\equiv\{T'_a,\,T'^a\}=\{T^a,\,T_a\}$ and $G$ is generated by $\{T'^a\}$.
We then take the coset. 
This redefinition can be expressed as $T_A\to T'_A=C_A{}^B\,T_B$ with $(C_A{}^B)=\bigl(\begin{smallmatrix} \bm{0} & \bm{1} \\ \bm{1} & \bm{0} \end{smallmatrix}\bigr)$, and the structure constants are transformed as
\begin{align}
 F_{ABC}\to F'_{ABC}=C_A{}^D\,C_B{}^E\,C_C{}^F\,F_{DEF}\,.
\end{align} 
All tensors associated with $T_A$ (such as $\hat{\cH}_{AB}$ and $E_A{}^M$ to be defined later) are transformed by the same matrix $C_A{}^B$. 
If we take another maximal isotropic subgroup $\tilde{G}'$ and consider the coset $\cD/\tilde{G}'$, we perform another redefinition $T_A\to T'_A=C_A{}^B\,T_B$ such that $\tilde{G}'$ is generated by $\{T'^a\}$.
To preserve the metric \eqref{eq:etaAB}, the matrix $C_A{}^B$ always should be an element of $\OO(D,D)$, and the PL $T$-plurality can be regarded as an $\OO(D,D)$ redefinition of the generators $T_A$\,. 

In the literature, some papers state that the structure constants $F_{AB}{}^C$ are invariant under the PL $T$-duality while some state that they transform covariantly under constant $\OO(D,D)$ transformations. 
The difference is only in the convention, and we take the latter in this paper. 
In the following discussion, we denote the generators of the gauge algebra $\mathfrak{f}$ as $T_{\sfi}\equiv \VV_{\sfi}{}^A\,T_A$\,. 
Under the PL $T$-duality, the constant vector $\VV_{\sfi}{}^A$ transforms as $\VV_{\sfi}{}^A\to \VV_{\sfi}{}^B\,(C^{-1})_B{}^A$ and one might get an impression that the gauge group $F$ is changed under the PL $T$-duality. 
However, this transformation is simply induced by our redefinition $T_A\to C_A{}^B\,T_B$\,, and the generators $T_{\sfi}= \VV_{\sfi}{}^A\,T_A$ and the structure constants $f_{\sfi\sfj}{}^{\sfk}$ of $\mathfrak{f}$ are invariant. 
Only the group $\tilde{G}$ is changed to another maximal isotropic subgroup $\tilde{G}'$ of $\cD$, and the information about the change in $\tilde{G}$ is recorded in the constant matrix $C_A{}^B$. 

\subsection{Gauged sigma model by Hull and Reid-Edwards}
\label{sec:Hull-Reid-Edwards}

Here we review how to derive the sigma model on $M=\cD/\tilde{G}$ from a duality-invariant action. 
Instead of the familiar first-order action of \cite{hep-th:9512040}, we consider a gauged sigma model of \cite{0902.4032,1001.2479} that has the manifest worldsheet diffeomorphism invariance. 
The gauged action is given by
\begin{align}
 4\pi\alpha'\,S = \int_\Sigma \Bigl(\frac{1}{2}\,\hat{\cH}_{AB}\,\hat{\cP}^A\wedge * \hat{\cP}^B + \langle \cP \overset{\wedge}{,}\, l\,\cC\,l^{-1}\rangle \Bigr)
 - \frac{1}{3!}\int_{\cB} \langle \cP\overset{\wedge}{,}\, [\cP,\, \cP]\rangle \,.
\label{eq:geugedDSM}
\end{align}
Here, $l(\sigma)\in \cD$ describes an embedding of the string worldsheet $\Sigma$ into the Drinfel'd double, $\cB$ is a three dimensional surface satisfying $\partial \cB=\Sigma$, and $\cP$ is defined by
\begin{align}
 \cP(\sigma) \equiv \cP^A(\sigma)\,T_A \equiv \rmd l \,l^{-1} \,.
\end{align}
In addition, $\hat{\cH}_{AB}$ is a ``constant'' symmetric matrix satisfying $\hat{\cH}_{A}{}^C\,\hat{\cH}_{C}{}^B =\delta_A^B$, $\cC(\sigma)\in \tilde{\mathfrak{g}}$ is a gauge field, and the 1-form field $\hat{\cP}$ is defined by
\begin{align}
 \hat{\cP}(\sigma) \equiv \rmd l\,l^{-1} + l\,\cC\,l^{-1} = \cP + l\,\cC\,l^{-1}\,. 
\end{align}
Furthermore, the Hodge star operator on $\Sigma$ is denoted as $*$ and we have defined
\begin{align}
 [\cP,\, \cP]\equiv [T_A,\,T_B]\,\cP^A\wedge \cP^B \,, \qquad \langle V \overset{\wedge}{,}\, W\rangle \equiv \eta_{AB}\,V^A\wedge W^B\,.
\end{align}
This gauged action is invariant under the gauge transformation
\begin{align}
 l(\sigma) \to l(\sigma)\,\tilde{h}(\sigma)\,, \qquad
 \cC(\sigma) \to \tilde{h}^{-1}(\sigma)\,\cC(\sigma)\,\tilde{h}(\sigma) - \tilde{h}^{-1}(\sigma)\,\rmd \tilde{h}(\sigma) \qquad \bigl[\tilde{h}(\sigma)\in \tilde{G} \bigr]\,.
\label{eq:gauge-transf}
\end{align}
Due to this gauge symmetry, this action describes string theory on the coset space $M=\cD/\tilde{G}$. 
If we assume that a group element $l(\sigma)\in\cD$ can be decomposed as
\begin{align}
 l(\sigma) = g(\sigma)\,\tilde{g}(\sigma) \qquad (g\in G,\quad \tilde{g}\in \tilde{G})\,,
\end{align}
we can use the gauge symmetry to eliminate $\tilde{g}$\,, yielding $l(\sigma) = g(\sigma)$. 
Then, after eliminating the auxiliary fields $\cC(\sigma)$, we get a sigma model on the group manifold of $G$. 
Namely, in this case, the $D$-dimensional space $M=\cD/\tilde{G}$ can be regarded as the group manifold of $G$. 

To compute the background fields on $M$, let us eliminate the gauge field $\cC(\sigma)$. 
For this purpose, it is convenient to rewrite the action \eqref{eq:geugedDSM}. 
By making the decomposition $l = g\,\tilde{g}$\,, we introduce the left-/right-invariant 1-forms for each subalgebra as
\begin{align}
\begin{split}
 \ell &\equiv \ell^a\,T_a \equiv g^{-1}\,\rmd g\,,\qquad
 r\equiv r^a\,T_a \equiv \rmd g\,g^{-1}\,,
\\
 \tilde{\ell}&\equiv \tilde{\ell}_a\,T^a \equiv \tilde{g}^{-1}\,\rmd \tilde{g}\,,\qquad
 \tilde{r}\equiv \tilde{r}_a\,T^a \equiv \rmd \tilde{g}\,\tilde{g}^{-1}\,.
\end{split}
\end{align}
We can then express the right-invariant 1-form $\cP= \cP^A\,T_A$ as
\begin{align}
 \cP = \rmd l \,l^{-1} = r + g\,\tilde{r}\,g^{-1} \,.
\end{align}
If we denote the adjoint action of $s\in \cD$ as
\begin{align}
 s\,T_A\,s^{-1} \equiv (\text{Ad}_s)_A{}^B\,T_B \,,
\end{align}
the matrix $\text{Ad}_{g}$ for the particular group element $g\in G$ can be parameterized as
\begin{align}
 (\text{Ad}_g)_A{}^B = \begin{pmatrix} (a^{-1})_a{}^b & 0 \\ a_c{}^a\,\pi^{cb} & a_b{}^a \end{pmatrix}.
\label{eq:Ad-g}
\end{align}
Then we have $g\,\tilde{r}\,g^{-1}=\tilde{r}_a\,\bigl(a_c{}^a\,\pi^{cb}\,T_b+ a_b{}^a\,T^b\bigr)$ and the right-invariant 1-form becomes
\begin{align}
 \cP^A = \begin{pmatrix} r^a - \pi^{ac}\,a_c{}^b\,\tilde{r}_b \\ a_a{}^b\,\tilde{r}_b \end{pmatrix}.
\end{align}
Since $\cC$ takes value in $\tilde{\mathfrak{g}}$, this can be parameterized as $(\cC^A)=(0,\,\cC_a)$, and then
\begin{align}
 (l\,\cC\,l^{-1})^A \equiv (g\,C\,g^{-1})^A = C^B\,(\text{Ad}_g)_B{}^A = \begin{pmatrix} - \pi^{ac}\,a_c{}^b\,C_b \\ a_a{}^b\,C_b \end{pmatrix} \qquad \bigl(C \equiv \tilde{g}\,\cC\,\tilde{g}^{-1}\in\tilde{\mathfrak{g}}\bigr)\,.
\end{align}
Using these parameterizations, we obtain
\begin{align}
 \hat{\cP}^A = \begin{pmatrix} r^a - \pi^{ab}\,P_b \\ P_a \end{pmatrix},\qquad
 P_a \equiv a_a{}^b\,(C_b + \tilde{r}_b) \,.
\label{eq:cP-hat-param}
\end{align}
We introduced the coordinates $(x^M)=(x^m,\,\tilde{x}_m)$ on $G$ and $\tilde{G}$ as $g=g(x^m)$ and $\tilde{g}=\tilde{g}(\tilde{x}_m)$. 
Then, denoting the right-invariant 1-form as $r^a= r^a_m(x)\,\rmd x^m$, we define a matrix $E_M{}^A$ as
\begin{align}
 E_M{}^A(x) \equiv \begin{pmatrix} r^a_m & 0 \\ -\pi^{ab}\,e_b^m & e_a^m \end{pmatrix}
 = \begin{pmatrix} r^a_m & 0 \\ \pi^{mn}\,r_n^a & e_a^m \end{pmatrix} ,
\label{eq:E-param}
\end{align}
where $(e_a^m)$ is the inverse matrix of $(r^a_m)$ and $\pi^{mn}\equiv e^m_a\,\pi^{ab}\,e_b^n$ is a bi-vector field called the Poisson--Lie structure. 
Then we can express $\hat{\cP}^A$ as
\begin{align}
 \hat{\cP}^A = E_M{}^A\,Z^M\,,\qquad 
 Z^M \equiv \begin{pmatrix} \rmd x^m \\ P_m \end{pmatrix},\qquad
 P_m \equiv r^a_m\,P_a = \ell^a_m\,(C_a + \tilde{r}_a)\,,
\label{eq:cP-Z}
\end{align}
where $\ell^a_m=a_b{}^a\,r^b_m$ are the components of the left-invariant 1-form, $\ell^a=\ell^a_m\,\rmd x^m$. 
Since $\ell^a_m$ and $a_a{}^b$ are invertible, we can treat $P_a$ or $P_m$ as independent fields instead of $C_a$ or $\cC_a$\,. 

The inverse of $E_M{}^A$ is called the generalized frame fields $E_A{}^M$. 
Using the properties \cite{hep-th:9710163}
\begin{align}
\begin{split}
 &f_{ab}{}^c = a_a{}^{d}\,a_b{}^{e}\,(a^{-1})_{f}{}^{c}\,f_{de}{}^{f}\,, \qquad
  f_{d}{}^{[ab}\,\pi^{c]d} + f_{de}{}^{[a}\,\pi^{b|d|}\,\pi^{c]e} = 0\,,
\\
 &f_a{}^{bc} = a_a{}^d\,(a^{-1})_e{}^b\,(a^{-1})_f{}^c\,f_d{}^{ef}
 + 2\,f_{ad}{}^{[b}\,\pi^{c]d} \,,
\\
 &D_a a_b{}^c = - f_{ab}{}^d\,a_d{}^c\,, \qquad
 D_a \pi^{bc} = f_a{}^{bc} + 2\,f_{ad}{}^{[b}\,\pi^{|d|c]} \quad \bigl(D_a\equiv e_a^m\,\partial_m\bigr)\,,
\end{split}
\label{eq:identities}
\end{align}
which follow from the definition \eqref{eq:Ad-g} of $a_a{}^b$ and $\pi^{ab}$, we can show that the generalized frame fields $E_A{}^M$ satisfy an important relation \cite{1707.08624,1810.11446}
\begin{align}
 [E_A,\, E_B]_{\text{D}} = - F_{AB}{}^C\,E_C \,, 
\label{eq:gLie-alg}
\end{align}
where $F_{AB}{}^C$ are the structure constants of the Drinfel'd double $\mathfrak{d}$\,. 

Now, the first term of the action \eqref{eq:geugedDSM} can be written as
\begin{align}
 \int_\Sigma \tfrac{1}{2}\,\cH_{MN}\,Z^M\wedge * Z^N \qquad 
 \bigl(\cH_{MN} \equiv E_M{}^A\,E_N{}^B\,\hat{\cH}_{AB}\bigr)\,.
\label{eq:cHMN}
\end{align}
Moreover, using
\begin{align}
 \langle \cP \overset{\wedge}{,}\, l\,\cC\,l^{-1}\rangle 
 &= r^a\wedge a_a{}^b\,C_b\,,
\\
 \tfrac{1}{3!} \langle \cP\overset{\wedge}{,}\, [\cP,\, \cP]\rangle
 &= \tfrac{1}{3!} \langle \ell + \tilde{r} \overset{\wedge}{,}\, [\ell + \tilde{r},\, \ell + \tilde{r}]\rangle
\nn\\
 &= \tfrac{1}{2}\,\langle \ell \overset{\wedge}{,}\, [\tilde{r},\, \tilde{r}]\rangle
 + \tfrac{1}{2}\,\langle \tilde{r} \overset{\wedge}{,}\, [\ell ,\, \ell]\rangle
 = - \rmd\langle \ell \overset{\wedge}{,}\, \tilde{r} \rangle
 = - \rmd\bigl(r^a\wedge a_a{}^b\,\tilde{r}_b\bigr)\,,
\end{align}
we obtain
\begin{align}
 \int_\Sigma \langle \cP \overset{\wedge}{,}\, l\,\cC\,l^{-1}\rangle 
 - \frac{1}{3!}\int_{\cB} \langle \cP\overset{\wedge}{,}\, [\cP,\, \cP]\rangle
 = \int_\Sigma r^a\wedge P_a 
 = \int_\Sigma \rmd x^m\wedge P_m \,.
\end{align}
Then the action \eqref{eq:geugedDSM} can be expressed as
\begin{align}
 S = \frac{1}{4\pi\alpha'}\int_\Sigma \bigl(\tfrac{1}{2}\,\cH_{MN}\,Z^M\wedge * Z^N + \rmd x^m\wedge P_m\bigr)\,.
\label{eq:DSM}
\end{align}
This is the familiar string action defined on a target space with the generalized metric $\cH_{MN}$. 
When $(\cH^{mn})$ is invertible, we can eliminate $P_m$ to obtain the usual action
\begin{align}
 S = \frac{1}{4\pi\alpha'}\int_\Sigma \bigl(g_{mn}\,\rmd x^m\wedge * \rmd x^n + B_{mn}\,\rmd x^m\wedge \rmd x^n\bigr)\,,
\label{eq:string-action}
\end{align}
where we have used the parameterization \eqref{eq:SUGRA-param}. 
If we parameterize the matrix $\hat{\cH}_{AB}$ as
\begin{align}
 \hat{\cH}_{AB} = \begin{pmatrix} \hat{G}_{ab} & - \hat{G}_{ac}\,\hat{\beta}^{cb} \\ \hat{\beta}^{ac}\,\hat{G}_{cb} & \hat{G}^{ab}-\hat{\beta}^{ac}\,\hat{G}_{cd}\,\hat{\beta}^{db} 
\end{pmatrix}\quad \bigl[\hat{G}_{ab}\equiv (\hat{G}^{-1})^{ab}\,,\ \hat{G}^{ab}=\hat{G}^{(ab)}\,,\ \hat{\beta}^{ab}=\hat{\beta}^{[ab]}\bigr],
\end{align}
by using $\cH_{MN}=E_M{}^A\,E_N{}^B\,\hat{\cH}_{AB}$\,, the background fields $E_{mn}\equiv g_{mn}+B_{mn}$ can be found as
\begin{align}
 E_{mn}=r_m^a\,\bigl[(\hat{E}+\pi)^{-1}\bigr]_{ab}\,r^b_n \qquad \bigl(\hat{E}^{ab} \equiv \hat{G}^{ab}+\hat{\beta}^{ab}\bigr)\,.
\label{eq:E-PL}
\end{align}
This is the PL-symmetric background given in Eq.~\eqref{eq:PL-symmetric}. 
Using Eq.~\eqref{eq:identities}, we can show that $\Lie_{v_a} \pi^{mn} = f_a{}^{bc} \,v_b^m\,v_c^n$ where $v_a^m$ are the left-invariant vector fields $v_a^m=(a^{-1})_a{}^b\,e_b^m$ satisfying $\Lie_{v_a}r^b_m=0$\,, and then we can show that $E_{mn}$ satisfies the dualizability condition \eqref{eq:PL-dualizability}. 

\subsection{Poisson--Lie $T$-plurality}
\label{sec:PL-T-DFT}

\subsubsection{A family of background fields}

As we have already mentioned, the PL $T$-plurality is the equivalence of multiple sigma models on various cosets of the form $\cD/\tilde{G}$. 
When we consider the duality between $\cD/\tilde{G}$ and $\cD/\tilde{G}'$, we redefine the generators as 
\begin{align}
 T_A\to T'_A = C_A{}^B\,T_B\qquad \bigl[C_A{}^B\in\OO(D,D)\bigr]\,,
\end{align}
such that $\{T'^a\}$ generate the Lie algebra $[T'^a,\,T'^b]=f'_c{}^{ab}\,T'^c$ of $\tilde{\mathfrak{g}}'$ that is associated with $\tilde{G}'$. 
We assume that $\{T'_a\}$ also generate a Lie subalgebra $[T'_a,\,T'_b]=f'_{ab}{}^c\,T'_c$ that is denoted as $\mathfrak{g}'$. 
Under this redefinition, the structure constants $F_{ABC}$ and the metric $\hat{\cH}_{AB}$ appearing in the gauged action \eqref{eq:geugedDSM} are transformed as
\begin{align}
 F_{ABC}\to F'_{ABC}=C_A{}^D\,C_B{}^E\,C_C{}^F\,F_{DEF}\,,\qquad \hat{\cH}_{AB} \to \hat{\cH}'_{AB} = C_A{}^C\,C_B{}^D\,\hat{\cH}_{CD}\,.
\label{eq:F-cH-transf}
\end{align}
Since $\{T'_A\}=\{T'_a,\,T'^a\}$ generate the Lie algebra of the Drinfel'd double $(\mathfrak{g}'|\tilde{\mathfrak{g}}')$, we can compute the generalized frame fields
\begin{align}
 E'_A{}^M \equiv \begin{pmatrix} e'^m_a & 0 \\ -\pi'^{ab}\,e'^m_b & r'^a_m \end{pmatrix},
\end{align}
in the same way as Eq.~\eqref{eq:E-param}.
We then find that they satisfy $[E'_A,\,E'_B]_{\text{D}} = - F'_{AB}{}^C\,E'_C$ by using the redefined structure constants $F'_{AB}{}^C$. 
Similar to Eq.~\eqref{eq:cHMN}, we can construct the generalized metric as $\cH'_{MN}\equiv E'_M{}^A\,E'_N{}^B\,\hat{\cH}'_{AB}$\,, and by following precisely the same procedure that led to Eq.~\eqref{eq:E-PL}, we obtain the background fields on $\cD/\tilde{G}'$ as
\begin{align}
 E'_{mn}=r'^a_m\,\bigl[(\hat{E}'+\pi')^{-1}\bigr]_{ab}\,r'^b_n \qquad \bigl(\hat{E}'^{ab} \equiv \hat{G}'^{ab}+\beta'^{ab}\bigr)\,,
\end{align}
where $\hat{G}'^{ab}$ and $\beta'^{ab}$ can be found from\footnote{If we parameterize $C_A{}^B$ as in Eq.~\eqref{eq:C-param}, we find $\hat{E}'^{ab}\equiv (\hat{G}'+\beta')^{ab}=[(\bm{c}+\bm{d}\,\hat{E})\,(\bm{a} + \bm{b}\,\hat{E})^{-1}]^{ab}$.}
\begin{align}
 \hat{\cH}'_{AB} = \begin{pmatrix} \hat{G}'_{ab} & - \hat{G}'_{ac}\,\hat{\beta}'^{cb} \\ \hat{\beta}'^{ac}\,\hat{G}'_{cb} & \hat{G}'^{ab}-\hat{\beta}'^{ac}\,\hat{G}'_{cd}\,\hat{\beta}'^{db} 
\end{pmatrix}= C_A{}^C\,C_B{}^D\,\hat{\cH}_{CD}\,.
\end{align}
If we choose another maximal isotropic subgroup $\tilde{G}''$ (which corresponds to choosing another constant matrix $C'_A{}^B$), we obtain new background fields $E''_{mn}$ (or $\cH''_{MN}$).
The PL $T$-plurality states that all of the background fields $E_{mn},\,E'_{mn},\cdots$ define equivalent sigma models. 

\subsubsection{Symmetry of DFT}

The PL $T$-plurality is known to be a symmetry of the DFT equations of motion \cite{1707.08624,1810.11446,1903.12175}. 
Namely, if the original background $\cH_{MN}$ is a solution of DFT, all of the backgrounds related through the PL $T$-plurality are solutions of DFT. 
Here we review the symmetry by ignoring the R--R fields and the spectator fields, but by introducing the DFT dilaton. 
We here use the formula \cite{hep-th:0205245} for the dilaton that can be expressed as
\begin{align}
 \Exp{-2d}= \Exp{-2\,\hat{d}} \abs{\det (\ell_m^a)}\,,
\label{eq:dilaton-ansatz}
\end{align}
where $\hat{d}$ is ``constant.''
Then the generalized fluxes defined in Eq.~\eqref{eq:gen-fluxes-def} become \cite{1707.08624,1810.11446}
\begin{align}
 \cF_{ABC} = F_{ABC}\,, \qquad 
 \cF_A = E_A{}^M\,\bigl(0,\,-f_{b}{}^{bc}\,v_c^m\bigr) \,. 
\end{align}
When $f_{b}{}^{ba}$ does not vanish (but $f_b{}^{ba}\,f_{ac}{}^c=0$), we modify the DFT dilaton as
\begin{align}
 \partial_M d \to \partial_M d + \bm{X}_M\,,\qquad \bm{X}_M\equiv \bigl(0,\,\tfrac{1}{2}\,f_{a}{}^{ab}\,v_b^m\bigr)\,.
\label{eq:dilaton-shift}
\end{align}
This modification corresponds to considering the generalized supergravity equations of motion \cite{1511.05795,1605.04884} with the Killing vector field $I^m=\tfrac{1}{2}\,f_{a}{}^{ab}\,v_b^m$. 
In terms of DFT, this corresponds to a modification of the ansatz for the DFT dilaton \cite{1703.09213}.
After the modification, we get $\cF_A=0$\,,\footnote{In general, $\hat{d}$ can depend on $x^M$ \cite{hep-th:0205245} if the resulting $\cF_A$ is a constant vector (see \cite{1903.12175}).} and then the DFT equations of motion (in the flux formulation) are simplified as
\begin{align}
\begin{split}
 \cR &= \tfrac{1}{12}\, \hat{\cH}^{AD}\,\bigl(3\,\eta^{BE}\,\eta^{CF} - \hat{\cH}^{BE}\, \hat{\cH}^{CF}\bigr) \,F_{ABC}\,F_{DEF} = 0\,,
\\
 \cG^{AB} &= \tfrac{1}{2}\,\bigl(\eta^{CE}\,\eta^{DF} - \hat{\cH}^{CE}\, \hat{\cH}^{DF}\bigr)\, \hat{\cH}^{G[A}\,F_{CD}{}^{B]}\,F_{EFG} = 0\,. 
\end{split}
\end{align}
These are manifestly covariant under the $\OO(D,D)$ PL $T$-plurality transformation \eqref{eq:F-cH-transf}, and this shows that the PL $T$-plurality is a solution generating transformation in DFT. 

\subsection{Left multiplication}
\label{sec:left-mult}

Here we show how the background fields $\cH_{MN}$\,, or the generalized frame fields $E_A{}^M$, transform under a constant left multiplication $l(\sigma)\to f\,l(\sigma)$ ($f\in \cD$). 
Supposing the gauge field $\cC$ to be invariant under the left multiplication, we find
\begin{align}
 (l\,\cC\,l^{-1}) \to f\,(l\,\cC\,l^{-1})\,f^{-1} \,,\qquad
 \cP \to f\, \cP \,f^{-1} \,,\qquad 
 \hat{\cP} \to f\, \hat{\cP} \,f^{-1} \,.
\label{eq:left-mult}
\end{align} 
Using the adjoint invariance of the inner product $\langle \cdot,\, \cdot \rangle$, we find that only the kinetic term $\tfrac{1}{2}\,\hat{\cH}_{AB}\,\hat{\cP}^A\wedge * \hat{\cP}^B$ of the action \eqref{eq:geugedDSM} is transformed:
\begin{align}
\begin{split}
 \tfrac{1}{2}\,\hat{\cH}_{AB}\,\hat{\cP}^A\wedge * \hat{\cP}^B &= \tfrac{1}{2}\,\hat{\cH}_{AB}\,E_M{}^A\,E_N{}^B\,\cZ^M\wedge * \cZ^N 
\\
 &\to \tfrac{1}{2}\,\hat{\cH}_{CD}\,(\text{Ad}_f)_A{}^C\,(\text{Ad}_f)_B{}^D\,E_M{}^A\,E_N{}^B\,\cZ^M\wedge * \cZ^N \,.
\end{split}
\end{align}
This can be regarded as a transformation of the vielbein $E_M{}^A$,\footnote{In Appendix \ref{app:left-mult}, we derive the same result from a different computation.}
\begin{align}
 E_M{}^A \to E_M{}^B\,(\text{Ad}_f)_B{}^A\,.
\end{align}
In particular, for an infinitesimal transformation $f=1+\epsilon^C\,T_C$\,, this can be expressed as
\begin{align}
 \delta E_M{}^A = \epsilon^C E_M{}^B\,F_{CB}{}^A\,,\qquad 
 \delta E_A{}^M = -\epsilon^C F_{CA}{}^B\,E_B{}^M\,.
\end{align}
Using the property \eqref{eq:gLie-alg}, this can be also expressed as
\begin{align}
 \delta E_A{}^M = \gLie_{\epsilon^B E_B} E_A{}^M\,.
\label{eq:left-gen-Lie}
\end{align}
Namely, the left multiplication is generated by $\gLie_{\epsilon^A E_A}$\,. 
Under the same left multiplication, the generalized metric transforms as
\begin{align}
 \delta \cH_{MN} = \gLie_{\epsilon^A E_A} \cH_{MN} = \epsilon^C E_M{}^A\,E_N{}^B\,\bigl(F_{CA}{}^D\,\hat{\cH}_{DB}+F_{CB}{}^D\,\hat{\cH}_{AD}\bigr)\,.
\end{align}
If the matrix $\hat{\cH}_{AB}$ satisfies the relation
\begin{align}
 (\text{Ad}_f)_A{}^C\,(\text{Ad}_f)_B{}^D\,\hat{\cH}_{CD} = \hat{\cH}_{AB}\,,
\label{eq:H-inv}
\end{align}
the gauged action is invariant under the global left multiplication. 
For an infinitesimal left multiplication by $f=1+\epsilon^C\,T_C$\,, the same condition can be expressed as
\begin{align}
 \epsilon^C \bigl(F_{CA}{}^D\,\hat{\cH}_{DB}+F_{CB}{}^D\,\hat{\cH}_{AD}\bigr) = 0\,.
\label{eq:H-inv-inf}
\end{align}
If this is satisfied, there is the corresponding conserved charge.
By defining $\VV^M\equiv \epsilon^A E_A{}^M$ and using its parameterization $(\VV^M)=(\kk^m,\,\tilde{\kk}_m)$, the conserved current satisfying $\rmd J=0$ can be found as
\begin{align}
 J = \kk^m\,(g_{mn}\,*\rmd x^n+B_{mn}\,\rmd x^n)+\tilde{\kk}_m\,\rmd x^m \,.
\end{align}
In section \ref{sec:Dressing-cosets}, in the case where the relation \eqref{eq:H-inv} or \eqref{eq:H-inv-inf} is satisfied, we promote this global symmetry to a local gauge symmetry by introducing an additional gauge field $\cA$\,. 

\section{Dressing cosets}
\label{sec:Dressing-cosets}

So far, we have studied string theory on the coset space $M = \cD/\tilde{G}$. 
Here, we consider a quotient of this space $M$ under the left multiplication by an $n$-dimensional subgroup $F\subset \cD$. 
As we explained in section \ref{sec:left-mult}, if $\hat{\cH}_{AB}$ satisfies Eq.~\eqref{eq:H-inv} for any element $f\in F$, the action \eqref{eq:geugedDSM} is invariant under a global left multiplication by $F$. 
In a recent paper \cite{1903.00439}, this global symmetry was promoted to a local gauge symmetry by introducing an additional gauge field. 
However, as we mentioned in the introduction, the gauged action proposed in \cite{1903.00439} is not manifestly covariant on the worldsheet. 
Here, we propose a worldsheet-covariant gauged action that is a gauged version of the gauged sigma model reviewed in section \ref{sec:Hull-Reid-Edwards}. 

\subsection{Our gauged action}

Let us denote the generators of $\mathfrak{f}$ as $T_{\sfi}$ ($\sfi=1,\dotsc,n\equiv \dim F$) and express these as $T_{\sfi}\equiv \VV_{\sfi}{}^A\,T_A$ by using a constant matrix $\VV_{\sfi}{}^A$. 
We assume that $\mathfrak{f}$ is isotropic with respect to $\langle \cdot,\,\cdot \rangle$\,, namely
\begin{align}
 \langle T_{\sfi},\, T_{\sfj} \rangle = 0\quad \Leftrightarrow \quad \eta_{AB}\,\VV_{\sfi}{}^A\,\VV_{\sfj}{}^B = 0\,.
\label{eq:F-isotropy}
\end{align}
Then, we propose a gauged sigma model\footnote{A similar gauged action that takes the form
\begin{align*}
 4\pi\alpha' S = \int_\Sigma\Bigl(\frac{1}{2}\,\langle \hat{\mathbb{H}}(\hat{\cP}) \overset{\wedge}{,} * \hat{\cP} \rangle + \langle\cP\overset{\wedge}{,}\, l\,\cC\,l^{-1}\rangle
 - \langle \hat{\cP} \overset{\wedge}{,}\, \cA\rangle \Bigr)
 - \frac{1}{3!}\int_{\cB} \langle \cP\overset{\wedge}{,}\, [\cP,\, \cP]\rangle \,,
\end{align*}
in our notation, has been studied by Jun-ichi Sakamoto. 
Here, the operator $\hat{\mathbb{H}}$ is degenerate similar to the operator $\hat{\cE}$ used in the degenerate $\hat{\cE}$-models \cite{1903.00439}. 
We thank Jun-ichi Sakamoto for sharing his private note.}
\begin{align}
\begin{split}
 4\pi\alpha' S &= \int_\Sigma \Bigl(\frac{1}{2}\,\hat{\cH}_{AB}\,\mathbb{P}^A \wedge * \mathbb{P}^B + \langle \cP \overset{\wedge}{,}\, l\,\cC\,l^{-1}\rangle 
 - \langle \hat{\cP} \overset{\wedge}{,}\, \cA\rangle \Bigr)
 - \frac{1}{3!}\int_{\cB} \langle \cP\overset{\wedge}{,}\, [\cP,\, \cP]\rangle \,,
\end{split}
\label{eq:geugedDSM2}
\end{align}
where we have defined $\mathbb{P} \equiv \mathbb{P}^A\,T_A \equiv \hat{\cP} +\cA$ and $\cA \equiv \cA^{\sfi}\,T_{\sfi}$\,.
This action is invariant under the gauge transformation
\begin{align}
 l(\sigma)\to f(\sigma)\,l(\sigma) \,,\quad
 \cA(\sigma) \to f(\sigma)\,\cA(\sigma)\,f^{-1}(\sigma) - \rmd f(\sigma)\,f^{-1}(\sigma) \,,\quad
 \cC(\sigma) \to \cC(\sigma) \,,
\label{eq:gauge-transf-L}
\end{align}
provided that the matrix $\hat{\cH}_{AB}$ satisfies the condition \eqref{eq:H-inv} for $\forall f\in F$. 
Indeed, we find 
\begin{align}
\begin{split}
 \mathbb{P} &\to f\,\mathbb{P}\,f^{-1}\,,
\qquad
 \langle \cP \overset{\wedge}{,}\, l\,\cC\,l^{-1}\rangle \to \langle \cP \overset{\wedge}{,}\, l\,\cC\,l^{-1}\rangle - \langle l\,\cC\,l^{-1} \overset{\wedge}{,}\, f^{-1}\,\rmd f\rangle\,,
\\
 \langle \hat{\cP} \overset{\wedge}{,}\, \cA\rangle &\to \langle \hat{\cP} \overset{\wedge}{,}\, \cA\rangle - \langle \hat{\cP} \overset{\wedge}{,}\, f^{-1} \,\rmd f \rangle \,,
\quad
 \tfrac{1}{3!}\,\langle \cP\overset{\wedge}{,}\, [\cP,\, \cP]\rangle \to \tfrac{1}{3!}\,\langle \cP\overset{\wedge}{,}\, [\cP,\, \cP]\rangle 
 + \rmd \langle\cP\wedge f^{-1}\,\rmd f\rangle\,,
\end{split}
\end{align}
and then it is easy to see that the action \eqref{eq:geugedDSM2} is invariant. 
This action is also invariant under the (local) right multiplication by $\tilde{G}$, by using the transformation rule \eqref{eq:gauge-transf} and the invariance of $\cA$\,. 
Then, after eliminating the auxiliary fields, we get the sigma model on a $d$-dimensional dressing coset $\check{M}= F\backslash \cD/\tilde{G}$, where $d=D-n$\,. 

We note that, after eliminating the gauge fields $\cC$ and $\cA$\,, the action is gauge invariant even if the condition \eqref{eq:H-inv} is slightly relaxed. 
This point is detailed later in section \ref{sec:gauge-inv}. 

\subsection{Gauged action by Hull and Spence}
\label{sec:Hull-Spence}

Here we take a straightforward approach to eliminate the gauge field $\cC$, or equivalently, the auxiliary fields $P_m$\,. 
To this end, we shall rewrite the action \eqref{eq:geugedDSM2} as
\begin{align}
 S = \frac{1}{4\pi\alpha'}\int_\Sigma \bigl(\tfrac{1}{2}\,\cH_{MN}\,\cZ^M \wedge * \cZ^N + \rmd x^m\wedge P_m - \eta_{MN}\,Z^M\wedge \cA^N \bigr)\,,
\end{align}
where
\begin{align}
 \cZ^M \equiv Z^M+ \cA^M\,,\qquad
 \cA^M\equiv \KK_{\sfi}{}^M\,\cA^{\sfi} \,,\qquad 
 \KK_{\sfi}{}^M \equiv \VV_{\sfi}{}^A\,E_A{}^M\,,
\label{eq:cZ-def}
\end{align}
and $Z^M$ is defined in Eq.~\eqref{eq:cP-Z}. 
Assuming that the original background $\cH_{MN}$ is Riemannian (i.e., $\cH^{mn}$ is non-degenerate), we can use the standard parameterizations
\begin{align}
 \cH_{MN} = \begin{pmatrix} (g-B\,g^{-1}\,B)_{mn} & (B\,g^{-1})_m{}^n \\ -(g^{-1}\,B)^m{}_n & g^{mn} \end{pmatrix},\qquad 
 \KK_{\sfi}{}^M = \begin{pmatrix} \kk_{\sfi}^m \\ \tilde{\kk}_{\sfi m} \end{pmatrix} = \begin{pmatrix} \kk_{\sfi}^m \\ \hat{\kk}_{\sfi m} + B_{mn}\,\kk_{\sfi}^n \end{pmatrix}.
\end{align}
Then we have
\begin{align}
 \cZ^M = \begin{pmatrix} \cD x^m \\ \cP_m \end{pmatrix},\qquad
 \cD x^m \equiv \rmd x^m + \kk_{\sfi}^m\,\cA^{\sfi}\,,\qquad 
 \cP_m\equiv P_m + (\hat{\kk}_{\sfi m} + B_{mn}\,\kk_{\sfi}^n)\,\cA^{\sfi}\,,
\end{align}
and the action is rewritten as
\begin{align}
\begin{split}
 S &= \frac{1}{4\pi\alpha'}\int_\Sigma \bigl[\tfrac{1}{2}\,\cH_{MN}\,\cZ^M \wedge * \cZ^N + \cD x^m\wedge \cP_m
\\
 &\qquad\qquad\quad + \kk_{\sfj}^m\,(\hat{\kk}_{\sfi m} + B_{mn}\,\kk_{\sfi}^n)\,\cA^{\sfi}\wedge \cA^{\sfj}
 - 2\,\rmd x^m \wedge (\hat{\kk}_{\sfi m} + B_{mn}\,\kk_{\sfi}^n)\,\cA^{\sfi} \bigr]
\\
 &= \frac{1}{4\pi\alpha'}\int_\Sigma \bigl[g_{mn}\,\cD x^m\wedge *\cD x^n + B_{mn}\,\cD x^m\wedge \cD x^n 
 +\tfrac{1}{2}\,g^{mn}\,\Lambda_m \wedge *\Lambda_n 
\\
 &\qquad\qquad\quad + \kk_{\sfj}^m\,(\hat{\kk}_{\sfi m} + B_{mn}\,\kk_{\sfi}^n)\,\cA^{\sfi}\wedge \cA^{\sfj}
 - 2\,\rmd x^m \wedge (\hat{\kk}_{\sfi m} + B_{mn}\,\kk_{\sfi}^n)\,\cA^{\sfi} \bigr] \,,
\end{split}
\end{align}
where
\begin{align}
 \Lambda_m \equiv \cP_m - B_{mn}\,\cD x^n - g_{mn}\,*\cD x^n \,.
\label{eq:Lambda-def}
\end{align}
Eliminating the auxiliary fields $P_m$\,, we obtain
\begin{align}
 S = \frac{1}{2\pi\alpha'} \int_\Sigma \bigl(\tfrac{1}{2}\,g_{mn}\,\cD x^m\wedge * \cD x^n - \hat{\kk}_{\sfi} \wedge \cA^{\sfi} + \tfrac{1}{2}\,\kk_{\sfj}^m\, \hat{\kk}_{\sfi m} \,\cA^{\sfi}\wedge \cA^{\sfj}\bigr)+ \frac{1}{2\pi\alpha'}\int_{\cB} H_3\,,
\label{eq:H-S-action}
\end{align}
where $\hat{\kk}_{\sfi}\equiv \hat{\kk}_{\sfi m}\,\rmd x^m$. 
This is precisely the gauged action studied by Hull and Spence \cite{Hull:1989jk}. 
Their gauged sigma model requires several conditions for the gauge invariance, which can be summarized as
\begin{align}
 \gLie_{\KK_{\sfi}}\cH_{MN}=0\,,\qquad \eta_{MN}\,\KK_{\sfi}^M\,\KK_{\sfj}^N = 0\,.
\end{align}
These are already assumed in the gauged sigma model \eqref{eq:geugedDSM2} and the action \eqref{eq:H-S-action} is invariant under the gauge transformation \eqref{eq:gauge-transf-L}, or equivalently,
\begin{align}
 \delta x^m(\sigma)= \epsilon^{\sfi}(\sigma)\,\kk_{\sfi}^m \,,\qquad
 \delta \cA^{\sfi}(\sigma) \to f_{\sfj\sfk}{}^{\sfi} \,\epsilon^{\sfj}(\sigma)\,\cA^{\sfk}(\sigma) - \rmd \epsilon^{\sfi}(\sigma) \,.
\end{align}
It is sometimes more convenient to rewrite the action \eqref{eq:H-S-action} as
\begin{align}
 S = \frac{1}{4\pi\alpha'} \int_\Sigma \bigl(g_{mn}\,\cD x^m\wedge * \cD x^n + B_{mn}\,\cD x^m\wedge \cD x^n - 2\,\tilde{\kk}_{\sfi} \wedge \cA^{\sfi} + \kk_{\sfj}^m\, \tilde{\kk}_{\sfi m} \,\cA^{\sfi}\wedge \cA^{\sfj}\bigr)\,.
\label{eq:H-S-action-2}
\end{align}

Let us consider two examples. 
The first is the case where the gauge group $F$ is a subgroup of $G$ while the second is the case where $F$ is a subgroup of $\tilde{G}$. 
\begin{enumerate}
\item 
In the first example, we choose the constant matrix $\VV_{\sfi}{}^A$ as $(\VV_{\sfi}{}^A) = (\delta_{\sfi}^a,\,0)$ and then
\begin{align}
 \KK_{\sfi}{}^M = \begin{pmatrix} \kk_{\sfi}^m \\ \tilde{\kk}_{\sfi m} \end{pmatrix} = E_{\sfi}{}^M = \begin{pmatrix} e_{\sfi}^m \\ 0 \end{pmatrix} .
\label{eq:V1}
\end{align}
The gauged action \eqref{eq:H-S-action-2} with $\cD x^m=\rmd x^m + e_{\sfi}^m\,\cA^{\sfi}$ is simplified as
\begin{align}
 S = \frac{1}{4\pi\alpha'} \int_\Sigma \bigl(g_{mn}\,\cD x^m\wedge * \cD x^n + B_{mn}\,\cD x^m\wedge \cD x^n \bigr) \,.
\end{align}

\item In the second example, we choose the constant matrix $\VV_{\sfi}{}^A$ as $(\VV_{\sfi}{}^A) = (0,\,\delta_{\sfi a})$ and then
\begin{align}
 \KK_{\sfi}{}^M = \begin{pmatrix} \kk_{\sfi}^m \\ \tilde{\kk}_{\sfi m} \end{pmatrix} = \delta_{\sfi a}\,E^{aM} = \delta_{\sfi a} \begin{pmatrix} -\pi^{ab}\,e_b^m \\ r^a_m \end{pmatrix} = \begin{pmatrix} \pi^{mn}\,r_n^a \\ r^a_m \end{pmatrix} \delta_{a\sfi}\,.
\label{eq:V2}
\end{align} 
The gauged action \eqref{eq:H-S-action-2} becomes
\begin{align}
 S = \frac{1}{4\pi\alpha'} \int_\Sigma \bigl(g_{mn}\,\cD x^m\wedge * \cD x^n + B_{mn}\,\cD x^m\wedge \cD x^n - 2\, r_{\sfi} \wedge \cA^{\sfi} + \pi_{\sfi\sfj} \,\cA^{\sfi}\wedge \cA^{\sfj}\bigr) \,,
\label{eq:dual-action-Stern}
\end{align}
where $r_{\sfi}\equiv \delta_{\sfi a}\,r^a$\,, $\pi_{\sfi\sfj}\equiv r_{\sfi m}\,\pi^{mn}\,r_{\sfj n}$\,, and $\cD x^m = \rmd x^m + \pi^{mn}\,r_{\sfi n}\,\cA^{\sfi}$\,. 
\end{enumerate}

The above two actions have been derived also in Eqs.~(3.48) and (3.58) of \cite{1903.00439} from the non-covariant gauged action (see Eq.~(3.34) of \cite{1903.00439}). 
This agreement suggests that our gauged action \eqref{eq:geugedDSM2} is equivalent to the one given in \cite{1903.00439}.
We also note that the action \eqref{eq:dual-action-Stern} seems to be the same as the dual action given in Eq.~(6.26) of the paper \cite{hep-th:9903170} by Stern. 

At this stage, the only assumption is that the background $\cH_{MN}$ is Riemannian. 
From the Hull--Spence action \eqref{eq:H-S-action}, the equations of motion for $\cA^{\sfi}$ can be found as
\begin{align}
 \kk_{\sfi}^m\, g_{mn}\,*Dx^n + \hat{\kk}_{\sfi m}\,\cD x^m = 0\,.
\label{eq:eom-A}
\end{align}
To solve this equation for the gauge fields $\cA^{\sfi}$, we additionally assume that the matrix
\begin{align}
 N_{\sfi\sfj} \equiv \kk_{\sfi}^m\,g_{mn}\,\kk_{\sfj}^n +\hat{\kk}_{\sfi m}\,\kk_{\sfj}^{m} 
 = \kk_{\sfi}^m\,E_{mn}\,\kk_{\sfj}^n + \tilde{\kk}_{\sfi m}\,\kk_{\sfj}^{m} \,,
\label{eq:Nij-def}
\end{align}
is invertible, and denote the inverse matrix as $N^{\sfi\sfj}$. 
Then we find
\begin{align}
 \cA^{\sfi} = - N^{(\sfi\sfj)}\,* J_{\sfj} - N^{[\sfi\sfj]}\,J_{\sfj} \qquad \bigl(J_{\sfi} \equiv \kk_{\sfi}^m\,g_{mn}\,*\rmd x^n + \hat{\kk}_{\sfi} \bigr)\,.
\end{align}
Substituting this into the gauged action \eqref{eq:H-S-action}, we obtain the action
\begin{align}
 S = \frac{1}{4\pi\alpha'} \int_\Sigma \bigl(\check{g}_{mn}\,\rmd x^m\wedge * \rmd x^n + \check{B}_{mn}\,\rmd x^m\wedge * \rmd x^n \bigr) \,,
\label{eq:reduced-action}
\end{align}
where $\check{g}_{mn}$ and $\check{B}_{mn}$ are symmetric and anti-symmetric parts of
\begin{align}
 \check{E}_{mn} = E_{mn} - \bigl(\kk_{\sfi m}-\hat{\kk}_{\sfi m}\bigr)\,N^{\sfi\sfj}\,\bigl(\kk_{\sfj n}+\hat{\kk}_{\sfj n}\bigr)\qquad \bigl(\kk_{\sfi m}\equiv g_{mn}\,\kk_{\sfi}^n\bigr)\,.
\label{eq:reduced-E}
\end{align}
Namely, this matrix plays the role of the background fields on the reduced space $\check{M}=F\backslash D/\tilde{G}$ (after making a certain gauge fixing). 
Accordingly, we call $\check{E}_{mn}$ the reduced background fields. 

In the following, we shall consider two examples that reproduce the Sfetsos model \cite{hep-th:9904188}. 
We also consider an example where $N_{\sfi\sfj}$ is degenerate. 

\subsubsection{Original background of the Sfetsos model}

When $(\KK_{\sfi}{}^A) \equiv (\mathsf{\kk}_{\sfi}^a,\,\tilde{\kk}_{\sfi a})= (\delta_{\sfi}^a,\,0)$, by using Eq.~\eqref{eq:V1}, the reduced background fields can be found as $\check{E}_{mn}= r_m^a\,\check{E}_{ab}\,r_n^b$ with
\begin{align}
 \check{E}_{ab} = E_{ab} - E_{ac}\,\delta_{\sfi}^c\,N^{\sfi\sfj}\,\delta_{\sfj}^d\,E_{db} \qquad \bigl(N_{\sfi\sfj}\equiv \delta_{\sfi}^a\,E_{ab}\,\delta_{\sfj}^b\bigr)\,,
\label{eq:cE-1}
\end{align}
where $E_{ab}\equiv e_a^m\,E_{mn}\,e_b^n = [(\hat{E}+\pi)^{-1}]_{ab}$\,. 
One can easily see that the reduced background fields satisfy $\check{E}_{ab}\,\mathsf{\kk}_{\sfi}^b=0=\mathsf{\kk}_{\sfi}^a\,\check{E}_{ab}$\,. 
Namely, after eliminating the gauge fields, the $\sfi$-components of $E_{ab}= [(\hat{E}+\pi)^{-1}]_{ab}$ are projected out. 
If we decompose the generators as $\{T_a\}=\{T_{\check{a}},\,T_{\sfi}\}$ $(\check{a}=1,\dotsc,d)$, the non-vanishing components of $\check{E}_{ab}$ are contained in the $d\times d$ block $\check{E}_{\check{a}\check{b}}$\,,
\begin{align}
 \check{E}_{ab} = \left(\begin{array}{c|c} \check{E}_{\check{a}\check{b}} & 0 \\\hline 0 & 0\end{array} \right).
\label{eq:E-block}
\end{align}
Then the action \eqref{eq:reduced-action} becomes
\begin{align}
 S = \frac{1}{4\pi\alpha'} \int_\Sigma \check{E}_{\check{a}\check{b}}\,\bigl(r^{\check{a}} \wedge * r^{\check{b}} + r^{\check{a}}\wedge r^{\check{b}} \bigr) \,.
\label{eq:reduced-action-o}
\end{align}
If we make a parameterization $g(\sigma)=\Exp{x^{\dot{m}}(\sigma)\,\delta_{\dot{m}}^{\sfi}\,T_{\sfi}}\Exp{x^{\check{m}}(\sigma)\,\delta_{\check{m}}^{\check{a}}\,T_{\check{a}}}$\,, we can remove the first factor $\Exp{x^{\dot{m}}(\sigma)\,\delta_{\dot{m}}^{\sfi}\,T_{\sfi}}\in F$ by using the gauge symmetry. 
Then the right-invariant 1-form $r^{\check{a}}$ becomes $r^{\check{a}}=r^{\check{a}}_{\check{m}}\,\rmd x^{\check{m}}$ and we can clearly see that the action \eqref{eq:reduced-action-o} describes string theory on a $d$-dimensional coset space $H\backslash G$ with coordinates $x^{\check{m}}$. 

Here, it is noted that the submatrix $\check{E}_{\check{a}\check{b}}$ of \eqref{eq:cE-1} takes the form of the Schur complement, and the $d\times d$ block $\check{E}_{\check{a}\check{b}}$ coincides with the inverse matrix of $(\hat{E}+\pi)^{\check{a}\check{b}}$. 
Namely, the non-vanishing components of the reduced background fields $\check{E}_{\check{a}\check{b}}$ are expressed as
\begin{align}
 \bigl(\check{E}_{\check{a}\check{b}}\bigr) = \bigl(\hat{E}^{\check{a}\check{b}}+\pi^{\check{a}\check{b}}\bigr)^{-1} \,.
\label{eq:sfetsos-E-1}
\end{align}
This reproduces the result \eqref{eq:E-limit} by Sfetsos \cite{hep-th:9904188}, which has been obtained by a singular limit. 

\subsubsection{Dual background of the Sfetsos model}

When $(\VV_{\sfi}{}^A) \equiv (\mathsf{\kk}_{\sfi}^a,\,\tilde{\kk}_{\sfi a}) = (0,\,\delta_{\sfi a})$, by using Eq.~\eqref{eq:V1}, the reduced background fields become
\begin{align}
\begin{split}
 \check{E}_{ab} &= [(\hat{E}+\pi)^{-1}]_{ab} + [(\hat{E}+\pi)^{-1}\,\hat{E}]_a{}^{c} \,\delta_{\sfi c}\,N^{\sfi\sfj}\,\delta_{\sfj d}\,\bigl[\hat{E}\,(\hat{E}+\pi)^{-1}\bigr]^d{}_b \,,
\\
 N_{\sfi\sfj}&= \delta_{\sfi a}\,\bigl[\hat{E} - \hat{E}\,(\hat{E}+\pi)^{-1}\,\hat{E}\bigr]^{ab}\,\delta_{b\sfj} \,.
\end{split}
\label{eq:S-d-EN}
\end{align}
To simplify this result, let us employ a matrix identity
\begin{align}
\begin{split}
 {\small\begin{pmatrix}
 \mathsf{A} & \mathsf{B} \\ \mathsf{C} & \mathsf{D}
\end{pmatrix}^{-1}} &= {\small\begin{pmatrix}\mathsf{A}^{-1}+\mathsf{A}^{-1}\mathsf{B} (\mathsf{D} -\mathsf{CA}^{-1}\mathsf{B} )^{-1}\mathsf{CA}^{-1}&-\mathsf{A}^{-1}\mathsf{B} (\mathsf{D} -\mathsf{CA}^{-1}\mathsf{B} )^{-1}\\-(\mathsf{D} -\mathsf{CA}^{-1}\mathsf{B} )^{-1}\mathsf{CA}^{-1}&(\mathsf{D} -\mathsf{CA}^{-1}\mathsf{B} )^{-1}\end{pmatrix}}
\\
 &= {\small\begin{pmatrix}(\mathsf{A} -\mathsf{BD}^{-1}\mathsf{C} )^{-1}&-(\mathsf{A} -\mathsf{BD}^{-1}\mathsf{C} )^{-1}\mathsf{BD}^{-1}\\-\mathsf{D}^{-1}\mathsf{C} (\mathsf{A} -\mathsf{BD}^{-1}\mathsf{C} )^{-1}&\quad \mathsf{D}^{-1}+\mathsf{D}^{-1}\mathsf{C} (\mathsf{A} -\mathsf{BD}^{-1}\mathsf{C} )^{-1}\mathsf{BD}^{-1}
 \end{pmatrix}}
\end{split},
\label{eq:formula}
\end{align}
whose 1--1 components give a non-trivial identity
\begin{align}
 \mathsf{A}^{-1}+\mathsf{A}^{-1}\,\mathsf{B}\,(\mathsf{D} -\mathsf{C}\,\mathsf{A}^{-1}\,\mathsf{B})^{-1}\,\mathsf{C}\,\mathsf{A}^{-1} = (\mathsf{A} -\mathsf{B}\,\mathsf{D}^{-1}\,\mathsf{C})^{-1}\,,
\end{align}
if various inverse matrices in the above expressions exist. 
Substituting
\begin{align}
 \mathsf{A} = (\hat{E}^{ab}+\pi^{ab})\,,\qquad 
 \mathsf{B} = (\hat{E}^{ac} \,\delta_{c\sfj})\,,\qquad
 \mathsf{C} = (\delta_{\sfi c}\,\hat{E}^{cb})\,,\qquad 
 \mathsf{D} = (\delta_{\sfi c}\,\hat{E}^{cd}\,\delta_{\sfj d})\,,
\end{align}
we can simplify the first line of Eq.~\eqref{eq:S-d-EN} as
\begin{align}
 \check{E}_{ab} = \bigl[(\widetilde{E} + \pi)^{-1}\bigr]_{ab} \qquad
 \bigl(\widetilde{E}^{ab} \equiv \hat{E}^{ab} - \hat{E}^{ac}\,\delta_{\sfi c}\,\cN^{\sfi\sfj}\,\delta_{\sfj d}\,\hat{E}^{db}\bigr)\,,
\end{align}
where $\cN^{\sfi\sfj}$ is the inverse of a constant matrix $\cN_{\sfi\sfj}\equiv \delta_{\sfi c}\,\hat{E}^{cd}\,\delta_{\sfj d}$\,. 
The non-vanishing components of the constant matrix $\widetilde{E}^{ab}$ are contained in the $d\times d$ matrix $\widetilde{E}^{\check{a}\check{b}}$\,, and again we find that $\widetilde{E}^{\check{a}\check{b}}$ is the inverse matrix of $\hat{E}_{\check{a}\check{b}}$\,. 
We then arrive at the simple expression
\begin{align}
 \bigl(\check{E}_{ab}\bigr) = \left(\begin{array}{c|c} \bigl[(\hat{E}_{\check{c}\check{d}})^{-1} + \pi\bigr]^{\check{a}\check{b}} & \pi^{\check{a}j} \\\hline \pi^{i\check{b}} & \pi^{ij} \end{array}\right)^{-1}\,,
\label{eq:sfetsos-E-2}
\end{align}
which takes the same form as the dual background \eqref{eq:E-limit-d} of the Sfetsos model. 
Under the PL $T$-duality, the matrix $\hat{E}^{ab}$ in the original model and in the dual model are related as
\begin{align}
 \bigl(\hat{E}_{\text{(orig)}}^{ab}\bigr) = \bigl(\hat{E}_{\text{(dual)}}^{ab}\bigr)^{-1}\,.
\end{align}
Thus the PL $T$-duality for the Sfetsos model can be expressed as
\begin{align}
 \bigl(\check{E}^{\text{(orig)}}_{ab}\bigr) =\left(\begin{array}{c|c} (\hat{E}_{\text{(orig)}}^{\check{a}\check{b}}+\pi^{\check{a}\check{b}})^{-1} & 0 \\\hline 0 & 0 \end{array}\right) 
 \ \leftrightarrow\ 
 \bigl(\check{E}^{\text{(dual)}}_{ab}\bigr) = \left(\begin{array}{c|c} (\hat{E}_{\text{(orig)}} + \tilde{\pi})^{\check{a}\check{b}} & \tilde{\pi}^{\check{a}j} \\\hline \tilde{\pi}^{i\check{b}} & \tilde{\pi}^{ij} \end{array}\right)^{-1},
\end{align}
where $\pi^{ab}$ in the original/dual model are denoted by $\pi^{ab}$ and $\tilde{\pi}^{ab}$, respectively. 

\subsubsection{A degenerate example}

Here we consider a case where the matrix $N_{\sfi\sfj}$ is degenerate. 
As an extreme case, let us suppose $N_{\sfi\sfj}=0$\,, namely, $g_{mn}\,\kk_{\sfi}^m\,\kk_{\sfj}^n=0$ and $\hat{\kk}_{\sfi m}\,\kk_{\sfj}^m=0$\,. 
This can be realized by choosing
\begin{align}
 \VV_{\sfi}{}^A = \begin{pmatrix} \hat{\beta}^{ab}\,\tilde{\kk}_{\sfi b} \\ \tilde{\kk}_{\sfi a} \end{pmatrix},\qquad \hat{G}^{ab}\,\tilde{\kk}_{\sfi b}=0\,.
\end{align}
In this case, the gauged action \eqref{eq:H-S-action} reduces to
\begin{align}
 S = \frac{1}{4\pi\alpha'} \int_\Sigma \bigl( g_{mn}\,\rmd x^m\wedge * \rmd x^n + 2\,\cA^{\sfi} \wedge J_{\sfi} \bigr)+ \frac{1}{2\pi\alpha'}\int_{\cB} H_3\,,
\label{eq:SM-KS}
\end{align}
where $J_{\sfi} = \kk_{\sfi}^m\,g_{mn}\,*\rmd x^n + \hat{\kk}_{\sfi}$\,. 
The quadratic term for the gauge field $\cA^{\sfi}$ vanishes and the equations of motion for $\cA^{\sfi}$ simply give $J_{\sfi}=0$\,.
Under the equations of motion, we have
\begin{align}
 S = \frac{1}{4\pi\alpha'} \int_\Sigma \bigl( g_{mn}\,\rmd x^m\wedge * \rmd x^n + B_{mn}\,\rmd x^m\wedge \rmd x^n \bigr) \,.
\end{align}
Namely, the background fields $E_{mn}=(g+B)_{mn}$ are exactly the same as the original one
\begin{align}
 \check{E}_{mn} = E_{mn} = r^a_m\,\bigl[(\hat{E} + \pi)^{-1}\bigr]_{ab}\,r^b_n\,. 
\label{eq:Emn-same-as-original}
\end{align}
However, after making a certain gauge fixing, this reduces to a $d$-dimensional sigma model. 

It is noted that the equations of motion for $P_m$ can be written as $\Lambda_m=0$ with $\Lambda_m$ defined in Eq.~\eqref{eq:Lambda-def}. 
Under $\Lambda_m=0$\,, $\cP_m$ is determined as
\begin{align}
 \cP_m = B_{mn}\,\cD x^n + g_{mn}* \cD x^n\,,
\end{align}
and by substituting this into $\cZ^M$, we find
\begin{align}
 \cZ^M = \begin{pmatrix} \cD x^m \\ B_{mn}\,\cD x^n + g_{mn} * \cD x^n \end{pmatrix}.
\label{eq:Z-on-shell}
\end{align}
Then the equations of motion $J_{\sfi}=0$ are rewritten as
\begin{align}
 \VV_{\sfi M}\,\cZ^M = \hat{\kk}_{\sfi m}\,\rmd x^m + \kk_{\sfi}^m\, g_{mn}* \rmd x^n = 0\,.
\end{align}
This can also be expressed as $\langle T_{\sfi},\,\mathbb{P}\rangle = 0$\,, or equivalently, $\mathbb{P} \in \mathfrak{f}^\perp$\,. 

We can regard Eq.~\eqref{eq:Emn-same-as-original} as a special case $\cR^{ab}=\hat{E}^{ab}$ of the model \eqref{eq:orig-model} by Klim\v{c}\'\i{}k and \v{S}evera. 
The required orthogonality condition \eqref{eq:r-f-orthogonal}, or equivalently $J_{\sfi}=0$\,, is reproduced as the equations of motion for $\cA^{\sfi}$\,.

Here we have considered a special case $N_{\sfi\sfj}=0$, but a more general case where $N_{\sfi\sfj}$ is degenerate is discussed in section \ref{sec:revisited}, or more generally in Appendix \ref{app:general-case}. 

\subsection{Simplifying the reduced geometry}
\label{sec:revisited}

In the previous subsection, we found the formula \eqref{eq:reduced-E} for $\check{E}_{mn}$\,. 
This can be applied to a wide class of examples, but the non-degeneracy of $N_{\sfi\sfj}$ has been assumed. 
In addition, this formula looks different from the one by Klim\v{c}\'\i{}k and \v{S}evera \eqref{eq:orig-model}.
Only in some specific cases, we found nice rewritings, Eq.~\eqref{eq:sfetsos-E-1} and Eq.~\eqref{eq:sfetsos-E-2}, that look similar to Eq.~\eqref{eq:orig-model}. 
In this subsection, to clarify the relation to the model by Klim\v{c}\'\i{}k and \v{S}evera, we find a nice rewriting of the formula for a general choice of the gauge group $F$. 
The approach in this section (which is inspired by \cite{1105.0162}) is somewhat technical, but thanks to this, it is no longer necessary to assume the non-degeneracy of $N_{\sfi\sfj}$\,. 

\subsubsection{Duality-covariant equations of motion}

Before performing a detailed analysis, let us make a few preparations.
In general, from the gauged action \eqref{eq:geugedDSM2}, the equations of motion for $\cA$ can be found as
\begin{align}
 \VV_{\sfi}^A\,\hat{\cH}_{AB}\,\mathbb{P}^B = - \VV_{\sfi A}\,* \mathbb{P}^A \,.
\label{eq:eom-A-cov}
\end{align}
In addition, when $\cH_{MN}$ is Riemannian, the equations of motion for $\cC$ are equivalent to Eq.~\eqref{eq:Z-on-shell}. 
They can also be expressed as the self-duality relation $\cZ^M = \cH^M{}_N \,* \cZ^N$, or
\begin{align}
 \mathbb{P}^A = \hat{\cH}^A{}_B\,* \mathbb{P}^B \,. 
\label{eq:self-duality}
\end{align}
Combining Eqs.~\eqref{eq:eom-A-cov} and \eqref{eq:self-duality}, the equations of motion can be summarized as\footnote{The second equation is equivalent to the equations of motion \eqref{eq:eom-A} that follows from the action \eqref{eq:H-S-action}.}
\begin{align}
 \mathbb{P}^A = \hat{\cH}^A{}_B\,* \mathbb{P}^B \,,\qquad 
 \eta_{AB}\,\VV_{\sfi}{}^A\,\mathbb{P}^B= 0\,.
\end{align}
An important relation that can be derived from these equations is
\begin{align}
 \hat{\cH}_{AB}\,\VV_{\sfi}{}^A\,\mathbb{P}^B = 0\,.
\end{align}
If we define an operator $\hat{\cH}(T_A)\equiv \hat{\cH}_A{}^B\,T_B$ that satisfies
\begin{align}
 \hat{\cH}^2 = 1\,,\qquad \langle \hat{\cH}(T_A),\, T_B\rangle = \langle T_A,\, \hat{\cH}(T_B)\rangle\,,
\end{align}
we can summarize the equations of motion for $\cC$ and $\cA$ as follows:
\begin{align}
 \mathbb{P},\,\hat{\cH}(\mathbb{P})\in \mathfrak{f}^\perp\,,\qquad *\mathbb{P} = \hat{\cH}(\mathbb{P})\,.
\label{eq:EOM-summary}
\end{align}

Under the equations of motion, we have $\hat{\cH}_{AB}\,\mathbb{P}^A\wedge * \mathbb{P}^B=\eta_{AB}\,\mathbb{P}^A\wedge \mathbb{P}^B=0$ and $\langle \mathbb{P} \overset{\wedge}{,}\, \cA\rangle=0$\,, and the gauged action \eqref{eq:geugedDSM2} reduces to
\begin{align}
 S = \frac{1}{4\pi\alpha'}\int_\Sigma r^a \wedge P_a \,.
\label{eq:r-p-action}
\end{align}
In the following, we rewrite this action such that the gauge fields do not appear. 

\subsubsection{Non-degenerate case}
\label{sec:non-deg}

For simplicity, we assume that the matrix $\hh_{\sfi\sfj}\equiv \langle T_{\sfi},\,\hat{\cH}(T_{\sfj})\rangle$ is non-degenerate, which has also been assumed in \cite{1903.00439}. 
This assumption is made throughout the main body of this paper.
More general cases, where the matrix $\hh_{\sfi\sfj}$ is degenerate are studied in Appendix \ref{app:general-case}. 

For each generator $T_{\sfi}\in \mathfrak{f}$, let us define $\tilde{T}_{\sfi}\equiv \hat{\cH}(T_{\sfi})$. 
Since the matrix $\hh_{\sfi\sfj}$ is non-degenerate, we shall denote the inverse matrix as $\hh^{\sfi\sfj}$. 
Then, we can easily see that the generators $\{\tilde{T}_{\sfi}\}$ span the $n$-dimensional subspace complement to $\mathfrak{f}^\perp$ \cite{1903.00439}. 
In the following, we may use the bases $T^{\sfi}\equiv \hh^{\sfi\sfj}\,\tilde{T}_{\sfj}$ instead of $\tilde{T}_{\sfi}$\,, for which we have $\langle T_{\sfi},\,T^{\sfj}\rangle =\delta_{\sfi}^{\sfj}$\,.
Using the isotropy of $\mathfrak{f}$ (i.e., $\langle T_{\sfi},\,T_{\sfj}\rangle=0$), we can show $\langle T^{\sfi},\,T^{\sfj}\rangle=0$\,.
For later convenience, we combine these generators as $\{T_{\sfI}\}=\{T_{\sfi},\,T^{\sfi}\}$ ($\sfI=1,\dotsc,2\,n$). 

We introduce the other independent generators $T_{\bcheck{A}}$ ($\bcheck{A}=1,\dotsc,2\,d$) of $\mathfrak{d}$ such that the linear subspace $\mathfrak{f}^\perp$ is spanned by $\{T_{\bcheck{A}},\,T_{\sfi}\}$\,. 
By the definition, they should satisfy
\begin{align}
 \langle T_{\bcheck{A}},\,T_{\sfi} \rangle = 0\,,\qquad \langle T_{\bcheck{A}},\,\hat{\cH}(T^{\sfi}) \rangle= 0\,.
\end{align}
Now, we assume that the $\hat{\cH}(T_{\bcheck{A}}) \in \mathfrak{f}^\perp$, and then we have
\begin{align}
 \langle \hat{\cH}(T_{\bcheck{A}}),\,T_{\sfi} \rangle = 0\,,\qquad 
 \langle T_{\bcheck{A}},\,T^{\sfi} \rangle = 0\,.
\end{align}
Under this situation, the two metrics $\langle \cdot ,\,\cdot \rangle$ and $\langle \cdot ,\,\hat{\cH}(\cdot) \rangle$ are block-diagonal:
\begin{align}
 \left(\begin{array}{c|c}
 \langle T_{\bcheck{A}},\,T_{\bcheck{B}} \rangle & \langle T_{\bcheck{A}},\,T_{\sfJ} \rangle \\\hline
 \langle T_{\sfI},\,T_{\bcheck{B}} \rangle & \langle T_{\sfI},\,T_{\sfJ} \rangle\end{array}\right)
 &= \left(\begin{array}{c|c} \langle T_{\bcheck{A}},\,T_{\bcheck{B}} \rangle & 0 \\\hline
 0 & \langle T_{\sfI},\,T_{\sfJ} \rangle \end{array}\right),
\\
 \left(\begin{array}{c|c} \langle T_{\bcheck{A}},\,\hat{\cH}(T_{\bcheck{B}}) \rangle & \langle T_{\bcheck{A}},\,\hat{\cH}(T_{\sfJ}) \rangle \\\hline
 \langle T_{\sfI},\,\hat{\cH}(T_{\bcheck{B}}) \rangle & \langle T_{\sfI},\,\hat{\cH}(T_{\sfJ}) \rangle \end{array}\right)
 &= \left(\begin{array}{c|c} \langle T_{\bcheck{A}},\,\hat{\cH}(T_{\bcheck{B}}) \rangle & 0 \\\hline
 0 & \langle T_{\sfI},\,\hat{\cH}(T_{\sfJ}) \rangle \end{array}\right).
\end{align}
Then, we can decompose the generators $T_{\bcheck{A}}$ as $\{T_{\bcheck{A}}\}=\{T^{\ubar{\bm{a}}},\,T^{\bar{\bm{a}}}\}$ ($\ubar{\bm{a}}=1,\dotsc,d$\,, $\bar{\bm{a}}=1,\dotsc,d$) such that the following relations are satisfied:\footnote{In general, the range of $\ubar{\bm{a}}$ and $\bar{\bm{a}}$ can be different (see for example section 2.3 of \cite{1707.03713}), but here we do not consider that possibility.}
\begin{align}
\begin{split}
 &\langle T^{\ubar{\bm{a}}},\,T^{\ubar{\bm{b}}} \rangle = \eta^{\ubar{\bm{a}}\ubar{\bm{b}}} = \eta^{\bar{\bm{a}}\bar{\bm{b}}} = - \langle T^{\bar{\bm{a}}},\,T^{\bar{\bm{b}}} \rangle \,,\quad 
 \langle T^{\ubar{\bm{a}}},\,T^{\bar{\bm{b}}} \rangle = 0\,,
\\
 &\hat{\cH}(T^{\ubar{\bm{a}}}) = +T^{\ubar{\bm{a}}}\,,\qquad 
 \hat{\cH}(T^{\bar{\bm{a}}}) =-T^{\bar{\bm{a}}}\,, 
\end{split}
\end{align}
where $(\eta^{\ubar{\bm{a}}\ubar{\bm{b}}})$ is a diagonal matrix with the elements $\pm 1$\,. 

We have introduced $2D$ generators $\{T_{\cA}\}\equiv \{T_{\bcheck{A}},\,T_{\sfI}\}=\{T^{\ubar{\bm{a}}},\,T^{\bar{\bm{a}}},\,T_{\sfi},\,T^{\sfi}\}$\,, and accordingly, we introduce constant vectors $\{\VV_{\cA}{}^A\}\equiv \{\VV_{\bcheck{A}}{}^A,\,\VV_{\sfI}{}^A\}$ such that $T_{\cA}= \VV_{\cA}{}^A\,T_A$ are realized. 
By the construction, $\{\VV_{\bcheck{A}}{}^A\}=\{\VV^{\ubar{\bm{a}}}{}^A,\,\bar{\VV}^{\bar{\bm{a}}}{}^A\}$ should satisfy
\begin{align}
 \hat{\cH}^A{}_B\,\VV^{\ubar{\bm{a}} B} = + \VV^{\ubar{\bm{a}} A}\,,\qquad 
 \hat{\cH}^A{}_B\,\bar{\VV}^{\bar{\bm{a}} B} = - \bar{\VV}^{\bar{\bm{a}} A}\,.
\end{align}
We parameterize $\VV^{\ubar{\bm{c}}}{}^A$, $\bar{\VV}^{\bar{\bm{c}}}{}^A$, and $\VV_{\sfi}{}^A$ as
\begin{align}
 \VV^{\ubar{\bm{c}}A} \equiv \frac{1}{\sqrt{2}}\begin{pmatrix} \hat{E}^{ab}\,\kk^{\ubar{\bm{c}}}_b \\ \kk^{\ubar{\bm{c}}}_a \end{pmatrix},\qquad
 \bar{\VV}^{\bar{\bm{c}}A} \equiv \frac{1}{\sqrt{2}}\begin{pmatrix} - \bar{\kk}^{\bar{\bm{c}}}_b\,\hat{E}^{ba} \\ \bar{\kk}^{\bar{\bm{c}}}_a \end{pmatrix},\qquad
 \VV_{\sfi}{}^A = \begin{pmatrix} \mathsf{\kk}_{\sfi}^a \\ \tilde{\kk}_{\sfi a} \end{pmatrix},
\label{eq:V-Vbar}
\end{align}
and then, for the orthogonality $\langle T_{\bcheck{A}},\,T_{\sfi}\rangle =0$ and $\langle T_{\sfi},\,T_{\sfj}\rangle =0$\,, $\mathsf{\kk}_{\sfi}^a$ and $\tilde{\kk}_{\sfi a}$ should satisfy
\begin{align}
 \tilde{\kk}_{\sfi c}\,\hat{E}^{cd}\,\kk_d^{\ubar{\bm{a}}} = - \mathsf{\kk}_{\sfi}^{c}\,\kk_c^{\ubar{\bm{a}}} \,,\qquad
 \bar{\kk}_c^{\bar{\bm{a}}}\,\hat{E}^{cd}\,\tilde{\kk}_{\sfi d} = + \mathsf{\kk}_{\sfi}^{c}\,\bar{\kk}_c^{\bar{\bm{a}}} \,,\qquad
 \mathsf{\kk}_{(\sfi}^{c}\,\tilde{\kk}_{\sfj) c} = 0\,.
\label{eq:kk-orthogonal}
\end{align}
Of course, the rank of the $d\times D$ matrices $\kk^{\ubar{\bm{a}}}_b$ and $\bar{\kk}^{\bar{\bm{a}}}_b$ must be $d$ for the linear independence. 
We can choose these as $\kk^{\ubar{\bm{a}}}_b=\bar{\kk}^{\bar{\bm{a}}}_b$ but they can be different. 

Under the above setup, we can expand the (on-shell) field $\mathbb{P}(\sigma)=\mathbb{P}^A\,T_A\in \mathfrak{f}^\perp$ as
\begin{align}
 \mathbb{P}(\sigma) = \VV^{\ubar{\bm{a}}}\,\alpha_{+\ubar{\bm{a}}}(\sigma) + \bar{\VV}^{\bar{\bm{a}}}\,\alpha_{-\bar{\bm{a}}}(\sigma) + \VV_{\sfi}\,\alpha^{\sfi}(\sigma) \,,
\label{eq:P-expand}
\end{align}
by using some 1-form fields $\alpha_{\pm\bm{a}}(\sigma)$ and $\alpha^{\sfi}(\sigma)$. 
The self-duality relation $*\mathbb{P} =\hat{\cH}(\mathbb{P})$ gives
\begin{align}
 \VV^{\ubar{\bm{a}}}\,*\alpha_{+\ubar{\bm{a}}} \, + \bar{\VV}^{\bar{\bm{a}}}\,*\alpha_{-\bar{\bm{a}}} + \VV_{\sfi}\, *\alpha^{\sfi} 
 = \VV^{\ubar{\bm{a}}}\,\alpha_{+\ubar{\bm{a}}} - \bar{\VV}^{\bar{\bm{a}}}\,\alpha_{-\bar{\bm{a}}} + \hh_{\sfi\sfj}\,\VV^{\sfi}\,\alpha^{\sfj} \,.
\end{align}
Since $\VV_{\sfi}$ and $\VV^{\sfi}$ are linearly independent, we have
\begin{align}
 \mathbb{P} = \VV^{\ubar{\bm{a}}}\,\alpha_{+\ubar{\bm{a}}} + \bar{\VV}^{\bar{\bm{a}}}\,\alpha_{-\bar{\bm{a}}} \qquad 
 \bigl(* \alpha_{+\ubar{\bm{a}}} = + \alpha_{+\ubar{\bm{a}}} \,,\quad 
 * \alpha_{-\bar{\bm{a}}} = - \alpha_{-\bar{\bm{a}}} \,,\quad 
 \alpha^{\sfi}=0\bigr)\,.
\label{eq:P-alpha}
\end{align}

Now, let us rewrite the action \eqref{eq:r-p-action}. 
Using the parameterization
\begin{align}
 \mathbb{P}^A = \begin{pmatrix} r^a - \pi^{ab}\,P_b + \mathsf{\kk}_{\sfi}^a\,\cA^{\sfi} \\ P_a + \tilde{\kk}_{\sfi a}\,\cA^{\sfi} \end{pmatrix} ,
\end{align}
we can rewrite Eq.\eqref{eq:P-alpha} as
\begin{align}
\begin{split}
 r^a - \pi^{ab}\,P_b + \mathsf{\kk}_{\sfi}^a \,\cA^{\sfi}&= \hat{E}^{ab}\,\kk^{\ubar{\bm{c}}}_b\,\alpha_{+\ubar{\bm{c}}} - \kk^{\bar{\bm{c}}}_b\,\hat{E}^{ba}\,\alpha_{-\bar{\bm{c}}} \,,
\\
 P_a + \tilde{\kk}_{\sfi a}\,\cA^{\sfi} &= \kk_a^{\ubar{\bm{b}}}\, \alpha_{+\ubar{\bm{b}}} + \bar{\kk}_a^{\bar{\bm{b}}}\,\alpha_{-\bar{\bm{b}}} \,.
\end{split}
\end{align}
These can be rewritten as
\begin{align}
\begin{split}
 r^a &= E^{ab}\,\kk^{\ubar{\bm{c}}}_b\,\alpha_{+\ubar{\bm{c}}} - \kk^{\bar{\bm{c}}}_b\,E^{ba}\,\alpha_{-\bar{\bm{c}}} - \kk_{\sfi}^a \,\cA^{\sfi} \,,
\\
 P_a &= \kk_a^{\ubar{\bm{b}}}\, \alpha_{+\ubar{\bm{b}}} + \bar{\kk}_a^{\bar{\bm{b}}}\,\alpha_{-\bar{\bm{b}}} - \tilde{\kk}_{\sfi a}\,\cA^{\sfi}\,,
\label{eq:r-P}
\end{split}
\end{align}
where we have used $E^{ab}=\hat{E}^{ab} + \pi^{ab}$ and have defined
\begin{align}
 \kk_{\sfi}^a \equiv \mathsf{\kk}_{\sfi}^a + \pi^{ab}\,\tilde{\kk}_{\sfi b}\,.
\end{align}
Then the action \eqref{eq:r-p-action} becomes
\begin{align}
 S = \frac{1}{4\pi\alpha'}\int_\Sigma \bigl[-2\,\bar{\kk}^{\bar{\bm{a}}}_c\,E^{cd} \kk^{\ubar{\bm{b}}}_d\,\alpha_{-\bar{\bm{a}}} \wedge \alpha_{+\ubar{\bm{b}}}
 +2\,\kk_{\sfi}^a\, \bigl(\kk^{\ubar{\bm{b}}}_a\,\alpha_{+\ubar{\bm{b}}} + \bar{\kk}^{\bar{\bm{b}}}_a\,\alpha_{-\bar{\bm{b}}}\bigr) \wedge \cA^{\sfi} 
 + \kk_{\sfi}^a\,\tilde{\kk}_{\sfj a} \,\cA^{\sfi} \wedge \cA^{\sfj} \bigr] \,,
\label{eq:r-p-action2}
\end{align}
where we have used Eq.~\eqref{eq:kk-orthogonal}. 

Using the relation
\begin{align}
\begin{split}
 r^a + * r^a &= {}+2\,\alpha_{+\ubar{\bm{c}}}\,E^{ab}\,\kk_b^{\ubar{\bm{c}}} - \kk_{\sfi}^a \,\bigl(\cA^{\sfi}+ *\cA^{\sfi}\bigr)\,,
\\
 r^a - * r^a &= {}-2\,\alpha_{-\bar{\bm{c}}}\,\bar{\kk}_b^{\bar{\bm{c}}}\,E^{ba} - \kk_{\sfi}^a \,\bigl(\cA^{\sfi}- *\cA^{\sfi}\bigr)\,,
\end{split}
\end{align}
which follows from \eqref{eq:r-P}, for an arbitrary matrix $\MM_{ab}$\,, we obtain
\begin{align}
 &\frac{1}{4\pi\alpha'}\int_\Sigma \MM_{ab}\,\bigl(r^a\wedge * r^b + r^a \wedge r^b \bigr)
 =\frac{1}{8\pi\alpha'}\int_\Sigma \MM_{ab}\,\bigl(r^a - * r^a\bigr)\wedge *\bigl(r^b + * r^b\bigr)
\nn\\
 &= - \frac{1}{2\pi\alpha'}\int_\Sigma \bar{\kk}_c^{\bar{\bm{a}}}\,\bigl(E\,\MM\,E\bigr)^{cd}\,\kk_d^{\ubar{\bm{b}}}\,\alpha_{-\bar{\bm{a}}} \wedge \alpha_{+\ubar{\bm{b}}}
\nn\\
 &\quad + \frac{1}{2\pi\alpha'}\int_\Sigma \bigl[\kk_{\sfi}^c\,(\MM\,E)_{c}{}^d\,\kk_d^{\ubar{\bm{a}}}\, \alpha_{+\ubar{\bm{a}}} + \bar{\kk}_c^{\bar{\bm{a}}}\,(E\,\MM)^c{}_d\,\kk_{\sfi}^d\,\alpha_{-\bar{\bm{a}}} \bigr] \wedge \cA^{\sfi}
\nn\\
 &\quad + \frac{1}{4\pi\alpha'}\int_\Sigma \kk_{\sfi}^a \,\MM_{ab}\, \kk_{\sfj}^b \, \bigl(\cA^{\sfi}\wedge \cA^{\sfj}+\cA^{\sfi}\wedge *\cA^{\sfj}\bigr) \,,
\end{align}
where we have used $\alpha_{\pm\bm{a}}\wedge *\cA^{\sfi}=\mp \alpha_{\pm\bm{a}}\wedge \cA^{\sfi}$ that follows from the (anti)self-duality of $\alpha_{\pm\bm{a}}$\,. 
This reproduces the action \eqref{eq:r-p-action2} if the matrix $\MM_{ab}$ satisfies the following conditions:
\begin{align}
 \bar{\kk}_b^{\bar{\bm{a}}}\,E^{bc}\,\MM_{cd} = \bar{\kk}_d^{\bar{\bm{a}}}\,,\qquad
 \MM_{ab}\,E^{bc}\,\kk_c^{\ubar{\bm{d}}} = \kk_a^{\ubar{\bm{d}}}\,,\qquad
 \kk_{\sfi}^a\,\MM_{ab}\,\kk_{\sfj}^b = \kk_{\sfi}^a\,\tilde{\kk}_{\sfj a} \,. 
\label{conds-general}
\end{align}
Then, by using such a matrix $\MM_{ab}$\,, the reduced background fields can be expressed as
\begin{align}
 \check{E}_{mn} = r_m^a\,\MM_{ab}\,r_n^b\,. 
\label{eq:E-M-simple}
\end{align}

Let us determine the explicit form of $\MM_{ab}$\,. 
It will be instructive to begin by considering a simple case where both $(\kk_{b}^{\hat{\bm{a}}}) = \bigl(\kk_{b}^{\ubar{\bm{a}}},\,\tilde{\kk}_{\sfi b}\bigr)$ and $(\bar{\kk}_{b}^{\hat{\bm{a}}}) = \bigl(\bar{\kk}_{b}^{\bar{\bm{a}}},\,\tilde{\kk}_{\sfi b}\bigr)$ span the linear subspace $\tilde{\mathfrak{g}}$\,.\footnote{If one of $(\kk_{b}^{\hat{\bm{a}}})$ and $(\bar{\kk}_{b}^{\hat{\bm{a}}})$ is non-degeneate, the other can also be made non-degenerate by choosing $\kk_{b}^{\hat{\bm{a}}} = \bar{\kk}_{b}^{\hat{\bm{a}}}$\,.} 
In this case, there is a matrix $\cR^{ab}$ (which gives a map $\tilde{\mathfrak{g}}\to \mathfrak{g}$), that realizes\footnote{If we use the operator $\hat{\mathbb{H}}$ to be given in Eq.~\eqref{eq:mathbbH-def}, $\text{Im}(\hat{\mathbb{H}}+\hat{\mathbb{H}}^2)$ is spanned by $\VV^{\ubar{\bm{a}}}$. Then the matrix $\cR^{ab}$ can be characterized as $\text{Im}(\hat{\mathbb{H}}+\hat{\mathbb{H}}^2)\oplus \mathfrak{f}=\text{span}\{T^a + T_b\,\cR^{ba}\}$, which corresponds to Eq.~(3.32) of \cite{1903.00439}.}
\begin{align}
 \VV^{\ubar{\bm{c}}A} \equiv \frac{1}{\sqrt{2}}\begin{pmatrix} \cR^{ab}\,\kk^{\ubar{\bm{c}}}_b \\ \kk^{\ubar{\bm{c}}}_a \end{pmatrix},\qquad
 \bar{\VV}^{\bar{\bm{c}}A} \equiv \frac{1}{\sqrt{2}}\begin{pmatrix} - \bar{\kk}^{\bar{\bm{c}}}_b\,\cR^{ba} \\ \bar{\kk}^{\bar{\bm{c}}}_a \end{pmatrix},\qquad
 \VV_{\sfi}{}^A = \begin{pmatrix} \cR^{ab}\,\tilde{\kk}_{\sfi b} \\ \tilde{\kk}_{\sfi a} \end{pmatrix}.
\end{align}
Recalling Eq.~\eqref{eq:V-Vbar} and the last condition of Eq.~\eqref{eq:kk-orthogonal}, this matrix $\cR^{ab}$ should satisfy
\begin{align}
 \cR^{cd}\,\kk^{\ubar{\bm{a}}}_d = \hat{E}^{cd}\,\kk^{\ubar{\bm{a}}}_d\,,\qquad 
 \kk^{\ubar{\bm{a}}}_c\,\cR^{cd} = \kk^{\ubar{\bm{a}}}_c\,\hat{E}^{cd} \,,\qquad 
 \mathsf{\kk}_{\sfi}^a=\cR^{ab}\,\tilde{\kk}_{\sfi b} = -\tilde{\kk}_{\sfi b}\,\cR^{ba} \,.
\end{align}
By denoting the inverse matrices of $\kk_{a}^{\hat{\bm{b}}}$ and $\bar{\kk}_{a}^{\hat{\bm{b}}}$ as $\kk^a_{\hat{\bm{b}}}$ and $\bar{\kk}^a_{\hat{\bm{b}}}$\,, such a matrix $\cR^{ab}$ is found as
\begin{align}
 \cR^{ab} = \mathbb{E}^{ab} - \mathsf{\kk}_{\sfi}^{a}\,\kk^{b\sfi} \qquad
 \bigl(\mathbb{E}^{ab} \equiv E^{ab} - \bar{\kk}^{a\sfi}\,\tilde{\kk}_{\sfi c}\,\hat{E}^{cd}\,\tilde{\kk}_{\sfj d}\,\kk^{b\sfj}\bigr)\,.
\label{eq:cR-1}
\end{align}
Using this, we can construct the matrix $\MM_{ab}$ as
\begin{align}
 \MM_{ab} = \bigl[(\cR+\pi)^{-1}\bigr]_{ab}\,. 
\end{align}
Indeed, the first condition of Eq.~\eqref{conds-general} can be checked as
\begin{align}
 \bar{\kk}_b^{\bar{\bm{a}}}\,E^{bc}\,\MM_{cd} 
 =\bar{\kk}_b^{\bar{\bm{a}}}\,\bigl(\hat{E}^{bc}+\pi^{bc}\bigr)\,\MM_{cd} 
 =\bar{\kk}_b^{\bar{\bm{a}}}\,\bigl(\cR^{bc}+\pi^{bc}\bigr)\,\MM_{cd} 
 = \bar{\kk}_d^{\bar{\bm{a}}}\,,
\end{align}
and the second condition is similarly shown. 
The third condition can be shown by using
\begin{align}
 \MM_{ab}\,\kk_{\sfj}^b = \MM_{ab}\,\bigl(\pi^{bc}\,\tilde{\kk}_{\sfj c}+\mathsf{\kk}_{\sfj}^b\bigr) = \MM_{ab}\,\bigl(\pi^{bc}+\cR^{bc}\bigr)\,\tilde{\kk}_{\sfj c} = \tilde{\kk}_{\sfj a} \,. 
\label{eq:MM-k}
\end{align}
Then we get a simple formula for the reduced background fields
\begin{align}
 \check{E}_{mn} = r_m^a\,\bigl[(\cR+\pi)^{-1}\bigr]_{ab}\,r_n^b \,.
\label{eq:KS-reproduced}
\end{align}
This reproduces the model by Klim\v{c}\'\i{}k and \v{S}evera \eqref{eq:orig-model}. 
The non-degeneracy of $(\kk_{b}^{\hat{\bm{a}}})$ and $(\bar{\kk}_{b}^{\hat{\bm{a}}})$ has been assumed in \cite{hep-th:9602162} (recall Eq.~\eqref{eq:KS-gauge}), and hence the equivalence between the original approach and the approach based on the gauged sigma model is shown. 
The non-degeneracy of $\langle T_{\sfi},\,\hat{\cH}(T_{\sfj})\rangle$ assumed here is relaxed in Appendix \ref{app:general-case}, and even in that general case, we obtain the same expression \eqref{eq:KS-reproduced}. 
We note that if the matrix $(\cR+\pi)^{ab}$ is not invertible, $\check{E}_{mn}$ will be singular. 
This kind of background may correspond to the non-Riemannian geometry in the sense of \cite{1307.8377,1707.03713}, where the usual supergravity fields are not well-defined. 

When $(\kk_{b}^{\hat{\bm{a}}})$ or $(\bar{\kk}_{b}^{\hat{\bm{a}}})$ does not span the subspace $\tilde{\mathfrak{g}}$, we introduce a constant $\OO(D,D)$ matrix
\begin{align}
 \bigl(\Xi^A{}_B\bigr) = \begin{pmatrix} \bm{p}^a{}_b & \bm{q}^{ab} \\ \bm{r}_{ab} & \bm{s}_a{}^b \end{pmatrix},
\end{align}
and change the parameterization \eqref{eq:V-Vbar} as
\begin{align}
\begin{split}
 \VV^{\ubar{\bm{c}}A} &= \Xi^A{}_B\,\VV'^{\ubar{\bm{c}}B} \,,\qquad
 \bar{\VV}^{\bar{\bm{c}}A} = \Xi^A{}_B\,\bar{\VV}'^{\bar{\bm{c}}B} \,,\qquad
 \VV_{\sfi}{}^A = \Xi^A{}_B\,\VV'_{\sfi}{}^B\,,
\\
 \VV'^{\ubar{\bm{c}}A} &\equiv \frac{1}{\sqrt{2}}\begin{pmatrix} \hat{E}'^{ab}\,\kk'^{\ubar{\bm{c}}}_b \\ \kk'^{\ubar{\bm{c}}}_a \end{pmatrix},\qquad
 \bar{\VV}'^{\bar{\bm{c}}A} \equiv \frac{1}{\sqrt{2}}\begin{pmatrix} - \bar{\kk}'^{\bar{\bm{c}}}_b\,\hat{E}'^{ba} \\ \bar{\kk}'^{\bar{\bm{c}}}_a \end{pmatrix},\qquad
 \VV'_{\sfi}{}^A = \begin{pmatrix} \mathsf{\kk}'^a_{\sfi} \\ \tilde{\kk}'_{\sfi a} \end{pmatrix},
\end{split}
\end{align}
such that $(\kk'^{\hat{\bm{a}}}_b) \equiv (\kk'^{\ubar{\bm{a}}}_b,\,\tilde{\kk}'_{\sfi b})$ and $(\bar{\kk}'^{\hat{\bm{a}}}_b)\equiv (\bar{\kk}'^{\bar{\bm{a}}}_b,\,\tilde{\kk}'_{\sfi b})$ span the linear subspace $\tilde{\mathfrak{g}}$. 
Then we can again find a matrix $\cR^{ab}$ that realizes
\begin{align}
 \VV'^{\ubar{\bm{c}}A} &\equiv \frac{1}{\sqrt{2}}\begin{pmatrix} \cR^{ab} \,\kk'^{\ubar{\bm{c}}}_b \\ \kk'^{\ubar{\bm{c}}}_a \end{pmatrix},\qquad
 \bar{\VV}'^{\bar{\bm{c}}A} \equiv \frac{1}{\sqrt{2}}\begin{pmatrix} - \bar{\kk}'^{\bar{\bm{c}}}_b\,\cR^{ab} \\ \bar{\kk}'^{\bar{\bm{c}}}_a \end{pmatrix},\qquad
 \VV'_{\sfi}{}^A = \begin{pmatrix} \cR^{ab}\,\tilde{\kk}'_{\sfi b} \\ \tilde{\kk}'_{\sfi a} \end{pmatrix},
\end{align}
and satisfies
\begin{align}
 \cR^{bc}\,\kk'^{\ubar{\bm{a}}}_c = \hat{E}'^{bc}\,\kk'^{\ubar{\bm{a}}}_c\,,\qquad 
 \kk'^{\ubar{\bm{a}}}_b\,\cR^{bc} = \kk'^{\ubar{\bm{a}}}_b\,\hat{E}'^{bc} \,,\qquad 
 \mathsf{\kk}'^a_{\sfi}=\cR^{ab}\,\tilde{\kk}'_{\sfi b}\,,\qquad
 \cR^{(ab)}\,\tilde{\kk}'_{\sfi b} = 0 \,.
\end{align}
Using the relation\footnote{The last equality follows from the $\OO(D,D)$ properties: $\bm{p}^{\rmT}\,\bm{r}+\bm{r}^{\rmT}\,\bm{p}=0$, $\bm{p}^{\rmT}\,\bm{s}+\bm{r}^{\rmT}\,\bm{q}=\bm{1}$, $\bm{q}^{\rmT}\,\bm{s}+\bm{s}^{\rmT}\,\bm{q}=0$\,.}
\begin{gather}
\begin{alignedat}{2}
 \kk^{\ubar{\bm{c}}}_a&=(\bm{r}\,\cR+\bm{s})_a{}^b\,\kk'^{\ubar{\bm{c}}}_b\,,\qquad&
 \bar{\kk}^{\bar{\bm{c}}}_a&=\bar{\kk}'^{\bar{\bm{c}}}_b\,(\bm{s}^{\rmT}-\cR\,\bm{r}^{\rmT})^b{}_a \,,
\\
 \mathsf{\kk}_{\sfi}^a &= (\bm{p}\,\cR +\bm{q})^{ab}\,\tilde{\kk}'_{\sfi b}\,,\qquad&
 \tilde{\kk}_{\sfi a}&= (\bm{r}\,\cR +\bm{s})_a{}^{b}\,\tilde{\kk}'_{\sfi b}\,,
\end{alignedat}
\\
 \hat{E}^{ab}= [(\bm{p}\,\hat{E}'+\bm{q})\,(\bm{r}\,\hat{E}'+\bm{s})^{-1}]^{ab} = [(\hat{E}'\,\bm{r}^{\rmT}-\bm{s}^{\rmT})^{-1}\,(\bm{q}^{\rmT}-\hat{E}'\,\bm{p}^{\rmT})]^{ab} \,,
\end{gather}
the condition \eqref{conds-general} becomes
\begin{align}
\begin{split}
 &\bar{\kk}'^{\bar{\bm{a}}}_b\, \bigl[\bm{q}^{\rmT}-\cR\,\bm{p}^{\rmT} + (\cR\,\bm{r}^{\rmT}-\bm{s}^{\rmT})\,\pi\bigr]^{bc}\,\MM_{cd} = \bar{\kk}'^{\bar{\bm{a}}}_b\,(\cR\,\bm{r}^{\rmT}-\bm{s}^{\rmT})^b{}_a\,,
\\
 &\MM_{ab}\,\bigl[\bm{p}\,\cR+\bm{q} +\pi\,(\bm{r}\,\cR+\bm{s})\bigr]^{bc}\, \kk'^{\ubar{\bm{d}}}_c = (\bm{r}\,\cR+\bm{s})_a{}^b\,\kk'^{\ubar{\bm{d}}}_b\,,
\\
 & \kk_{\sfi}^a\,\bigl\{\MM_{ab}\,[\bm{p}\,\cR+\bm{q}+\pi\,(\bm{r}\,\cR+\bm{s})]^{bc} -(\bm{r}\,\cR+\bm{s})_a{}^c\bigr\}\,\tilde{\kk}'_{\sfj c}=0 \,. 
\end{split}
\end{align}
By choosing
\begin{align}
 \MM_{ab} = \bigl\{(\bm{r}\,\cR+\bm{s})\,\bigl[\bm{p}\,\cR+\bm{q}+\pi\,(\bm{r}\,\cR+\bm{s})\bigr]^{-1}\bigr\}_{ab}\,,
\label{eq:MM-general}
\end{align}
we can readily see that the second and the third are satisfied. 
The first condition can also be checked by using the identity $(\cR\,\bm{p}^{\rmT}-\bm{q}^{\rmT})\,(\bm{r}\,\cR+\bm{s}) = (\bm{s}^{\rmT}-\cR\,\bm{r}^{\rmT})\,(\bm{p}\,\cR+\bm{q})$.
We thus obtain a general formula for the reduced background fields
\begin{align}
\begin{split}
 \check{E}_{mn} &= r_m^a\,\bigl\{(\bm{r}\,\cR+\bm{s})\,\bigl[\bm{p}\,\cR+\bm{q}+\pi\,(\bm{r}\,\cR+\bm{s})\bigr]^{-1}\bigr\}_{ab}\,r_n^b 
\\
 &= r_m^a\,\bigl\{\bigl[\bm{q}^{\rmT} - \cR\,\bm{p}^{\rmT} + (\cR\,\bm{r}^{\rmT}-\bm{s}^{\rmT})\,\pi\bigr]^{-1}\,(\cR\,\bm{r}^{\rmT}-\bm{s}^{\rmT})\bigr\}_{ab}\,r_n^b \,.
\end{split}
\label{eq:gauge-formula}
\end{align}
The previous formula \eqref{eq:KS-reproduced} is reproduced as the specific case, $\bm{p}=\bm{s}=\bm{1}$ and $\bm{q}=\bm{r}=\bm{0}$\,.

The choice of $\Xi^A{}_B$ is arbitrary as long as the matrices $(\kk'^{\hat{\bm{a}}}_{b})$ and $(\bar{\kk}'^{\hat{\bm{a}}}_{b})$ are non-degenerate. 
The matrix $\Xi^A{}_B$ is only a tool to avoid the singularity, and if we get a finite result, it does not depend on the choice of $\Xi^A{}_B$\,. 
In practice, when $(\kk^{\hat{\bm{a}}}_{b})$ and $(\bar{\kk}^{\hat{\bm{a}}}_{b})$ are degenerate, we usually find $(\mathsf{\kk}^{\hat{\bm{a}}b})$ and $(\mathsf{\bar{\kk}}^{\hat{\bm{a}}b})$ are non-degenerate. 
Then, we can choose the constant matrix as
\begin{align}
 \bigl(\Xi^A{}_B\bigr) = \begin{pmatrix} \bm{0} & \bm{1} \\ \bm{1} & \bm{0} \end{pmatrix}.
\end{align}
In this case, the formula \eqref{eq:gauge-formula} reduces to
\begin{align}
 \check{E}_{mn} = r_m^a\,\bigl\{\tilde{\cR}\,\bigl[(\bm{1}+\pi\,\tilde{\cR})\bigr]^{-1}\bigr\}_{ab}\,r_n^b = r_m^a\,\bigl\{\bigl[(\bm{1}+\tilde{\cR}\,\pi)\,\tilde{\cR}\bigr]^{-1}\bigr\}_{ab}\,r_n^b \,,
\label{eq:KSd}
\end{align}
where $\tilde{\cR}_{ab}$ is defined by
\begin{align}
 \tilde{\cR}_{ab} = \widetilde{\mathbb{E}}_{ab} - \tilde{\kk}_{\sfi a}\,\kk_{b}^{\sfi} \qquad
 \bigl[\widetilde{E}_{ab} \equiv (\hat{E}^{-1})_{ab} - \bar{\kk}_{a}^{\sfi}\,\mathsf{\kk}_{\sfi}^c \,(\hat{E}^{-1})_{cd}\,\mathsf{\kk}_{\sfj}^d \,\kk_{b}^{\sfj}\bigr]\,.
\label{eq:cR-2}
\end{align}

The original and the dual backgrounds of the Sfetsos model can be reproduced from Eqs.~\eqref{eq:KSd} and \eqref{eq:KS-reproduced}, respectively. 
All of our examples studied in section \ref{sec:examples} can also be reproduced from the simplified formula Eq.~\eqref{eq:KS-reproduced} or Eq.~\eqref{eq:KSd}. 

If we consider the consistency with Sfetsos's procedure, the formula \eqref{eq:gauge-formula} should coincides with $\check{E}_{mn}={\displaystyle\lim_{\lambda\to 0}}E_{mn}$\,. 
This shows that $\cR^{ab}$ can also be expressed as
\begin{align}
 \cR^{ab} = \lim_{\lambda\to 0} \bigl[(\bm{p}-\hat{E}\,\bm{r})^{-1}\,(\hat{E}\,\bm{s}-\bm{q})\bigr]^{ab}\,.
\end{align}
In particular, for Eqs.~\eqref{eq:cR-1} and \eqref{eq:cR-2}, we have $\cR^{ab} = {\displaystyle\lim_{\lambda\to 0}}\hat{E}^{ab}$ and $\tilde{\cR}_{ab} = {\displaystyle\lim_{\lambda\to 0}}(\hat{E}^{-1})_{ab}$\,. 

\subsection{Properties of the reduced background fields}
\label{sec:cE-prop}

So far, using the gauged action \eqref{eq:geugedDSM2}, we obtained the formulas for the reduced background fields $\check{E}_{mn}$\,. 
Here, we find several properties of the reduced background fields $\check{E}_{mn}$ by using these formulas. 
They help us to clarify the structure of dressing cosets. 

Let us parameterize the vector fields $\VV_{\sfi}{}^M$ as $(\VV_{\sfi}{}^M)=(\kk_{\sfi}^m,\,\tilde{\kk}_{\sfi m})=(\kk_{\sfi}^m,\,\hat{\kk}_{\sfi m}+B_{mn}\,\kk_{\sfi}^n)$, and introduce some generalized vector fields, $H_{\ubar{\bm{a}}}{}^M$, $\bar{H}_{\bar{\bm{a}}}{}^M$, and $\widetilde{\VV}_{\sfi}{}^M \equiv \cH^M{}_N\,\VV_{\sfi}{}^N$, as
\begin{align}
 H_{\ubar{\bm{a}}}{}^M \equiv \frac{1}{\sqrt{2}}\begin{pmatrix} h_{\ubar{\bm{a}}}^{m} \\ E_{mn}\, h_{\ubar{\bm{a}}}^n \end{pmatrix} ,\quad
 \bar{H}_{\bar{\bm{a}}}{}^M \equiv \frac{1}{\sqrt{2}}\begin{pmatrix} \bar{h}_{\bar{\bm{a}}}^m \\ -\bar{h}_{\bar{\bm{a}}}^n\,E_{nm} \end{pmatrix},\quad
 \widetilde{\VV}_{\sfi}{}^M = \begin{pmatrix} \hat{\kk}_{\sfi}^m \\
 \kk_{\sfi m} + B_{mn}\, \hat{\kk}_{\sfi}^n \end{pmatrix},
\label{eq:HHbVt}
\end{align}
where $\hat{\kk}_{\sfi}^m\equiv g^{mn}\,\hat{\kk}_{\sfi n}$\,.
For $H_{\ubar{\bm{a}}}{}^M$ and $\bar{H}_{\bar{\bm{a}}}{}^M$ to be orthogonal to $\VV_{\sfi}{}^M$, $h_{\ubar{\bm{a}}}^{m}$ and $\bar{h}_{\bar{\bm{a}}}^{m}$ should satisfy
\begin{align}
 \bar{h}_{\bar{\bm{a}}}^{m}\,\bigl(\kk_{\sfi m} - \hat{\kk}_{\sfi m}\bigr) = 0\,,\qquad
 \bigl(\kk_{\sfi m}+ \hat{\kk}_{\sfi m}\bigr)\, h_{\ubar{\bm{a}}}^{m} = 0\qquad \bigl(\kk_{\sfi m}= g_{mn}\,\kk_{\sfi}^n\bigr)\,.
\label{eq:k-w-orthogonal}
\end{align}
Then, $\{H_{\ubar{\bm{a}}}{}^M,\,\bar{H}_{\bar{\bm{a}}}{}^M,\,\VV_{\sfi}{}^M,\,\widetilde{\VV}_{\sfi}{}^M\}$ span the $2D$-dimensional generalized tangent space. 

Under the above setup, we find that the reduced background fields $\check{E}_{mn}$ satisfy
\begin{align}
 \bar{h}_{\bar{\bm{a}}}^m\,\check{E}_{mn} = \bar{h}_{\bar{\bm{a}}}^m\,E_{mn}\,, \qquad
 \check{E}_{mn}\,h_{\ubar{\bm{a}}}^n = E_{mn}\,h_{\ubar{\bm{a}}}^n\,,\qquad 
 \check{E}_{mn}\,\kk_{\sfi}^n &= \tilde{\kk}_{\sfi m}\,.
\label{eq:checkE}
\end{align}
We can show these properties, for example, by using the formula \eqref{eq:reduced-E} for $\check{E}_{mn}$\,,
\begin{align}
 \check{E}_{mn} = E_{mn} - \bigl(\kk_{\sfi m}-\hat{\kk}_{\sfi m}\bigr)\,N^{\sfi\sfj}\,\bigl(\kk_{\sfj n}+\hat{\kk}_{\sfj n}\bigr)\,.
\end{align}
The first two properties can be easily shown by using Eq.~\eqref{eq:k-w-orthogonal}.
The third property is also shown by recalling the definition of $N^{\sfi\sfj}$\,,
\begin{align}
 \check{E}_{mn}\,\kk_{\sfi}^n = E_{mn}\,\kk_{\sfi}^n - \bigl(\kk_{\sfi m}-\hat{\kk}_{\sfi m}\bigr) = B_{mn}\,\kk_{\sfi}^n + \hat{\kk}_{\sfi m} = \tilde{\kk}_{\sfi m}\,.
\end{align}
The same properties can be also shown by using the formula \eqref{eq:KS-reproduced}. 
To show the first two, we use the expressions $H_{\ubar{\bm{a}}}{}^M=H_{\ubar{\bm{a}}}{}^B\,E_B{}^M$ and $\bar{H}_{\bar{\bm{a}}}{}^M=\bar{H}_{\bar{\bm{a}}}{}^B\,E_B{}^M$ with
\begin{align}
 H_{\ubar{\bm{a}}}{}^B = \begin{pmatrix} \mathsf{h}_{\ubar{\bm{a}}}^b \\ \tilde{h}_{\ubar{\bm{a}}b} \end{pmatrix} ,\qquad
 \bar{H}_{\bar{\bm{a}}}{}^B = \begin{pmatrix} \bar{\mathsf{h}}_{\bar{\bm{a}}}^b \\ \tilde{\bar{h}}_{\bar{\bm{a}}b} \end{pmatrix} .
\end{align}
By the construction, $H_{\ubar{\bm{a}}}{}^B$ and $\bar{H}_{\bar{\bm{a}}}{}^B$ are spanned by $\VV^{\ubar{\bm{a}}B}$ and $\bar{\VV}^{\bar{\bm{a}}B}$ of Eq.~\eqref{eq:V-Vbar}, respectively. 
Then the defining properties \eqref{conds-general} of $\MM_{ab}$ give
\begin{align}
 \bigl(-\bar{\mathsf{h}}_{\bar{\bm{a}}}^b+\tilde{\bar{h}}_{\bar{\bm{a}}d}\,\pi^{db}\bigr)\,\MM_{bc} = \tilde{\bar{h}}_{\bar{\bm{a}}b}\,,\qquad
 \MM_{ab}\,\bigl(\mathsf{h}_{\ubar{\bm{a}}}^b+\pi^{bc}\,\tilde{h}_{\ubar{\bm{a}}c}\bigr) = \tilde{h}_{\ubar{\bm{a}}b}\,.
\end{align}
Using $r_m^a\,\MM_{ab}\,r^b_n=\check{E}_{mn}$\,, $e_b^m\,(\mathsf{h}_{\ubar{\bm{a}}}^b+\pi^{bc}\,\tilde{h}_{\ubar{\bm{a}}c})=\frac{1}{\sqrt{2}}\,h^m_{\ubar{\bm{a}}}$\,, and $r^b_m\,\tilde{h}_{\ubar{\bm{a}}b}=\frac{1}{\sqrt{2}}\,E_{mn}\,h^n_{\ubar{\bm{a}}}$\,, we obtain the first property. 
The second one can also shown similarly. 
The third property readily follows from $\MM_{ab}\,\kk_{\sfj}^b = \tilde{\kk}_{\sfj a}$ that is shown in Eq.~\eqref{eq:MM-k}. 

Using the above properties, we can express the generalized vector fields as%%%%%%%%%%%%%%%%%%%%%%%%%%%%%%%%%%%%%%%%%%
\footnote{This is reminiscent of non-Riemannian geometry studied in \cite{1307.8377,1707.03713}. In a special case $\hat{\kk}_{\sfi}^m=0$, we find $\widetilde{\VV}_{\sfi}{}^M = (0,\, \hh_{\sfi\sfj}\,x^{\sfj}_m)$ with $x^{\sfi}_m = \hh^{\sfi\sfj}\,\kk_{\sfi m}$ ($\hh_{\sfi\sfj}=\kk_{\sfi}^m\,g_{mn}\,\kk_{\sfj}^n$), and then $\{H_{\ubar{\bm{a}}}{}^M,\, \bar{H}_{\bar{\bm{a}}}{}^M,\,\tfrac{1}{\sqrt{2}}(\VV_{\sfi}{}^M\pm\widetilde{\VV}_{\sfi}^M)\}$ can be identified with the DFT-vielbeins for non-Riemannian geometry.}
%%%%%%%%%%%%%%%%%%%%%%%%%%%%%%%%%%%%%%%%%%
\begin{align}
 H_{\ubar{\bm{a}}}{}^M = \frac{1}{\sqrt{2}}\begin{pmatrix} h_{\ubar{\bm{a}}}^{m} \\ \check{E}_{mn}\, h_{\ubar{\bm{a}}}^n \end{pmatrix} ,\qquad
 \bar{H}_{\bar{\bm{a}}}{}^M = \frac{1}{\sqrt{2}}\begin{pmatrix} \bar{h}_{\bar{\bm{a}}}^m \\ -\bar{h}_{\bar{\bm{a}}}^n\,\check{E}_{nm} \end{pmatrix} ,\qquad
 \VV_{\sfi}{}^M = \begin{pmatrix} \kk_{\sfi}^{m} \\ \check{E}_{mn}\, \kk_{\sfi}^n \end{pmatrix}.
\label{eq:frame-check}
\end{align}
For later convenience, we shall choose $h_{\ubar{\bm{a}}}^{m}$ and $\bar{h}_{\bar{\bm{a}}}^m$ such that they satisfy
\begin{align}
 \eta_{\ubar{\bm{a}}\ubar{\bm{b}}} = h_{\ubar{\bm{a}}}^{m}\,\check{g}_{mn}\,h_{\ubar{\bm{b}}}^{n} \,,\qquad 
 \eta_{\bar{\bm{a}}\bar{\bm{b}}} = \bar{h}_{\bar{\bm{a}}}^{m}\,\check{g}_{mn}\,\bar{h}_{\bar{\bm{b}}}^{n} \,. 
\label{eq:w-vielbein}
\end{align}
We can easily find $\check{g}_{mn}\, \kk_{\sfi}^n=0$ and then $\VV_{\sfi}{}^M$ can also be expressed as
\begin{align}
 \VV_{\sfi}{}^M = \begin{pmatrix} \kk_{\sfi}^{m} \\ \check{E}_{mn}\, \kk_{\sfi}^n \end{pmatrix} = \begin{pmatrix} \kk_{\sfi}^{m} \\ -\kk_{\sfi}^n\,\check{E}_{nm} \end{pmatrix} = \begin{pmatrix} \kk_{\sfi}^{m} \\ \check{B}_{mn}\, \kk_{\sfi}^n \end{pmatrix}.
\label{eq:k-g-degenerate}
\end{align}
Then the generalized vector fields realize the relations
\begin{align}
\begin{split}
 \hat{\cH}_{AB} &= \eta_{\ubar{\bm{c}}\ubar{\bm{d}}}\, H^{\ubar{\bm{c}}}{}_A\,H^{\ubar{\bm{d}}}{}_B + \eta_{\bar{\bm{c}}\bar{\bm{d}}}\,\bar{H}^{\bar{\bm{c}}}{}_A\,\bar{H}^{\bar{\bm{d}}}{}_B + \hh^{\sfi\sfj}\,\bigl(\VV_{\sfi A}\,\VV_{\sfj B} + \widetilde{\VV}_{\sfi A}\,\widetilde{\VV}_{\sfj B}\bigr)\,,
\\
 \eta_{AB} &= \eta_{\ubar{\bm{c}}\ubar{\bm{d}}}\, H^{\ubar{\bm{c}}}{}_A\,H^{\ubar{\bm{d}}}{}_B - \eta_{\bar{\bm{c}}\bar{\bm{d}}}\,\bar{H}^{\bar{\bm{c}}}{}_A\,\bar{H}^{\bar{\bm{d}}}{}_B + \hh^{\sfi\sfj}\,\bigl(\VV_{\sfi A}\,\widetilde{\VV}_{\sfj B} + \widetilde{\VV}_{\sfi A}\,\VV_{\sfj B}\bigr)\,.
\end{split}
\label{eq:frame-param}
\end{align}
These generalized vector fields play an important role in the following discussion. 

\subsection{Reduction of the generalized metric}
\label{sec:reduction-GM}

Here we discuss Sfetsos's approach in the perspective of our gauged sigma model, and extend the singular limit to the case where $F$ is an arbitrary isotropic subgroup of $\cD$. 
This extension results in a new approach to compute the reduced background fields $\check{E}_{mn}$\,. 

\subsubsection{Sfetsos's singular limit}
\label{sec:Sfetsos-limit}

Here we review the details of the singular limit proposed by Sfetsos \cite{hep-th:9904188}. 
For this purpose, we shall consider the standard coset, where $(\VV_{\sfi}{}^A)=(\kk_{\sfi}^a,\,\tilde{\kk}_{\sfi a}) =(\delta_{\sfi}^a,\,0)$. 
We can compute $\widetilde{\VV}_{\sfi} \equiv \hat{\cH}(\VV_{\sfi})$ as $(\widetilde{\VV}_{\sfi}{}^A) = \bigl( \hat{\beta}^{ab} \,\hat{G}_{b\sfi},\, \hat{G}_{a\sfi} \bigr)$. 
We also introduce
\begin{align}
 (\VV^{\ubar{\bm{a}} B}) = \frac{1}{\sqrt{2}} \begin{pmatrix} \hat{E}^{bc}\,\delta_c^{\ubar{\bm{a}}}\\ \delta_{b}^{\ubar{\bm{a}}}\end{pmatrix} ,\qquad
 (\bar{\VV}^{\bar{\bm{a}} B}) = \frac{1}{\sqrt{2}} \begin{pmatrix} -\delta^{\bar{\bm{a}}}_c\,\hat{E}^{cb}\\ \delta^{\bar{\bm{a}}}_b \end{pmatrix},
\label{eq:VV-coset}
\end{align}
and express the matrix $\hat{\cH}_{AB}$ as
\begin{align}
 \hat{\cH}_{AB} = \hh_{\ubar{\bm{c}}\ubar{\bm{d}}}\, \VV^{\ubar{\bm{c}}}{}_A\,\VV^{\ubar{\bm{d}}}{}_B + \hh_{\bar{\bm{c}}\bar{\bm{d}}}\,\bar{\VV}^{\bar{\bm{c}}}{}_A\,\bar{\VV}^{\bar{\bm{d}}}{}_B + \hh^{\sfi\sfj}\,\bigl(\VV_{\sfi A}\,\VV_{\sfj B} + \widetilde{\VV}_{\sfi A}\,\widetilde{\VV}_{\sfj B}\bigr)\,,
\label{eq:cH-expand}
\end{align}
where $(\hh_{\ubar{\bm{a}}\ubar{\bm{b}}})=(\hh_{\bar{\bm{a}}\bar{\bm{b}}})$ and $(\hh^{\sfi\sfj})$ are inverse of $(\hat{G}^{\check{a}\check{b}})$ ($\check{a},\check{b}=1,\dotsc,d$) and $(\hat{G}_{\sfi\sfj})$, respectively. 
Then, Sfetsos's limit is defined by a replacement
\begin{align}
 (\hat{E}^{ab}) = \begin{pmatrix} \hat{E}^{\check{a}\check{b}} & \hat{E}^{\check{a}\sfj} \\ \hat{E}^{\sfi\check{b}} & \hat{E}^{\sfi\sfj}\end{pmatrix} \to \begin{pmatrix} \hat{E}^{\check{a}\check{b}} & \hat{E}^{\check{a}\sfj} \\ \hat{E}^{\sfi\check{b}} & \lambda^{-1}\,\hat{E}^{\sfi\sfj}\end{pmatrix}, 
\label{eq:sfetsos-limit}
\end{align}
and the limit $\lambda\to 0$\,. 
As we can see from the formula \eqref{eq:formula}, under this limit, we have
\begin{align}
 (\hat{E}_{ab}) = \begin{pmatrix} \hat{E}_{\check{a}\check{b}} & \hat{E}_{\check{a}\sfj} \\ \hat{E}_{\sfi\check{b}} & \hat{E}_{\sfi\sfj} \end{pmatrix} = \begin{pmatrix} \hat{E}_{\check{a}\check{b}} & 0 \\ 0 & 0 \end{pmatrix} + \cO(\lambda) \,,
\end{align}
and then we find
\begin{align}
 \widetilde{\VV}_{\sfk}{}^A &= \bigl( \widetilde{\VV}_{\sfk}{}^{\check{a}}, \,\widetilde{\VV}_{\sfk}{}^{\sfi} , \, \widetilde{\VV}_{\sfk \check{a}} , \, \widetilde{\VV}_{\sfk\sfi} \bigr) 
 = \bigl( 0, \,\hat{\beta}^{\sfi\sfj} \,\hat{G}_{\sfj\sfk} , \, 0 , \, 0 \bigr) + \cO(\lambda)
\nn\\
 &= -\hat{G}_{\sfk\sfj}\,\hat{\beta}^{\sfj\sfi} \,\VV_{\sfi}{}^A + \cO(\lambda)\,.
\label{eq:Vt-limit}
\end{align}
The $\OO(D,D)$ matrix $\hat{\cH}_{AB}$ becomes
\begin{align}
 \hat{\cH}_{AB} = \hh_{\ubar{\bm{c}}\ubar{\bm{d}}}\, \VV^{\ubar{\bm{c}}}{}_A\,\VV^{\ubar{\bm{d}}}{}_B + \hh_{\bar{\bm{c}}\bar{\bm{d}}}\,\bar{\VV}^{\bar{\bm{c}}}{}_A\,\bar{\VV}^{\bar{\bm{d}}}{}_B + \lambda^{-1}\,\bigl(\hh^{\sfi\sfj}-\hat{\beta}^{\sfi\sfk}\,\hat{G}_{\sfk\sfl}\,\hat{\beta}^{\sfl\sfj}\bigr)\, \VV_{\sfi A}\,\VV_{\sfj B} +\cO(\lambda)\,,
\end{align}
and diverges under the limit. 
Then one might wonder why such a singular limit would yield the non-singular background fields $\check{E}_{mn}$\,, which we shall explain below. 

Under the equations of motion \eqref{eq:EOM-summary}, $\mathbb{P} \in \mathfrak{f}^\perp$ satisfies the self-duality relation $* \mathbb{P} = \hat{\cH}(\mathbb{P})$. 
If we decompose the generators of $\mathfrak{f}^\perp$ as $\{T_{\bcheck{A}},\,T_{\sfi}\}$ and define an operator $\hat{\mathbb{H}}:\mathfrak{f}^\perp \to \mathfrak{f}^\perp$ as
\begin{align}
 \hat{\mathbb{H}}(T_{\ubar{\bm{a}}}) = \hat{\cH}(T_{\ubar{\bm{a}}})=+T_{\ubar{\bm{a}}}\,,\qquad 
 \hat{\mathbb{H}}(T_{\bar{\bm{a}}}) = \hat{\cH}(T_{\bar{\bm{a}}})=-T_{\bar{\bm{a}}}\,,\qquad 
 \hat{\mathbb{H}}(T_{\sfi}) = 0\,,
\label{eq:bbH-def-prop}
\end{align}
by recalling Eq.~\eqref{eq:P-alpha}, the self-duality relation can also be expressed as
\begin{align}
 * \mathbb{P} = \hat{\mathbb{H}}(\mathbb{P})\,. 
\label{eq:self-duality-bbH}
\end{align}
The merit of this rewriting is that, unlike the divergent operator $\hat{\cH}$, the operator $\hat{\mathbb{H}}$ can be made non-singular under the limit $\lambda\to 0$\,. 
Indeed, if we define $\hat{\mathbb{H}}_{AB} \equiv \langle T_A,\,\hat{\mathbb{H}}(T_B)\rangle$ as
\begin{align}
 \hat{\mathbb{H}}_{AB} \equiv \hh_{\ubar{\bm{c}}\ubar{\bm{d}}}\, \VV^{\ubar{\bm{c}}}{}_A\,\VV^{\ubar{\bm{d}}}{}_B + \hh_{\bar{\bm{c}}\bar{\bm{d}}}\,\bar{\VV}^{\bar{\bm{c}}}{}_A\,\bar{\VV}^{\bar{\bm{d}}}{}_B\,,
\label{eq:bbH-def}
\end{align}
it satisfies the defining properties \eqref{eq:bbH-def-prop} and is independent of $\lambda$\,. 
Under the condition \eqref{eq:bbH-def-prop}, there is still an arbitrariness to redefine $\hat{\mathbb{H}}_{AB}$ as\footnote{In \cite{2107.05607}, the operator $\hat{\mathbb{H}}$ is extended to a map $\hat{\mathbb{H}}:\mathfrak{d}\to \mathfrak{d}$\,, and there, $c^{\sfi\sfj}$ is chosen as $c^{\sfi\sfj}=\hh^{\sfi\sfj}$.}
\begin{align}
 \hat{\mathbb{H}}_{AB}\to \hat{\mathbb{H}}_{AB} + c^{\sfi\sfj}\, \VV_{\sfi A}\,\VV_{\sfj B}\,,
\end{align}
by using an arbitrary constant symmetric matrix $c^{\sfi\sfj}$.
However, this redefinition does not have any physical consequence because $c^{\sfi\sfj}$ disappears from the self-duality relation \eqref{eq:self-duality-bbH} under the condition $\mathbb{P} \in \mathfrak{f}^\perp$. 
If we choose $c^{\sfi\sfj}$ as a divergent one $c^{\sfi\sfj}=\lambda^{-1}\,\bigl(\hh^{\sfi\sfj}-\hat{\beta}^{\sfi\sfk}\,\hat{G}_{\sfk\sfl}\,\hat{\beta}^{\sfl\sfj}\bigr)$, the matrix $\hat{\cH}_{AB}$ coincides with $\hat{\mathbb{H}}_{AB}$ after taking the singular limit $\lambda\to 0$\,, and all physical information of $\hat{\cH}_{AB}$ is contained in the finite matrix $\hat{\mathbb{H}}_{AB}$ given in Eq.~\eqref{eq:bbH-def}. 
In the following, we show that the reduced background fields $\check{E}_{\check{m}\check{n}}$ can be reproduced from the finite matrix $\hat{\mathbb{H}}_{AB}$\,. 

\subsubsection{Reduced background fields on the standard coset}
\label{sec:Sfetsos-revisted}

Here, in the simple case of the standard coset, we explain how to compute the reduced background fields $\check{E}_{\check{m}\check{n}}$ from $\hat{\mathbb{H}}_{AB}$\,. 
Before taking the singular limit, using a parameterization $g=\Exp{x^{\dot{m}}\,\delta_{\dot{m}}^{\sfi}\,T_{\sfi}}\Exp{x^{\check{m}}\,\delta_{\check{m}}^{\bm{a}}\,T_{\bm{a}}}$\,, the right-invariant vector fields/1-form fields take the form
\begin{align}
 (e_a^m)= \begin{pmatrix} e_{\bm{a}}^{\check{m}} & e_{\bm{a}}^{\dot{m}} \\ e_{\sfi}^{\check{m}} & e_{\sfi}^{\dot{m}}\end{pmatrix} 
 =\begin{pmatrix} e_{\bm{a}}^{\check{m}} & e_{\bm{a}}^{\dot{m}} \\ 0 & e_{\sfi}^{\dot{m}}\end{pmatrix} ,\qquad
 (r_m^a)= \begin{pmatrix} r_{\check{m}}^{\bm{a}} & r_{\check{m}}^{\sfi} \\ r_{\dot{m}}^{\bm{a}} & r_{\dot{m}}^{\sfi}\end{pmatrix} 
 =\begin{pmatrix} r_{\check{m}}^{\bm{a}} & r_{\check{m}}^{\sfi} \\ 0 & r_{\dot{m}}^{\sfi}\end{pmatrix}.
\label{eq:e-param}
\end{align}
Then we have
\begin{align}
 (\VV_{\sfi}{}^M)=(\kk_{\sfi}^{\check{m}},\, \kk_{\sfi}^{\dot{m}}\,;\, \tilde{\kk}_{\sfi \check{m}},\,\tilde{\kk}_{\sfi \dot{m}})=(0,\,e_{\sfi}^{\dot{m}}\,;\,0,\,0)\,, \qquad
 (\widetilde{\VV}_{\sfi}{}^M)=\bigl( \beta^{mn} \,G_{np}\, e_{\sfi}^p\,;\, G_{mn}\,e_{\sfi}^n \bigr) \,,
\label{eq:V-param}
\end{align}
where $G_{mn}$ is the inverse matrix of $G^{mn}=e_{a}^m\,\hat{G}^{ab}\,e_b^n$ and $\beta^{mn}\equiv e_{a}^m\,(\hat{\beta}+\pi)^{ab}\,e_b^n$\,. 
Using the generalized vector fields $\VV^{\bm{a}}{}_M\equiv \VV^{\bm{a}}{}_B\,E_M{}^B$ and $\bar{\VV}^{\bar{\bm{a}}}{}_M\equiv \bar{\VV}^{\bar{\bm{a}}}{}_B\,E_M{}^B$ associated with the vectors \eqref{eq:VV-coset},
\begin{align}
 (\VV^{\ubar{\bm{a}}}{}_M) = \frac{1}{\sqrt{2}} \begin{pmatrix}
r^{\ubar{\bm{a}}}_m\\ E^{mn}\,r^{\ubar{\bm{a}}}_n \end{pmatrix} ,\quad
 (\bar{\VV}^{\bar{\bm{a}}}{}_M)= \frac{1}{\sqrt{2}} \begin{pmatrix}
r^{\bar{\bm{a}}}_m\\ -r^{\bar{\bm{a}}}_n\,E^{nm}\end{pmatrix} \quad \bigl(E^{mn}\equiv G^{mn}+\beta^{mn}\bigr)\,,
\end{align}
we can express the generalized metric as
\begin{align}
 \cH_{MN} = \hh_{\ubar{\bm{a}}\ubar{\bm{b}}}\, \VV^{\ubar{\bm{a}}}{}_M\,\VV^{\ubar{\bm{b}}}{}_N + \hh_{\bar{\bm{a}}\bar{\bm{b}}}\,\bar{\VV}^{\bar{\bm{a}}}{}_M\,\bar{\VV}^{\bar{\bm{b}}}{}_N + \hh^{\sfi\sfj}\,\bigl(\VV_{\sfi M}\,\VV_{\sfj N} + \widetilde{\VV}_{\sfi M}\,\widetilde{\VV}_{\sfj N}\bigr)\,.
\end{align}
After taking the limit $\lambda\to 0$\,, by subtracting the irrelevant divergent part, the generalized metric $\cH_{MN}$ reduces to the degenerate generalized metric
\begin{align}
 \mathbb{H}_{MN} \equiv \,E_M{}^A\,E_N{}^B\,\hat{\mathbb{H}}_{AB} = \hh_{\ubar{\bm{a}}\ubar{\bm{b}}}\, \VV^{\ubar{\bm{a}}}{}_M\,\VV^{\ubar{\bm{b}}}{}_N + \hh_{\bar{\bm{a}}\bar{\bm{b}}}\,\bar{\VV}^{\bar{\bm{a}}}{}_M\,\bar{\VV}^{\bar{\bm{b}}}{}_N\,.
\label{eq:degenerate-GM}
\end{align}
By recalling $\mathbb{P}^A =\cZ^M\,E_M{}^A$ ($\cZ^{M}$ has been defined in Eq.~\eqref{eq:cZ-def}), the self-duality relation \eqref{eq:self-duality-bbH} can be expressed as
\begin{align}
 *\cZ^{M} = \mathbb{H}^M{}_N \,\cZ^{N}\,.
\label{eq:degenerate-self-dual}
\end{align}
Using Eqs.~\eqref{eq:e-param} and \eqref{eq:V-param}, we find that $\cZ^M$ and $\mathbb{H}^M{}_N$ take the form,
\begin{align}
 \cZ^M = \begin{pmatrix} \rmd x^{\check{m}} \\ \rmd x^{\dot{m}}+ e_{\sfi}^{\dot{m}}\,\cA^{\sfi} \\\hline P_{\check{m}} \\ P_{\dot{m}} \end{pmatrix} ,\qquad
 \mathbb{H}^M{}_N = \left(\begin{array}{cc|cc}
 \mathbb{H}^{\check{m}}{}_{\check{n}}&0&\mathbb{H}^{\check{m}\check{n}}&\mathbb{H}^{\check{m}\dot{n}} \\
 \mathbb{H}^{\dot{m}}{}_{\check{n}}&0&\mathbb{H}^{\dot{m}\check{n}}&\mathbb{H}^{\dot{m}\dot{n}} \\\hline
 \mathbb{H}_{\check{m}\check{n}}&~0~&\mathbb{H}_{\check{m}}{}^{\check{n}}&\mathbb{H}_{\check{m}}{}^{\dot{n}} \\
 0&0&0&0 
\end{array}\right).
\end{align}
We make a gauge fixing $x^{\dot{m}}=\text{const.}$, and then the fourth and the second lines of the self-duality relation give
\begin{align}
 P_{\dot{m}}=0\,,\qquad e_{\sfi}^{\dot{m}}\,*\cA^{\sfi} = \bigl(\mathbb{H}^{\dot{m}}{}_{\check{n}}\,\rmd x^{\check{n}}+\mathbb{H}^{\dot{m}\check{n}}\,P_{\check{n}}\bigr)\rvert_{x^{\dot{m}}=\text{const.}}\,.
\label{eq:eom-Sfetsos}
\end{align}
They determines the on-shell values of the auxiliary fields $P_{\dot{m}}$ and $\cA^{\sfi}$. 
The other components can be expressed as a reduced self-duality relation for the first and the third components,
\begin{align}
 *\check{Z}^{\check{M}} = \check{\cH}^{\check{M}}{}_{\check{N}} \,\check{Z}^{\check{N}} \,,\qquad
 \check{Z}^{\check{M}} \equiv \begin{pmatrix} \rmd x^{\check{m}} \\ P_{\check{m}} \end{pmatrix},\qquad
 \check{\cH}^{\check{M}}{}_{\check{N}} \equiv \begin{pmatrix}
 \mathbb{H}^{\check{m}}{}_{\check{n}}&\mathbb{H}^{\check{m}\check{n}} \\
 \mathbb{H}_{\check{m}\check{n}}&\mathbb{H}_{\check{m}}{}^{\check{n}} \end{pmatrix}\Biggr\rvert_{x^{\dot{m}}=\text{const.}} .
\label{eq:reduced-self-duality}
\end{align}
Under the first equation of Eq.~\eqref{eq:eom-Sfetsos}, the action \eqref{eq:r-p-action} reduces to
\begin{align}
 S = \frac{1}{4\pi\alpha'}\int_\Sigma \rmd x^{m} \wedge P_{m} = \frac{1}{4\pi\alpha'}\int_\Sigma \rmd x^{\check{m}} \wedge P_{\check{m}} \,,
\label{eq:string-action-d}
\end{align}
where $(\rmd x^{\check{m}},\, P_{\check{m}})=\check{Z}^{\check{M}}$ satisfies the self-duality relation \eqref{eq:reduced-self-duality}. 
Then, Eq.~\eqref{eq:string-action-d} is precisely the string action in a $d$-dimensional target space with the reduced generalized metric $\check{\cH}_{\check{M}\check{N}}$\,. 
Then, recalling the definition \eqref{eq:degenerate-GM} of $\mathbb{H}_{MN}$, we can express $\check{\cH}_{\check{M}\check{N}}$ as
\begin{align}
 \check{\cH}_{\check{M}\check{N}} = \hh_{\ubar{\bm{c}}\ubar{\bm{d}}}\,\check{\VV}^{\ubar{\bm{c}}}{}_{\check{M}}\,\check{\VV}^{\ubar{\bm{d}}}{}_{\check{N}} + \hh_{\bar{\bm{c}}\bar{\bm{d}}}\,\check{\bar{\VV}}^{\bar{\bm{c}}}{}_{\check{M}}\,\check{\bar{\VV}}^{\bar{\bm{d}}}{}_{\check{N}}\,, 
\label{eq:Odd-reduced}
\end{align}
where $\check{\VV}^{\ubar{\bm{a}}}{}_{\check{M}}$ and $\check{\bar{\VV}}^{\bar{\bm{a}}}{}_{\check{M}}$ can be expressed as follows by using Eq.~\eqref{eq:e-param}:
\begin{align}
 \check{\VV}^{\ubar{\bm{a}}}{}_{\check{M}}\equiv \frac{1}{\sqrt{2}} \begin{pmatrix}
r^{\ubar{\bm{a}}}_{\check{m}}\\ E^{\check{m}\check{n}}\,r^{\ubar{\bm{a}}}_{\check{n}}\end{pmatrix} ,\qquad
 \check{\bar{\VV}}^{\bar{\bm{a}}}{}_{\check{M}}\equiv \frac{1}{\sqrt{2}} \begin{pmatrix} r^{\bar{\bm{a}}}_{\check{m}}\\ -r^{\bar{\bm{a}}}_{\check{n}}\,E^{\check{n}\check{m}}\end{pmatrix} .
\end{align}
This expression shows that $\check{\cH}_{\check{M}\check{N}}$ is an $\OO(d,d)$ matrix, and the reduced background fields $\check{E}_{\check{m}\check{n}}$ are found as the inverse matrix of
\begin{align}
 E^{\check{m}\check{n}}=e^{\check{m}}_a\,(\hat{E}+\pi)^{ab}\,e_b^{\check{n}}=e^{\check{m}}_{\check{a}}\,(\hat{E}+\pi)^{\check{a}\check{b}}\,e_{\check{b}}^{\check{n}}\,.
\end{align}
In this way, the result \eqref{eq:sfetsos-E-1} by Sfetsos is exactly reproduced. 

To summarize, by considering a special case of the standard coset, we have demonstrated that the reduced background fields $\check{E}_{\check{m}\check{n}}$ can be found by truncating the degenerate generalized metric $\mathbb{H}_{MN}$ defined in Eq.~\eqref{eq:degenerate-GM}. 
In the following, considering a general gauge group $F$, we explain how to take the Sfetsos limit, and how to construct the degenerate generalized metric $\mathbb{H}_{MN}$. 
We also explain how to reduce $\mathbb{H}_{MN}$ to the reduced generalized metric $\check{\cH}_{\check{M}\check{N}}$\,. 

\subsubsection{Generalization of the Sfetsos limit}
\label{sec:g-Sfetsos}

Similar to Eq.~\eqref{eq:cH-expand}, we introduce a set of vectors that realize
\begin{align}
\begin{split}
 \eta_{AB} &= \eta_{\ubar{\bm{c}}\ubar{\bm{d}}}\,\VV^{\ubar{\bm{c}}}{}_A\,\VV^{\ubar{\bm{d}}}{}_B - \eta_{\ubar{\bm{c}}\ubar{\bm{d}}}\,\bar{\VV}^{\bar{\bm{c}}}{}_A\,\bar{\VV}^{\bar{\bm{d}}}{}_B
 + \hh^{\sfi\sfj}\,\bigl(\VV_{\sfi A}\,\widetilde{\VV}_{\sfj B} + \VV_{\sfi A}\,\widetilde{\VV}_{\sfj B}\bigr)\,,
\\
 \hat{\cH}_{AB} &= \eta_{\ubar{\bm{c}}\ubar{\bm{d}}}\,\VV^{\ubar{\bm{c}}}{}_A\,\VV^{\ubar{\bm{d}}}{}_B + \eta_{\bar{\bm{c}}\bar{\bm{d}}}\,\bar{\VV}^{\bar{\bm{c}}}{}_A\,\bar{\VV}^{\bar{\bm{d}}}{}_B
 + \hh^{\sfi\sfj}\,\bigl(\VV_{\sfi A}\,\VV_{\sfj B} + \widetilde{\VV}_{\sfi A}\,\widetilde{\VV}_{\sfj B}\bigr)\,.
\end{split}
\end{align}
Unlike the previous subsections, we here consider general vectors $\VV_{\sfi}{}^A$ satisfying $\langle \VV_{\sfi},\,\VV_{\sfj}\rangle =0$\,.
In addition, for simplicity, we have normalized $\VV^{\ubar{\bm{c}}}{}_A$ and $\bar{\VV}^{\bar{\bm{c}}}{}_A$ such that $(\hh_{\ubar{\bm{a}}\ubar{\bm{b}}})=(\hh_{\bar{\bm{a}}\bar{\bm{b}}})$ becomes the diagonal metric $(\eta_{\ubar{\bm{a}}\ubar{\bm{b}}})=(\eta_{\bar{\bm{a}}\bar{\bm{b}}})=\diag(\pm1,\dotsc,\pm1)$. 

Let us consider a rescaling
\begin{align}
 \VV^{\ubar{\bm{b}}}{}_A\to \VV^{\ubar{\bm{b}}}{}_A\,,\quad
 \bar{\VV}^{\bar{\bm{b}}}{}_A\to \bar{\VV}^{\bar{\bm{b}}}{}_A\,,\quad
 \VV_{\sfi A} \to \VV_{\sfi A} \,,\quad
 \widetilde{\VV}_{\sfi A} \to \lambda\,\widetilde{\VV}_{\sfi A} \,,\quad
 \hh_{\sfi\sfj}\to \lambda\,\hh_{\sfi\sfj}\,.
\end{align}
This does not change the $\OO(D,D)$ metric $\eta_{AB}$ and introduce the parameter $\lambda$ into $\hat{\cH}_{AB}$\,. 
Under the limit $\lambda\to 0$, the metric $\hat{\cH}_{AB}$ is reduced to the degenerate one
\begin{align}
 \hat{\mathbb{H}}_{AB} &= \eta_{\ubar{\bm{c}}\ubar{\bm{d}}}\,\VV^{\ubar{\bm{c}}}{}_A\,\VV^{\ubar{\bm{d}}}{}_B + \eta_{\bar{\bm{c}}\bar{\bm{d}}}\,\bar{\VV}^{\bar{\bm{c}}}{}_A\,\bar{\VV}^{\bar{\bm{d}}}{}_B\,,
\end{align}
up to the divergent term $\hh^{\sfi\sfj}\,\VV_{\sfi A}\,\VV_{\sfj B}$\,, and this plays the same role as the Sfetsos limit. 
However, this is not the unique way to introduce $\lambda$\,. 
The rescaling $\widetilde{\VV}_{\sfi A} \to \lambda\,\widetilde{\VV}_{\sfi A}$ is somewhat specific, and here we suppose $\widetilde{\VV}_{\sfi}$ to have a more general form,
\begin{align}
 \widetilde{\VV}_{\sfi} = \sfK_{\sfi} + \tilde{\sfK}_{\sfi} \,, \qquad
 \sfK_{\sfi} = \zeta_{\sfi}{}^{\sfj}\,\VV_{\sfj}\,,\qquad \tilde{\sfK}_{\sfi}=\cO(\lambda)\,,
\label{eq:VVt-decomp}
\end{align}
where $\zeta_{\sfi}{}^{\sfj}$ are constants of $\cO(\lambda^0)$. 
We find that $\tilde{\sfK}_{\sfi}$ satisfies
\begin{align}
 \langle \tilde{\sfK}_{\sfi},\,\VV_{\sfj}\rangle=\hh_{\sfi\sfj}\,,\qquad
 \langle \tilde{\sfK}_{\sfi},\,\VV^{\ubar{\bm{a}}}\rangle=0\,,\qquad
 \langle \tilde{\sfK}_{\sfi},\,\bar{\VV}^{\bar{\bm{a}}}\rangle=0\,,
\end{align}
and then we have $\hh_{\sfi\sfj}=\cO(\lambda)$.
Under the limit $\lambda\to 0$\,, we find
\begin{align}
 \hat{\cH}_{AB} = \eta_{\ubar{\bm{c}}\ubar{\bm{d}}}\,\VV^{\ubar{\bm{c}}}{}_A\,\VV^{\ubar{\bm{d}}}{}_B + \eta_{\bar{\bm{c}}\bar{\bm{d}}}\,\bar{\VV}^{\bar{\bm{c}}}{}_A\,\bar{\VV}^{\bar{\bm{d}}}{}_B
 + \bigl(\hh^{\sfi\sfj}+\hh^{\sfk\sfl}\,\zeta_{\sfk}{}^{\sfi}\,\zeta_{\sfl}^{\sfj}\bigr)\, \VV_{\sfi A}\,\VV_{\sfj B} + \cO(\lambda)\,,
\end{align}
and again this can be reduced to $\mathbb{H}_{AB}$\,. 
In section \ref{sec:DFT-EOM}, we discuss the DFT equations of motion by using this generalized Sfetsos limit that can contain the non-vanishing $\zeta_{\sfi}{}^{\sfj}$\,. 

The constants $\zeta_{\sfi}{}^{\sfj}$ do not affect the reduced background fields and may be chosen freely. 
For example, in the standard coset, Eq.~\eqref{eq:Vt-limit} shows that we have $\zeta_{\sfi}{}^{\sfj}= -\hat{G}_{\sfk\sfj}\,\hat{\beta}^{\sfj\sfi}$\,. 
However, if we consider a non-standard scaling
\begin{align}
 (\hat{E}^{ab}) = \begin{pmatrix} \hat{E}^{\bm{a}\bm{b}} & \hat{E}^{\bm{a}\sfj} \\ \hat{E}^{\sfi\bm{b}} & \hat{E}^{\sfi\sfj}\end{pmatrix} \to \begin{pmatrix} \hat{E}^{\bm{a}\bm{b}} & \hat{E}^{\bm{a}\sfj} \\ \hat{E}^{\sfi\bm{b}} & \lambda^{-1}\,\hat{G}^{\sfi\sfj} + \hat{\beta}^{\sfi\sfj}\end{pmatrix}, 
\end{align}
we find $\zeta_{\sfi}{}^{\sfj}=0$\,. 
This limit also gives the same result as Eq.~\eqref{eq:E-limit}, and the change in the way to introduce $\lambda$ is not important. 
As long as $\lambda$ does not enter in $\{\VV^{\ubar{\bm{b}}}{}_A,\, \bar{\VV}^{\bar{\bm{b}}}{}_A,\, \VV_{\sfi A}\}$ and the matrix $\hh_{\sfi\sfj}$ is degenerate after the limit, we will obtain the same reduced background fields. 

Under the PL $T$-plurality, $\widetilde{\VV}_{\sfi}{}^A$ and $\VV_{\sfi}{}^A$ are transformed as
\begin{align}
 \widetilde{\VV}_{\sfi}{}^A \to \widetilde{\VV}_{\sfi}{}^B\,(C^{-1})_B{}^A\,,\qquad
 \VV_{\sfi}{}^A \to \VV_{\sfi}{}^B\,(C^{-1})_B{}^A\,,
\end{align}
and the constants $\zeta_{\sfi}{}^{\sfj}$\,, $\hh_{\sfi\sfj}=\langle \widetilde{\VV}_{\sfi},\,\VV_{\sfj}\rangle$\,, and the structure constants $f_{\sfi\sfj}{}^{\sfk}$ are all duality invariant. 
Therefore, if $\zeta_{\sfi}{}^{\sfj}=0$ in the initial duality frame, it vanishes in all of the duality frames connected by the PL $T$-plurality (which is the case in the examples given in section \ref{sec:examples}). 

Let us also comment on additional properties of $\sfK_{\sfi}$ and $\tilde{\sfK}_{\sfi}$\,. 
Thanks to $\gLie_{\VV_{\sfi}}\cH_{MN}=0$\,, $\widetilde{\VV}_{\sfi}{}^M=\cH^M{}_N\,\VV_{\sfi}{}^N$ satisfies $[\VV_{\sfi},\,\widetilde{\VV}_{\sfj}]_{\text{D}} = - f_{\sfi\sfj}{}^{\sfk}\,\widetilde{\VV}_{\sfk}$\,.
This can be decomposed as
\begin{align}
 [\VV_{\sfi},\,\sfK_{\sfj}]_{\text{D}}=-f_{\sfi\sfj}{}^{\sfk}\,\sfK_{\sfk}\,,\qquad
 [\VV_{\sfi},\,\tilde{\sfK}_{\sfj}]_{\text{D}}=-f_{\sfi\sfj}{}^{\sfk}\,\tilde{\sfK}_{\sfk}\,.
\label{eq:V-sfU-tildeU}
\end{align}
By recalling $\sfK_{\sfi} = \zeta_{\sfi}{}^{\sfj}\,\VV_{\sfj}$\,, the first relation gives a non-trivial relation $\zeta_{\sfj}{}^{\sfl}\,f_{\sfi\sfl}{}^{\sfk}=f_{\sfi\sfj}{}^{\sfl}\,\zeta_{\sfl}{}^{\sfk}$\,. 

\subsubsection{Degenerate generalized metric}
\label{sec:dGM}

For later convenience, let us introduce an $n$-dimensional subbundle $\mathbb{F}$ spanned at each point by the generalized vector fields $\{\VV_{\sfi}{}^M\}$ and an $(2d+n)$-dimensional subbundle $\mathbb{F}^\perp$ spanned at each point by $\{\VV_{\bcheck{A}}{}^M,\,\VV_{\sfi}{}^M\}$\,. 
Sections of each bundle are denoted as $\Gamma$ as usual, and in particular, $\mathfrak{f}$-invariant sections (i.e., sections satisfying $\gLie_{\VV_{\sfi}}=0$) are denoted as $\Gamma^{\mathfrak{f}}$\,. 
For example, a section $\cK\in \Gamma^{\mathfrak{f}}(\mathbb{F})$ can be expressed as $\cK=\cK^{\sfi}\,\VV_{\sfi}$ and satisfies $\gLie_{\VV_{\sfi}}\cK=0$\,. 

We also introduce the operator $\hat{\mathbb{H}}:\mathfrak{f}^\perp\to \mathfrak{f}^\perp$ in the same way as Eq.~\eqref{eq:bbH-def-prop}, i.e.,
\begin{align}
 \hat{\mathbb{H}}(T_{\ubar{\bm{a}}}) = \hat{\cH}(T_{\ubar{\bm{a}}})=+T_{\ubar{\bm{a}}}\,,\qquad 
 \hat{\mathbb{H}}(T_{\bar{\bm{a}}}) = \hat{\cH}(T_{\bar{\bm{a}}})=-T_{\bar{\bm{a}}}\,,\qquad 
 \hat{\mathbb{H}}(T_{\sfi}) = 0\,.
\label{eq:mathbbH}
\end{align}
This operator satisfies
\begin{align}
 \hat{\mathbb{H}}^2(T_{\bcheck{A}}) =T_{\bcheck{A}}\,, \qquad
 \hat{\mathbb{H}}_{AB}\equiv \langle T_{\bcheck{A}},\,\hat{\mathbb{H}}(T_{\bcheck{B}}) \rangle
 =\langle \hat{\mathbb{H}}(T_{\bcheck{A}}),\,T_{\bcheck{B}} \rangle\,,
\end{align}
and thus the matrix $\hat{\mathbb{H}}_{\bcheck{A}\bcheck{B}}$ is a symmetric $\OO(d,d)$ matrix. 
This matrix can be constructed as
\begin{align}
 \hat{\mathbb{H}}_{AB} = \eta_{\ubar{\bm{c}}\ubar{\bm{d}}}\,\VV^{\ubar{\bm{c}}}{}_A\,\VV^{\ubar{\bm{d}}}{}_B + \eta_{\bar{\bm{c}}\bar{\bm{d}}}\,\bar{\VV}^{\bar{\bm{c}}}{}_A\,\bar{\VV}^{\bar{\bm{d}}}{}_B\,.
\label{eq:mathbbH-def}
\end{align}
Using this matrix, we define the degenerate generalized metric as
\begin{align}
 \mathbb{H}_{MN} \equiv E_M{}^A\,E_N{}^B\,\hat{\mathbb{H}}_{AB}
 = \eta_{\ubar{\bm{a}}\ubar{\bm{b}}}\,\VV^{\ubar{\bm{a}}}{}_M\,\VV^{\ubar{\bm{b}}}{}_N + \eta_{\bar{\bm{a}}\bar{\bm{b}}}\,\bar{\VV}^{\bar{\bm{a}}}{}_M\,\bar{\VV}^{\bar{\bm{b}}}{}_N\,.
\end{align}
In the case of standard cosets, this was easily reduced to the reduced generalized metric $\check{\cH}_{\check{M}\check{N}}$ (recall Eq.~\eqref{eq:Odd-reduced}), but in general, the reduction procedure is not straightforward. 

In fact, in general, it is useful to consider a redefinition of $\{\VV^{\ubar{\bm{a}}}{}_M,\,\bar{\VV}^{\bar{\bm{a}}}{}_M\}$ such that $\mathbb{H}_{MN}$ becomes manifestly gauge invariant. 
Namely, we introduce $\mathfrak{f}$-invariant generalized vector fields $\{H_{\bcheck{A}}{}^M\}\equiv \{H_{\ubar{\bm{a}}}{}^M,\,\bar{H}_{\bar{\bm{a}}}{}^M\}$, which are spanned at each point by $\{\VV_{\bcheck{A}}{}^M\}$ and satisfy
\begin{align}
 \gLie_{\VV_{\sfi}}H_{\ubar{\bm{a}}}{}^M = 0\,,\qquad 
 \gLie_{\VV_{\sfi}}\bar{H}_{\bar{\bm{a}}}{}^M = 0\,,
\label{eq:Lie-K-0}
\end{align}
and reconstruct the degenerate generalized metric as
\begin{align}
 \mathbb{H}_{MN} = \eta^{\ubar{\bm{a}}\ubar{\bm{b}}}\,H_{\ubar{\bm{a}} M}\,H_{\ubar{\bm{b}} N} + \eta^{\bar{\bm{a}}\bar{\bm{b}}}\,\bar{H}_{\bar{\bm{a}} M}\,\bar{H}_{\bar{\bm{b}} N}\,.
\label{eq:H-KK}
\end{align}
In the following, we show that this rewriting is always possible under a mild assumption. 

The assumption is that the gauge group $F$ acts on the $D$-dimensional space $M$ freely and transitively. 
Under this assumption, the $D$-dimensional space $M$ can be decomposed into a product of a $d$-dimensional surface $\cN$ and the gauge group $F$. 
Accordingly, we introduce the coordinates on $\cN$ as $(x^{\check{m}})$ and also introduce $(x^{\dot{m}})$ that parameterize the group action of $F$. 
In this adapted coordinate system, the generators of the group action take the form
\begin{align}
 \kk_{\sfi}^m = \begin{pmatrix} \kk_{\sfi}^{\check{m}} \\ \kk_{\sfi}^{\dot{m}} \end{pmatrix} = \begin{pmatrix} 0 \\ \kk_{\sfi}^{\dot{m}} \end{pmatrix},
\label{eq:kk-good-coordinates}
\end{align}
where the $d\times d$ matrix $(\kk_{\sfi}^{\dot{m}})$ should be non-degenerate for the group action to be transitive. 
The gauge-fixing condition can be chosen as $x^{\dot{m}}=\text{const.}$ 
This gauge fixing reduces the $D$-dimensional space $M$ to the $d$-dimensional surface $\cN$, which plays the role of the dressing coset $\check{M}$ discussed in the previous sections. 

Let us recall the generalized vector fields $\{H_{\ubar{\bm{a}}}{}^M,\,\bar{H}_{\bar{\bm{a}}}{}^M\}$ that are defined in Eq.~\eqref{eq:frame-check} and satisfy Eq.~\eqref{eq:H-KK}. 
Thanks to Eq.~\eqref{eq:kk-good-coordinates}, the property $\check{g}_{mn}\,\kk_{\sfi}^n = 0$ (recall Eq.~\eqref{eq:k-g-degenerate}) gives
\begin{align}
 \check{g}_{mn} = \begin{pmatrix} \check{g}_{\check{m}\check{n}} & \check{g}_{\check{m}\dot{n}} \\ \check{g}_{\dot{m}\check{n}} & \check{g}_{\dot{m}\dot{n}} \end{pmatrix} = \begin{pmatrix} \check{g}_{\check{m}\check{n}} & 0 \\ 0 & 0 \end{pmatrix}. 
\end{align}
Moreover, the Killing equation $\Lie_{\kk_{\sfi}}\check{g}_{\check{m}\check{n}}=\kk_{\sfi}\cdot\check{g}_{\check{m}\check{n}} + 2\,\partial_{(\check{m}}k^{\check{p}}\,\check{g}_{\check{n})\check{p}}=\kk_{\sfi}\cdot\check{g}_{\check{m}\check{n}}=0$ shows that $\check{g}_{\check{m}\check{n}}$ depends only on $x^{\check{m}}$. 
Then, parameterizing $h_{\ubar{\bm{a}}}^m$ and $\bar{h}_{\bar{\bm{a}}}^m$ as
\begin{align}
 h_{\ubar{\bm{a}}} = \bigl(h_{\ubar{\bm{a}}}^{\check{m}},\,h_{\ubar{\bm{a}}}^{\dot{m}}\bigr)\,,\qquad
 \bar{h}_{\bar{\bm{a}}} = \bigl(\bar{h}_{\bar{\bm{a}}}^{\check{m}},\,\bar{h}_{\bar{\bm{a}}}^{\dot{m}}\bigr)\,, 
\end{align}
we find that the condition Eq.~\eqref{eq:w-vielbein} gives
\begin{align}
 \eta_{\ubar{\bm{a}}\ubar{\bm{b}}} = h_{\ubar{\bm{a}}}^{\check{m}}\,\check{g}_{\check{m}\check{n}}\,h_{\ubar{\bm{b}}}^{\check{n}} \,,\qquad 
 \eta_{\bar{\bm{a}}\bar{\bm{b}}} = \bar{h}_{\bar{\bm{a}}}^{\check{m}}\,\check{g}_{mn}\,\bar{h}_{\bar{\bm{b}}}^{\check{n}} \,. 
\end{align}
Namely, the $d\times d$ matrices $h_{\ubar{\bm{a}}}^{\check{m}}$ and $\bar{h}_{\bar{\bm{a}}}^{\check{m}}$ play the role of certain vielbeins associated with $\check{g}_{\check{m}\check{n}}$\,. 
Since $\check{g}_{\check{m}\check{n}}$ depend only on $x^{\check{m}}$, $h_{\ubar{\bm{a}}}^{\check{m}}$ and $\bar{h}_{\bar{\bm{a}}}^{\check{m}}$ also can be chosen such that they depend only on $x^{\check{m}}$. 
Then we can show the Killing equation for the $\check{m}$ components, $\Lie_{\kk_{\sfi}} h_{\ubar{\bm{a}}}^{\check{m}} = 0$ and $\Lie_{\kk_{\sfi}} \bar{h}_{\bar{\bm{a}}}^{\check{m}} = 0$, for any choices of the other components $\{h_{\ubar{\bm{a}}}^{\dot{m}},\,\bar{h}_{\bar{\bm{a}}}^{\dot{m}}\}$\,. 
The components $h_{\ubar{\bm{a}}}^{\dot{m}}$ and $\bar{h}_{\bar{\bm{a}}}^{\dot{m}}$ can be chosen arbitrarily as long as the orthogonality condition \eqref{eq:k-w-orthogonal} is satisfied. 
We choose these such that the condition \eqref{eq:k-w-orthogonal} is satisfied on the $d$-dimensional surface $\cN$. 
We then Lie transport $h_{\ubar{\bm{a}}}^m$ and $\bar{h}_{\bar{\bm{a}}}^m$ along the flow of $\kk_{\sfi}^m$, which generates the group action.
Then $h_{\ubar{\bm{a}}}^m$ and $\bar{h}_{\bar{\bm{a}}}^m$ are defined over the $D$-dimensional space. 
By the construction, they satisfy
\begin{align}
 \Lie_{\kk_{\sfi}}h_{\ubar{\bm{a}}}^m =0\,,\qquad
 \Lie_{\kk_{\sfi}}\bar{h}_{\bar{\bm{a}}}^m =0\,.
\label{eq:h-Lie}
\end{align}
Using $\Lie_{\kk_{\sfi}}\bigl(\kk_{\sfj m}\pm \hat{\kk}_{\sfj m}\bigr)=-f_{\sfi\sfj}{}^{\sfk}\,\bigl(\kk_{\sfk m}\pm \hat{\kk}_{\sfk m}\bigr)$, we can see that the orthogonality condition \eqref{eq:k-w-orthogonal} is preserved under the flow of $\kk_{\sfi}^m$.
Namely, $\{h_{\ubar{\bm{a}}}^m,\,\bar{h}_{\bar{\bm{a}}}^m\}$ satisfy the orthogonality over the entire $D$-dimensional space. 
In this way, we can construct the vector fields $h_{\ubar{\bm{a}}}^m$ and $\bar{h}_{\bar{\bm{a}}}^m$. 
Then, using Eq.~\eqref{eq:h-Lie} and the generalized Killing equation $\Lie_{\kk_{\sfi}}E_{mn} + 2\,\partial_{[m|}\tilde{\kk}_{\sfi|n]}=0$\,, we find that $H_{\ubar{\bm{a}}}{}^{M}$ defined in Eq.~\eqref{eq:HHbVt} satisfies
\begin{align}
 \gLie_{\VV_{\sfi}} H_{\ubar{\bm{a}}}{}^{M} = \frac{1}{\sqrt{2}}\begin{pmatrix} \Lie_{\kk_{\sfi}}h_{\ubar{\bm{a}}}^{\check{m}} \\ \Lie_{\kk_{\sfi}}E_{mn}\, h_{\ubar{\bm{a}}}^{\check{n}} + 2\,\partial_{[m|}\tilde{\kk}_{\sfi|n]}\,h_{\ubar{\bm{a}}}^{n} \end{pmatrix}=0\,.
\label{eq:Lie-H}
\end{align}
We can similarly check $\gLie_{\VV_{\sfi}}\bar{H}_{\bar{\bm{a}}}{}^{M}=0$\,. 
Thus, the degenerate generalized metric $\mathbb{H}_{MN}$ can be constructed by using the $\mathfrak{f}$-invariant generalized vector fields $H_{\bcheck{A}}{}^{M}$ as in Eq.~\eqref{eq:H-KK}.

\subsubsection{Reduced generalized metric}
\label{sec:reduced-GM}

In the previous subsection, we have constructed the degenerate generalized metric $\mathbb{H}_{MN}$ by using the $\mathfrak{f}$-invariant generalized vector fields $H_{\bcheck{A}}{}^M$. 
Here, we construct the reduced generalized metric $\check{\cH}_{\check{M}\check{N}}\in\OO(d,d)$ from the degenerate generalized metric $\mathbb{H}_{MN}$. 
A naive expectation is that a simple truncation of $H_{\bcheck{A}}{}^M$ will give the generalized vielbeins associated with the $2d$-dimensional reduced doubled space $\check{\cM}$\,. 
However, if we simply truncate the $(x^{\dot{m}},\, \tilde{x}_{\dot{m}})$-components of $H_{\ubar{\bm{a}}}{}^M$ and construct a $2d\times 2d$ matrix
\begin{align}
 \mathbb{H}_{\check{M}\check{N}} = \eta^{\ubar{\bm{a}}\ubar{\bm{b}}}\,H_{\ubar{\bm{a}} \check{M}}\,H_{\ubar{\bm{b}} \check{N}} + \eta^{\bar{\bm{a}}\bar{\bm{b}}}\,\bar{H}_{\bar{\bm{a}} \check{M}}\,\bar{H}_{\bar{\bm{b}} \check{N}}\,,
\end{align}
this is not necessarily an $\OO(d,d)$ matrix. 
This is because that the vector components of $H_{\ubar{\bm{a}}}{}^M$, namely $\{h_{\bcheck{A}}^m\}\equiv \{h_{\ubar{\bm{a}}}^m,\,\check{h}_{\bar{\bm{a}}}^m\}$, are not tangent vectors on the surface $\cN$. 
To construct generalized vector fields whose vector components are tangent to $\cN$, we introduce
\begin{align}
 \cK_{\ubar{\bm{a}}}{}^M\equiv -\bigl[h_{\ubar{\bm{a}}}^{\dot{n}}\,(\kk^{-1})_{\dot{n}}^{\sfi}\bigr]\,\VV_{\sfi}{}^M\,,\qquad
 \bar{\cK}_{\bar{\bm{a}}}{}^M\equiv -\bigl[\bar{h}_{\bar{\bm{a}}}^{\dot{n}}\,(\kk^{-1})_{\dot{n}}^{\sfi}\bigr]\,\VV_{\sfi}{}^M\,,
\label{eq:cK-def}
\end{align}
where we have used the invertibility of the matrix $(\kk_{\sfi}^{\dot{m}})$ (see below Eq.~\eqref{eq:kk-good-coordinates}).
We then redefine the generalized vector fields $H_{\bcheck{A}}{}^M$ as
\begin{align}
 \CE_{\ubar{\bm{a}}}{}^M \equiv H_{\ubar{\bm{a}}}{}^M + \cK_{\ubar{\bm{a}}}{}^M\,,\qquad
 \bar{\CE}_{\bar{\bm{a}}}{}^M \equiv \bar{H}_{\bar{\bm{a}}}{}^M + \bar{\cK}_{\bar{\bm{a}}}{}^M\,.
\label{eq:CE-def}
\end{align}
Using the property $\tilde{\kk}_{\sfi m}=\check{E}_{mn}\,\kk_{\sfi}^n$ given in Eq.~\eqref{eq:checkE}, $\CE_{\ubar{\bm{a}}}{}^M$ and $\bar{\CE}_{\bar{\bm{a}}}{}^M$ are found as
\begin{align}
 \CE_{\ubar{\bm{a}}}{}^M = \frac{1}{\sqrt{2}}\begin{pmatrix} \check{h}_{\ubar{\bm{a}}}^m \\ \check{E}_{mn}\,\check{h}_{\ubar{\bm{a}}}^n \end{pmatrix},\qquad
 \bar{\CE}_{\bar{\bm{a}}}{}^M = \frac{1}{\sqrt{2}}\begin{pmatrix} \check{\bar{h}}_{\bar{\bm{a}}}^m \\ -\check{\bar{h}}_{\bar{\bm{a}}}^n\,\check{E}_{nm} \end{pmatrix},
\label{eq:CE-check0}
\end{align}
where we have defined the following vector fields:
\begin{align}
 \check{h}_{\ubar{\bm{a}}}^m \equiv \begin{pmatrix} h_{\ubar{\bm{a}}}^{\check{m}} \\ 0 \end{pmatrix} ,\qquad \check{\bar{h}}_{\bar{\bm{a}}}^m\equiv \begin{pmatrix} \bar{h}_{\bar{\bm{a}}}^{\check{m}} \\ 0 \end{pmatrix}.
\end{align}
Thanks to this redefinition, the vector components of $\{\CE_{\bcheck{A}}{}^M\}\equiv \{\CE_{\ubar{\bm{a}}}{}^M,\,\bar{\CE}_{\bar{\bm{a}}}{}^M\}$ are tangential to the $d$-dimensional surface $\cN$.
Then we can simply truncate their $(x^{\dot{m}},\,\tilde{x}_{\dot{m}})$-components and obtain the reduced generalized vector fields $\{\check{\CE}_{\bcheck{A}}{}^{\check{M}}\}\equiv \{\check{\CE}_{\ubar{\bm{a}}}{}^{\check{M}},\,\bar{\check{\CE}}_{\bar{\bm{a}}}{}^{\check{M}}\}$ as
\begin{align}
 \check{\CE}_{\ubar{\bm{a}}}{}^{\check{M}} = \frac{1}{\sqrt{2}}\begin{pmatrix} h_{\ubar{\bm{a}}}^{\check{m}} \\ \check{E}_{\check{m}\check{n}}\,h_{\ubar{\bm{a}}}^{\check{n}} \end{pmatrix} ,\qquad
 \bar{\check{\CE}}_{\bar{\bm{a}}}{}^{\check{M}} = \frac{1}{\sqrt{2}}\begin{pmatrix} \bar{h}_{\bar{\bm{a}}}^{\check{m}} \\ - \bar{h}_{\bar{\bm{a}}}^{\check{n}}\,\check{E}_{\check{n}\check{m}} \end{pmatrix} .
\label{eq:CE-check}
\end{align}
They plays the role of the generalized vielbeins for the reduced generalized metric
\begin{align}
 \check{\cH}_{\check{M}\check{N}} \equiv \eta^{\ubar{\bm{a}}\ubar{\bm{b}}}\,\check{\CE}_{\ubar{\bm{a}} \check{M}}\,\check{\CE}_{\ubar{\bm{b}} \check{N}} + \eta^{\bar{\bm{a}}\bar{\bm{b}}}\,\bar{\check{\CE}}_{\bar{\bm{a}} \check{M}}\,\bar{\check{\CE}}_{\bar{\bm{b}} \check{N}}\,.
\label{eq:reduced-GM}
\end{align}
This is always an $\OO(d,d)$ matrix associated with the reduced background fields $\check{E}_{\check{m}\check{n}}$\,. 

Below, we shall show three important properties of $\CE_{\bcheck{A}}{}^M$. 
The first is that they form a closed algebra under the D-bracket. 
Since $\CE_{\bcheck{A}}{}^M$ are sections of $\mathbb{F}^\perp$, their D-brackets are also orthogonal to $\VV_{\sfi}{}^M$. 
Namely, we can expand $[\CE_{\bcheck{A}},\,\CE_{\bcheck{B}}]_{\text{D}}$ as a linear combination of $\CE_{\bcheck{A}}$ and $\VV_{\sfi}$\,. 
If we compute the D-bracket, the vector components become
\begin{align}
 [\CE_{\bcheck{A}},\,\CE_{\bcheck{B}}]^m_{\text{D}}
 = \frac{1}{2}\,[\check{h}_{\bcheck{A}},\,\check{h}_{\bcheck{B}}]^{m} 
 = \frac{1}{2}\begin{pmatrix} \Lie_{\check{h}_{\bcheck{A}}}\check{h}_{\bcheck{B}}^{\check{m}} \\ \Lie_{\check{h}_{\bcheck{A}}} \check{h}_{\bcheck{B}}^{\dot{m}} \end{pmatrix} 
 = \frac{1}{2}\begin{pmatrix} \Lie_{\check{h}_{\bcheck{A}}}\check{h}_{\bcheck{B}}^{\check{m}} \\ 0 \end{pmatrix} ,
\label{eq:HH-alg}
\end{align}
where $\{\check{h}_{\bcheck{A}}^{\check{m}}\}\equiv \{\check{h}_{\ubar{\bm{a}}}^{\check{m}},\,\check{h}_{\bar{\bm{a}}}^{\check{m}}\}$\,. 
The absence of the second row shows that we can expand $[\CE_{\bcheck{A}},\,\CE_{\bcheck{B}}]_{\text{D}}$ only by $\CE_{\bcheck{A}}$\,.
Then, using certain $\mathfrak{f}$-invariant functions $\bm{f}_{\bcheck{A}\bcheck{B}}{}^{\bcheck{C}}$, we obtain
\begin{align}
 [\CE_{\bcheck{A}},\,\CE_{\bcheck{B}}]_{\text{D}} = - \bm{f}_{\bcheck{A}\bcheck{B}}{}^{\bcheck{C}}\,\CE_{\bcheck{C}} \,.
\label{eq:H-algebra}
\end{align}
The second property is that the reduced fields $\check{\CE}_{\bcheck{A}}{}^{\check{M}}$ satisfy the same algebra as $\CE_{\bcheck{A}}{}^{M}$,
\begin{align}
 [\check{\CE}_{\bcheck{A}},\,\check{\CE}_{\bcheck{B}}]_{\text{D}}= -\bm{f}_{\bcheck{A}\bcheck{B}}{}^{\bcheck{C}}\,\check{\CE}_{\bcheck{C}}\,.
\end{align}
This is because the second component of $\CE_{\bcheck{A}}{}^M$ is absent and the D-bracket reduces to
\begin{align}
\begin{split}
 [\CE_{\bcheck{A}},\, \CE_{\bcheck{B}}]^{\check{M}}_{\text{D}} 
 &= 2\,\CE_{[\bcheck{A}}{}^N\,\partial_N \CE_{\bcheck{B}]}{}^{\check{M}} + \CE_{\bcheck{B}N}\,\partial^{\check{M}} \CE_{\bcheck{A}}{}^N
 = 2\,\CE_{[\bcheck{A}}{}^{\check{N}}\,\partial_{\check{N}} \CE_{\bcheck{B}]}{}^{\check{M}} + \CE_{\bcheck{B}{\check{N}}}\,\partial^{\check{M}} \CE_{\bcheck{A}}{}^{\check{N}}\,.
\\
 &= 2\,\check{\CE}_{[\bcheck{A}}{}^{\check{N}}\,\partial_{\check{N}} \check{\CE}_{\bcheck{B}]}{}^{\check{M}} + \check{\CE}_{\bcheck{B}{\check{N}}}\,\partial^{\check{M}} \check{\CE}_{\bcheck{A}}{}^{\check{N}}
 = [\check{\CE}_{\bcheck{A}},\, \check{\CE}_{\bcheck{B}}]^{\check{M}}_{\text{D}} \,.
\end{split}
\end{align}

To consider the third property, let us recall that $x^{\check{m}}$ are defined as the coordinates on the $d$-dimensional surface $\cN$. 
Under the group action, this surface $\cN$ is mapped to another $d$-dimensional surface $\cN'$, and at the same time, the vector fields $h_{\ubar{\bm{a}}}^m$ and $\bar{h}_{\bar{\bm{a}}}^m$ tangential to $\cN$ are mapped to those tangential to $\cN'$. 
Namely, the $\dot{m}$-components of $\check{h}_{\ubar{\bm{a}}}^m$ and $\bar{\check{h}}_{\bar{\bm{a}}}^m$ do not appear under the group action.
This can be expressed as
\begin{align}
 \Lie_{\kk_{\sfi}} \check{h}_{\ubar{\bm{a}}}^{\dot{m}}=\check{h}_{\ubar{\bm{a}}}^{\check{n}}\,\partial_{\check{n}}\kk_{\sfi}^{\dot{m}}=0\,,\qquad
 \Lie_{\kk_{\sfi}} \bar{\check{h}}_{\bar{\bm{a}}}^{\dot{m}}=\bar{\check{h}}_{\bar{\bm{a}}}^{\check{n}}\,\partial_{\check{n}}\kk_{\sfi}^{\dot{m}}=0\,.
\end{align}
Namely, we find $\partial_{\check{m}}\kk_{\sfi}^{\dot{m}}=0$\,.
Then we can rewrite $\Lie_{\kk_{\sfi}}h_{\ubar{\bm{a}}}^{\dot{m}}=0$ and $\Lie_{\kk_{\sfi}}\bar{h}_{\bar{\bm{a}}}^{\dot{m}}=0$ as
\begin{align}
 \kk_{\sfi}^{\dot{n}}\,\partial_{\dot{n}} h_{\ubar{\bm{a}}}^{\dot{m}} = h_{\ubar{\bm{a}}}^{\dot{n}}\,\partial_{\dot{n}} \kk_{\sfi}^{\dot{m}}\,,\qquad
 \kk_{\sfi}^{\dot{n}}\,\partial_{\dot{n}} \bar{h}_{\bar{\bm{a}}}^{\dot{m}} = \bar{h}_{\bar{\bm{a}}}^{\dot{n}}\,\partial_{\dot{n}} \kk_{\sfi}^{\dot{m}}\,.
\label{eq:assum}
\end{align}
The gauge algebra $[\kk_{\sfi},\, \kk_{\sfj}]^{\dot{m}} = - f_{\sfi\sfj}{}^{\sfk}\,\kk_{\sfk}^{\dot{m}}$ gives
\begin{align}
 \kk_{\sfi}^{\dot{n}}\,\partial_{\dot{n}} (\kk^{-1})_{\dot{m}}^{\sfj} = - (\kk^{-1})_{\dot{p}}^{\sfj}\,\partial_{\dot{m}} \kk_{\sfi}^{\dot{p}} + \kk_{\dot{m}}^{\sfk}\,f_{\sfi\sfk}{}^{\sfj}\,,
\end{align}
and this can be combined with Eq.~\eqref{eq:assum} as
\begin{align}
 \kk_{\sfi}^{\dot{n}}\,\partial_{\dot{n}} \bigl[h_{\ubar{\bm{a}}}^{\dot{m}}\,(\kk^{-1})_{\dot{m}}^{\sfj}\bigr] 
 = \bigl[h_{\ubar{\bm{a}}}^{\dot{m}}\,\kk_{\dot{m}}^{\sfk}\bigr]\,f_{\sfi\sfk}{}^{\sfj} \,,\qquad 
 \kk_{\sfi}^{\dot{n}}\,\partial_{\dot{n}} \bigl[\bar{h}_{\bar{\bm{a}}}^{\dot{m}}\,(\kk^{-1})_{\dot{m}}^{\sfj}\bigr] 
 = \bigl[\bar{h}_{\bar{\bm{a}}}^{\dot{m}}\,\kk_{\dot{m}}^{\sfk}\bigr]\,f_{\sfi\sfk}{}^{\sfj} \,.
\end{align}
Then we find that $\cK_{\ubar{\bm{a}}}{}^M$ and $\bar{\cK}_{\bar{\bm{a}}}{}^M$ are $\mathfrak{f}$-invariant. 
Indeed, we can show
\begin{align}
 [\VV_{\sfi},\, \cK_{\ubar{\bm{a}}}]_{\text{D}} = -\bigl[h_{\ubar{\bm{a}}}^{\dot{m}}\,(\kk^{-1})_{\dot{m}}^{\sfj}\bigr]\,f_{\sfi\sfj}{}^{\sfk}\,\VV_{\sfk} 
 + \bigl[h_{\ubar{\bm{a}}}^{\dot{m}}\,(\kk^{-1})_{\dot{m}}^{\sfj}\bigr]\,f_{\sfi\sfj}{}^{\sfk}\,\VV_{\sfk} = 0\,,
\end{align}
and similarly for $\bar{\cK}_{\bar{\bm{a}}}{}^M$. 
Recalling the definition \eqref{eq:CE-def}, we find the third property
\begin{align}
 \bigl[\VV_{\sfi},\,\CE_{\bcheck{A}}\bigr]_{\text{D}} = 0\,.
\end{align}

\subsubsection{A short summary}

Let summarize the results of section \ref{sec:reduction-GM}, where we defined various generalized vector fields. 
We have introduced a complete basis $\{K^{\ubar{\bm{a}} A},\,\bar{K}^{\bar{\bm{a}} A},\,\VV_{\sfi}{}^A,\,\widetilde{\VV}_{\sfi}{}^A\}$ and expressed $\hat{\cH}_{AB}$ as
\begin{align}
 \hat{\cH}_{AB} = \eta_{\ubar{\bm{c}}\ubar{\bm{d}}}\,\VV^{\ubar{\bm{c}}}{}_A\,\VV^{\ubar{\bm{d}}}{}_B + \eta_{\bar{\bm{c}}\bar{\bm{d}}}\,\bar{\VV}^{\bar{\bm{c}}}{}_A\,\bar{\VV}^{\bar{\bm{d}}}{}_B
 + \hh^{\sfi\sfj}\,\bigl(\VV_{\sfi A}\,\VV_{\sfj B} + \widetilde{\VV}_{\sfi A}\,\widetilde{\VV}_{\sfj B}\bigr)\,.
\end{align}
The generalized Sfetsos limit makes $\hh_{\sfi\sfj}$ degenerate and then $\widetilde{\VV}_{\sfi}$ are linear combination of $\VV_{\sfi}$\,, and then we find
\begin{align}
 \hat{\cH}_{AB} = \eta_{\ubar{\bm{c}}\ubar{\bm{d}}}\,\VV^{\ubar{\bm{c}}}{}_A\,\VV^{\ubar{\bm{d}}}{}_B + \eta_{\bar{\bm{c}}\bar{\bm{d}}}\,\bar{\VV}^{\bar{\bm{c}}}{}_A\,\bar{\VV}^{\bar{\bm{d}}}{}_B
 + \bigl(\hh^{\sfi\sfj}+\hh^{\sfk\sfl}\,\zeta_{\sfk}{}^{\sfi}\,\zeta_{\sfl}^{\sfj}\bigr)\, \VV_{\sfi A}\,\VV_{\sfj B} + \cO(\lambda)\,,
\end{align}
The term including $\VV_{\sfi A}\,\VV_{\sfj B}$ is divergent, but this term does not contribute the string equations of motion and can be ignored. 
Then we define the degenerate matrix
\begin{align}
 \hat{\mathbb{H}}_{AB} = \eta_{\ubar{\bm{c}}\ubar{\bm{d}}}\,\VV^{\ubar{\bm{c}}}{}_A\,\VV^{\ubar{\bm{d}}}{}_B + \eta_{\bar{\bm{c}}\bar{\bm{d}}}\,\bar{\VV}^{\bar{\bm{c}}}{}_A\,\bar{\VV}^{\bar{\bm{d}}}{}_B\,,
\end{align}
and construct the degenerate generalized metric $\hat{\mathbb{H}}_{MN}=E_M{}^A\,E_N{}^B\,\hat{\mathbb{H}}_{AB}$\,. 
We can always find a set of $\mathfrak{f}$-invariant generalized vector fields $H_{\bcheck{A}}{}^M$, and using these, we reconstruct the degenerate generalized metric as
\begin{align}
 \hat{\mathbb{H}}_{MN} = \eta^{\ubar{\bm{a}}\ubar{\bm{b}}}\,H_{\ubar{\bm{a}}M}\,H_{\ubar{\bm{b}}M} + \eta^{\bar{\bm{a}}\bar{\bm{b}}}\,\bar{H}_{\bar{\bm{a}}M}\,\bar{H}_{\bar{\bm{b}}N} \,.
\end{align}
We then introduce $\cK_{\hat{A}}{}^M$ and construct $\cE_{\hat{A}}{}^M=H_{\hat{A}}{}^M+\cK_{\hat{A}}{}^M\in \Gamma^{\mathfrak{f}}(\mathbb{F})$. 
Considering the gauge-fixing $x^{\dot{m}}=\text{const.}$, we truncate the $(x^{\dot{m}},\,\tilde{x}_{\dot{m}})$-components of $\cE_{\hat{A}}{}^M$, which are denoted as $\check{\cE}_{\hat{A}}{}^{\check{M}}$.
Using these, we obtain the reduced generalized metric
\begin{align}
 \check{\cH}_{\check{M}\check{N}} = \eta^{\ubar{\bm{a}}\ubar{\bm{b}}}\,\check{\CE}_{\ubar{\bm{a}} \check{M}}\,\check{\CE}_{\ubar{\bm{b}} \check{N}} + \eta^{\bar{\bm{a}}\bar{\bm{b}}}\,\bar{\check{\CE}}_{\bar{\bm{a}} \check{M}}\,\bar{\check{\CE}}_{\bar{\bm{b}} \check{N}}\,.
\end{align}
This metric gives the reduced background fields $\check{E}_{\check{m}\check{n}}$\,. 
The above steps do not rely on the gauged sigma model, and provide an independent approach to compute $\check{E}_{\check{m}\check{n}}$\,. 

This approach is not so useful as a method to compute $\check{E}_{\check{m}\check{n}}$\,, but helps us to understand how the $d$-dimensional space is embedded into the $D$-dimensional space. 
Moreover, this allows us to clearly discuss the symmetry of the DFT equations of motion under the PL $T$-plurality. 
In section \ref{sec:PL-DFT}, we consider the equations of motion for $\check{\cH}_{\check{M}\check{N}}$ in $d$-dimensions. 
This $d$-dimensional generalized metric is naturally uplifted to a $D$-dimensional metric
\begin{align}
 \bm{H}_{MN} \equiv \eta^{\ubar{\bm{a}}\ubar{\bm{b}}}\,\CE_{\ubar{\bm{a}} M}\,\CE_{\ubar{\bm{b}} N} + \eta^{\bar{\bm{a}}\bar{\bm{b}}}\,\bar{\CE}_{\bar{\bm{a}} M}\,\bar{\CE}_{\bar{\bm{b}} N}\,.
\label{eq:bmH-def}
\end{align}
By using the relations between the metric $\bm{H}_{MN}$, the degenerate generalized metric $\mathbb{H}_{MN}$, and the generalized metric $\cH_{MN}$ in $D$-dimensions, the $d$-dimensional equations of motion can be expressed by using the algebraic quantities of the Drinfel'd double. 
Then we can show that the PL $T$-plurality is a symmetry of the DFT equations of motion. 

\subsection{Comments on the gauge invariance of the action}
\label{sec:gauge-inv}

Before closing this section, let us comment on the condition \eqref{eq:H-inv-inf} for the gauge invariance. 
As we show in Appendix \ref{app:left-mult}, when the gauge symmetry is generated by $T_{\sfi}=\mathsf{\kk}_{\sfi}^a\,T_a+\tilde{\kk}_{\sfi a}\,T^a$\,, the invariance condition can also be expressed as
\begin{align}
 \Omega_{\sfi}^{ab} \equiv \mathsf{\kk}_{\sfi}^c\,\bigl(f_c{}^{ab} - f_{cd}{}^a\,\hat{E}^{db} - f_{cd}{}^b\,\hat{E}^{ad}\bigr)
 - \tilde{\kk}_{\sfi c}\,\bigl(f_{de}{}^c\,\hat{E}^{ad}\,\hat{E}^{eb} - f_d{}^{ca}\,\hat{E}^{db} - f_d{}^{cb}\,\hat{E}^{ad} \bigr) = 0\,.
\end{align}
For example, if we consider the case of the standard coset $(\VV_{\sfi}{}^A)=(\delta_{\sfi}^a,\,0)$, this becomes
\begin{align}
 \Omega_{\sfi}^{ab} = f_{\sfi}{}^{ab} - f_{\sfi c}{}^a\,\hat{E}^{cb} - f_{\sfi c}{}^b\,\hat{E}^{ac} = 0\,,
\label{eq:inv-orig}
\end{align}
which corresponds to Eq.~(2.16) of \cite{hep-th:9904188}. 
If we consider its dual, $(\VV_{\sfi}{}^A)=(0,\,\delta_{\sfi a})$, we have
\begin{align}
 \Omega_{\sfi}^{ab} = - \delta_{\sfi c}\,\bigl(f_{de}{}^c\,\hat{E}^{ad}\,\hat{E}^{eb} - f_d{}^{ca}\,\hat{E}^{db} - f_d{}^{cb}\,\hat{E}^{ad} \bigr) = 0\,.
\end{align}
Similarly, if we consider $\VV_{\sfi}=(\hat{E}^{ab}\,\delta_{\sfi b},\,\delta_{\sfi a})$, we find
\begin{align}
\begin{split}
 \Omega_{\sfi}^{ab} = \delta_{\sfi c}\,\bigl(\hat{E}^{db}\,f_{de}{}^c\,\hat{E}^{ae}
 - \hat{E}^{dc}\,f_{de}{}^a\,\hat{E}^{eb}
 - \hat{E}^{dc}\,f_{de}{}^b\,\hat{E}^{ae} &
\\
 + f_d{}^{ab}\,\hat{E}^{dc}
 + f_d{}^{ca}\,\hat{E}^{db}
 + f_d{}^{cb}\,\hat{E}^{ad} &\bigr) = 0\,,
\end{split}
\end{align}
which corresponds to the condition discussed around Eq.~(110) of \cite{1105.0162}.

We note that the condition \eqref{eq:inv-orig} is slightly stronger than the condition found in \cite{hep-th:9904188}:
\begin{align}
 f_{\sfi}{}^{\check{a}\check{b}} = f_{\sfi\check{d}}{}^{\check{a}}\,\hat{E}^{\check{d}\check{b}} + f_{\sfi \check{d}}{}^b\,\hat{E}^{\check{a}\check{d}} \,.
\label{eq:Sfetsos-cond}
\end{align}
In fact, this is the $\check{a}$--$\check{b}$ component of the condition \eqref{eq:inv-orig} (note that $f_{\sfi\sfj}{}^{\check{a}}=0$). 
Below, we explain why Eq.~\eqref{eq:Sfetsos-cond} is enough for the gauge invariance in the approach of \cite{hep-th:9904188}. 

In the case of the standard coset studied in \cite{hep-th:9904188}, the easiest way is to take the variation of the standard action \eqref{eq:string-action}
\begin{align}
 \delta_{\sfi} S = \frac{1}{4\pi\alpha'}\int_\Sigma \delta_{\sfi} E_{mn}\,\bigl(\rmd x^m\wedge * \rmd x^n + \rmd x^m\wedge \rmd x^n\bigr)\,.
\end{align}
Using the result \eqref{eq:delta-E} given in Appendix \ref{app:left-mult}, we find (up to a boundary term)
\begin{align}
 \delta_{\sfi} S = - \frac{1}{4\pi\alpha'}\int_\Sigma E_{ac}\,\Omega_{\sfi}^{cd}\,E_{db}\,\bigl(r^a\wedge * r^b + r^a\wedge r^b\bigr)\,.
\end{align}
The equations of motion $\mathbb{P}\in\mathfrak{f}^\perp$, namely, $\eta_{MN}\,\VV_{\sfi}^M\,\cZ^N=0$ gives
\begin{align}
 \tilde{k}_{\sfi a}\,* r^a + (\mathsf{\kk}_{\sfi}^a+\pi^{ac}\,\tilde{k}_{\sfi c}\bigr)\,E_{ab}\,r^b = 0\,,
\end{align}
and for the standard coset ($\mathsf{\kk}_{\sfi}^a=\delta_{\sfi}^a$ and $\tilde{k}_{\sfi a}=0$), they reduce to $\delta_{\sfi}^a\,E_{ab}\,r^b = 0$\,. 
Then, for the invariance of the action under $\mathbb{P}\in\mathfrak{f}^\perp$, it is enough to impose $\Omega_{\sfi}^{\check{a}\check{b}}=0$\,. 
In the original gauged action \eqref{eq:geugedDSM2}, $\mathbb{P}\in\mathfrak{f}^\perp$ is not satisfied (because it appears as the equations of motion), and we need to impose the sufficient condition $\Omega_{\sfi}^{ab}=0$\,. 
In all of the examples studied in this paper, the sufficient condition is satisfied but it may be also interesting to study the case where only $\Omega_{\sfi}^{\check{a}\check{b}}=0$ is satisfied. 
Such a model may not be described by the gauged action \eqref{eq:geugedDSM2}. 

\section{Poisson--Lie $T$-plurality in DFT}
\label{sec:PL-DFT}

In this section, we discuss the symmetry of the DFT equations of motion under the PL $T$-plurality by using a similar idea to the one given in \cite{1810.07763} (see also \cite{1502.04517}). 
The content of this section is technical, and the readers can skip this section and go to section \ref{sec:examples} immediately (although section \ref{sec:PL-AdS2-RR} requires section \ref{sec:R-R}). 

\subsection{Assumptions and properties}
\label{sec:assumptions}

\subsubsection{Assumptions on the dilaton}

The discussion in this section is based on the generalized vector fields $\CE_{\bcheck{A}}$ constructed in section \ref{sec:reduced-GM}. 
They are defined to satisfy
\begin{align}
 \gLie_{\VV_{\sfi}} \CE_{\bcheck{A}} = 0\,,\qquad 
 [\CE_{\bcheck{A}},\,\CE_{\bcheck{B}}]_{\text{D}} = - \bm{f}_{\bcheck{A}\bcheck{B}}{}^{\bcheck{C}}\,\CE_{\bcheck{C}}\,,
\label{eq:CE-setup1}
\end{align}
and we have found that they can be reduced to $\check{\CE}_{\bcheck{A}}{}^{\check{M}}$, which also satisfy
\begin{align}
 [\check{\CE}_{\bcheck{A}},\,\check{\CE}_{\bcheck{B}}]_{\text{D}} = - \bm{f}_{\bcheck{A}\bcheck{B}}{}^{\bcheck{C}}\,\check{\CE}_{\bcheck{C}}\,. 
\end{align}
The reduced fields $\check{\CE}_{\bcheck{A}}$ play the role of the generalized frame fields in the $2d$-dimensional reduced doubled space $\check{\cM}$ with coordinates $(x^{\check{M}})=(x^{\check{m}},\,\tilde{x}_{\check{m}})$\,. 
Since we are interested in $d$-dimensional physics, any fields are supposed to depend on the $d$-dimensional coordinates $x^{\check{m}}$\,. 
An arbitrary generalized tensor field on $\check{\cM}$ can be constructed as $\check{T}=T^{\bcheck{A}\bcheck{B}\cdots}\,\check{\CE}_{\bcheck{A}}\otimes \check{\CE}_{\bcheck{B}}\otimes \cdots$ with the coefficients $T^{\bcheck{A}\bcheck{B}\cdots}$ depending only on $x^{\check{m}}$\,. 
This can be naturally uplifted to the $2D$-dimensional doubled space $\cM$ as $T=T^{\bcheck{A}\bcheck{B}\cdots}\,\CE_{\bcheck{A}}\otimes \CE_{\bcheck{B}}\otimes \cdots$ that satisfies $\gLie_{\VV_{\sfi}}T=0$\,. 

To discuss the DFT equations of motion, we shall consider the DFT dilaton. 
We suppose that $d(x)$ is also defined on the $d$-dimensional space and is a function of $x^{\check{m}}$. 
Then, since the components $\CE_{\bcheck{A}}{}^m=\frac{1}{\sqrt{2}}\,h_{\bcheck{A}}^m$ depend only on the coordinates $x^{\check{m}}$, we find
\begin{align}
 \div \CE_{\bcheck{A}} = \Exp{2d} \partial_M\bigl(\Exp{-2d} \CE_{\bcheck{A}}{}^M\bigr) = \Exp{2d} \partial_{\check{M}}\bigl(\Exp{-2d} \CE_{\bcheck{A}}{}^{\check{M}}\bigr)
 = \Exp{2d} \partial_{\check{M}}\bigl(\Exp{-2d} \check{\CE}_{\bcheck{A}}{}^{\check{M}}\bigr) 
 \equiv \widetilde{\div}\, \check{\CE}_{\bcheck{A}}\,.
\label{eq:CE-setup2}
\end{align}
We also find that, for an arbitrary generalized vector field $W^M$, the divergence satisfies
\begin{align}
 \div [\VV_{\sfi},\,W]_{\text{D}} = \VV_{\sfi} \cdot \div W - W\cdot \div \VV_{\sfi} \,.
\end{align}
Since we are assuming that the dilaton flux $\cF_{\sfi}\equiv \VV_{\sfi}{}^A\,\cF_A=-\div \VV_{\sfi}$ is constant, we have
\begin{align}
 \div [\VV_{\sfi},\,W]_{\text{D}} = \VV_{\sfi} \cdot \div W \,. 
\label{eq:V-div1}
\end{align}
We additionally assume that the dilaton flux is given by\footnote{In this paper, we employ the ansatz \eqref{eq:dilaton-ansatz} and we have $\cF_{A}=0$. Then $f_{\sfi\sfj}{}^{\sfj}=0$ needs to be satisfied.}
\begin{align}
 \cF_{\sfi} = - f_{\sfi\sfj}{}^{\sfj}\,,
\end{align}
where $f_{\sfi\sfj}{}^{\sfk}$ are the structure constants of the gauge algebra. 
This ensures that an arbitrary generalized vector field $\cK=\cK^{\sfi}\,\VV_{\sfi} \in \Gamma^{\mathfrak{f}}(\mathbb{F})$ satisfies \cite{1810.07763}
\begin{align}
 \div \cK = \Exp{2d} \partial_M\bigl(\Exp{-2d} \cK^M\bigr) = \partial_M \cK^M = \VV_{\sfi}\cdot \cK^{\sfi} + \cK^{\sfi}\,\partial_M \VV_{\sfi}^M
 = - \cK^{\sfi}\,f_{\sfi\sfk}{}^{\sfk} - \cK^{\sfi}\, \cF_{\sfi} = 0\,,
\label{eq:V-div2}
\end{align}
where we used $\cF_{\sfi} = - \partial_M \VV_{\sfi}{}^M$ and $\VV_{\sfi}\cdot\cK^{\sfj}=-\cK^{\sfk}\,f_{\sfk\sfi}{}^{\sfj}$ that follows from $[\VV_{\sfi},\,\cK]_{\text{D}}=0$\,. 
When the properties \eqref{eq:V-div1} and \eqref{eq:V-div2} are satisfied, $\div$ is called an equivariant divergence \cite{1810.07763} and then the $\div$ defined in $\cM$ can be reduced to the divergence $\widetilde{\div}$ defined in $\check{\cM}$\,. 

\subsubsection{Important properties}

For convenience, when two generalized vector fields $A$ and $B$ are related as
\begin{align}
 A = B + \cK\,,
\end{align}
by using a generalized vector field $\cK\in \Gamma^{\mathfrak{f}}(\mathbb{F})$, we denote
\begin{align}
 A\sim B\,,
\end{align}
because the difference is not important in our discussion. 
Then, we find
\begin{align}
 \mathbb{H}(X) \sim \cH(X)\,,\qquad
 P_{\pm}(X) \sim \mathbb{P}_{\pm}(X)\qquad \bigl[\forall X\in \Gamma^{\mathfrak{f}}(\mathbb{F}^\perp)\bigr]\,,
\end{align}
where we have defined
\begin{align}
 P_{\pm}(X) \equiv \frac{X\pm \cH(X)}{2}\,,\qquad 
 \mathbb{P}_{\pm}(X) \equiv \frac{X\pm \mathbb{H}(X)}{2}\,.
\end{align}
In particular, we have
\begin{align}
 P_{\pm}(\CE_{\bcheck{A}}) \sim \mathbb{P}_{\pm}(\CE_{\bcheck{A}}) \,.
\end{align}
For arbitrary $X\in \Gamma^{\mathfrak{f}}(\mathbb{F}^\perp)$ and $\cK\equiv \cK^{\sfi}\,\VV_{\sfi}\in \Gamma^{\mathfrak{f}}(\mathbb{F})$, we find
\begin{align}
\begin{split}
 [\cK,\,X]_{\text{D}}^M &= \cK^{\sfi}\,[\VV_{\sfi},\,X]_{\text{D}}^M - (X\cdot \cK^{\sfi})\, \VV_{\sfi}^M + \partial^M \cK^{\sfi}\,\langle \VV_{\sfi},\,X\rangle
 =- (X\cdot \cK^{\sfi})\, \VV_{\sfi}^M \,,
\\
 [X,\,\cK]_{\text{D}}^M&= \cK^{\sfi}\,[X,\,\VV_{\sfi}]_{\text{D}}^M + (X\cdot \cK^{\sfi})\,\VV_{\sfi}^M = (X\cdot \cK^{\sfi})\,\VV_{\sfi}^M\,.
\end{split}
\end{align}
Then, for $A\sim C$ and $B\sim D$ which are $\mathfrak{f}$-invariant sections of $\mathbb{F}^\perp$, we have
\begin{align}
 [A,\,B]_{\text{D}} \sim [C,\,D]_{\text{D}}\,,\qquad \langle A,\,B\rangle = \langle C,\,D\rangle \,.
\end{align}

Let us also summarize properties associated with the divergence. 
Using $\div \cK = 0$ for an arbitrary $\cK\equiv \cK^{\sfi}\,\VV_{\sfi} \in \Gamma^{\mathfrak{f}}(\mathbb{F})$, for two sections of $\mathbb{F}^\perp$ related by $X\sim Y$, we find
\begin{align}
 \div X \sim \div Y\,.
\end{align}
In addition, using Eq.~\eqref{eq:V-div1}, for an arbitrary $X\in \Gamma^{\mathfrak{f}}(\mathbb{F}^\perp)$, we have
\begin{align}
 \VV_{\sfi}\cdot \div X = \div [\VV_{\sfi},\,X]_{\text{D}} = 0\,.
\label{eq:V-divA-0}
\end{align}
Then for sections of $\mathbb{F}^\perp$, $X$, $Y$, and $Z$ satisfying $X\sim Y$ and $\gLie_{\VV_{\sfi}}Z=0$\,, we have
\begin{align}
 X\cdot \div Z = Y\cdot \div Z\,.
\end{align}

\subsection{Invariance of the DFT equations of motion}
\label{sec:DFT-EOM}

Here we consider the DFT equations of motion for the reduced generalized metric $\check{\cH}_{\check{M}\check{N}}$ and the DFT dilaton $d$ defined in $\check{\cM}$\,.

\subsubsection{Generalized Ricci tensor}
\label{sec:DFT-EOM-1}

The equations of motion for $\check{\cH}_{\check{M}\check{N}}$ can be expressed as
\begin{align}
\begin{split}
 \widetilde{\text{GRic}}_{+}(\check{\CE}_{\bcheck{A}},\,\check{\CE}_{\bcheck{B}})
 &= \widetilde{\div} \bigl[\check{P}_+([\check{P}_-(\check{\CE}_{\bcheck{B}}),\,\check{P}_+(\check{\CE}_{\bcheck{A}})]_{\text{D}})\bigr] - \check{P}_-(\check{\CE}_{\bcheck{B}})\cdot \widetilde{\div} \check{P}_+(\check{\CE}_{\bcheck{A}})
\\
 &\quad - \langle [\check{P}_-([\check{P}_+(\check{\CE}_{\bcheck{C}}),\,\check{P}_-(\check{\CE}_{\bcheck{B}})]_{\text{D}}),\,\check{P}_+(\check{\CE}_{\bcheck{A}})]_{\text{D}},\,\check{P}_+(\check{\CE}^{\bcheck{C}}) \rangle = 0\,.
\end{split}
\end{align}
Let us rewrite these equations of motion in terms of the $D$-dimensional fields. 
If we uplift $\check{\cH}_{\check{M}\check{N}}$ to the metric $\bm{H}_{MN}$ defined in Eq.~\eqref{eq:bmH-def} and define $\bm{H}(X)\equiv \bm{H}^M{}_N\,X^N$ for arbitrary $X \in \Gamma(\mathbb{F}^\perp)$\,, we find
\begin{align}
 \bm{H}(\CE_{\ubar{\bm{a}}}) = +\CE_{\ubar{\bm{a}}}\,,\qquad
 \bm{H}(\CE_{\bar{\bm{a}}}) = -\CE_{\bar{\bm{a}}}\,,\qquad
 \bm{H}(\VV_{\sfi}) = 0\,. 
\end{align}
For the operator $\mathbb{H}$ defined in Eq.~\eqref{eq:mathbbH}, we have
\begin{align}
 \mathbb{H}(\CE_{\ubar{\bm{a}}}) = +H_{\ubar{\bm{a}}}\,,\qquad
 \mathbb{H}(\CE_{\bar{\bm{a}}}) = -H_{\bar{\bm{a}}}\,,\qquad
 \mathbb{H}(\VV_{\sfi}) = 0\,.
\end{align}
Then, defining projectors $\bm{P}_{\pm}(X) \equiv \frac{1}{2}\,\bigl[X\pm \bm{H}(X)\bigr]$\,, we have
\begin{align}
 \bm{H}(X) \sim \mathbb{H}(X)\,,\qquad 
 \bm{P}_\pm(X) \sim \mathbb{P}_\pm(X)\qquad (\forall X\in\Gamma(\mathbb{F}^\perp))\,.
\end{align}
For arbitrary $X,\,Y\in \Gamma(\mathbb{F}^\perp)$, we define
\begin{align}
\begin{split}
 \bm{G}\text{Ric}_{+}(X,\,Y)
 &\equiv \div \bigl[\bm{P}_+([\bm{P}_-(Y),\,\bm{P}_+(X)]_{\text{D}})\bigr] - \bm{P}_-(Y)\cdot \div \bm{P}_+(X)
\\
 &\quad - \langle [\bm{P}_-([\bm{P}_+(\CE_{\bcheck{A}}),\,\bm{P}_-(Y)]_{\text{D}}),\,\bm{P}_+(X)]_{\text{D}},\,\bm{P}_+(\CE^{\bcheck{A}}) \rangle \,,
\end{split}
\end{align}
and then it is easy to see that $\bm{G}\text{Ric}_{+}$ is the $D$-dimensional uplift of $\widetilde{\text{GRic}}_{+}$,
\begin{align}
 \widetilde{\text{GRic}}_{+}(\check{\CE}_{\bcheck{A}},\,\check{\CE}_{\bcheck{B}}) = \bm{G}\text{Ric}_{+}(\CE_{\bcheck{A}},\,\CE_{\bcheck{B}})\,.
\end{align}

For arbitrary $X,\,Y\in \Gamma(\mathbb{F}^\perp)$, we also define
\begin{align}
\begin{split}
 \text{$\mathbb{G}$Ric}_{+}(X,\,Y)
 &= \div \bigl[\mathbb{P}_+([\mathbb{P}_-(Y),\,\mathbb{P}_+(X)]_{\text{D}})\bigr] - \mathbb{P}_-(Y)\cdot \div \mathbb{P}_+(X)
\\
 &\quad - \langle [\mathbb{P}_-([\mathbb{P}_+(\CE_{\bcheck{A}}),\,\mathbb{P}_-(Y)]_{\text{D}}),\,\mathbb{P}_+(X)]_{\text{D}},\,\mathbb{P}_+(\CE^{\bcheck{A}}) \rangle \,,
\end{split}
\end{align}
which is a generalized tensor, and additionally, is independent of the choice of the $\mathfrak{f}$-invariant frame fields $\CE_{\bcheck{A}}$ (see Appendix \ref{app:GRic}). 
Then we find a non-trivial relation
\begin{align}
 \bm{G}\text{Ric}_{+}(X,\,Y) = \text{$\mathbb{G}$Ric}_{+}(X,\,Y)\qquad \bigl[\forall X,\,Y\in \Gamma(\mathbb{F}^\perp)\bigr]\,.
\end{align}
Indeed, using $\bm{P}_-(Y)\sim \mathbb{P}_-(Y)$ and $\bm{P}_+(X)\sim \mathbb{P}_+(X)$, we can show $[\bm{P}_-(Y),\,\bm{P}_+(X)]_{\text{D}}=[\mathbb{P}_-(Y),\,\mathbb{P}_+(X)]_{\text{D}}$\,.
We then find $\bm{P}_+([\bm{P}_-(Y),\,\bm{P}_+(X)]_{\text{D}}) \sim \mathbb{P}_+([\mathbb{P}_-(Y),\,\mathbb{P}_+(X)]_{\text{D}})$, and the first terms of the two tensors are the same
\begin{align}
 \div \bigl[\bm{P}_+([\bm{P}_-(Y),\,\bm{P}_+(X)]_{\text{D}})\bigr]_{\text{D}} 
 = \div \bigl[\mathbb{P}_+([\mathbb{P}_-(Y),\,\mathbb{P}_+(X)]_{\text{D}})\bigr] \,.
\end{align}
Similarly, we can easily show that other terms are also the same. 
Thus, the equations of motion for the generalized metric $\bm{G}\text{Ric}_{+}(\CE_{\bcheck{A}},\,\CE_{\bcheck{B}})=0$ can be expressed as $\text{$\mathbb{G}$Ric}_{+}(\CE_{\bcheck{A}},\,\CE_{\bcheck{B}}) = 0$\,.
Using $\mathbb{P}_\pm(\VV_{\sfj})=\frac{1}{2}\,\VV_{\sfj}$\,, $[\mathbb{P}_\pm(\CE_{\bcheck{A}}),\,\VV_{\sfi}]=[\VV_{\sfi},\,\mathbb{P}_\pm(\CE_{\bcheck{A}})]=0$\,, Eq.~\eqref{eq:V-divA-0}, $[\VV_{\sfi},\,\VV_{\sfj}]_{\text{D}}=-f_{\sfi\sfj}{}^{\sfk}\,\VV_{\sfk}$\,, and that $\div \VV_{\sfi}=-\cF_{\sfi}$ is constant, we can show the identities
\begin{align}
 \text{$\mathbb{G}$Ric}_{+}(\CE_{\bcheck{A}},\,\VV_{\sfj}) = 0\,,\qquad 
 \text{$\mathbb{G}$Ric}_{+}(\VV_{\sfi},\,\CE_{\bcheck{B}}) = 0\,,\qquad
 \text{$\mathbb{G}$Ric}_{+}(\VV_{\sfi},\,\VV_{\sfj}) = 0 \,.
\end{align}
Thus the equations of motion can be expressed as
\begin{align}
 \text{$\mathbb{G}$Ric}_{+}(X,\,Y) = 0\qquad \bigl[\forall X,\,Y\in \Gamma^{\mathfrak{f}}(\mathbb{F}^\perp)\bigr]\,.
\end{align}

An additional non-trivial point is to show that $\text{$\mathbb{G}$Ric}_{+}(X,\,Y)$ can be expressed as
\begin{align}
 \text{$\mathbb{G}$Ric}_{+}(X,\,Y) = \lim_{\lambda\to 0} \text{GRic}_{+}(X,\,Y)\qquad \bigl[\forall X,\,Y\in \Gamma^{\mathfrak{f}}(\mathbb{F}^\perp)\bigr]\,.
\label{eq:GGRic}
\end{align}
Under Sfetsos's singular limit, $P_\pm$ reduces to $\mathbb{P}_\pm$ for any $\mathfrak{f}$-invariant sections of $\mathbb{F}^\perp$.
Thus, the only non-trivial point in showing the relation \eqref{eq:GGRic} is the equality of the last term:
\begin{align}
\begin{split}
 &\langle [\mathbb{P}_-([\mathbb{P}_+(\CE_{\bcheck{A}}),\,\mathbb{P}_-(Y)]_{\text{D}}),\,\mathbb{P}_+(X)]_{\text{D}},\,\mathbb{P}_+(\CE^{\bcheck{A}}) \rangle
\\
 &= \lim_{\lambda\to 0} \langle [P_-([P_+(E_{A}),\,P_-(Y)]_{\text{D}}),\,P_+(X)]_{\text{D}},\,P_+(E^{A}) \rangle\,.
\end{split}
\label{eq:last-term}
\end{align}
The choice of the $2D$-dimensional frame fields $E_{\bcheck{A}}$ is arbitrary and we can choose these as
\begin{align}
 \{E_{A}\} = \{\CE_{\ubar{\bm{a}}},\,\bar{\CE}_{\bar{\bm{a}}},\,\VV_{\sfi},\,\widetilde{\VV}_{\sfi}\}\,,\qquad 
 \{E^{A}\} = \{ \eta^{\ubar{\bm{a}}\ubar{\bm{b}}}\,\CE_{\ubar{\bm{b}}},\, \eta^{\bar{\bm{a}}\bar{\bm{b}}}\,\bar{\CE}_{\bar{\bm{b}}},\, \hh^{\sfi\sfj}\,\widetilde{\VV}_{\sfj},\, \hh^{\sfi\sfj}\,\VV_{\sfj}\}\,.
\label{eq:EA-decomp}
\end{align}
Since we have $P_+(\widetilde{\VV}_{\sfi})=P_+(\VV_{\sfi})=\frac{1}{2}\,\bigl(\VV_{\sfi}+\widetilde{\VV}_{\sfi}\bigr)$, the third and the fourth components of $E_{A}$ or $E^{A}$ give the same contribution, and we have
\begin{align}
\begin{split}
 &\langle [P_-([P_+(E_{A}),\,P_-(Y)]_{\text{D}}),\,P_+(X)]_{\text{D}},\,P_+(E^{A}) \rangle
\\
 &= \langle [P_-([P_+(\CE_{\bcheck{A}}),\,P_-(Y)]_{\text{D}}),\,P_+(X)]_{\text{D}},\,P_+(\CE^{\bcheck{A}}) \rangle
\\
 &\quad+ \tfrac{1}{2}\,\hh^{\sfi\sfj}\,\langle [P_-([\VV_{\sfi}+\widetilde{\VV}_{\sfi},\, P_-(Y)]_{\text{D}}),\,P_+(X)]_{\text{D}},\, \widetilde{\VV}_{\sfj} + \VV_{\sfj} \rangle\,.
\end{split}
\label{eq:expand}
\end{align}
In the last line, we have $[\VV_{\sfi},\, P_-(Y)]_{\text{D}}=0$ and $[\widetilde{\VV}_{\sfi},\, P_-(Y)]_{\text{D}}\in \Gamma(\mathbb{F}^\perp)$ because $P_-(Y)\in\Gamma(\mathbb{F}^\perp)$. 
We also find $\chi\equiv [P_-([\widetilde{\VV}_{\sfi},\, P_-(Y)]_{\text{D}}),\,P_+(X)]_{\text{D}}\in \Gamma(\mathbb{F}^\perp)$ and thus $\langle \chi ,\, \widetilde{\VV}_{\sfj} + \VV_{\sfj} \rangle = \langle \chi ,\, \tilde{\sfK}_{\sfj} \rangle$, where we used the decomposition \eqref{eq:VVt-decomp}. 
Under the limit $\lambda\to 0$\,, $\hh^{\sfi\sfj}=\cO(\lambda^{-1})$ but $[\widetilde{\VV}_{\sfi},\, P_-(Y)]_{\text{D}}=[\sfK_{\sfi},\, P_-(Y)]_{\text{D}}+\cO(\lambda)=\cO(\lambda)$.
Then, the last line of Eq.~\eqref{eq:expand} is of order $\cO(\lambda)$ and negligible. 
This gives Eq.~\eqref{eq:last-term} and thus Eq.~\eqref{eq:GGRic}. 
The tensor $\text{GRic}_{+}(X,\,Y)$ is invariant under the PL $T$-plurality, and then $\text{$\mathbb{G}$Ric}_{+}(X,\,Y)= \lim_{\lambda\to 0} \text{GRic}_{+}(X,\,Y)$ is also invariant. 
In this way, we can show the invariance of the equations of motion for $\check{\cH}_{\check{M}\check{N}}$ under the PL $T$-plurality. 

\subsubsection{Generalized Ricci scalar}
\label{sec:DFT-EOM-2}

Let us also consider the dilaton equations of motion in $d$-dimensions:
\begin{align}
\begin{split}
 \widetilde{\text{GR}}_+ &= -8\Exp{d}\gLie_{\check{P}_+(\check{\CE}_{\bcheck{A}})}\gLie_{\check{P}_+(\check{\CE}^{\bcheck{A}})} \Exp{-d}
\\
 &\quad + \tfrac{1}{3}\,\langle [\check{P}_+(\check{\CE}_{\bcheck{A}}),\,\check{P}_+(\check{\CE}_{\bcheck{B}})]_{\text{D}},\,\check{P}_+(\check{\CE}_{\bcheck{C}})\rangle\,\langle [\check{P}_+(\check{\CE}^{\bcheck{A}}),\,\check{P}_+(\check{\CE}^{\bcheck{B}})]_{\text{D}},\,\check{P}_+(\check{\CE}^{\bcheck{C}})\rangle
\\
 &\quad + \langle [\check{P}_+(\check{\CE}_{\bcheck{A}}),\,\check{P}_+(\check{\CE}_{\bcheck{B}})]_{\text{D}},\,\check{P}_-(\check{\CE}_{\bcheck{C}})\rangle\,\langle [\check{P}_+(\check{\CE}^{\bcheck{A}}),\,\check{P}_+(\check{\CE}^{\bcheck{B}})]_{\text{D}},\,\check{P}_-(\check{\CE}^{\bcheck{C}})\rangle = 0 \,.
\end{split}
\end{align}
The generalized Ricci scalar $\widetilde{\text{GR}}_+$ can be readily uplifted to $D$-dimensions as
\begin{align}
\begin{split}
 \bm{G}\text{R}_+ &\equiv -8\Exp{d}\gLie_{\bm{P}_+(\CE_{\bcheck{A}})}\gLie_{\bm{P}_+(\CE^{\bcheck{A}})} \Exp{-d}
\\
 &\quad + \tfrac{1}{3}\,\langle [\bm{P}_+(\CE_{\bcheck{A}}),\,\bm{P}_+(\CE_{\bcheck{B}})]_{\text{D}},\,\bm{P}_+(\CE_{\bcheck{C}})\rangle\,\langle [\bm{P}_+(\CE^{\bcheck{A}}),\,\bm{P}_+(\CE^{\bcheck{B}})]_{\text{D}},\,\bm{P}_+(\CE^{\bcheck{C}})\rangle
\\
 &\quad + \langle [\bm{P}_+(\CE_{\bcheck{A}}),\,\bm{P}_+(\CE_{\bcheck{B}})]_{\text{D}},\,\bm{P}_-(\CE_{\bcheck{C}})\rangle\,\langle [\bm{P}_+(\CE^{\bcheck{A}}),\,\bm{P}_+(\CE^{\bcheck{B}})]_{\text{D}},\,\bm{P}_-(\CE^{\bcheck{C}})\rangle \,.
\end{split}
\end{align}
Again, by making the replacement $\bm{P}_\pm\to\mathbb{P}_\pm$\,, we define
\begin{align}
\begin{split}
 \text{$\mathbb{G}$R}_+ &\equiv -8\Exp{d}\gLie_{\mathbb{P}_+(\CE_{\bcheck{A}})}\gLie_{\mathbb{P}_+(\CE^{\bcheck{A}})} \Exp{-d}
\\
 &\quad + \tfrac{1}{3}\,\langle [\mathbb{P}_+(\CE_{\bcheck{A}}),\,\mathbb{P}_+(\CE_{\bcheck{B}})]_{\text{D}},\,\mathbb{P}_+(\CE_{\bcheck{C}})\rangle\,\langle [\mathbb{P}_+(\CE^{\bcheck{A}}),\,\mathbb{P}_+(\CE^{\bcheck{B}})]_{\text{D}},\,\mathbb{P}_+(\CE^{\bcheck{C}})\rangle
\\
 &\quad + \langle [\mathbb{P}_+(\CE_{\bcheck{A}}),\,\mathbb{P}_+(\CE_{\bcheck{B}})]_{\text{D}},\,\mathbb{P}_-(\CE_{\bcheck{C}})\rangle\,\langle [\mathbb{P}_+(\CE^{\bcheck{A}}),\,\mathbb{P}_+(\CE^{\bcheck{B}})]_{\text{D}},\,\mathbb{P}_-(\CE^{\bcheck{C}})\rangle \,,
\end{split}
\label{eq:GGR}
\end{align}
and then we can show that $\bm{G}\text{R}_+=\text{$\mathbb{G}$R}_+$ in a similar way to the generalized Ricci tensor. 
The non-trivial point is to relate $\text{$\mathbb{G}$R}_+$ and the generalized Ricci scalar in $D$-dimensions,
\begin{align}
\begin{split}
 \text{GR}_+ &= -8\Exp{d}\gLie_{P_+(E_A)}\gLie_{P_+(E^A)} \Exp{-d}
\\
 &\quad + \tfrac{1}{3}\,\langle [P_+(E_A),\,P_+(E_B)]_{\text{D}},\,P_+(E_C)\rangle\,\langle [P_+(E^A),\,P_+(E^B)]_{\text{D}},\,P_+(E^C)\rangle
\\
 &\quad + \langle [P_+(E_A),\,P_+(E_B)]_{\text{D}},\,P_-(E_C)\rangle\,\langle [P_+(E^A),\,P_+(E^B)]_{\text{D}},\,P_-(E^C)\rangle \,.
\end{split}
\label{eq:GR-to-be-reduced}
\end{align}
Again, using Eq.~\eqref{eq:EA-decomp}, we decompose the summation over $E_A$ and $E^A$. 
The first line becomes
\begin{align}
 -8\Exp{d}\gLie_{P_+(E_A)}\gLie_{P_+(E^A)} \Exp{-d}
 = -8\Exp{d}\gLie_{P_+(\CE_{\bcheck{A}})}\gLie_{P_+(\CE^{\bcheck{A}})} \Exp{-d}
 -4\,\hh^{\sfi\sfj}\,\Exp{d}\gLie_{\VV_{\sfi}+\widetilde{\VV}_{\sfi}}\gLie_{\VV_{\sfj}+\widetilde{\VV}_{\sfj}} \Exp{-d}.
\end{align}
By using $\gLie_{\VV_{\sfj}} \Exp{-d}=0$ and $\gLie_{\sfK_{\sfj}} \Exp{-d}=0$\,, the second term becomes
\begin{align}
 -4\,\hh^{\sfi\sfj} \Exp{d}\gLie_{\VV_{\sfi}+\widetilde{\VV}_{\sfi}}\gLie_{\VV_{\sfj}+\widetilde{\VV}_{\sfj}} \Exp{-d}
 = -4\,\hh^{\sfi\sfj} \Exp{d}\gLie_{\VV_{\sfi}+\widetilde{\VV}_{\sfi}}\gLie_{\tilde{\sfK}_{\sfj}} \Exp{-d}.
\end{align}
Further using $[\gLie_V,\,\gLie_W]=\gLie_{[V,W]_{\text{D}}}$ and Eq.~\eqref{eq:V-sfU-tildeU}, we find
\begin{align}
 -4\,\hh^{\sfi\sfj}\Exp{d}\gLie_{\VV_{\sfi}+\widetilde{\VV}_{\sfi}}\gLie_{\tilde{\sfK}_{\sfj}} \Exp{-d}
 &= -4\,\hh^{\sfi\sfj}\Exp{d}\gLie_{\tilde{\sfK}_{\sfj}} \gLie_{\VV_{\sfi}+\widetilde{\VV}_{\sfi}}\Exp{-d}
 -4\,\hh^{\sfi\sfj}\Exp{d}\gLie_{[\VV_{\sfi}+\widetilde{\VV}_{\sfi},\, \tilde{\sfK}_{\sfj}]_{\text{D}}} \Exp{-d} 
\nn\\
 &= -4\,(\delta_{\sfi}^{\sfk}+\zeta_{\sfi}{}^{\sfk})\,\hh^{\sfi\sfj}\Exp{d}\gLie_{[\VV_{\sfk},\, \tilde{\sfK}_{\sfj}]_{\text{D}}} \Exp{-d} + \cO(\lambda) = \bm{\delta} + \cO(\lambda) \,,
\end{align}
where $\bm{\delta} \equiv -2\,(\delta_{\sfi}^{\sfk}+\zeta_{\sfi}{}^{\sfk})\,\hh^{\sfi\sfj}\,f_{\sfk\sfj}{}^{\sfl}\,\tilde{\sfK}_{\sfl}{}^A \cF_A=\cO(\lambda^0)$\,.
Namely, we have shown that the first line of Eq.~\eqref{eq:GR-to-be-reduced} is related to that of Eq.~\eqref{eq:GGR} as
\begin{align}
 \lim_{\lambda\to 0}\bigl[-8\Exp{d}\gLie_{P_+(E_A)}\gLie_{P_+(E^A)} \Exp{-d}\,\bigr]
 = -8\Exp{d}\gLie_{\mathbb{P}_+(\CE_{\bcheck{A}})}\gLie_{\mathbb{P}_+(\CE^{\bcheck{A}})} \Exp{-d} +\bm{\delta}\,.
\end{align}
Let us also decompose the second and the third lines of Eq.~\eqref{eq:GR-to-be-reduced}. 
Using\footnote{The first two follow from $\widetilde{\VV}_{\sfi}= \sfK_{\sfj} + \cO(\lambda)$, $[\VV_{\sfi},\,P_+(\CE_{\bcheck{B}})]_{\text{D}}=0$\,, and $\langle [P_+(\CE_{\bcheck{A}}),\,P_+(\CE_{\bcheck{B}})]_{\text{D}},\, \VV_{\sfi}\rangle=0$\,. 
The third can be shown by using $[\VV_{\sfi}+\widetilde{\VV}_{\sfi},\,P_+(\CE_{\bcheck{B}})]_{\text{D}}=[\widetilde{\VV}_{\sfi},\,P_+(\CE_{\bcheck{B}})]_{\text{D}}\in \Gamma(\mathbb{F}^\perp)$, $[\widetilde{\VV}_{\sfi},\,P_+(\CE_{\bcheck{B}})]_{\text{D}}=\cO(\lambda)$, and thus $\langle [\widetilde{\VV}_{\sfi},\,P_+(\CE_{\bcheck{B}})]_{\text{D}},\, \VV_{\sfk}\pm \widetilde{\VV}_{\sfk} \rangle=\cO(\lambda^{2})$. 
The fourth is shown from the third by using $\langle [\VV_{\sfi}+\widetilde{\VV}_{\sfi},\,\VV_{\sfk}+\widetilde{\VV}_{\sfk}]_{\text{D}},\, P_\pm(\CE_{\bcheck{C}}) \rangle = -\langle [\VV_{\sfi}+\widetilde{\VV}_{\sfi},\,P_\pm(\CE_{\bcheck{C}})]_{\text{D}},\,\VV_{\sfk}+\widetilde{\VV}_{\sfk} \rangle$\,.}
\begin{align}
\begin{split}
 \hh^{\sfi\sfj}\,\langle [\VV_{\sfi}^+,\,P_+(\CE_{\bcheck{B}})]_{\text{D}},\, P_\pm(\CE_{\bcheck{C}}) \rangle\,\langle [\VV_{\sfj}^+,\,P_+(\CE^{\bcheck{B}})]_{\text{D}},\,P_\pm(\CE^{\bcheck{C}}) \rangle &= \cO(\lambda)\,,
\\
 \hh^{\sfi\sfj}\,\langle [P_+(\CE_{\bcheck{A}}),\,P_+(\CE_{\bcheck{B}})]_{\text{D}},\, \VV_{\sfi}^{\pm} \rangle\,\langle [P_+(\CE^{\bcheck{A}}),\,P_+(\CE^{\bcheck{B}})]_{\text{D}},\,\VV_{\sfj}^{\pm}\rangle &= \cO(\lambda)\,,
\\
 \hh^{\sfi\sfj}\,\hh^{\sfk\sfl}\,\langle [\VV_{\sfi}^+ ,\,P_+(\CE_{\bcheck{B}})]_{\text{D}},\, \VV_{\sfk}^{\pm} \rangle\,\langle [\VV_{\sfj}^+ ,\,P_+(\CE^{\bcheck{B}})]_{\text{D}},\,\VV_{\sfl}^{\pm}\rangle &= \cO(\lambda)\,,
\\
 \hh^{\sfi\sfj}\,\hh^{\sfk\sfl}\,\langle [\VV_{\sfi}^+,\,\VV_{\sfk}^+]_{\text{D}},\, P_\pm(\CE_{\bcheck{C}}) \rangle\,\langle [\VV_{\sfj}^+,\,\VV_{\sfl}^+]_{\text{D}},\,P_\pm(\CE^{\bcheck{C}}) \rangle &= \cO(\lambda)\,,
\end{split}
\end{align}
where $\VV_{\sfi}^\pm\equiv \VV_{\sfi}\pm \widetilde{\VV}_{\sfi}$\,, we can show
\begin{align}
 &\tfrac{1}{3}\,\langle [P_+(E_A),\,P_+(E_B)]_{\text{D}},\,P_+(E_C)\rangle\,\langle [P_+(E^A),\,P_+(E^B)]_{\text{D}},\,P_+(E^C)\rangle
\nn\\
 &\quad + \langle [P_+(E_A),\,P_+(E_B)]_{\text{D}},\,P_-(E_C)\rangle\,\langle [P_+(E^A),\,P_+(E^B)]_{\text{D}},\,P_-(E^C)\rangle 
\\
 &\overset{\lambda\to 0}{\to} \tfrac{1}{3}\,\langle [\mathbb{P}_+(\CE_{\bcheck{A}}),\,\mathbb{P}_+(\CE_{\bcheck{B}})]_{\text{D}},\,\mathbb{P}_+(\CE_{\bcheck{C}})\rangle\,\langle [\mathbb{P}_+(\CE^{\bcheck{A}}),\,\mathbb{P}_+(\CE^{\bcheck{B}})]_{\text{D}},\,\mathbb{P}_+(\CE^{\bcheck{C}})\rangle
\nn\\
 &\qquad + \langle [\mathbb{P}_+(\CE_{\bcheck{A}}),\,\mathbb{P}_+(\CE_{\bcheck{B}})]_{\text{D}},\,\mathbb{P}_-(\CE_{\bcheck{C}})\rangle\,\langle [\mathbb{P}_+(\CE^{\bcheck{A}}),\,\mathbb{P}_+(\CE^{\bcheck{B}})]_{\text{D}},\,\mathbb{P}_-(\CE^{\bcheck{C}})\rangle 
 + \Delta \,,
\end{align}
where $\Delta\equiv \tfrac{1}{24}\,\Delta_+ + \frac{1}{8}\,\Delta_-$ with
\begin{align}
 \Delta_\pm &\equiv \lim_{\lambda\to 0} \hh^{\sfi\sfi'}\,\hh^{\sfj\sfj'}\,\hh^{\sfk\sfk'}\,\langle [\VV_{\sfi}^+,\,\VV_{\sfj}^+]_{\text{D}},\,\VV_{\sfk}^{\pm}\rangle\,\langle [\VV_{\sfi'}^+,\,\VV_{\sfj'}^+]_{\text{D}},\,\VV_{\sfk'}^{\pm}\rangle\,.
\end{align}
Using, $\langle [\widetilde{\VV}_{\sfi},\,\widetilde{\VV}_{\sfj}]_{\text{D}},\, \VV_{\sfk} \rangle = - \langle \widetilde{\VV}_{\sfj},\, [\widetilde{\VV}_{\sfi},\,\VV_{\sfk}]_{\text{D}} \rangle=0$\,, we can show
\begin{align}
 \langle [\VV_{\sfi}^+,\,\VV_{\sfj}^+]_{\text{D}},\, \VV_{\sfk}^{\pm} \rangle
 &= \pm \langle [\VV_{\sfi},\,\VV_{\sfj}]_{\text{D}},\, \widetilde{\VV}_{\sfk} \rangle
 + 2\,\langle [\VV_{\sfi},\,\widetilde{\VV}_{\sfj}]_{\text{D}},\, \VV_{\sfk} \rangle
 \pm \langle [\widetilde{\VV}_{\sfi},\,\widetilde{\VV}_{\sfj}]_{\text{D}},\, \widetilde{\VV}_{\sfk} \rangle
\nn\\
 &= -(2\pm 1)\,f_{\sfi\sfj}{}^{\sfl}\,\hh_{\sfl\sfk} 
 \mp 3\,\zeta_{[\sfi}{}^{\sfi'}\,\zeta_{\sfj}{}^{\sfj'}\,\hh_{\sfk]\sfk'}\,f_{\sfi'\sfj'}{}^{\sfk'} 
 + \cO(\lambda^{3}) \,,
\end{align}
where, in the second equality, we have used
\begin{align}
 \langle [\widetilde{\VV}_{\sfi},\,\widetilde{\VV}_{\sfj}]_{\text{D}},\, \widetilde{\VV}_{\sfk} \rangle
 = 3\,\langle [\sfK_{[\sfi},\,\sfK_{\sfj}]_{\text{D}},\, \tilde{\sfK}_{\sfk]} \rangle
 + \cO(\lambda^{3}) 
 = -3\,\zeta_{[\sfi}{}^{\sfi'}\,\zeta_{\sfj}{}^{\sfj'}\,\hh_{\sfk]\sfk'}\,f_{\sfi'\sfj'}{}^{\sfk'}
 + \cO(\lambda^{3}) \,.
\end{align}
Thus, in general, we find that $\Delta_{\pm}=\cO(\lambda^{-1})$ and it may diverge under the limit $\lambda\to 0$\,. 
However, this divergence is not important because it is associated with an unphysical space that is orthogonal to the reduced $2d$-dimensional doubled space $\check{\cM}$\,. 
Fortunately, $\Delta$ is composed of the duality-invariant quantities $\{f_{\sfi\sfj}{}^{\sfk},\,\zeta_{\sfi}{}^{\sfj},\,\hh_{\sfi\sfj}\}$, and is invariant under the PL $T$-plurality.\footnote{$\Delta$ does not contain an $\cO(\lambda^0)$ term and only the duality-invariant $\cO(\lambda^{-1})$ term survives under the limit.} 
The finite term $\bm{\delta}$ is also invariant because $\tilde{\sfK}_{\sfl}{}^A \cF_A$ is also invariant.
Then, using the quantities that are invariant under the PL $T$-plurality, we can express the finite quantity $\widetilde{\text{GR}}_+$ as
\begin{align}
 \widetilde{\text{GR}}_+ = \text{$\mathbb{G}$R}_+ = \lim_{\lambda\to 0}\bigl(\text{GR}_+ - \Delta - \bm{\delta}\bigr)\,.
\end{align}
Thus the dilaton equations of motion $\widetilde{\text{GR}}_+=0$ is invariant under the PL $T$-plurality. 

\subsubsection{Spectator fields}

Here we discuss the extension of the previous discussion to the case with spectator fields. 
Namely, so far we have assumed that the metric $\hat{\cH}_{AB}$ (or $\hat{E}^{ab}$) is constant, but here we allow these to depend on the spectator fields $y^\mu$ ($\mu=1,\dots,m$), which are also called the external coordinates. 
Accordingly, we introduce the coordinates $(\hat{x}^{\hat{m}})=(y^\mu,\,x^m)$ $(\hat{m}=1,\dotsc, \hat{D}\equiv D+m)$. 
The doubled coordinates are denoted as $x^{\hat{M}}$ ($\hat{M}=1,\dots,2\hat{D}$), and we construct the generalized metric as
\begin{align}
 \cH_{\hat{M}\hat{N}}(\hat{x}) = E_{\hat{M}}{}^{\hat{A}}(x)\,E_{\hat{N}}{}^{\hat{B}}(x)\,\hat{\cH}_{\hat{A}\hat{B}}(y)\,,\qquad
 E_{\hat{M}}{}^{\hat{A}}(x) = \left(\begin{array}{c|c} \bm{1}_{2m} & 0 \\ \hline 0 & E_M{}^A(x)\end{array}\right) ,
\label{eq:H-E-def}
\end{align}
where $\{\hat{A}\}=\{\alpha,\,a\}$ and $\alpha=1,\dotsc,m$\,. 
We also introduce the vielbein $V_{\hat{A}}{}^{\cB}(y)\in\OO(\hat{D},\hat{D})$ as
\begin{align}
 \hat{\cH}_{\hat{A}\hat{B}}(y) = V_{\hat{A}}{}^{\cC}(y)\,V_{\hat{B}}{}^{\cD}(y)\,H_{\cC\cD}\,,\qquad
 \eta_{\hat{A}\hat{B}} = V_{\hat{A}}{}^{\cC}(y)\,V_{\hat{B}}{}^{\cD}(y)\,\,\eta_{\cC\cD}\,, 
\end{align}
where we have defined
\begin{align}
 H_{\cA\cB}\equiv \begin{pmatrix} \eta_{\dot{a}\dot{b}} & 0 \\ 0 & \eta^{\dot{a}\dot{b}} \end{pmatrix} ,\qquad
 \eta_{\cA\cB} \equiv \begin{pmatrix} \eta_{\dot{a}\dot{b}} & 0 \\ 0 & -\eta^{\dot{a}\dot{b}} \end{pmatrix},
\end{align}
and $\eta_{\dot{a}\dot{b}}$ ($\dot{a}, \dot{b}=1,\dotsc,\hat{D}$) are diagonal matrix with elements $\pm 1$ and $\eta^{\dot{a}\dot{b}}$ is its inverse. 
We also denote the inverse of $V_{\hat{A}}{}^{\cB}$ as $V_{\cA}{}^{\hat{B}}$. 
In addition to the metric, we also allow the DFT dilaton to depend on the spectator fields as
\begin{align}
 \Exp{-2d}= \Exp{-2\,\hat{d}(y)} \abs{\det (\ell_m^a)}\,.
\end{align}
We then find that the equations of motion, in the flux formulation, can be expressed as $\bm{\cR}=0$ and $\bm{\cG}^{\hat{A}\hat{B}}=0$ with
\begin{align}
 \bm{\cR} &= H^{\cA\cB}\, \bigl(2\,\bm{\cD}_{\cA} \bm{\cF}_{\cB} - \bm{\cF}_{\cA}\,\bm{\cF}_{\cB}\bigr) 
 + \frac{1}{12}\, H^{\cA\cD}\,\bigl(3\,\eta^{\cB\cE}\,\eta^{\cC\cF} - H^{\cB\cE}\, H^{\cC\cF}\bigr) \,\bm{\cF}_{\cA\cB\cC}\,\bm{\cF}_{\cD\cE\cF} \,,
\\
 \bm{\cG}^{\cA\cB} &= 2\,H^{\cD[\cA}\, \bm{\cD}^{\cB]} \bm{\cF}_{\cD} 
 -\frac{1}{2}\,H^{\cD\cE}\,(\eta^{\cA\cF}\,\eta^{\cB\cG}-H^{\cA\cF}\,H^{\cB\cG})\,\bigl(\bm{\cF}_{\cD}-\bm{\cD}_{\cD}\bigr)\, \bm{\cF}_{\cE\cF\cG} 
\nn\\
 &\quad -H_{\cE}{}^{[\cA}\,\bigl(\bm{\cF}_{\cD}-\bm{\cD}_{\cD}\bigr)\, \bm{\cF}^{\cB]\cD\cE} 
  +\frac{1}{2}\,\bigl(\eta^{\cC\cE}\,\eta^{\cD\cF} - H^{\cC\cE}\, H^{\cD\cF}\bigr)\, H^{\cG[\cA}\,\bm{\cF}_{\cC\cD}{}^{\cB]}\,\bm{\cF}_{\cE\cF\cG} \,,
\end{align}
where $\bm{\cD}_{\cA}\equiv \hat{V}_{\cA}{}^{\hat{B}}\,E_{\hat{B}}{}^{\hat{M}}\,\partial_{\hat{M}}$ and the generalized fluxes are given by
\begin{align}
\begin{split}
 \bm{\cF}_{\cA} &= \hat{\cF}_{\cA}(y) + V_{\cA}{}^{\hat{B}}(y) \, \cF_{\hat{B}}\,,
\\
 \bm{\cF}_{\cA\cB\cC} &= \hat{\cF}_{\cA\cB\cC}(y) + V_{\cA}{}^{\hat{D}}(y)\,V_{\cB}{}^{\hat{E}}(y)\,V_{\cC}{}^{\hat{F}}(y)\,F_{\hat{D}\hat{E}\hat{F}}\,,
\end{split}
\end{align}
where $\cF_{\hat{A}}$ and $F_{\hat{A}\hat{B}\hat{C}}$ are extensions of the dilaton flux and the structure constants of the Drinfel'd double whose non-vanishing components are $\cF_{A}$ and $F_{ABC}$\,. 
We have also defined
\begin{align}
 \hat{\cF}_{\cA\cB\cC} \equiv 3\,\hat{\Omega}_{[\cA\cB\cC]}\,, \qquad 
 \hat{\cF}_{\cA} \equiv \hat{\Omega}^{\cB}{}_{\cA\cB} + 2\,\bm{\cD}_{\cA} \hat{d} \,, \qquad
 \hat{\Omega}_{\cA\cB\cC} \equiv -\bm{\cD}_{\cA} V_{\cB}{}^{\hat{D}}\, V_{\hat{D}\cC} \,.
\end{align}

In this general setup, the PL $T$-plurality is an $\OO(D,D)$ transformation $C_{A}{}^{B}$ that is embedded into the $\OO(\hat{D},\hat{D})$ matrix $\hat{C}_{\hat{A}}{}^{\hat{B}}$ with the components
\begin{align}
 \hat{C}_{\alpha}{}^{\beta}=\hat{C}^\beta{}_\alpha =\delta_{\alpha}^\beta\,,\quad \hat{C}^{\alpha\beta}=\hat{C}_{\alpha\beta}=\hat{C}_{\alpha}{}^B=\hat{C}^{\alpha B}=\hat{C}_A{}^{\beta} = \hat{C}_{A\beta}=0\,,\quad \hat{C}_A{}^B=C_A{}^B\,.
\end{align}
Using this matrix, we can express the PL $T$-plurality transformation as
\begin{align}
 H_{\cA\cB}\to H_{\cA\cB}\,,\quad 
 \hat{d} \to \hat{d}\,,\quad
 F_{ABC} \to C_A{}^D\,C_B{}^E\,C_C{}^F\, F_{DEF}\,,\quad
 V_{\cA}{}^{\hat{B}}\to (\hat{C}^{-1})_{\hat{C}}{}^{\hat{B}}\,V_{\cA}{}^{\hat{C}}\,.
\label{eq:PL-T-spectator}
\end{align}
We can easily see that the generalized fluxes $\bm{\cF}_{\cA}$ and $\bm{\cF}_{\cA\cB\cC}$ are invariant. 
Moreover, in the expression for $\cR$ or $\cG^{\cA\cB}$\,, the derivative $\bm{\cD}_{\cA}$ is contained but all quantities appearing there depend only on $y^\mu$ and we find
\begin{align}
 \bm{\cD}_{\cA} = \hat{V}_{\cA}{}^{\hat{B}}\,E_{\hat{B}}{}^{\mu}\,\partial_{\mu} = \hat{V}_{\cA}{}^{\beta}\,\delta_{\beta}^{\mu}\,\partial_{\mu}\,.
\end{align}
Since the PL $T$-plurality \eqref{eq:PL-T-spectator} does not transform $\hat{V}_{\cA}{}^{\beta}$\,, the derivative $\bm{\cD}_{\cA}$ is invariant under the transformation. 
Then we can show the invariance of the equations of motion 
\begin{align}
 \bm{\cR}' =\bm{\cR} \,,\qquad \bm{\cG}'^{\cA\cB}=\bm{\cG}^{\cA\cB}\,.
\end{align}
Recalling the relations $\bm{\cR}=\text{\textbf{GR}}_+$ and $\tfrac{1}{2}\,\bm{\cG}(X,\,Y) = \text{\textbf{GR}}_+(X,\,Y) - \text{\textbf{GR}}_+(Y,\,X)$, we can also express these as the invariance of the $(m+D)$-dimensional generalized Ricci tensors
\begin{align}
 \text{\textbf{GR}}'_+ =\text{\textbf{GR}}_+\,,\qquad \text{\textbf{GR}}'_+(X,\,Y)=\text{\textbf{GR}}_+(X,\,Y)\,.
\label{eq:EOM-invariance}
\end{align}

The above discussion is only a review of \cite{1903.12175}, but to consider the dressing cosets, we need to perform the reduction of the $(m+D)$-dimensional tensors to the $(m+d)$-dimensional ones. 
When we show the invariance \eqref{eq:EOM-invariance}, the coordinates $x^m$ on the group manifold $G$ and the spectator fields $y^\mu$ have played different roles.
However, when we consider their reductions, it is important to notice that $y^\mu$ play the same role as the $d$-dimensional coordinates $x^{\check{m}}$ on the dressing coset. 
In Eq.~\eqref{eq:CE-setup1} we have introduced the generalized vector fields $\CE_{\bcheck{A}}$\,, but here we consider their extension by including the external directions. 
We shall keep using the same notation for the enlarged ones $\CE_{\bcheck{A}}$ $(\bcheck{A}=1,\dotsc,2m+2d)$ and we suppose that they are orthogonal to $\VV_{\sfi}$ and $\widetilde{\VV}_{\sfi}$ at any points. 
We also suppose that they are gauge invariant $\gLie_{\VV_{\sfi}}\CE_{\bcheck{A}}=0$\,, form a closed algebra $[\CE_{\bcheck{A}},\,\CE_{\bcheck{B}}]_{\text{D}}=-\bm{f}_{\bcheck{A}\bcheck{B}}{}^{\bcheck{C}}\,\CE_{\bcheck{C}}$\,, and satisfy Eq.~\eqref{eq:CE-setup2}. 
If the spectator fields can be introduced in that way, we only repeat the same discussion as sections \ref{sec:DFT-EOM-1} and \ref{sec:DFT-EOM-2}. 
By taking the Sfetsos limit, $2n$ dimensions spanned by $\VV_{\sfi}$ and $\widetilde{\VV}_{\sfi}$ can be eliminated and the $(m+D)$-dimensional tensors $\bm{\cG}^{\cA\cB}$ and $\bm{\cR}$ reduce to the $(m+d)$-dimensional tensors on the dressing coset. 
Using the invariance \eqref{eq:EOM-invariance} of the higher-dimensional tensors, the $(m+d)$-dimensional tensors, namely, the equations of motion are the same in all of the duality frames that are related by the PL $T$-plurality. 

\subsubsection{Ramond--Ramond fields}
\label{sec:R-R}

In this subsection, we review the R--R sector of DFT, and find the transformation rule for the R--R field strength under the PL $T$-plurality. 
We here follow the approach of \cite{1107.0008} that is based on \cite{hep-th:9907132} and describe the R--R fields as an $\OO(10,10)$ spinor. 
This is useful when we consider only the bosonic fields, but when we also consider fermions, it may be more useful to take the approach of \cite{1206.3478} that is based on \cite{hep-th:9907152,hep-th:9912236} and describes the R--R fields as a bi-spinor of the double Lorentz group $\OO(1,9)\times \OO(9,1)$. 

To make the indices simple, only in this subsection, we shall use the notation $m,n=0,\dotsc,9$\,, $a,b=0,\dotsc,9$\,, $({}^M)=({}^m,\,{}_m)$, and $({}^A)=({}^a,\,{}_a)$. 
We introduce the real gamma matrix $(\Gamma_{M})=(\Gamma_m,\,\Gamma^m)$ satisfying $(\Gamma_M)^\rmT=\Gamma^M$ and
\begin{align}
 \{\Gamma_{M},\,\Gamma_{N}\} = \eta_{MN}\qquad 
 \bigl(\Leftrightarrow \ 
 \{\Gamma^m,\,\Gamma_n\}=\delta^m_n\,,\quad \{\Gamma^m,\,\Gamma^n\}=0=\{\Gamma_m,\,\Gamma_n\}\ \bigr)\,.
\end{align}
We then regard $\Gamma^m$ and $\Gamma_m$ as the creation and the annihilation operators, respectively. 
The Clifford vacuum $\ket{0}$\,, which is normalized as $\langle 0\ket{0}=1$ for $\bra{0}\equiv \ket{0}^\rmT$, is defined as $\Gamma_m\,\ket{0}=0$ and an $\OO(10,10)$ spinor can be expanded as
\begin{align}
 \ket{S} = \sum_{p=0}^9 \frac{1}{p!}\,S_{m_1\cdots m_p}\,\Gamma^{m_1\cdots m_p}\ket{0}\,,\qquad 
 \Gamma^{m_1\cdots m_p}\equiv \Gamma^{[m_1}\cdots\Gamma^{m_p]}\,.
\end{align}
Under an $\OO(10,10)$ transformation by $(h_M{}^N)\in\OO(10,10)$,
\begin{align}
 \Gamma^M\to \Gamma'^M = (h^{\rmT})^M{}_N \, \Gamma^N\,,\qquad
 \Gamma_M\to \Gamma'_M = (h^{-1})_M{}^N \, \Gamma_N\,,
\end{align}
an $\OO(10,10)$ spinor transforms as
\begin{align}
 \ket{S} \to \ket{S'} = S_h\,\ket{F}\,,
\end{align}
where $S_h$ is defined through
\begin{align}
 S_h\,\Gamma_M\,S_h^{-1} = (h^{-1})_M{}^N \, \Gamma_N \,. 
\end{align}
For example, the operator $S_h$ associated with the $\GL(10)$ transformation $h_a$\,, $B$-shift $h_b$\,, or the factorized $T$-duality $h_z$\,,
\begin{align}
 h_a \equiv \begin{pmatrix} a_m{}^n &0 \\ 0& (a^{-1})_n{}^m \end{pmatrix},\quad
 h_b \equiv \begin{pmatrix} \delta_m^n & b_{mn} \\ 0& \delta^m_n \end{pmatrix},\quad
 h_z \equiv \begin{pmatrix}
 \delta_m^n -\delta_m^z\,\delta^n_z & \delta_m^z\,\delta_n^z \\
 \delta^m_z\,\delta^n_z & \delta^m_n -\delta^m_z\,\delta_n^z
\end{pmatrix},
\end{align}
are found as
\begin{align}
 S_{h_a} = \Exp{\frac{1}{2}\,\rho_m{}^n\,[\Gamma^m,\,\Gamma_n]} \quad \bigl(\rho \equiv \ln a \bigr)\,,\qquad
 S_{h_b} = \Exp{\frac{1}{2}\,b_{mn}\,\Gamma^{mn}} \,,\qquad 
 S_{h_z} = \bigl(\Gamma^z - \Gamma_z\bigr)\,\Gamma^{11}\,,
\end{align}
where we have defined the chirality operator
\begin{align}
 \Gamma^{11} \equiv (-1)^{\Gamma^m\,\Gamma_m} \qquad
 \bigl[(\Gamma^{11})^{\rmT}=\Gamma^{11}\,,\quad (\Gamma^{11})^{-1}=\Gamma^{11}\,,\quad \{\Gamma^{11},\,\Gamma_M\}=0\bigr]\,. 
\end{align}
It is noted that, under a $\GL(10)$ transformation, the spinor transforms as
\begin{align}
 \ket{S}\to \ket{S'} = \sum_{p=0}^9 \frac{1}{p!}\,S'_{m_1\cdots m_p}\,\Gamma^{m_1\cdots m_p}\ket{0}\,,\qquad 
 S'_{m_1\cdots m_p}=\frac{a_{m_1}{}^{n_1}\cdots a_{m_p}{}^{n_p}}{\sqrt{\abs{\det (a_m{}^n)}}}\,S_{n_1\cdots n_p}\,,
\end{align}
where the determinant factor comes from $S_{h_a}\ket{0}=\Exp{-\frac{1}{2} \tr \rho}\Exp{\rho_m{}^n\, \Gamma^m\,\Gamma_n}\ket{0}= \abs{\det (a_m{}^n)}^{-\frac{1}{2}}\ket{0}$\,. 
Namely, the components $S_{m_1\cdots m_p}$ of the $\OO(10,10)$ spinor behaves as tensor densities with weight $-\frac{1}{2}$\,. 
We also note that when $a_m{}^n$ has a negative determinant, we decompose it into a $\GL(10)$ transformation with a positive determinant and a parity transformation. 
For example, the parity transformation along the $x^i$-direction is generated by
\begin{align}
 S_{h_i} = [\Gamma^i,\,\Gamma_i] = (S_{h_i})^{-1}\,. 
\end{align}

The R--R field strength is defined as an $\OO(10,10)$ spinor $\ket{\cF}$ that transforms as
\begin{align*}
 \gLie_V \ket{\cF} = \bigl(V^M\,\partial_M + \partial_M V_N \,\Gamma^{MN}\bigr)\,\ket{\cF} \,,
\end{align*}
under the generalized Lie derivative. 
This is defined to have a definite chirality
\begin{align}
 \Gamma^{11}\,\ket{\cF} = \pm \ket{\cF} \,,
\label{eq:chirality}
\end{align}
in type IIA/IIB theory. 
To write down the action, we define the charge conjugation matrix as
\begin{align}
 \cC \equiv (\Gamma^0+\Gamma_0) \cdots (\Gamma^9+\Gamma_9) \qquad
 \bigl[\cC\,\Gamma^M\,\cC^{-1} = -(\Gamma^M)^\rmT \,,\quad \cC^{-1}= -\cC = \cC^\rmT\bigr]\,,
\end{align}
and introduce the notation
\begin{align}
 \overline{\bra{\cF}} \equiv \bra{\cF}\,\cC^\rmT\,,
\end{align}
where
\begin{align}
 \bra{\cF} \equiv (\ket{\cF})^\rmT = \sum_p \frac{1}{p!}\, \cF_{m_1\cdots m_p}\,\bra{0}\,(\Gamma^{m_p})^\rmT \cdots (\Gamma^{m_1})^\rmT \,.
\end{align}
We also define $\cK_{\bm{g}}$ and $\cK_{\cH}$ as
\begin{align}
\begin{split}
 &\cK_{\bm{g}}\,\Gamma_M\, \cK_{\bm{g}}^{-1} = -\bm{g}_{M}{}^N \, \Gamma_N \,,\qquad 
 (\bm{g}_{MN})\equiv \begin{pmatrix} g_{mn} & 0 \\ 0 & g^{mn} \end{pmatrix},
\\
 &\cK_{\cH}\,\Gamma_M\, \cK_{\cH}^{-1} = - \cH_{M}{}^N\, \Gamma_N \,,\qquad 
 \cK_{\cH} = \Exp{\bm{B}} \cK_{\bm{g}} \Exp{-\bm{B}}\,, \qquad \bm{B} \equiv \frac{1}{2}\,B_{mn}\,\Gamma^{mn}\,. 
\end{split}
\end{align}
When the metric $g_{mn}$ takes the form $g_{mn}=e_m{}^a\,e_n{}^b\,\eta_{ab}$ with $(\eta_{ab})=\diag(-1,+1,\dotsc,+1)$, we can construct $\cK_{\bm{g}}$ as
\begin{align}
 \cK_{\bm{g}}=\cC\,(S_{h_e})^\rmT\,\cK_\eta\,S_{h_e}\,,\qquad
 \cK_\eta\equiv [\Gamma^0,\,\Gamma_0]\,,
\end{align}
where $\cK_\eta$ and $S_{h_e}$ are associated with the Minkowski metric and the $\GL(10)$ matrix $e_m{}^a$\,, respectively. 
For arbitrary $\OO(D,D)$ spinors $\ket{\alpha}$ and $\ket{\beta}$\,, we find
\begin{align}
 \overline{\bra{\alpha}}\,\cK_{\bm{g}}\,\ket{\beta}
 = -\sqrt{\abs{g}}\,\sum_p \frac{1}{p!}\,g^{m_1n_1}\cdots g^{m_pn_p}\,\alpha_{m_1\cdots m_p}\,\beta_{n_1\cdots n_p} = \overline{\bra{\beta}}\,\cK_{\bm{g}}\,\ket{\alpha}\,.
\end{align}
Then the Lagrangian for the R--R sector is given by
\begin{align}
 \cL = \frac{1}{4}\Exp{-2\,d}\overline{\bra{\cF}}\, S_{\cH}\, \ket{\cF} \,.
\end{align}
We can relate $\ket{\cF}$ with the standard supergravity fields as
\begin{align}
 \ket{\cF} = \Exp{d} \ket{F} = \Exp{d} \Exp{\bm{B}} \ket{G}\,,
\end{align}
where
\begin{align}
 \ket{F}\equiv \sum_p \frac{1}{p!}\,F_{m_1\cdots m_p}\,\Gamma^{m_1\cdots m_p}\ket{0}\,,\qquad
 \ket{G}\equiv \sum_p \frac{1}{p!}\,G_{m_1\cdots m_p}\,\Gamma^{m_1\cdots m_p}\ket{0}\,,
\end{align}
and $F_{m_1\cdots m_p}$ and $G_{m_1\cdots m_p}$ are components of the polyform $F = \Exp{B_2\wedge} G$\,.
Using these fields, the Lagrangian can be expressed in the standard form
\begin{align}
 \cL = \frac{1}{4} \overline{\bra{G}}\, S_{\bm{g}}\, \ket{G} = -\frac{1}{4} \sqrt{\abs{g}}\, \sum_p \frac{1}{p!}\,G_{m_1\cdots m_p}\,G^{m_1\cdots m_p}\,.
\end{align}
Due to the chirality \eqref{eq:chirality}, the summation over $p$ is restricted to even/odd number in type IIA/IIB theory. 
In addition to the chirality constraint, we also impose the self-duality relation
\begin{align}
 \ket{\cF} = \cK_{\cH} \, \ket{\cF} \,,
\label{eq:RR-self-dual}
\end{align}
at the level of the equations of motion. 
The R--R field strength $\ket{\cF}$ is defined as
\begin{align}
 \ket{\cF} = \bigl(\sla{\partial} - \sla{\partial}d\bigr)\,\ket{\cA} \qquad \bigl(\sla{\partial} \equiv \Gamma^M\,\partial_M\bigr)\,,
\end{align}
by using the R--R potential $\ket{\cA}$, and then we can check the Bianchi identity
\begin{align}
 \bigl(\sla{\partial} - \sla{\partial}d\bigr)\,\ket{\cF} = 0 \,,
\label{eq:RR-Bianchi}
\end{align}
by using the nilpotency $(\sla{\partial} + \sla{\partial}d)^2 =0$ that is ensured by the section condition. 
When we consider the generalized supergravity, the dilaton derivative receives a shift
\begin{align}
 \partial_M d \to \partial_M d + \bm{X}_M\,, \qquad \bm{X}_M\equiv (0,\,I^m)\,,
\label{eq:GSE-shift}
\end{align}
and the modified Bianchi identity becomes
\begin{align}
 \bigl(\sla{\partial} - \sla{\partial}d - I^m\,\Gamma_m \bigr)\,\ket{\cF} = 0 \,.
\label{eq:RR-Bianchi-modified}
\end{align}
This is equivalent to the equations of motion for the R--R field under the self-duality relation. 

In the presence of the R--R fields, the equations of motion for the DFT dilaton $\cS=0$ is not modified but those for the generalized metric is modified as $\cS_{MN} = \cE_{MN}$ where $\cE_{MN}$ is the energy-momentum tensor
\begin{align}
 \cE_{MN} \equiv -\frac{1}{4}\,\Bigl[\overline{\bra{\cF}} \,\Gamma_{(M}\,S_{\cH}\,\Gamma_{N)}\, \ket{\cF} + \frac{1}{2}\,\cH_{MN}\,\overline{\bra{\cF}}\,S_{\cH}\,\ket{\cF}\Bigr]\,.
\end{align}

For convenience, let us express the equations of motion in (generalized) type II supergravity \cite{1511.05795,1605.04884,1703.09213} in terms of the standard supergravity fields in our convention. 
For the consistency (that is related to the section condition of DFT \cite{1611.05856,1703.09213}), $I^m$ must be a Killing vector field satisfying $\Lie_I g_{mn}=\Lie_I H_3=\Lie_I \Phi=\Lie_I F=0$ \cite{1511.05795}. 
Under this constraint, the shift \eqref{eq:GSE-shift} modifies the dilaton equations of motion $\cS=0$ as
\begin{align}
 R + 4\,\nabla^m Z_m - 4\,Z^m\,Z_m -4\,I^m\,I_m - \tfrac{1}{12}\,H_{mnp}\,H^{mnp} = 0 \,,
\end{align}
where $Z_m\equiv \partial_m \Phi + I^n\, B_{nm}$\,, $I_m\equiv g_{mn}\,I^n$\,, and $Z^m\equiv g^{mn}\,Z_n$\,. 
The equations of motion $\cS_{MN} = \cE_{MN}$ for the generalized metric becomes
\begin{align}
 R_{mn}-\tfrac{1}{4}\,H_{mpq}\,H_n{}^{pq} + \nabla_m Z_n + \nabla_n Z_m &= T_{mn}\,,
\\
 - \tfrac{1}{2}\,\nabla^p H_{pmn} + Z^p\,H_{pmn} + \nabla_m I_n - \nabla_n I_m &= K_{mn}\,,
\end{align}
where we have defined
\begin{align}
\begin{split}
 T_{mn} &\equiv \frac{1}{4}\sum_{p}\bigl[\tfrac{1}{(p-1)!}\,G_{(m}{}^{q_1\cdots q_{p-1}}\,G_{n) q_1\cdots q_{p-1}} -\tfrac{1}{2\,p!}\,g_{mn}\,G_{q_1\cdots q_p}\,G^{q_1\cdots q_p}\bigr]\,,
\\
 K_{mn} &\equiv \frac{1}{4}\sum_{p} \tfrac{1}{(p-2)!}\,G_{q_1\cdots q_{p-2}}\,G_{mn}{}^{q_1\cdots q_{p-2}} \,,
\end{split}
\end{align}
and the summation is over even/odd numbers in type IIA/IIB theory. 
The modified Bianchi identity \eqref{eq:RR-Bianchi-modified} can be expressed as
\begin{align}
 \rmd F - \iota_I F = 0 \,,\qquad
 \rmd G + H_3\wedge G - \iota_I B_2 \wedge G - \iota_I G= 0 \,,
\end{align}
and the self-duality relation \eqref{eq:RR-self-dual}, or $\ket{G} = \cK_{\bm{g}} \, \ket{G}$\,, can be expressed as
\begin{align}
 G_p = (-1)^{\frac{p(p-1)}{2}} * G_{10-p}\,.
\label{eq:G-SD}
\end{align}
As is discussed in \cite{1107.0008}, if we perform a timelike $T$-duality, the overall sign of $\cK_\eta$\,, $\cK_{\bm{g}}$\,, and $S_{\cH}$ is flipped, and the kinetic term of the R--R fields has the wrong sign. 
Accordingly, a solution of (generalized) type IIA/IIB supergravity is mapped to a solution of (generalized) type IIB${}^*$/IIA${}^*$ supergravity. 
Due to the sign flip, the sign of the self-duality relation \eqref{eq:RR-self-dual} is also changed. 
In the examples studied in section \ref{sec:PL-AdS2-RR}, the relation \eqref{eq:G-SD} may have an additional minus sign, but that is due to the timelike $T$-duality. 
It is noted that we use the convention $\epsilon_{0\cdots 9}=+1$ in ten dimensions.
However, if we perform a $\GL(10)$ transformation with negative determinant (which changes the orientation), the sign becomes $\epsilon_{0\cdots 9}=-1$\,. 

Now, we shall discuss the symmetry under the PL $T$-plurality. 
The invariance of the generalized Ricci tensor $\widetilde{\text{GRic}}_{+}(\check{\CE}_{\bcheck{A}},\,\check{\CE}_{\bcheck{B}})$ indicates that $\cS_{MN}$ takes the form
\begin{align}
 \cS_{MN} = \check{\CE}_{M}{}^{\bcheck{A}}\,\check{\CE}_{N}{}^{\bcheck{B}}\,\hat{\cS}_{\bcheck{A}\bcheck{B}}\,,
\end{align}
where $\hat{\cS}_{\bcheck{A}\bcheck{B}}$ is a duality-invariant matrix that can depend only on $y^\mu$. 
Then, under the PL $T$-plurality, the generalized Ricci tensor $\cS_{MN}$ is mapped as
\begin{align}
 \cS'_{MN} = \Lambda_M{}^P\,\Lambda_N{}^Q\,\cS_{PQ}\,,\qquad \Lambda_M{}^N\equiv \check{\CE}'_M{}^{\bcheck{B}}\,\check{\CE}_{\bcheck{B}}{}^N\in\OO(10,10)\,.
\end{align}
To satisfy the equation of motion, $\cE_{MN}$ also should be transformed by the same matrix $\Lambda_M{}^N$\,, and this requirement determines the transformation rule of the R--R field strength,
\begin{align}
 \ket{\cF'} = S_\Lambda \ket{\cF}\,.
\label{eq:RR-rule}
\end{align}
We can rewrite this rule as follows. 
We introduce a constant matrix $T_A{}^{\bcheck{B}}$ that realizes
\begin{align}
 D_{AB} = T_A{}^{\ubar{\bm{c}}}\,T_B{}^{\ubar{\bm{d}}}\,\eta_{\ubar{\bm{c}}\ubar{\bm{d}}} + T_A{}^{\bar{\bm{c}}}\,T_B{}^{\bar{\bm{d}}}\,\eta_{\bar{\bm{c}}\bar{\bm{d}}}\,,\qquad 
 \eta_{AB} = T_A{}^{\ubar{\bm{c}}}\,T_B{}^{\ubar{\bm{d}}}\,\eta_{\ubar{\bm{c}}\ubar{\bm{d}}} - T_A{}^{\bar{\bm{c}}}\,T_B{}^{\bar{\bm{d}}}\,\eta_{\bar{\bm{c}}\bar{\bm{d}}}\,,
\label{eq:def-T}
\end{align}
where $D_{AB}\equiv\diag(\eta_{ab},\,\eta^{ab})$, and define $\check{\CE}_A{}^M\equiv T_A{}^{\bcheck{B}}\,\check{\CE}_{\bcheck{B}}{}^M\in \OO(10,10)$.\footnote{The matrix $T_A{}^{\bcheck{B}}$ is a tool to transform $\check{\CE}_{\bcheck{A}}{}^M$ into an $\OO(10,10)$ matrix $\check{\CE}_A{}^M$, and the choice can be arbitrary. A different choice may change $\ket{\hat{\cF}}$ but the dual R--R field strength $\ket{\cF'}$ is independent of $T_A{}^{\bcheck{B}}$.} 
Then, in the original duality frame, we construct the R--R field strength as
\begin{align}
 \ket{\cF} = S_{\check{\CE}}^{-1}\ket{\hat{\cF}}\,, \qquad \check{\CE}\equiv (\check{\CE}_A{}^M),
\label{eq:cF-dualizability}
\end{align}
where $\ket{\hat{\cF}}$ depends only on $y^\mu$. 
If it is not possible to find such an $\ket{\hat{\cF}}$, we cannot perform the PL $T$-plurality.
Namely, Eq.~\eqref{eq:cF-dualizability} can be understood as a dualizability condition. 
After performing a PL $T$-plurality, we find the dual field $\check{\CE}'_{\bcheck{A}}{}^M$, and again define $\check{\CE}'_A{}^M\equiv T_A{}^{\bcheck{B}}\,\check{\CE}'_{\bcheck{B}}{}^M\in \OO(10,10)$ by using the same $T_A{}^{\bcheck{B}}$\,. 
We then construct the R--R field strength as
\begin{align}
 \ket{\cF'} = S_{\check{\CE}'}^{-1}\ket{\hat{\cF}}\,.
\end{align}
This construction is consistent with the rule \eqref{eq:RR-rule}. 
The self-duality relation \eqref{eq:RR-self-dual} is a covariant under $\OO(10,10)$ transformations, and if it is satisfied in the original configuration $\{\cH_{MN},\,\ket{\cF}\}$, it is also satisfied in the dual one $\{\cH'_{MN},\,\ket{\cF'}\}$\,.
The only non-trivial relation is the Bianchi identity \eqref{eq:RR-Bianchi}. 
Under the ansatz \eqref{eq:cF-dualizability}, this can be expressed as
\begin{align}
 \sla{D} \ket{\hat{\cF}} = 0 \qquad
 \bigl(\sla{D}\equiv \sla{\partial} + \tfrac{1}{3!}\,\bm{f}_{ABC}\,\Gamma^{ABC} - \tfrac{1}{2}\,\bm{f}_A\,\Gamma^A\bigr)\,,
\end{align}
where
\begin{align}
 \bm{f}_{ABC} \equiv -3\,\check{\CE}_{[A}\cdot \check{\CE}_B{}^N\, \check{\CE}_{|N|C]}\,, \qquad 
 \bm{f}_A \equiv - \partial_M \check{\CE}_A{}^M + 2\, \check{\CE}_A{}^M\,(\partial_M d+\bm{X}_M) \,.
\label{eq:bm-f}
\end{align}
Unfortunately, we have failed to show the covariance of this identity under the PL $T$-plurality. 
Consequently, we need to check the Bianchi identity in each duality frame.\footnote{It seems that the covariance of the Bianchi identity has been generally shown in \cite{1810.07763}.}

\section{Examples}
\label{sec:examples}

In this section, we study specific examples of the PL $T$-plurality for dressing cosets and explain in detail how to compute the reduced background fields $\check{E}_{mn}$\,. 
In the previous sections, we have discussed the following four procedures to compute $\check{E}_{mn}$\,:
\begin{enumerate}
\item When the matrix $N_{\sfi\sfj}$ defined in \eqref{eq:Nij-def} is non-degenerate, the reduced background fields are given by Eq.~\eqref{eq:reduced-E}, namely,
\begin{align}
 \check{E}_{mn} = E_{mn} - \bigl(\kk_{\sfi m}-\hat{\kk}_{\sfi m}\bigr)\,N^{\sfi\sfj}\,\bigl(\kk_{\sfj n}+\hat{\kk}_{\sfj n}\bigr)\,.
\label{eq:former-formula}
\end{align}
When $N_{\sfi\sfj}=0$\,, we simply have $\check{E}_{mn} = E_{mn}$\,. 
In all of the examples studied in this paper, the matrix $N_{\sfi\sfj}$ is non-degenerate, and Eq.~\eqref{eq:former-formula} can be applied. 

\item The gauged sigma model gives another formula \eqref{eq:gauge-formula} for the reduced background fields. 
In our examples, depending on the situation, this formula reduces to one of the two simple expressions, i.e., Eqs.~\eqref{eq:KS-reproduced} and \eqref{eq:KSd},
\begin{align}
 \check{E}_{mn} = r_m^a\,\bigl[(\cR+\pi)^{-1}\bigr]_{ab}\,r_n^b \,,\qquad 
 \check{E}_{mn} = r_m^a\,\bigl[(1+\tilde{\cR}\,\pi)^{-1}\,\tilde{\cR}\bigr]_{ab}\,r_n^b \,,
\label{eq:2nd-formula}
\end{align}
where $\cR^{ab}$ and $\tilde{\cR}_{ab}$ are defined in Eqs.~\eqref{eq:cR-1} and \eqref{eq:cR-2}, respectively. 

\item In our examples, the initial configuration is a standard coset. 
In that case, the standard Sfetsos limit can be applied, $\check{E}_{mn} = {\displaystyle\lim_{\lambda\to 0}} E_{mn}$\,. 
Under the PL $T$-plurality, the matrix $\hat{E}^{ab}$ is covariantly transformed as $\hat{E}^{ab}\to \hat{E}'^{ab}=[(\bm{c}+\bm{d}\,\hat{E})\,(\bm{a}+\bm{b}\,\hat{E})^{-1}]^{ab}$ and the $\lambda$-dependence of $\hat{E}'^{ab}$ becomes non-trivial. 
However, since $\lambda$ has been introduced such that $\widetilde{\VV}_{\sfi}$ are reduced to linear combinations of $\VV_{\sfi}$\,, the Sfetsos limit works even after the PL $T$-plurality. 
Namely, we can obtain the dual geometries as $\check{E}'_{mn} = {\displaystyle\lim_{\lambda\to 0}} E'_{mn}$\,, where $E'_{mn}=r'^a_m\,[(\hat{E}'+\pi')^{-1}]_{ab}\,r'^b_n$ and $\{r'^a_m,\,\pi'^{ab}\}$ are associated with the new Manin triple. 

\item In the geometric approach discussed in section \ref{sec:reduced-GM}, we look for certain frame fields $\CE_{\hat{A}}\in\Gamma^{\mathfrak{f}}(\mathbb{F}^\perp)$ that form a closed algebra under the D-bracket. 
We then consider their reduction and obtain $\check{\CE}_{\hat{A}}$.
Then, using Eq.~\eqref{eq:reduced-GM}, we obtain the reduced generalized metric $\check{\cH}_{\check{M}\check{N}}$ and this gives the reduced background fields $\check{E}_{\check{m}\check{n}}$\,. 
\end{enumerate}
As a hybrid of the second and the third approaches, we can consider the following one as well. 
In the second approach, $\cR^{ab}$ and $\tilde{\cR}_{ab}$ are found from Eqs.~\eqref{eq:cR-1} and \eqref{eq:cR-2}, but if we know the non-degenerate matrix $\hat{E}^{ab}$ that contains the parameter $\lambda$\,, we can also compute these as $\cR^{ab}= {\displaystyle\lim_{\lambda\to 0}} \hat{E}^{ab}$ or $\tilde{\cR}_{ab}= {\displaystyle\lim_{\lambda\to 0}} (\hat{E}^{-1})_{ab}$\,. 
This may reduce the amount of the computation. 

In section \ref{sec:PL-AdS2}, we consider various PL $T$-pluralities, and there we take a shortcut approach.
In the original configuration, we take Sfetsos's approach, and the reduced background fields are obtained as $\check{E}_{mn} = r_m^a\,[(\bm{1}+\tilde{\cR}\,\pi)^{-1}\,\tilde{\cR}]_{ab}\,r^b_n$ where $\tilde{\cR}_{ab}= {\displaystyle\lim_{\lambda\to 0}}(\hat{E}^{-1})_{ab}$\,. 
We then perform the PL $T$-pluralities, that is the $\OO(D,D)$ rotations of the form
\begin{align}
 C_A{}^B = \begin{pmatrix} \bm{a}_a{}^b & \bm{b}_{ab} \\ \bm{c}^{ab} & \bm{d}^a{}_b \end{pmatrix}.
\label{eq:C-param}
\end{align}
Under this transformation, $\hat{E}^{ab}$ is transformed as
\begin{align}
 \hat{E}^{ab} \to \bigl[(\bm{c}+\bm{d}\,\hat{E})\,(\bm{a}+\bm{b}\,\hat{E})^{-1}\bigr]^{ab}\,,
\end{align}
and the $D$-dimensional background fields become
\begin{align}
\begin{split}
 \check{E}'_{mn} &= r'^a_m\,\bigl\{[(\bm{c}+\bm{d}\,\hat{E})\,(\bm{a}+\bm{b}\,\hat{E})^{-1}+\pi' ]^{-1}\bigr\}_{ab}\,r'^b_n
\\
 &= r'^a_m\,\bigl\{(\bm{a}\,\hat{E}^{-1}+\bm{b})\, [\bm{c}\,\hat{E}^{-1}+\bm{d} +\pi'\,(\bm{a}\,\hat{E}^{-1}+\bm{b})]^{-1}\bigr\}_{ab}\,r'^b_n \,.
\end{split}
\end{align}
By taking the limit $\lambda\to 0$\,, we find a formula
\begin{align}
 \check{E}'_{mn} = r'^a_m\,\bigl\{(\bm{a}\,\tilde{\cR}+\bm{b})\,[\bm{c}\,\tilde{\cR}+\bm{d}+\pi'\,(\bm{a}\,\tilde{\cR}+\bm{b})]^{-1}\bigr\}_{ab}\,r'^b_n\,. 
\label{eq:formula-fast}
\end{align}
In each duality frame, we only need to compute $\{r'^a_m,\,\pi'^{ab}\}$\,. 
Using the matrix $\tilde{\cR}_{ab}$ computed in the original frame, we can easily obtain $\check{E}'_{mn}$\,. 
This will be the fastest way to compute $\check{E}'_{mn}$\,. 

In section \ref{sec:S2}, to make it easy to understand various approaches, we begin with a well-known example: the PL $T$-duality of S$^2$. 
There, the PL $T$-plurality is nothing but the non-Abelian $T$-duality. 
We find that all formulas give the same results. 
Considering a spherically symmetric supergravity solution, we obtain its dual supergravity solution. 

In section \ref{sec:hYB}, we discuss a class of PL $T$-pluralities called the homogeneous Yang--Baxter deformations. 
Here, we start with a standard coset, where the gauge group $F$ is a subgroup of $G$. 
Since the homogeneous Yang--Baxter deformations do not change $G$, even after the deformation, the background is a standard coset. 
The important point is that this is a special class of the PL $T$-pluralities, and thus the homogeneous Yang--Baxter deformations of coset spaces are solutions-generating transformations. 

In section \ref{sec:PL-AdS2}, we study the PL $T$-plurality of AdS$_2$\,. 
In this example, the Drinfel'd double has ten Manin triples, and accordingly, we obtain a chain of ten supergravity backgrounds. 
In some of the Manin triples, $F$ is not a subgroup of $G$ nor $\tilde{G}$, and their PL $T$-dualities give some pairs of non-trivial dressing cosets. 
In this non-trivial example, we again find that all approaches always give the same results. 
Moreover, we consider an $\text{AdS}_2\times \text{S}^2\times \text{T}^6$ solution with $H$-flux, and find that the dualized geometries are solutions of DFT. 

In section \ref{sec:PL-AdS2-RR}, we apply the results of section \ref{sec:PL-AdS2} to another $\text{AdS}_2\times \text{S}^2\times \text{T}^6$ solution with R--R fields and study its PL $T$-plurality. 
When the timelike $T$-duality is involved, we find a solution of (generalized) type II$^*$ supergravity equations of motion. 

\subsection{Poisson--Lie $T$-duality of S$^2$}
\label{sec:S2}

We start with a simple example of the usual coset $\text{S}^2=\UU(1)\backslash\SO(3)$. 
We choose the Drinfel'd double as a six-dimensional one (i.e., $D=3$) with non-vanishing structure constants
\begin{align}
 f_{12}{}^3 = 1\,,\qquad
 f_{23}{}^1 = 1\,,\qquad
 f_{13}{}^2 = -1\,.
\label{eq:9-1}
\end{align}
This Drinfel'd double is called DD5 in the classification of \cite{math:0202210,hep-th:0403164}. 
We introduce the local coordinates $(x^m)=(\theta,\,\phi,\,\psi)$ on $\SO(3)$ and parameterize the group element as 
\begin{align}
 g=\Exp{\psi\,T_3}\Exp{\theta\,T_1}\Exp{-\phi\,T_3}.
\end{align}
The right-invariant vector fields/1-form fields become
\begin{align}
 (e_a{}^m) = {\footnotesize \begin{pmatrix} \cos \psi & -\frac{\sin \psi}{\sin\theta} & - \frac{\sin \psi}{\tau\theta} \\
 \sin \psi & \frac{\cos \psi}{\sin\theta} & \frac{\cos \psi}{\tan \theta} \\ 0 & 0 & 1\end{pmatrix}},\qquad 
 (r^a{}_m) = {\footnotesize \begin{pmatrix}
 \cos \psi & -\sin \theta \sin \psi & 0 \\
 \sin \psi & \sin \theta \cos \psi & 0 \\
 0 & -\cos \theta & 1 \end{pmatrix}},
\end{align}
and we have $\pi^{ab}=0$ because $f_a{}^{bc}=0$\,. 
We choose the matrix $\hat{E}^{ab}$ and its inverse matrix as
\begin{align}
 \hat{E}^{ab} = \begin{pmatrix}
 \frac{a-b}{a^2+b^2} & -\frac{a+b}{a^2+b^2} & 0 \\
 \frac{a+b}{a^2+b^2} & \frac{a-b}{a^2+b^2} & 0 \\
 0 & 0 & \lambda^{-1} \end{pmatrix},\qquad 
 \hat{E}_{ab} = \begin{pmatrix} \frac{a-b}{2} & \frac{a+b}{2} & 0 \\ -\frac{a+b}{2} & \frac{a-b}{2} & 0 \\ 0 & 0 & \lambda
\end{pmatrix},
\label{eq:S2-Ehat}
\end{align}
which may depend on the spectator fields. 
The parameter $\lambda$ is supposed to be $\lambda=1$\,, but this plays an important role when we take the Sfetsos limit $\lambda\to 0$\,. 
We can check that the matrix $\hat{E}^{ab}$ satisfies the invariance condition \eqref{eq:inv-orig} for infinitesimal transformation $T_{\sfi} = T_3$ ($\sfi=1$), and we choose the gauge algebra as $\mathfrak{f}=\text{span}\{T_3\}$\,. 

Using Eq.~\eqref{eq:E-PL}, we obtain the $D$-dimensional background fields $E_{mn}=g_{mn}+B_{mn}$ as
\begin{align}
 \rmd s^2 = \frac{a-b}{2}\,\bigl(\rmd\theta^2+\sin^2 \theta\,\rmd\phi^2\bigr) + \lambda\,\bigl(\rmd\psi - \cos\theta\,\rmd\phi\bigr)^2\,,
\qquad
 B_2 = \frac{a+b}{2}\sin \theta\,\rmd\theta\wedge\rmd \phi\,.
\label{eq:S2-original}
\end{align}
They are defined on the coset space $\text{DD5}/\tilde{G}_{\bm{1}}$ where $\tilde{G}_{\bm{1}}$ is the three-dimensional Abelian group associated with the Abelian algebra generated by $\{T^a\}$\,. 
From the metric and the $B$-field, we can compute the generalized metric $\cH_{MN}$\,. 
Then, we can check $\gLie_{E_3}\cH_{MN}=0$\,. 
Since $E_3 = \partial_\psi$\,, this gauge symmetry generates the $\UU(1)$ symmetry that shifts the angular coordinate $\psi$. 
Here we choose the gauge-fixing condition as $\psi=0$\,. 

Let us compute the reduced background fields $\check{E}_{\check{m}\check{n}}$ on the $d\,(=2)$-dimensional coset space $\check{M}=\UU(1)\backslash\text{DD5}/A_3$\,. 
If we denote the generator of the gauge algebra $\mathfrak{f}$ as $T_{\sfi}= \VV_{\sfi}{}^A\,T_A$ and denote the corresponding generalized vector field as $\VV_{\sfi}{}^M= \VV_{\sfi}{}^A\,E_A{}^M$, we find
\begin{align}
\begin{split}
 \VV_{\sfi=1}{}^A &\equiv \VV^A = \bigl(\mathsf{\kk}^a\,;\, \tilde{\kk}_{a}\bigr) = ( 0,\, 0,\, 1\,;\, 0,\, 0,\, 0)\,,
\\
 \VV_{\sfi=1}{}^M &\equiv \VV^M = \bigl(\kk^m,\, \tilde{\kk}_{m}\bigr) = ( 0,\, 0,\, 1\,;\, 0,\, 0,\, 0)\,.
\end{split}
\end{align}
Then the $n\times n$ matrix \eqref{eq:Nij-def} becomes
\begin{align}
 (N_{\sfi\sfj}) = (N_{11}) = \lambda\,,
\end{align}
and the formula \eqref{eq:reduced-E} gives
\begin{align}
 \check{E}_{mn} = E_{mn} - \lambda^{-1}\,\kk_{m}\, \kk_{n} \,,\qquad \kk_{m}\,\rmd x^m = \lambda\,(\rmd \psi - \cos\theta\,\rmd\phi) \,.
\label{eq:subtract}
\end{align}
Namely, we find the reduced background fields as
\begin{align}
 \rmd \check{s}^2 = \frac{a-b}{2}\,\bigl(\rmd\theta^2+\sin^2 \theta\,\rmd\phi^2\bigr) \,, \qquad
 \check{B}_2 = \frac{a+b}{2}\sin \theta\,\rmd\theta\wedge\rmd \phi\,.
\label{eq:S2-limit}
\end{align}
We note that the subtraction of Eq.~\eqref{eq:subtract} has removed the $\lambda$-dependent part of $E_{mn}$\,.

Sfetsos's approach is much faster. 
Starting with the background given in Eq.~\eqref{eq:S2-original}, we only take the limit $\lambda\to 0$\,. 
Without doing the computation of Eq.~\eqref{eq:subtract}, we can quickly obtain the result \eqref{eq:S2-limit} from Eq.~\eqref{eq:S2-original}. 

To consider the geometric approach, let us introduce $\VV^{\ubar{\bm{a}}}$, $\VV^{\bar{\bm{a}}}$, and $\widetilde{\VV}\equiv \widetilde{\VV}_{\sfi=1}$ as
\begin{align}
 \VV^{\ubar{\bm{a}}} &= \frac{1}{\sqrt{2}}\begin{pmatrix} \hat{E}^{ab}\,\kk^{\ubar{\bm{a}}}_{b} \\ \kk^{\ubar{\bm{a}}}_{a} \end{pmatrix},\qquad
 \bar{\VV}^{\bar{\bm{a}}} = \frac{1}{\sqrt{2}}\begin{pmatrix} - \bar{\kk}^{\bar{\bm{a}}}_{b}\,\hat{E}^{ba} \\ \bar{\kk}^{\bar{\bm{a}}}_{a} \end{pmatrix},
\\
 \widetilde{\VV} &= \hat{\cH}(\VV) = (0,\, 0,\, 0\,;\, 0,\, 0,\, \lambda) \,,
\end{align}
where
\begin{align}
 \kk^{\ubar{\bm{1}}}_{a} = \bar{\kk}^{\bar{\bm{1}}}_{a} = \sqrt{\frac{a^2+b^2}{a-b}}\, (1,\,0,\,0)\,,\qquad 
 \kk^{\ubar{\bm{2}}}_{a} = \bar{\kk}^{\bar{\bm{2}}}_{a} = \sqrt{\frac{a^2+b^2}{a-b}}\,(0,\,1,\,0)\,.
\end{align}
By using the flat metric $(\eta_{\ubar{\bm{a}}\ubar{\bm{b}}})=(\eta_{\bar{\bm{a}}\bar{\bm{b}}})= \diag(+1,+1)$, they realize
\begin{align}
\begin{split}
 \eta_{AB} &= \eta_{\ubar{\bm{c}}\ubar{\bm{d}}}\,\VV^{\ubar{\bm{c}}}_{A}\,\VV^{\ubar{\bm{d}}}_{B} - \eta_{\bar{\bm{c}}\bar{\bm{d}}}\,\bar{\VV}^{\bm{c}}_{A}\, \bar{\VV}^{\bm{d}}_{B} + \lambda^{-1}\,\bigl(\VV_{A}\,\widetilde{\VV}_{B}+\widetilde{\VV}_{A}\,\VV_{B}\bigr)\,,
\\
 \hat{\cH}_{AB} &= \eta_{\ubar{\bm{c}}\ubar{\bm{d}}}\,\VV^{\bm{c}}_{A}\,\VV^{\bm{d}}_{B} + \eta_{\bar{\bm{c}}\bar{\bm{d}}}\,\bar{\VV}^{\bm{c}}_{A}\, \bar{\VV}^{\bm{d}}_{B} + \lambda^{-1}\,\bigl(\VV_{A}\,\VV_{B}+\widetilde{\VV}_{A}\,\widetilde{\VV}_{B}\bigr)\,.
\end{split}
\end{align}
We find that $\VV^{\ubar{\bm{a}}M}\equiv \VV^{\ubar{\bm{a}}A}\,E_A{}^M$ and $\bar{\VV}^{\bar{\bm{a}}M}\equiv \bar{\VV}^{\bar{\bm{a}}A}\,E_A{}^M$ are not $\mathfrak{f}$-invariant. 
However, if we consider a local rotation that preserves the Euclidean metrics $\eta_{\ubar{\bm{a}}\ubar{\bm{b}}}$ and $\eta_{\bar{\bm{a}}\bar{\bm{b}}}$\,, 
\begin{align}
\begin{split}
 \begin{pmatrix} G^{\ubar{1}}\\ G^{\ubar{2}} \end{pmatrix}
 = \begin{pmatrix} \cos \psi & \sin \psi \\ -\sin \psi & \cos \psi \end{pmatrix}
 \begin{pmatrix} \VV^{\ubar{1}}\\ \VV^{\ubar{2}} \end{pmatrix} ,
 \qquad
 \begin{pmatrix} \bar{G}^{\bar{1}}\\ \bar{G}^{\bar{2}} \end{pmatrix}
 = \begin{pmatrix} \cos \psi & \sin \psi \\ -\sin \psi & \cos \psi \end{pmatrix}
 \begin{pmatrix} \bar{\VV}^{\bar{1}}\\ \bar{\VV}^{\bar{2}} \end{pmatrix} ,
\end{split}
\end{align}
we find that $G^{\ubar{\bm{a}}}$ and $\bar{G}^{\bar{\bm{a}}}$ are $\mathfrak{f}$-invariant: $\gLie_{\VV} G^{\ubar{\bm{a}}}=0$ and $\gLie_{\VV} \bar{G}^{\bar{\bm{a}}} = 0$\,.
They can be related to $H_{\bcheck{A}}$ introduced around Eq.~\eqref{eq:Lie-K-0} as $H_{\ubar{\bm{a}}}=\eta_{\ubar{\bm{a}}\ubar{\bm{b}}}\,G^{\ubar{\bm{b}}}$ and $\bar{H}_{\bar{\bm{a}}}=\eta_{\bar{\bm{a}}\bar{\bm{b}}}\,\bar{G}^{\bar{\bm{b}}}$. 
Their explicit components are as follows:
\begin{align}
\begin{split}
 H_{\ubar{\bm{1}}}{}^M &= \tfrac{1}{\sqrt{2}}\,\bigl(\sqrt{\tfrac{a-b}{a^2+b^2}},\, \tfrac{a+b}{\sqrt{(a-b)(a^2+b^2)} \sin \theta},\,\tfrac{a+b}{\sqrt{(a-b)(a^2+b^2)} \tan\theta}\,;\, \sqrt{\tfrac{a^2+b^2}{a-b}},\, 0,\,0 \bigr),
\\
 H_{\ubar{\bm{2}}}{}^M &= \tfrac{1}{\sqrt{2}}\,\bigl( -\tfrac{a+b}{\sqrt{(a-b) (a^2+b^2)}},\, \sqrt{\tfrac{a-b}{a^2+b^2}}\tfrac{1}{\sin \theta},\,\sqrt{\tfrac{a-b}{a^2+b^2}}\tfrac{1}{\tan\theta}\,;\, 0,\, \sqrt{\tfrac{a^2+b^2}{a-b}} \sin \theta,\,0 \bigr)\,,
\\
 \bar{H}_{\bar{\bm{1}}}{}^M &= \tfrac{1}{\sqrt{2}}\,\bigl( - \sqrt{\tfrac{a-b}{a^2+b^2}},\, \tfrac{a+b}{\sqrt{(a-b) (a^2+b^2)} \sin \theta},\, \tfrac{a+b}{\sqrt{(a-b) (a^2+b^2)} \tan \theta}\,;\, \sqrt{\tfrac{a^2+b^2}{a-b}} ,\, 0,\,0 \bigr)\,,
\\
 \bar{H}_{\bar{\bm{2}}}{}^M &= \tfrac{1}{\sqrt{2}}\,\bigl(-\tfrac{a+b}{\sqrt{(a-b) (a^2+b^2)}},\, -\sqrt{\tfrac{a-b}{a^2+b^2}}\tfrac{1}{\sin \theta},\, -\sqrt{\tfrac{a-b}{a^2+b^2}}\tfrac{1}{\tan \theta} \,;\, 0,\, \sqrt{\tfrac{a^2+b^2}{a-b}} \sin \theta,\,0 \bigr)\,.
\end{split}
\end{align}
Although they are $\mathfrak{f}$-invariant, they do not form a closed algebra under the D-bracket. 
To close the algebra, similar to Eq.~\eqref{eq:CE-def}, we redefine these as
\begin{align}
\begin{split}
\begin{alignedat}{2}
 \CE_{\ubar{\bm{1}}} &\equiv H_{\ubar{\bm{1}}} - \tfrac{a+b}{\sqrt{2\,(a-b)(a^2+b^2)} \tan \theta}\,\VV\,,
\qquad&
 \CE_{\ubar{\bm{2}}} &\equiv H_{\ubar{\bm{2}}} - \tfrac{a+b}{\sqrt{2\,(a-b)(a^2+b^2)} \tan \theta}\,\VV\,,
\\
 \bar{\CE}_{\bar{\bm{1}}} &\equiv \bar{H}_{\bar{\bm{1}}} - \sqrt{\tfrac{a-b}{2(a^2+b^2)}} \,\tfrac{1}{\tan \theta} \,\VV \,,
\qquad&
 \bar{\CE}_{\bar{\bm{2}}} &\equiv \bar{H}_{\bar{\bm{2}}} + \sqrt{\tfrac{a-b}{2(a^2+b^2)}} \,\tfrac{1}{\tan \theta} \,\VV \,.
\end{alignedat}
\end{split}
\end{align}
This redefinition simply removes the $\psi$-components of these vectors
\begin{align}
\begin{split}
 \CE_{\ubar{\bm{1}}}{}^M &= \tfrac{1}{\sqrt{2}}\,\bigl(\sqrt{\tfrac{a-b}{a^2+b^2}},\, \tfrac{a+b}{\sqrt{(a-b)(a^2+b^2)} \sin \theta},\,0\,;\, \sqrt{\tfrac{a^2+b^2}{a-b}},\, 0,\,0 \bigr),
\\
 \CE_{\ubar{\bm{2}}}{}^M &= \tfrac{1}{\sqrt{2}}\,\bigl( -\tfrac{a+b}{\sqrt{(a-b) (a^2+b^2)}},\, \sqrt{\tfrac{a-b}{a^2+b^2}}\tfrac{1}{\sin \theta},\,0\,;\, 0,\, \sqrt{\tfrac{a^2+b^2}{a-b}} \sin \theta,\,0 \bigr)\,,
\\
 \bar{\CE}_{\bar{\bm{1}}}{}^M &= \tfrac{1}{\sqrt{2}}\,\bigl( - \sqrt{\tfrac{a-b}{a^2+b^2}},\, \tfrac{a+b}{\sqrt{(a-b) (a^2+b^2)} \sin \theta},\, 0\,;\, \sqrt{\tfrac{a^2+b^2}{a-b}} ,\, 0,\,0 \bigr)\,,
\\
 \bar{\CE}_{\bar{\bm{2}}}{}^M &= \tfrac{1}{\sqrt{2}}\,\bigl(-\tfrac{a+b}{\sqrt{(a-b) (a^2+b^2)}},\, -\sqrt{\tfrac{a-b}{a^2+b^2}}\tfrac{1}{\sin \theta},\, 0\,;\, 0,\, \sqrt{\tfrac{a^2+b^2}{a-b}} \sin \theta,\,0 \bigr)\,.
\end{split}
\end{align}
Then we find that they are $\mathfrak{f}$-invariant and also form a closed algebra $[\CE_{\bcheck{A}},\,\CE_{\bcheck{B}}]_{\text{D}}=-\bm{f}_{\bcheck{A}\bcheck{B}}{}^{\bcheck{C}}\,\CE_{\bcheck{C}}$ under the D-bracket. 
Now, we consider the gauge-fixing condition $\psi=0$\,, and define the reduced fields $\check{\CE}_{\bcheck{A}}$ by truncating the $(\psi,\,\tilde{\psi})$ components of $\CE_{\bcheck{A}}$\,,
\begin{align}
\begin{split}
 \check{\CE}_{\ubar{\bm{1}}}{}^{\check{M}} &= \tfrac{1}{\sqrt{2}}\,\bigl(\sqrt{\tfrac{a-b}{a^2+b^2}},\, \tfrac{a+b}{\sqrt{(a-b)(a^2+b^2)} \sin \theta} \,;\, \sqrt{\tfrac{a^2+b^2}{a-b}},\, 0 \bigr)\,,
\\
 \check{\CE}_{\ubar{\bm{2}}}{}^{\check{M}} &= \tfrac{1}{\sqrt{2}}\,\bigl( -\tfrac{a+b}{\sqrt{(a-b) (a^2+b^2)}},\, \sqrt{\tfrac{a-b}{a^2+b^2}}\tfrac{1}{\sin \theta} \,;\, 0,\, \sqrt{\tfrac{a^2+b^2}{a-b}} \sin \theta \bigr)\,,
\\
 \bar{\check{\CE}}_{\bar{\bm{1}}}{}^{\check{M}} &= \tfrac{1}{\sqrt{2}}\,\bigl( - \sqrt{\tfrac{a-b}{a^2+b^2}},\, \tfrac{a+b}{\sqrt{(a-b) (a^2+b^2)} \sin \theta} \,;\, \sqrt{\tfrac{a^2+b^2}{a-b}} ,\, 0 \bigr)\,,
\\
 \bar{\check{\CE}}_{\bar{\bm{2}}}{}^{\check{M}} &= \tfrac{1}{\sqrt{2}}\,\bigl(-\tfrac{a+b}{\sqrt{(a-b) (a^2+b^2)}},\, -\sqrt{\tfrac{a-b}{a^2+b^2}}\tfrac{1}{\sin \theta} \,;\, 0,\, \sqrt{\tfrac{a^2+b^2}{a-b}} \sin \theta \bigr)\,.
\end{split}
\end{align}
They indeed satisfy the same algebra as that of $\CE_{\bcheck{A}}$\,: $[\check{\CE}_{\bcheck{A}},\,\check{\CE}_{\bcheck{B}}]_{\text{D}}=-\bm{f}_{\bcheck{A}\bcheck{B}}{}^{\bcheck{C}}\,\check{\CE}_{\bcheck{C}}$\,. 
We then construct the reduced generalized metric as
\begin{align}
 \check{\cH}_{\check{M}\check{N}} &= \eta^{\ubar{\bm{a}}\ubar{\bm{b}}}\, \check{\CE}_{\bm{a}\check{M}}\,\check{\CE}_{\bm{b}\check{N}} + \eta^{\bar{\bm{a}}\bar{\bm{b}}}\,\bar{\check{\CE}}_{\bar{\bm{a}}\check{M}}\,\bar{\check{\CE}}_{\bar{\bm{b}}\check{N}} 
\nn\\
 &= \begin{pmatrix}
\frac{a^2+b^2}{a-b} & 0 & 0 & \frac{a+b}{(a-b) \sin \theta} \\
 0 & \frac{(a^2+b^2) \sin^2 \theta}{a-b} & -\frac{(a+b) \sin \theta}{a-b} & 0 \\
 0 & -\frac{(a+b) \sin \theta}{a-b} & \frac{2}{a-b} & 0 \\
 \frac{a+b}{(a-b) \sin \theta} & 0 & 0 & \frac{2}{(a-b) \sin^2\theta}\end{pmatrix} .
\end{align}
This exactly corresponds to the background obtained in Eq.~\eqref{eq:S2-limit}. 

Let us also consider the formulas \eqref{eq:KS-reproduced} and \eqref{eq:KSd}. 
Here, \eqref{eq:KS-reproduced} does not apply and we shall use the formula \eqref{eq:KSd},
\begin{align}
 \check{E}_{mn} = r_m^a\,\bigl[(1+\tilde{\cR}\,\pi)^{-1}\,\tilde{\cR}\bigr]_{ab}\,r_n^b = r_m^a\,\tilde{\cR}_{ab}\,r_n^b \,.
\label{eq:KSd-reproduced-2}
\end{align}
We can compute the matrices $(\mathsf{\kk}^{\hat{\bm{b}}a})\equiv (\mathsf{\kk}^{\ubar{\bm{1}}a},\,\mathsf{\kk}^{\ubar{\bm{2}}a},\,\mathsf{\kk}^a)$ and $(\mathsf{\bar{\kk}}^{\hat{\bm{b}}a})\equiv (\mathsf{\bar{\kk}}^{\ubar{\bm{1}}a},\,\mathsf{\bar{\kk}}^{\ubar{\bm{2}}a},\,\mathsf{\kk}^a)$ by using
\begin{align}
\begin{split}
\begin{alignedat}{2}
 (\mathsf{\kk}^{\ubar{\bm{1}}a})&= \tfrac{1}{\sqrt{(a-b)\,(a^2+b^2)}}\,(a-b,\, a+b,\, 0)\,,\qquad&
 (\mathsf{\kk}^{\ubar{\bm{2}}a})&= \tfrac{-1}{\sqrt{(a-b)\,(a^2+b^2)}}\,(a+b,\, b-a,\, 0)\,,
\\
 (\mathsf{\bar{\kk}}^{\ubar{\bm{1}}a})&=\tfrac{1}{\sqrt{(a-b)\,(a^2+b^2)}}\,(b-a,\, a+b,\, 0)\,,\qquad&
 (\mathsf{\bar{\kk}}^{\ubar{\bm{2}}a})&=\tfrac{-1}{\sqrt{(a-b)\,(a^2+b^2)}}\,(a+b,\, a-b,\, 0)\,,
\end{alignedat}
\end{split}
\end{align}
and $(\mathsf{\kk}^a)=(0,\,0,\,1)$. 
We then find $(\bar{\kk}_{a}^{\sfi=1})=(\kk_{a}^{\sfi=1})=(0,\,0,\,1)$, and Eq.~\eqref{eq:cR-2} gives
\begin{align}
 (\tilde{\cR}_{ab}) = (\widetilde{\mathbb{E}}_{ab}) = \begin{pmatrix} \frac{a-b}{2} & \frac{a+b}{2} & 0 \\
 -\frac{a+b}{2} & \frac{a-b}{2} & 0 \\
 0 & 0 & 0 \end{pmatrix}.
\label{eq:tilde-R-91}
\end{align}
Substituting this into Eq.~\eqref{eq:KSd-reproduced-2}, we obtain
\begin{align}
 (\check{E}_{mn}) = \begin{pmatrix}
 \frac{a-b}{2} & \frac{a+b}{2} \sin \theta & 0 \\
 -\frac{a+b}{2} \sin \theta & \frac{a-b}{2} \sin^2 \theta & 0 \\
 0 & 0 & 0
\end{pmatrix},
\end{align}
which is exactly the result \eqref{eq:S2-limit}. 
We can also check that $\tilde{\cR}_{ab} = {\displaystyle\lim_{\lambda\to 0}} (\hat{E}^{-1})_{ab}$\,. 

For the Drinfel'd double DD5, the only Manin triples, i.e., the decomposition of a Drinfel'd double into the pair $(\mathfrak{g}|\tilde{\mathfrak{g}})$, are the $(\bm{9}|\bm{1})$ and its dual $(\bm{1}|\bm{9})$ \cite{math:0202210}.
Here, $\bm{9}$ denotes the Lie algebra \eqref{eq:9-1} of $\mathfrak{so}(3)$ and $\bm{1}$ denotes the three-dimensional Abelian Lie algebra (see \cite{math:0202210,hep-th:0403164} for the details of the notation). 
The two Manin triples $(\bm{9}|\bm{1})$ and $(\bm{1}|\bm{9})$ are related by $T_a\leftrightarrow T^a$\,, which corresponds to the traditional non-Abelian $T$-duality. 
Namely, for DD5, the non-Abelian $T$-duality is the only possible PL $T$-plurality. 

\subsubsection{A supergravity solution}

Before performing the non-Abelian $T$-duality, let us embed the obtained $\text{S}^2$ background into a solution of supergravity. 
As it has been studied in \cite{hep-th:9210021}, if we choose $a$ and $b$ as
\begin{align}
 a = r^2\,,\qquad b = - r^2\,,
\end{align}
the reduced background fields \eqref{eq:S2-limit} become
\begin{align}
 \rmd \check{s}^2 = r^2\,\bigl(\rmd\theta^2+\sin^2 \theta\,\rmd\phi^2\bigr) \,, \qquad
 \check{B}_2 = 0\,.
\end{align}
Using this, we can consider, for example, the Schwarzschild solution \cite{hep-th:9210021}. 
Here, instead, we consider a spherically symmetric solution with $B$-field studied in \cite{hep-th:9410142,1606.09307}
\begin{align}
\begin{split}
 \rmd s^2 &= \frac{-\rmd t^2 + f^{-1}(r) \,\rmd r^2}{2\,f(r) - 1 + 2\,v f^{\frac{1}{2}}(r)[1 - f(r)]^{\frac{1}{2}}} + r^2\,\bigl(\rmd \theta^2+\sin^2\theta\,\rmd\phi^2\bigr)\,,\quad 
 f(r)\equiv 1-\frac{h^2}{4\,r^2}\,,
\\
 B_2&= h\,t\sin\theta\,\rmd\theta\wedge\rmd\phi\,,\qquad 
 \Exp{-2d}= r^2 f^{-\frac{1}{2}}(r) \abs{\det(\ell_m^a)}\qquad \bigl(\,\det(\ell_m^a)=\sin\theta\,\bigr)\,,
\end{split}
\label{eq:S2-solution}
\end{align}
where $v$ is an arbitrary parameter. 
This can be reproduced from \eqref{eq:S2-limit} by choosing
\begin{align}
 a = r^2 + h\,t\,,\qquad b = -r^2 + h\, t\,,
\label{eq:S2-a-b}
\end{align}
where $h$ is an arbitrary constant. 
Below, we consider the non-Abelian $T$-dual of this solution. 

\subsubsection{NATD of S$^2$}
\label{sec:NATDofS2}

The non-Abelian $T$-duality can be defined as a redefinition of generators, $T_a\leftrightarrow T^a$, namely
\begin{align}
 T_A \to T'_A = C_A{}^B\,T_B\,,\qquad C = \begin{pmatrix} \bm{0} & \bm{1} \\ \bm{1} & \bm{0} \end{pmatrix}. 
\label{eq:NATD}
\end{align}
Under this redefinition, the structure constants become
\begin{align}
 f_3{}^{12} = 1\,,\qquad
 f_1{}^{23} = 1\,,\qquad
 f_2{}^{13} = -1\,,
\end{align}
and the Manin triple becomes $(\bm{1}|\bm{9})$. 
Using the parameterization $g=\Exp{z\,T_3}\Exp{\rho\sin\theta\,T_2}\Exp{\rho\cos\theta\,T_1}$, the right-invariant vector fields $e_a^m$\,, the matrix $\pi^{ab}$, and $\det(\ell_m^a)$ become
\begin{align}
 e_a{}^m = {\footnotesize \begin{pmatrix} 
 \cos \theta & -\frac{\sin \theta}{\rho} & 0 \\
 \sin \theta & \frac{\cos \theta}{\rho} & 0 \\
 0 & 0 & 1 \end{pmatrix}},\qquad 
 \pi^{ab} = {\footnotesize \begin{pmatrix}
 0 & z & -\rho \sin \theta \\
 -z & 0 & \rho \cos \theta \\
 \rho \sin \theta & -\rho \cos \theta & 0 \end{pmatrix}},\qquad 
 \det(\ell_m^a)=\rho\,,
\end{align}
where we use the coordinate system $(x^m)=(\rho,\,\theta,\,z)$. 
The matrix $\hat{E}^{ab}$ can be obtained as the inverse matrix of $\hat{E}^{ab}$ given in Eq.~\eqref{eq:S2-Ehat}, and we have
\begin{align}
 \hat{E}^{ab} = \begin{pmatrix} \frac{a-b}{2} & \frac{a+b}{2} & 0 \\ -\frac{a+b}{2} & \frac{a-b}{2} & 0 \\ 0 & 0 & \lambda \end{pmatrix},\qquad 
 \hat{E}_{ab} = \begin{pmatrix}
 \frac{a-b}{a^2+b^2} & -\frac{a+b}{a^2+b^2} & 0 \\
 \frac{a+b}{a^2+b^2} & \frac{a-b}{a^2+b^2} & 0 \\
 0 & 0 & \lambda^{-1} \end{pmatrix}.
\end{align}
Then we obtain
\begin{align}
 E_{mn} = {\footnotesize \frac{1}{(a-b)\,\rho^2 +\lambda\,\Sigma}\begin{pmatrix}
 2\,\rho^2 + \lambda\,(a-b) & -\lambda\,(a+b+2\,z)\,\rho & (a+b+2\,z)\,\rho \\
 \lambda\,(a+b+2\,z)\,\rho & \lambda\,(a-b)\,\rho^2 & -(a-b)\,\rho^2 \\
 (a+b+2\,z)\,\rho & (a-b)\,\rho^2 & \Sigma
\end{pmatrix}},
\label{eq:E-NATD}
\end{align}
where $\Sigma \equiv a^2+b^2+2\,(a+b)\,z+2\,z^2$\,. 

Under the duality \eqref{eq:NATD}, the index $A$ is covariantly transformed and, for example, the vectors $\VV^A$ and $\widetilde{\VV}^A$ are transformed as $\VV^A \to \VV^B\,(C^{-1})_B{}^A$ and $\widetilde{\VV}^A \to \widetilde{\VV}^B\,(C^{-1})_B{}^A$. 
The resulting $\VV^A$ and $\widetilde{\VV}^A$, and the corresponding generalized vector fields $\VV^M\equiv \VV^A\,E_A{}^M$ and $\widetilde{\VV}^M\equiv \widetilde{\VV}^A\,E_A{}^M$, take the form
\begin{align}
\begin{split}
\begin{alignedat}{2}
 \VV^A&= \bigl(0,\, 0,\, 0\,;\, 0,\, 0,\, 1\bigr) ,\qquad& \VV^M&= \bigl( 0,\, 1,\, 0\,;\, 0,\, 0,\, 1\bigr)\,,
\\
 \widetilde{\VV}^A&= \bigl(0,\, 0,\, \lambda\,;\, 0,\, 0,\, 0\bigr) ,\qquad& \widetilde{\VV}^M&= \bigl( 0,\, 0,\, \lambda\,;\, 0,\, 0,\, 0\bigr)\,.
\end{alignedat}
\end{split}
\end{align}
If we construct the generalized metric $\cH_{MN}$ from the above $E_{mn}=g_{mn}+B_{mn}$\,, we can check the $\mathfrak{f}$-invariance $\gLie_{\VV}\cH_{MN}=0$\,. 
This should be satisfied because the invariance condition \eqref{eq:H-inv-inf} is manifestly covariant under the PL $T$-plurality. 
Here, $\VV^M$ shifts the $\theta$-coordinate and we take $\theta=0$ as the gauge-fixing condition. 

Now, let us compute the reduced background fields by using the formula \eqref{eq:reduced-E}. 
By using
\begin{align}
\begin{split}
 (N_{\sfi\sfj}) &= (N_{11}) = \tfrac{(a-b)\,\lambda\,\rho^2}{(a-b)\,\rho^2 + \lambda\,\Sigma}\,,\qquad
 \kk_{m}=\bigl(0,\,\tfrac{(a-b)\,\lambda\,\rho^2}{(a-b)\,\rho^2 + \lambda\,\Sigma},\,0\bigr)\,,
\\
 \hat{\kk}_{m} &= \bigl( \tfrac{(a+b+2\,z)\,\lambda\,\rho}{(a-b)\,\rho^2 + \lambda\,\Sigma},\, 0,\, 1-\tfrac{(a-b)\,\rho^2}{(a-b)\,\rho^2 + \lambda\,\Sigma} \bigr)\,,
\end{split}
\end{align}
the reduced background fields are found as
\begin{align}
 \check{E}_{mn} = \begin{pmatrix}
 \frac{2}{a-b} & 0 & \frac{a+b+2\,z}{(a-b)\,\rho} \\
 0 & 0 & -1 \\
 \frac{a+b+2\,z}{(a -b)\,\rho} & 1 & \frac{\Sigma}{(a-b)\,\rho^2}
\end{pmatrix} .
\label{eq:NATD-E}
\end{align}
It is worth noting that the complicated $\lambda$-dependence appearing in $E_{mn}$ of Eq.~\eqref{eq:E-NATD} has been completely subtracted. 
One can easily check that this is exactly the same as taking the Sfetsos limit $\lambda\to 0$ in Eq.~\eqref{eq:E-NATD}. 

After the gauge fixing $\theta=0$\,, we obtain the reduced background fields as\footnote{If we change the coordinates as $\rho=\sqrt{x^2-y^2}$ and $z=y$\,, and choose $b=-a$\,, the metric becomes $\rmd \check{s}^2 = \frac{a^2\rmd y^2 + x^2\rmd x^2}{a\,(x^2-y^2)}$ which reproduces the dual geometry of \cite{hep-th:9210021}.}
\begin{align}
 \check{g}_{\check{m}\check{n}} = \begin{pmatrix} \frac{2}{a-b} & \frac{a+b+2\,z}{(a-b)\, \rho} \\
 \frac{a+b+2\,z}{(a-b)\,\rho} & \frac{\Sigma}{(a-b)\,\rho^2} 
\end{pmatrix} ,\qquad \check{B}_{\check{m}\check{n}} = 0\,.
\label{eq:NATD-bg}
\end{align}
Then, substituting Eq.~\eqref{eq:S2-a-b} to $a$ and $b$\,, we obtain the non-Abelian $T$-dual of Eq.~\eqref{eq:S2-solution} as
\begin{align}
\begin{split}
 \rmd s^2 &= \tfrac{-\rmd t^2 + f^{-1}(r) \,\rmd r^2}{2\,f(r) - 1 + 2\,v f^{\frac{1}{2}}(r)[1 - f(r)]^{\frac{1}{2}}} 
 + r^{-2}\,\Bigl[ \rmd^2\rho + \frac{2\,(z+h\,t)}{\rho}\,\rmd \rho\,\rmd z + \frac{r^4 + (z+h\,t)^2}{\rho^2} \,\rmd^2 z \Bigr]\,, 
\\
 B_2&= 0\,,\qquad 
 \Exp{-2d}= r^2 f^{-\frac{1}{2}}(r)\,\rho\,,
\end{split}
\end{align}
where only the $\text{S}^2$ part of the background fields are replaced. 
We can easily check that this satisfies the usual supergravity equations of motion. 

Let us also consider the geometric approach based on $\CE_{\bcheck{A}}{}^M$. 
The set of vectors become
\begin{align}
 \VV^{\ubar{\bm{a}}} = \frac{1}{\sqrt{2}}\begin{pmatrix} \hat{E}^{ab}\,\kk^{\ubar{\bm{a}}}_{b} \\ \kk^{\ubar{\bm{a}}}_{a} \end{pmatrix},\qquad
 \bar{\VV}^{\bar{\bm{a}}} = \frac{1}{\sqrt{2}}\begin{pmatrix} - \bar{\kk}^{\bar{\bm{a}}}_{b}\,\hat{E}^{ba} \\ \bar{\kk}^{\bar{\bm{a}}}_{a} \end{pmatrix},
\end{align}
where
\begin{align}
 \kk^{\ubar{\bm{1}}}_{a} = \bar{\kk}^{\bar{\bm{1}}}_{a} = \sqrt{\frac{2}{a-b}}\, (1,\,0,\,0)\,,\qquad 
 \kk^{\ubar{\bm{2}}}_{a} = \bar{\kk}^{\bar{\bm{2}}}_{a} = \sqrt{\frac{2}{a-b}}\,(0,\,1,\,0)\,.
\end{align}
They are obtained by acting the $\OO(3,3)$ transformation \eqref{eq:NATD} to those of the original duality frame. 
By the construction, it is obvious that they realize
\begin{align}
\begin{split}
 \eta_{AB} &= \eta_{\ubar{\bm{c}}\ubar{\bm{d}}}\,\VV^{\ubar{\bm{c}}}_{A}\,\VV^{\ubar{\bm{d}}}_{B} - \eta_{\bar{\bm{c}}\bar{\bm{d}}}\,\bar{\VV}^{\bm{c}}_{A}\, \bar{\VV}^{\bm{d}}_{B} + \lambda^{-1}\,\bigl(\VV_{A}\,\widetilde{\VV}_{B}+\widetilde{\VV}_{A}\,\VV_{B}\bigr)\,,
\\
 \hat{\cH}_{AB} &= \eta_{\ubar{\bm{c}}\ubar{\bm{d}}}\,\VV^{\bm{c}}_{A}\,\VV^{\bm{d}}_{B} + \eta_{\bar{\bm{c}}\bar{\bm{d}}}\,\bar{\VV}^{\bm{c}}_{A}\, \bar{\VV}^{\bm{d}}_{B} + \lambda^{-1}\,\bigl(\VV_{A}\,\VV_{B}+\widetilde{\VV}_{A}\,\widetilde{\VV}_{B}\bigr)\,,
\end{split}
\end{align}
where $(\eta_{\ubar{\bm{a}}\ubar{\bm{b}}})=(\eta_{\bar{\bm{a}}\bar{\bm{b}}})= \diag(+1,+1)$.
Again, $\VV^{\ubar{\bm{a}}M}= \VV^{\ubar{\bm{a}}A}\,E_A{}^M$ and $\bar{\VV}^{\bar{\bm{a}}M}=\bar{\VV}^{\bar{\bm{a}}A}\,E_A{}^M$ are not $\mathfrak{f}$-invariant, and we perform a redefinition
\begin{align}
\begin{split}
 \begin{pmatrix} H_{\ubar{1}}\\ H_{\ubar{2}} \end{pmatrix}
 = \begin{pmatrix} \cos \theta & \sin \theta \\ -\sin \theta & \cos \theta \end{pmatrix}
 \begin{pmatrix} \VV^{\ubar{1}}\\ \VV^{\ubar{2}} \end{pmatrix} ,
 \qquad
 \begin{pmatrix} \bar{H}_{\bar{1}}\\ \bar{H}_{\bar{2}} \end{pmatrix}
 = \begin{pmatrix} \cos \theta & \sin \theta \\ -\sin \theta & \cos \theta \end{pmatrix}
 \begin{pmatrix} \bar{\VV}^{\bar{1}}\\ \bar{\VV}^{\bar{2}} \end{pmatrix} .
\end{split}
\end{align}
Then we find that $H_{\ubar{\bm{a}}}$ and $\bar{H}_{\bar{\bm{a}}}$ are $\mathfrak{f}$-invariant.
However, they do not form a closed algebra. 
We thus redefine these as
\begin{align}
\begin{split}
\begin{alignedat}{2}
 \CE_{\ubar{\bm{1}}} &\equiv H_{\ubar{\bm{1}}} + \tfrac{a+b+2\,z}{2\,\rho\sqrt{a-b}}\,\VV \,,
\qquad&
 \CE_{\ubar{\bm{2}}} &\equiv H_{\ubar{\bm{2}}} - \tfrac{\sqrt{a-b}}{2\,\rho}\,\VV \,,
\\
 \bar{\CE}_{\bar{\bm{1}}} &\equiv \bar{H}_{\bar{\bm{1}}} + \tfrac{a+b+2\,z}{2\,\rho\sqrt{a-b}}\,\VV \,,
\qquad&
 \bar{\CE}_{\bar{\bm{2}}} &\equiv \bar{H}_{\bar{\bm{2}}} + \tfrac{\sqrt{a-b}}{2\,\rho} \,\VV \,.
\end{alignedat}
\end{split}
\end{align}
Writing explicitly, we obtain
\begin{align}
\begin{split}
 \CE_{\ubar{\bm{1}}}{}^M &= \bigl(\tfrac{\sqrt{a-b}}{2},\, 0,\,0 \,;\, \tfrac{1}{\sqrt{a-b}},\, 0,\, \tfrac{a+b+2\,z}{2\,\rho\sqrt{a-b}} \bigr),
\\
 \CE_{\ubar{\bm{2}}}{}^M &=\bigl(\tfrac{a+b+2\,z}{2\sqrt{a-b}},\, 0,\, -\tfrac{\rho}{\sqrt{a-b}} \,;\, 0,\, \tfrac{\rho}{\sqrt{a-b}},\, -\tfrac{\sqrt{a-b}}{2\,\rho}\bigr)\,,
\\
 \bar{\CE}_{\bar{\bm{1}}}{}^M &= \bigl(-\tfrac{\sqrt{a-b}}{2},\,0,\,0 \,;\, \tfrac{1}{\sqrt{a-b}},\, 0,\, \tfrac{a+b+2\,z}{2\,\rho\sqrt{a-b}}\bigr)\,,
\\
 \bar{\CE}_{\bar{\bm{2}}}{}^M &=\bigl(\tfrac{a+b+2\,z}{2\sqrt{a-b}},\, 0 ,\, -\tfrac{\rho}{\sqrt{a-b}} \,;\, 0,\, \tfrac{\rho}{\sqrt{a-b}},\, \tfrac{\sqrt{a-b}}{2\,\rho}\bigr)\,.
\end{split}
\end{align}
Then truncating the $(\theta,\,\tilde{\theta})$-components, we obtain
\begin{align}
 \check{\cH}_{\check{M}\check{N}} &= \eta^{\ubar{\bm{a}}\ubar{\bm{b}}}\, \check{\CE}_{\bm{a}\check{M}}\,\check{\CE}_{\bm{b}\check{N}} + \eta^{\bar{\bm{a}}\bar{\bm{b}}}\,\bar{\check{\CE}}_{\bar{\bm{a}}\check{M}}\,\bar{\check{\CE}}_{\bar{\bm{b}}\check{N}} 
\nn\\
 &= \begin{pmatrix}
 \frac{2}{a-b} & \frac{a+b+2\,z}{(a-b)\,\rho} & 0 & 0 \\
 \frac{a+b+2\,z}{(a-b)\,\rho} & \frac{\Sigma}{(a-b)\,\rho^2} & 0 & 0 \\
 0 & 0 & \frac{\Sigma}{a-b} & -\frac{(a+b+2\,z)\,\rho}{a-b} \\
 0 & 0 & -\frac{(a+b+2\,z)\,\rho}{a-b} & \frac{2\,\rho^2}{a-b}
\end{pmatrix},
\end{align}
and the background \eqref{eq:NATD-bg} is reproduced as expected. 

Let us also consider the formula \eqref{eq:KS-reproduced},
\begin{align}
 \check{E}_{mn} = r_m^a\,\bigl[(\cR+\pi)^{-1}\bigr]_{ab}\,r_n^b \,.
\label{eq:KS-reproduced-2}
\end{align}
We can easily compute
\begin{align}
 (\cR^{ab})=(\mathbb{E}^{ab})=\begin{pmatrix}
 \frac{a-b}{2} & \frac{a+b}{2} & 0 \\
 -\frac{a+b}{2} & \frac{a-b}{2} & 0 \\
 0 & 0 & 0 \end{pmatrix},
\label{eq:R-19}
\end{align}
and then Eq.~\eqref{eq:KS-reproduced-2} gives the same result as Eq.~\eqref{eq:NATD-E}. 
Again, we find $\cR^{ab}={\displaystyle\lim_{\lambda\to 0}}\hat{E}^{ab}$\,. 
The same result can also be obtained from the formula \eqref{eq:formula-fast} with $\bm{b}=\bm{c}=\bm{1}$\,, $\bm{a}=\bm{d}=\bm{0}$\,, and $(\tilde{\cR}_{ab})$ given in Eq.~\eqref{eq:tilde-R-91}. 
We note that $(\tilde{\cR}_{ab})$ is the same as $(\cR^{ab})$ obtained in Eq.~\eqref{eq:R-19}. 

\subsection{Homogeneous Yang--Baxter deformations of standard cosets}
\label{sec:hYB}

Let us consider a standard coset that solves the supergravity equations of motion. 
We suppose that the Drinfel'd double is semi-Abelian (i.e., $f_a{}^{bc}=0$)\,, and decompose the generators as $\{T_a\}=\{T_{\check{a}},\,T_{\sfi}\}$ and $\{T^a\}=\{T^{\check{a}},\,T^{\sfi}\}$ such that $\mathfrak{f}$ is generated by $T_{\sfi}$\,. 
We then find
\begin{align}
 \check{E}_{mn} = r_m^a\,\tilde{\cR}_{ab}\,r^b_n\qquad \Bigl[\,\tilde{\cR}_{ab}\equiv \lim_{\lambda\to 0} (\hat{E}^{-1})_{ab} \,\Bigr] \,.
\end{align}
Now, we perform a specific PL $T$-plurality transformation $T_A\to T'_A\equiv C_A{}^B\,T_B$ with
\begin{align}
 \bigl(C_A{}^B\bigr) = \begin{pmatrix} \delta_a^b & 0 \\ r^{ab} & \delta^a_b \end{pmatrix} \in \OO(D,D)\qquad \bigl(r^{ab}=r^{[ab]}\bigr)\,.
\label{eq:YB-C}
\end{align}
If the constant matrix $r^{ab}$ satisfies the homogeneous classical Yang--Baxter equation
\begin{align}
 r^{ad}\,r^{be}\,f_{de}{}^c + r^{bd}\,r^{ce}\,f_{de}{}^a + r^{cd}\,r^{ae}\,f_{de}{}^b = 0\,,
\end{align}
this maps the semi-Abelian double to another Manin triple with the structure constants
\begin{align}
 f'_{ab}{}^c = f_{ab}{}^c\,,\qquad 
 f'_a{}^{bc} = r^{bd}\,f_{ad}{}^c - r^{cd}\,f_{ad}{}^b\,.
\end{align}
This type of PL $T$-plurality is called the homogeneous Yang--Baxter deformation \cite{hep-th:0210095,0802.3518,1308.3581,1501.03665}. 
The generators $T_a$ are not transformed $T'_a=T_a$\,, and the group $G$ is not changed $G'=G$. 
Then the generators of the gauge algebra is also unchanged $T'_{\sfi}=T_{\sfi}$ and we find $F\subset G'$. 
Consequently, the deformed background $\check{M}'=F\backslash \cD/\tilde{G}'$ is also a standard coset. 

After the deformation, the non-vanishing dual structure constants $f'_a{}^{bc}$ produces the Poisson--Lie structure $\pi'^{mn}$. 
It can be found to take the form
\begin{align}
 \pi'^{mn} = r^{ab}\,\bigl(v_a^m\,v_b^n - e_a^m\,e_b^n\bigr) \,,
\end{align}
where $e'^m_a=e_a^m$ and $v'^m_a=v_a^m$ because of $T'_a=T_a$\,. 
Using $\hat{E}'^{ab}=\hat{E}^{ab}+r^{ab}$, we find $\tilde{\cR}'=\tilde{\cR}\,(\bm{1} + r\,\tilde{\cR})^{-1}$, and then the reduced background fields become
\begin{align}
\begin{split}
 \check{E}'_{mn} &= r_m^a\,\bigl[\tilde{\cR}'\,(\bm{1}+\pi\,\tilde{\cR}')^{-1}\bigr]_{ab}\,r^b_n
 = r_m^a\,\bigl[\tilde{\cR}\,(\bm{1} + r\,\tilde{\cR} + \pi\,\tilde{\cR})^{-1} \bigr]_{ab}\,r^b_n
\\
 & = E_{mp}\,[(\bm{1}+\bm{r}\,\check{E})^{-1}]^p{}_n\qquad \bigl(\bm{r}^{mn}\equiv r^{ab}\,v_a^m\,v_b^n\bigr) \,.
\end{split}
\end{align}
The deformed background can be easily obtained once we found a solution $r^{ab}$ of the homogeneous classical Yang--Baxter equation and the left-invariant vector fields $v_{a}^{m}$\,. 

The dilaton should be invariant under the Yang--Baxter deformation $d'=d$ because the left-invariant 1-form $\ell^a_m$ is invariant. 
However, when the dual structure constants have non-vanishing trace $f'_b{}^{ba} = r^{bd}\,f_{bd}{}^a - r^{ad}\,f_{bd}{}^b\neq 0$\,, due to the shift of the dilaton flux \eqref{eq:dilaton-shift}, the deformed background satisfies the generalized supergravity equations of motion with the Killing vector field $I'^{\check{m}}=\tfrac{1}{2}\,f'_{b}{}^{ba}\,v_a^{\check{m}}$\,. 
The R--R fields are also transformed covariantly under the Yang--Baxter deformation (see for example \cite{1705.07116}).

\subsection{Poisson--Lie $T$-plurality of AdS$_2$}
\label{sec:PL-AdS2}

As a non-trivial example, we here study the PL $T$-pluralities of $\text{AdS}_2=\SO(1,1)\backslash\SL(2)$.
Here we employ a six-dimensional Drinfel'd double, called the DD6 \cite{math:0202210,hep-th:0403164}.
This has the following five Manin triples and their PL $T$-duals \cite{math:0202210}:
\begin{align}
 \text{DD6:}\qquad (\bm{8}|\bm{1}) \cong (\bm{8}|\bm{5.iii}) \cong (\bm{7_0}|\bm{5.i}) \cong (\bm{6_0}|\bm{5.i}) \cong (\bm{5}|\bm{2.ii}) \,,
\end{align}
where $\bm{8}$ corresponds to the Lie algebra $\mathfrak{sl}(2)$ and the Bianchi type V algebra (denoted as $\bm{5}$, $\bm{5.i}$, or $\bm{5.iii}$) is non-unimodular. 
Accordingly, in total, ten supergravity backgrounds are related under PL $T$-plurality. 
The constant $\OO(D,D)$ matrices $C_A{}^B$ that are used for the redefinitions of generators $T_A\to T'_A=C_A{}^B\,T_B$ are given in \cite{math:0202210} (though some of them may have slightly different forms in our convention). 

In the following, we determine the two-dimensional reduced background fields $\{\check{g}_{\check{m}\check{n}},\,\check{B}_{\check{m}\check{n}}\}$ for each of the ten Manin triples. 
We also consider the embedding of each background into a solution of a ten-dimensional supergravity. 
In the initial Manin triple $(\bm{8}|\bm{1})$, we consider an $\text{AdS}_2\times \text{S}^2\times \text{T}^6$ solution with $H$-flux,
\begin{align}
\begin{split}
 \rmd s^2 &= \frac{-\rmd t^2 + \rmd z^2}{z^2} + \bigl(\rmd\theta^2 +\sin^2\theta\,\rmd\phi^2\bigr) + \bigl(\rmd w^2 +\rmd s^2_{\text{T}^5}\bigr) \,,
\\
 H_3 &= \rmd \hat{w}\wedge \Bigl(\frac{\rmd t \wedge \rmd z}{z^2} + \sin \theta\,\rmd \theta \wedge\rmd \phi\Bigr) \,,\qquad \Exp{-2\Phi}=1 \,,
\end{split}
\label{eq:AdS2-sol}
\end{align}
where we have defined the rescaled coordinate
\begin{align}
 \hat{w}\equiv \sqrt{2}\,w\,,
\end{align}
which is used throughout this subsection. 
This supergravity solution does not involve the R--R fields and the dualized backgrounds contain only the supergravity fields in the NS--NS sector. 
Similar to the standard PL $T$-plurality (where the group $G$ acts on $M$ without fixed points), when the dual algebra $\tilde{\mathfrak{g}}$ is unimodular/non-unimodular, we find that backgrounds satisfy the standard/generalized supergravity equations of motion. 

Here we have computed the reduced background fields $\check{E}_{mn}$ by taking the four approaches described at the beginning of this section, and have checked that all approaches give the same results. 
To make the presentation short, in most of the examples, we obtain the results by using the formula \eqref{eq:formula-fast}. 
Namely, we start with the Manin triple $(\vect{8}|\vect{1})$ and introduce a certain matrix $\hat{E}^{ab}$. 
This matrix contains the parameter $\lambda$ and by taking the Sfetsos limit, we obtain $\tilde{\cR}_{ab}\equiv {\displaystyle\lim_{\lambda\to 0}}(\hat{E}^{-1})_{ab}$\,. 
The reduced background fields are obtained as $\check{E}_{mn} = r_m^a\,\tilde{\cR}_{ab}\,r^b_n$\,. Then, we perform an $\OO(D,D)$ rotations of the form \eqref{eq:C-param}, and consider another Manin triple. 
There, using a certain parameterization of the group element $g\in G$, we compute $r'^a_m$ and $\pi'^{ab}$\,, and then the formula \eqref{eq:formula-fast} gives the reduced background fields $\check{E}'_{mn}$\,.

\subsubsection{$(\vect{8}|\vect{1})$}

Let us consider a semi-Abelian double, $\mathfrak{g}=\bm{8}=\text{sl}(2)$ and $\tilde{\mathfrak{g}}=\bm{1}$\,,
\begin{align}
 f_{12}{}^3 = -1\,,\qquad
 f_{23}{}^1 = 1\,,\qquad
 f_{13}{}^2 = 1\,.
\end{align}
Using the coordinates $(x^m)=(t,\,z,\,y)$ and the parameterization
\begin{align}
 g = \Exp{y\,T_3}\Exp{\ln z\,T_2}\Exp{t\,(T_1 - T_3)},
\label{eq:81-param}
\end{align}
we can compute the right-invariant vector fields $e_a^m$\,, $\det(\ell^a_m)$, and $\pi^{ab}$ as
\begin{align}
 e_a{}^m= {\footnotesize\begin{pmatrix}
 z \cosh y & z \sinh y & \cosh y \\
 z \sinh y & z \cosh y & \sinh y \\
 0 & 0 & 1\end{pmatrix}},\qquad \det(\ell^a_m)=z^{-2}\,,\qquad \pi^{ab}=0\,. 
\end{align}
To construct the background fields, we introduce the matrix $\hat{E}^{ab}=\hat{G}^{ab}+\hat{\beta}^{ab}$ as
\begin{align}
 \hat{E}^{ab} = {\footnotesize\begin{pmatrix} -\alpha & p & 0 \\ -p & \alpha & 0 \\ 0 & 0 & \lambda^{-1} \end{pmatrix}}.
\label{eq:Eab-81}
\end{align}
Here, we suppose $\lambda=1$ but this vanishes under the Sfetsos limit $\lambda\to 0$\,. 
The invariance condition \eqref{eq:inv-orig} is satisfied for $T_{\sfi}=T_3$\,, and here we take the gauge algebra as $\mathfrak{f}=\text{span}\{T_3\}$\,. 
Correspondingly, we introduce a vector
\begin{align}
 \VV_{\sfi=1}^A \equiv \VV^A \equiv (\mathsf{\kk}^a\,;\,\tilde{\kk}_a) \equiv (0,\,0,\,1\,;\,0,\,0,\,0)\,,
\end{align}
and the corresponding generalized vector field $\VV^M = \VV^A\,E_A{}^M$ as
\begin{align}
 \VV^M = \VV^A\,E_A{}^M = E_3{}^M = (0,\,0,\,1\,;\,0,\,0,\,0)\,.
\end{align}
This generates a translation in the $y$-coordinate, and the gauge-fixing condition can be chosen as $y=0$\,. 
Before reducing the dimensions of the target space, the three-dimensional background fields $E_{mn}=r_m^a\,[(\hat{E}+\pi)^{-1}]_{ab}\,r^b_n$ can be found as
\begin{align}
 E_{mn} = \frac{1}{z^2\,(\alpha^2-p^2)} {\footnotesize\begin{pmatrix}
 -\alpha & p & 0 \\
 -p & \alpha & 0 \\
 0 & 0 & 0 \end{pmatrix}}
 + \lambda {\footnotesize\begin{pmatrix}
 \frac{1}{z^2} & 0 & -\frac{1}{z} \\
 0 & 0 & 0 \\
 -\frac{1}{z} & 0 & 1 \end{pmatrix}}.
\end{align}
Here, the second term vanishes under the Sfetsos limit $\lambda\to 0$\,. 
By choosing $y=0$\,, the last row/column are truncated, and the reduced background fields become
\begin{align}
 \check{g}_{mn} = \frac{\alpha}{\alpha^2-p^2} \begin{pmatrix}
 -\frac{1}{z^2} & 0 \\
 0 & \frac{1}{z^2} \end{pmatrix} ,\qquad 
 \check{B}_{mn} = \frac{p}{\alpha^2-p^2} \begin{pmatrix}
 0 & \frac{1}{z^2} \\
 -\frac{1}{z^2} & 0 \end{pmatrix} .
\label{eq:DG-81}
\end{align}
If $\alpha$ and $p$ are constants, this corresponds to an AdS$_2$ background with $B$-field. 

Here we choose $\alpha$ and $p$ as the following functions of the spectator field $w$\,:
\begin{align}
 \alpha = \frac{1}{1-\hat{w}^2}\,,\qquad 
 p= \frac{\hat{w}}{1-\hat{w}^2}\,.
\label{eq:AdS2-param}
\end{align}
Then the reduced background fields can be embedded into the solution \eqref{eq:AdS2-sol} as
\begin{align}
 \rmd s^2 = \frac{- \rmd t^2 + \rmd z^2}{z^2} + \rmd s^2_{\text{S}^2\times \text{T}^6} \,,\qquad
 B_2 = \hat{w}\,\frac{\rmd t \wedge \rmd z}{z^2} + B_{\text{S}^2} \,,\qquad 
 \Exp{-2\,d}= \Exp{-2\,\hat{d}} z^{-2}\,,
\end{align}
where
\begin{align}
\begin{split}
 \rmd s^2_{\text{S}^2\times \text{T}^6}&\equiv \rmd\theta^2 +\sin^2\theta\,\rmd\phi^2 + \rmd w^2 + (\rmd z^1)^2 + \cdots + (\rmd z^5)^2\,,
\\
 B_{\text{S}^2} &\equiv \hat{w} \sin \theta\,\rmd \theta \wedge\rmd \phi\,,\qquad 
 \Exp{-2\,\hat{d}} \equiv \sin\theta\,.
\end{split}
\end{align}
In the following, we study various duality transformations, but $\rmd s^2_{\text{S}^2\times \text{T}^6}$\,, $B_{\text{S}^2}$\,, and $\Exp{-2\,\hat{d}}$ are invariant. 
Only the two-dimensional part of the background fields are transformed. 

For later convenience, we compute the degenerate matrix $\tilde{\cR}_{ab}$ as
\begin{align}
 (\tilde{\cR}_{ab}) = \lim_{\lambda\to 0} (\hat{E}^{-1})_{ab} = \frac{1}{\alpha^2-p^2}\, {\footnotesize\begin{pmatrix} - \alpha & p & 0 \\
 -p & \alpha & 0 \\
 0 & 0 & 0 \end{pmatrix}}\,.
\label{eq:cRt-ads2}
\end{align}
Using this matrix and the formula \eqref{eq:formula-fast}, we obtain various dual geometries.

\subsubsection{$(\vect{1}|\vect{8})$}

Let us perform a PL $T$-duality, $T_A\to T'_A=C_A{}^B\,T_B$ with
\begin{align}
 C_A{}^B = \begin{pmatrix} 
 \bm{0}_3 & \bm{1}_3 \\
 \bm{1}_3 & \bm{0}_3 \end{pmatrix}.
\label{eq:C-18}
\end{align}
The structure constants are transformed as
\begin{align}
 f_3{}^{12} = -1\,,\qquad
 f_1{}^{23} = 1\,,\qquad
 f_2{}^{13} = 1\,,
\end{align}
and this corresponds to the Manin triple $(\vect{1}|\vect{8})$. 
Using the coordinates $(x^m)=(r,\,\theta,\, z)$ and the parameterization
\begin{align}
 g=\Exp{r\cosh\theta\,T_1}\Exp{r\sinh\theta\,T_2}\Exp{z\,T_3},
\end{align}
we obtain
\begin{align}
\begin{split}
 e_a{}^m &= {\footnotesize\begin{pmatrix}
 \cosh \theta & -\frac{\sinh \theta}{r} & 0 \\
 -\sinh \theta & \frac{\cosh \theta}{r} & 0 \\
 0 & 0 & 1\end{pmatrix}} ,\quad
 \pi^{ab} = {\footnotesize\begin{pmatrix}
 0 & -z & r \sinh \theta \\
 z & 0 & r \cosh \theta \\
 -r \sinh \theta & -r \cosh \theta & 0\end{pmatrix}},\quad
 \det(\ell^a_m)=r\,.
\end{split}
\end{align}
Under the PL $T$-duality, $\VV^A$ is covariantly transformed and we find
\begin{align}
 \VV^A = (0,\,0,\,0\,;\,0,\,0,\,1)\,,\qquad 
 \VV^M = (0,\,1,\,0\,;\,0,\,0,\,1)\,.
\end{align}
Then the gauge-fixing condition can be taken as $\theta=0$\,. 

The reduced background fields can be easily found by substituting Eqs.~\eqref{eq:cRt-ads2} and \eqref{eq:C-18} into the formula \eqref{eq:formula-fast}. 
We find
\begin{align}
 \rmd \check{s}^2 = \frac{(p^2-\alpha^2)\,\rmd r^2}{\alpha}-\frac{2\,[p-(\alpha^2-p^2)\,z]\, \rmd r\,\rmd z}{\alpha\,r}+\frac{[1 + 2\,p\,z - (\alpha^2 - p^2)\,z^2]\,\rmd z^2}{\alpha\,r^2}\,,
\label{eq:DG-18}
\end{align}
and $\check{B}_2=0$\,.
Substituting the parameters \eqref{eq:AdS2-param}, we obtain the dual background
\begin{align}
\begin{split}
 \rmd s^2 &= -\rmd r^2
 +\frac{2\,(z-\hat{w})\,\rmd r\,\rmd z}{r}
 +\frac{(1-z^2+2\,z\,\hat{w}-\hat{w}^2)\,\rmd z^2}{r^2}
 + \rmd s^2_{\text{S}^2\times \text{T}^6}\,,
\\
 B_2 &= B_{\text{S}^2}\,,\qquad
 \Exp{-2d}=\Exp{-2\,\hat{d}} r\,. 
\end{split}
\end{align}
We can check that this satisfies the supergravity equations of motion. 

\subsubsection{$(\vect{8}|\vect{5.iii})$}

We again start with $(\vect{8}|\vect{1})$, and consider an $\OO(3,3)$ transformation
\begin{align}
 C_A{}^B = \begin{pmatrix} \delta_a^b & 0 \\ r^{ab} & \delta^a_b \end{pmatrix},\qquad
 r^{ab} \equiv {\footnotesize\begin{pmatrix}
 0 & -\eta & 0 \\
 \eta & 0 & \eta \\
 0 & -\eta & 0\end{pmatrix}},
\end{align}
which corresponds to the homogeneous Yang--Baxter deformation. 
The structure constants are found as
\begin{align}
\begin{split}
 f_{12}{}^3 &= -1\,,\quad
 f_{23}{}^1 = 1\,,\quad
 f_{13}{}^2 = 1\,, 
\\
 f_2{}^{12} &= -\eta\,,\quad
 f_2{}^{23} = \eta\,,\quad
 f_1{}^{13} = \eta\,,\quad
 f_3{}^{13} = -\eta\,,
\end{split}
\end{align}
and we obtain $(\vect{8}|\vect{5.iii})$ of \cite{math:0202210} (up to a trivial $\GL(3)$ redefinition of the generators). 
Using the coordinates $(x^m)=(t,\,z,\,y)$ and the parameterization
\begin{align}
 g = \Exp{y\,T_3}\Exp{\ln z\,T_2}\Exp{t\,(T_1 - T_3)},
\end{align}
we find
\begin{align}
\begin{split}
 e_a{}^m &= {\footnotesize\begin{pmatrix}
 z \cosh y & z \sinh y & \cosh y \\
 z \sinh y & z \cosh y & \sinh y \\
 0 & 0 & 1\end{pmatrix}},\qquad \det(\ell^a_m)=z^{-2}\,,
\\
 \pi^{ab} &= \eta{\footnotesize\begin{pmatrix}
 0 & 1-z+\frac{t^2}{z} & 2\,t \cosh y -\frac{(t^2+z^2) \sinh y}{z} \\
 * & 0 & (\frac{t^2}{z}+z) \cosh y -2\,t \sinh y-1 \\
 * & * & 0 \end{pmatrix}}.
\end{split}
\end{align}
The gauge generator is found as
\begin{align}
 \VV^A = (0,\,0,\,1\,;\,0,\,0,\,0)\,,\qquad 
 \VV^M = (0,\,0,\,1\,;\,0,\,0,\,0)\,,
\end{align}
and the gauge-fixing condition can be chosen as $y=0$\,. 

Then, the formula \eqref{eq:formula-fast} gives
\begin{align}
 \rmd\check{s}^2 = \frac{\alpha\,(-\rmd t^2 + \rmd z^2)}{\alpha^2\,z^2 - [p\,z + \eta\,(t^2-z^2)]^2}\,, \qquad
 \check{B}_2 = \frac{p + \eta\,z^{-1}\,(t^2-z^2)}{\alpha^2\,z^2 - [p\,z + \eta\,(t^2-z^2)]^2}\, \rmd t\wedge\rmd z\,.
\end{align}
If we take the limit $\eta\to 0$\,, the original geometry \eqref{eq:DG-81} is reproduced as expected. 
This can be embedded to the ten-dimensional background,
\begin{align}
\begin{split}
 \rmd s^2 &= \frac{-\rmd t^2+\rmd z^2}{z^2 + 2\,\eta\,\hat{w}\,z\,(z^2-t^2) - \eta^2\,(1-\hat{w}^2)\,(z^2-t^2)^2}
 + \rmd s^2_{\text{S}^2\times \text{T}^6}\,,
\\
 B_2 &= \frac{[\hat{w} - \eta\,z^{-1}\,(z^2-t^2)\,(1-\hat{w}^2)]\,\rmd t\wedge\rmd z}{z^2 + 2\,\eta\,\hat{w}\,z\,(z^2-t^2) - \eta^2\,(1-\hat{w}^2)\,(z^2-t^2)^2} + B_{\text{S}^2}\,,\qquad
 \Exp{-2d}=\Exp{-2\,\hat{d}} z^{-2}\,. 
\end{split}
\end{align}
In this case, the dual algebra $\tilde{\mathfrak{g}}=\vect{5.iii}$ is non-unimodular, and this background is a solution of generalized supergravity equations of motion. 
The Killing vector field is
\begin{align}
 I^m = \tfrac{1}{2}\,f_{b}{}^{ba}\,v_a^m = \eta\,(t^2+z^2,\, 2\,t\,z,\, 2\,z)\,,
\end{align}
and, under the gauge fixing $y=0$\,, this reduces to
\begin{align}
 I = \eta\,\bigl[(t^2+z^2)\,\partial_t + 2\,t\,z\,\partial_z \bigr]\,.
\end{align}
We can check the generalized supergravity equations of motion by using this $I$\,. 

\subsubsection{$(\vect{5.iii}|\vect{8})$}

Here we consider a PL $T$-plurality given by
\begin{align}
 C_A{}^B= {\footnotesize\left(\begin{array}{ccc|ccc}
 0 & -\eta & 0 & 1 & 0 & 0 \\
 \eta & 0 & \eta & 0 & 1 & 0 \\
 0 & -\eta & 0 & 0 & 0 & 1 \\ \hline
 1 & 0 & 0 & 0 & 0 & 0 \\
 0 & 1 & 0 & 0 & 0 & 0 \\
 0 & 0 & 1 & 0 & 0 & 0\end{array}\right) }.
\end{align}
The structure constants are found as
\begin{align}
\begin{split}
 f_{12}{}^2 &= -\eta\,,\quad
 f_{23}{}^2 = \eta\,,\quad
 f_{13}{}^1 = \eta\,,\quad
 f_{13}{}^3 = -\eta\,,
\\
 f_3{}^{12} &= -1\,,\quad
 f_1{}^{23} = 1\,,\quad
 f_2{}^{13} = 1\,.
\end{split}
\end{align}
This corresponds to the PL $T$-dual of the homogeneous Yang--Baxter deformation. 
A naive parameterization of $g$ makes it difficult to find a gauge-fixing condition. 
Accordingly, using the coordinates $(x^m)=(x,\,y,\,z)$, we make the following complicated parameterization:
\begin{align}
\begin{split}
 g &= \exp\bigl[\eta^{-1} \ln (\Exp{\eta\, y -z} - \Exp{2\,\eta\, y-z} - \Exp{z} + 2\,x - \Exp{-z}x^2)\,T_1\bigr]
\\
 &\quad \cdot \exp\bigl[\eta^{-1}\,( \Exp{2\,\eta\,y-z} -\Exp{\eta\,y-z} - \Exp{z} + \Exp{-z} x^2)\,T_2\bigr] \exp\bigl[-(y + \eta^{-1} \ln 2)\,T_3\bigr]\,.
\end{split}
\end{align}
This is singular in the limit $\eta\to 0$ and we cannot see the connection to the case $(\vect{1}|\vect{8})$. 
However, thanks to this parameterization, we obtain
\begin{align}
 \VV^A = (0,\,0,\,0\,;\,0,\,0,\,1)\,,\qquad 
 \VV^M = \bigl(0,\,0,\,1\,;\, 
 0,\, \tfrac{\Exp{z}}{\Exp{2\,\eta\,y} - \Exp{\eta\,y} + x^2 -2\Exp{z}x+\Exp{2 z}},\, 0 \bigr)\,.
\end{align}
Then, we can choose the gauge-fixing condition as $z=z_0$ ($z_0:$ constant)\,. 

The formula \eqref{eq:formula-fast} gives
\begin{align}
\begin{split}
 \rmd\check{s}^2 &= -\tfrac{\bigl[2\,(\alpha + p -\eta)\,\rmd x +\eta\,(\alpha -2\,p\,x + p + 2\,\eta\,x - 2\,\alpha\,x)\,\rmd y\bigr]\bigl[2\,(\alpha -p+\eta)\,\rmd x + \eta\,(\alpha +2\, p\,x-p-2\,\eta\,x -2\,\alpha\,x)\,\rmd y\bigr]}{4\,\alpha\,\eta^2\, (\Exp{2\,\eta\,y}-\Exp{\eta y}+x^2)}\,, 
\\
 \check{B}_2 &= \frac{\rmd x \wedge \rmd y}{\Exp{2\,\eta\,y}-\Exp{\eta\,y} + x^2 - 2\Exp{z_0} x +\Exp{2\,z_0}} - \frac{\rmd x \wedge \rmd y}{2\,(\Exp{2\,\eta\,y}-\Exp{\eta\,y}+x^2)} \,.
\end{split}
\end{align}
This can be embedded into a ten-dimensional configuration,
\begin{align}
\begin{split}
 \rmd s^2 &= -\tfrac{\bigl\{1 + \eta\,\bigl[\hat{w}\,(\eta\,\hat{w}+2)-\eta\bigr] \bigr\}\,\rmd x^2 + \eta\,\bigl\{1 + \eta\,\hat{w} - 2\,x\, \bigl[1-\eta^2 + \eta\,\hat{w}\, (\eta\,\hat{w} + 2)\bigr] \bigr\}\,\rmd x\,\rmd y}{\eta^2\,(x^2-\Exp{\eta\,y}+\Exp{2\,\eta\,y})}
\\
 &\quad - \tfrac{\bigl\{1 - 4\,x\,\bigl[ 1 - x + \eta\,\hat{w} + \eta^2\,x - \eta\,x\,\hat{w}\,(\eta\,\hat{w} + 2) \bigr] \bigr\}\,\rmd y^2}{4\,(x^2-\Exp{\eta\,y}+\Exp{2\,\eta\,y})}
  + \rmd s^2_{\text{S}^2\times \text{T}^6}\,,
\\
 B_2 &= \tfrac{\rmd x \wedge \rmd y}{\Exp{2\,\eta\, y}-\Exp{\eta\, y}+x^2-2\Exp{z_0}x +\Exp{2\,z_0}}-\tfrac{\rmd x \wedge \rmd y}{2\,(\Exp{2\,\eta\, y}-\Exp{\eta\, y}+x^2)} 
 + B_{\text{S}^2}\,,\qquad
 \Exp{-2d}=\Exp{-2\,\hat{d}} \tfrac{\Exp{-2\,\eta\,y}}{2\,\eta^2} \,. 
\end{split}
\end{align}
This expression is not so illuminating, but we can check that this indeed satisfies the standard supergravity equations of motion. 
The $B$-field can be slightly simplified by choosing $z_0=\infty$\,. 
Since the two-dimensional part of the $B$-field does not produce the $H$-flux, the equations of motion are satisfied even if we truncate the two-dimensional part of the $B$-field.

\subsubsection{$(\vect{6_0}|\vect{5.i})$}

Here we consider
\begin{align}
 C_A{}^B= {\footnotesize\left(\begin{array}{ccc|ccc}
 0 & 0 & 0 & 0 & 1 & 0 \\
 0 & 0 & 0 & -1 & 0 & 0 \\
 0 & 0 & 1 & 0 & 0 & 0 \\ \hline
 0 & 1 & 0 & 0 & 0 & 0 \\
 -1 & 0 & 1 & 0 & 0 & 0 \\
 0 & 0 & 0 & 1 & 0 & 1 \end{array}\right)},
\end{align}
which gives
\begin{align}
 f_{23}{}^1 = 1\,,\qquad
 f_{13}{}^2 = 1\,,\qquad
 f_2{}^{12} = -1\,,\qquad
 f_3{}^{13} = -1\,.
\label{eq:f-65}
\end{align}
Using the coordinates $(x^m)=(x,\,y,\,z)$ and
\begin{align}
 g= \Exp{x\,T_1} \Exp{y\,T_2} \Exp{z\,T_3},
\end{align}
we obtain
\begin{align}
\begin{split}
 e_a{}^m &= {\footnotesize\begin{pmatrix}
 1 & 0 & 0 \\
 0 & 1 & 0 \\
 -y & -x & 1 
\end{pmatrix}},\qquad \det(\ell^a_m)=1\,,
\\
 \pi^{ab} &= \eta{\footnotesize\begin{pmatrix}
 0 & -x \sinh z -y \cosh z & -\sinh z \\
 x \sinh z +y \cosh z & 0 & \cosh z -1 \\
 \sinh z & 1-\cosh z & 0 \end{pmatrix}},
\end{split}
\end{align}
We also find
\begin{align}
 \VV^A = (0,\,0,\,1\,;\,0,\,0,\,0)\,,\qquad 
 \VV^M = \bigl( -y,\,-x,\,1\,;\, 0,\,0,\,0 \bigr)\,,
\label{eq:VM-65}
\end{align}
and we choose the gauge-fixing condition as $z=z_0$ ($z_0:$ const.). 
We note that the coordinates $(x,y)$ are not invariant under the gauge transformation and are different from the adapted coordinates discussed in section \ref{sec:dGM}.

We then find the reduced background fields as
\begin{align}
\begin{split}
 \rmd\check{s}^2 &= \frac{\alpha\,(\rmd x^2-\rmd y^2)}{[1-(\alpha-p) (\sinh z_0\,x+\cosh z_0\,y)][1+(\alpha +p) (\sinh z_0\,x+\cosh z_0\,y)]}\,,
\\
 \check{B}_2 &= \frac{p -(\alpha^2-p^2) (\sinh z_0\,x+\cosh z_0\,y)}{[1-(\alpha-p) (\sinh z_0\,x+\cosh z_0\,y)][1+(\alpha +p)(\sinh z_0\,x+\cosh z_0\,y)]}\,\rmd x \wedge \rmd y \,.
\end{split}
\label{eq:65-1}
\end{align}
By choosing $z_0=0$\,, this is simplified as
\begin{align}
 \rmd\check{s}^2 = \frac{\alpha\,(\rmd x^2-\rmd y^2)}{1+ 2\,p\,y -(\alpha^2-p^2)\,y^2}\,, \qquad
 \check{B}_2 = \frac{p-(\alpha^2-p^2)\,y}{1+2\,p\,y -(\alpha^2-p^2)\,y^2}\,\rmd x \wedge \rmd y \,.
\label{eq:65-2}
\end{align}
We note that the geometry \eqref{eq:65-1} can be mapped to \eqref{eq:65-2} through a Lorentz transformation
\begin{align}
 x'=\cosh z_0\,x-\sinh z_0\,y\,,\qquad
 y'=-\sinh z_0\,x+\cosh z_0\,y\,.
\end{align}
This can be also understood from that the gauge generator \eqref{eq:VM-65} generates an infinitesimal Lorentz transformation as well as the translation along $z$\,. 

We can embed the configuration \eqref{eq:65-2} into the ten-dimensional background
\begin{align}
\begin{split}
 \rmd s^2 &= \frac{\rmd x^2-\rmd y^2}{1-y^2+2\,y\,\hat{w} - \hat{w}^2}
 + \rmd s^2_{\text{S}^2\times \text{T}^6}\,,
\\
 B_2 &= \frac{(\hat{w}-y)\,\rmd x \wedge \rmd y}{1-y^2+2\,y\,\hat{w}-\hat{w}^2} + B_{\text{S}^2}\,,\qquad
 \Exp{-2d}=\Exp{-2\,\hat{d}} \,. 
\end{split}
\end{align}
The dual algebra is non-unimodular, and we need to introduce the Killing vector field as
\begin{align}
 I^m = (\cosh z,\, -\sinh z,\, 0) \quad \overset{z=0}{\to}\quad I = \partial_x \,.
\end{align}
Using this, we can check that the generalized supergravity equations of motion are satisfied.

\subsubsection{$(\vect{5.i}|\vect{6_0})$}

Let us consider
\begin{align}
 C_A{}^B= {\footnotesize\left(\begin{array}{ccc|ccc}
 0 & 1 & 0 & 0 & 0 & 0 \\
 -1 & 0 & 1 & 0 & 0 & 0 \\
 0 & 0 & 0 & 1 & 0 & 1 \\\hline
 0 & 0 & 0 & 0 & 1 & 0 \\
 0 & 0 & 0 & -1 & 0 & 0 \\
 0 & 0 & 1 & 0 & 0 & 0 \end{array}\right) },
\end{align}
which gives
\begin{align}
 f_{12}{}^2 = -1\,,\qquad
 f_{13}{}^3 = -1\,,\qquad
 f_1{}^{23} = 1\,,\qquad
 f_2{}^{13} = 1\,.
\end{align}
We choose the coordinates as $(x^m)=(x,\,y,\,z)$ and use
\begin{align}
 g= \Exp{\frac{\cosh z\sin x}{\cos x + \sin x\sinh z}\,T_2} \Exp{\ln(\cos x + \sin x\sinh z)\, T_1} \Exp{y\,T_3}.
\end{align}
We then find
\begin{align}
\begin{split}
 e_a{}^m &= {\footnotesize\begin{pmatrix}
 -\sinh z & 0 & \frac{\cosh z}{\tan x} \\
 \cosh z & 0 & 1-\frac{\sinh z}{\tan x} \\
 0 & \sin x \sinh z +\cos x & 0 \end{pmatrix}},\qquad
 \det (\ell^a_m) = \sin x\,,
\\
 \pi^{ab} &= {\footnotesize\begin{pmatrix}
 0 & 0 & \frac{\tan x \cosh z}{1+\tan x \sinh z} \\
 0 & 0 & \frac{\tan x \sinh z}{1+\tan x \sinh z} \\
 -\frac{\tan x \cosh z}{1+\tan x \sinh z} & -\frac{\tan x \sinh z}{1+\tan x \sinh z} & 0 \end{pmatrix}} ,
\end{split}
\end{align}
and also
\begin{align}
 \VV^A = (0,\,0,\,0\,;\,0,\,0,\,1)\,,\quad 
 \VV^M = \bigl( 0,\,0,\,1\,;\,
 0,\,\tfrac{1}{\sin x \sinh z +\cos x},\,0 \bigr)\,.
\end{align}
The gauge-fixing condition we take is $z=0$\,. 
The reduced background fields are simple,
\begin{align}
\begin{split}
 \rmd\check{s}^2 = -\frac{\rmd x^2}{\alpha} + \frac{2\,p\,\rmd x\,\rmd y}{\alpha\sin x} + \frac{(\alpha^2-p^2)\,\rmd y^2}{\alpha\sin^2 x}\,, \qquad
 \check{B}_2 = \frac{\rmd x \wedge \rmd y}{\cos y}\,.
\end{split}
\end{align}
This can be embedded into the ten-dimensional supergravity solution
\begin{align}
\begin{split}
 \rmd s^2 &= - \rmd x^2 + \Bigl(\frac{\rmd y}{\sin x}+\hat{w}\, \rmd x\Bigr)^2 + \rmd s^2_{\text{S}^2\times \text{T}^6}\,,
\\
 B_2 &= \frac{\rmd x \wedge \rmd y}{\cos x} + B_{\text{S}^2}\,,\qquad
 \Exp{-2d}=\Exp{-2\,\hat{d}}\sin x\,. 
\end{split}
\end{align}

\subsubsection{$(\vect{7_0}|\vect{5.i})$}

Here we consider
\begin{align}
 C_A{}^B= {\footnotesize\left(\begin{array}{ccc|ccc}
 0 & 0 & 0 & 0 & -1 & 0 \\
 0 & 0 & 0 & 0 & 0 & 1 \\
 1 & 0 & 0 & 0 & 0 & 0 \\\hline
 0 & -1 & 0 & 0 & 0 & 0 \\
 1 & 0 & 1 & 0 & 0 & 0 \\
 0 & 0 & 0 & 1 & 0 & -1\end{array}\right)},
\end{align}
and the structure constants become
\begin{align}
 f_{23}{}^1 = 1\,,\qquad
 f_{13}{}^2 = -1\,,\qquad
 f_2{}^{12} = -1\,,\qquad
 f_3{}^{13} = -1\,.
\end{align}
The only difference from Eq.~\eqref{eq:f-65} is in the sign of $f_{13}{}^2$\,. 
We use the coordinates $(x^m)=(x,\,y,\,z)$ and
\begin{align}
 g= \Exp{z\,T_3} \Exp{x\,T_1} \Exp{y\,T_2} .
\end{align}
Then we find
\begin{align}
 e_a{}^m= {\footnotesize\begin{pmatrix}
 \cos z & -\sin z & 0 \\
 \sin z & \cos z & 0 \\
 0 & p & 1 \end{pmatrix}} ,\qquad
 \pi^{ab} = {\footnotesize\begin{pmatrix}
 0 & -y & -\sin z \\
 y & 0 & \cos z -1 \\
 \sin z & 1-\cos z & 0 \end{pmatrix}} ,\qquad
 \det(\ell^a_m)=1\,,
\end{align}
and the gauge generator is
\begin{align}
 \VV^A = (0,\,0,\,-1\,;\,0,\,1,\,0)\,,\qquad 
 \VV^M = (-y \cos z,\, y \sin z\,, -\cos z\,;\, \sin z,\, \cos z,\, 0)\,.
\end{align}
Again, the coordinates $(x,y)$ are not gauge-invariant, but we can choose the gauge $z=0$\,. 
We also note that the gauge algebra $\mathfrak{f}$ is generated by $T_{\sfi=1}=T^2-T_3$\,, which is neither a subgroup of $\mathfrak{g}$ nor of $\tilde{\mathfrak{g}}$\,. 
Thus this gives a dressing coset that is neither the usual coset nor the PL $T$-dual of the usual coset. 

The reduced background fields become
\begin{align}
\begin{split}
 \rmd\check{s}^2 = \frac{\alpha\, (\rmd x^2-\rmd y^2)}{1+2\,p\,y-(\alpha^2-p^2)\,y^2}\,, \qquad
 \check{B}_2 = \frac{p-(\alpha^2-p^2)\, y}{1+2\,p\,y -(\alpha^2-p^2)\, y^2}\,\rmd x \wedge \rmd y \,,
\end{split}
\label{eq:Echeck-75}
\end{align}
If we perform a coordinate transformation
\begin{align}
 x'=x-\tan z_0\,y\,,\qquad y'=\frac{y}{\cos z_0}\qquad (z_0:\text{ const.})\,,
\end{align}
this configuration is mapped to that of a more general gauge $z=z_0$\,. 
Namely, the choice of the value $z$ is not important. 
We note that the reduced background fields \eqref{eq:Echeck-75} are exactly the same as Eq.~\eqref{eq:65-2} associated with the Manin triple $(\vect{6_0}|\vect{5.i})$. 
This will be accidental and may be due to the similarity of the structure constants. 
The dilaton is also the same due to and the Killing vector field is also the same:
\begin{align}
 I = v_1 = \partial_x\,.
\end{align}
Thus the ten-dimensional uplift is also the same as the case of the Manin triple $(\vect{6_0}|\vect{5.i})$. 
The only difference is how the gauge group $F$ acts on $M$. 

\subsubsection{$(\vect{5.i}|\vect{7_0})$}

Let us consider the PL $T$-dual of the previous example
\begin{align}
 C_A{}^B= {\footnotesize\left(\begin{array}{ccc|ccc}
 0 & -1 & 0 & 0 & 0 & 0 \\
 1 & 0 & 1 & 0 & 0 & 0 \\
 0 & 0 & 0 & 1 & 0 & -1 \\\hline
 0 & 0 & 0 & 0 & -1 & 0 \\
 0 & 0 & 0 & 0 & 0 & 1 \\
 1 & 0 & 0 & 0 & 0 & 0 \end{array}\right)},
\label{eq:C-57}
\end{align}
where
\begin{align}
 f_{12}{}^2 = -1\,,\qquad
 f_{13}{}^3 = -1\,,\qquad
 f_1{}^{23} = 1\,,\qquad
 f_2{}^{13} = -1\,.
\end{align}
Using $(x^m)=(x,\,y,\,z)$ and
\begin{align}
 g= \Exp{\ln(\cosh y\cosh z + \sinh y)\,T_1}\Exp{x\,T_3}\Exp{\cosh y\sinh z\,T_2},
\end{align}
we obtain
\begin{align}
\begin{split}
 e_a{}^m&= {\footnotesize\begin{pmatrix}
 0 & \cosh z & -\tanh y \sinh z \\
 0 & -\sinh z & 1+\tanh y \cosh z \\
 \cosh y \cosh z +\sinh y & 0 & 0 \end{pmatrix}} ,
\\
 \pi^{ab} &= {\footnotesize\begin{pmatrix}
 0 & 0 & -\frac{\sinh z}{\tanh y +\cosh z} \\
 0 & 0 & \frac{\tanh y}{\tanh y + \cosh z} \\
 \frac{\sinh z}{\tanh y +\cosh z} & -\frac{\tanh y}{\tanh y + \cosh z} & 0 \end{pmatrix}} ,\qquad
 \det(\ell^a_m)=\cosh y\,,
\end{split}
\end{align}
and
\begin{align}
 \VV^A = (0,\,1,\,0\,;\,0,\,0,\,-1)\,,\quad 
 \VV^M = \bigl( 0,\,0,\,1\,;\,
 -\tfrac{1}{\cosh y \cosh z +\sinh y},\,0,\,0\bigr)\,.
\end{align}
The gauge-fixing condition can be chosen as $z=0$\,, and then we find
\begin{align}
 \rmd \check{s}^2 = -\frac{(\alpha^2-p^2)\,\rmd x^2}{\cosh^2y}-\frac{2\,p\,\rmd x\,\rmd y}{\cosh y} +\frac{\rmd y^2}{\alpha}\,,
\qquad
 \check{B}_2 = 0\,.
\label{eq:Echeck-57}
\end{align}
This is an intriguing background with the Ricci scalar $\check{R}=\frac{2\,(\alpha^2-p^2)}{\alpha}\,\frac{2-\cosh^2 y}{\cosh^2 y}$\,. 
For large $y$\,, we find $\check{R}\to -\frac{2\,(\alpha^2-p^2)}{\alpha}$ and this is asymptotically AdS$_2$/dS$_2$ for $\alpha^2\gtrless p^2$\,, but the sign of the Ricci scalar is opposite, $\check{R}=\frac{2\,(\alpha^2-p^2)}{\alpha}$\,, at $y=0$\,.
Accordingly, this is a vacuum solution (without cosmological constant). 
This is embedded into a ten-dimensional background
\begin{align}
\begin{split}
 \rmd s^2 &= - \frac{\rmd x^2}{\cosh^2 y}
 -\frac{2\,\hat{w}\,\rmd x\,\rmd y}{\cosh y}
 +(1-\hat{w}^2)\,\rmd y^2 + \rmd s^2_{\text{S}^2\times \text{T}^6}\,,
\\
 B_2 &= B_{\text{S}^2}\,,\qquad
 \Exp{-2d}=\Exp{-2\,\hat{d}}\cosh y\,,
\end{split}
\end{align}
which is a solution of the usual supergravity.

Since this is the PL $T$-dual of the previous example, the gauge algebra $\mathfrak{f}$ is neither a subalgebra of $\mathfrak{g}$ nor $\tilde{\mathfrak{g}}$\,.
Then, the standard procedure of Sfetsos does not apply. 
Namely, if we set $\lambda=1$ in $\hat{E}^{ab}$ given in Eq.~\eqref{eq:Eab-81} and bring this into the present duality frame through the $\OO(3,3)$ transformation \eqref{eq:C-57}, it is not clear how to recover the parameter $\lambda$\,. 
However, by using a prescription discussed in section \ref{sec:g-Sfetsos}, we can introduce the parameter $\lambda$ as
\begin{align}
 \hat{G}^{ab} &= {\footnotesize\begin{pmatrix}
 \frac{\alpha -\lambda\,(\alpha^2-p^2)}{1-\lambda\,\alpha} & 0 & \frac{\lambda\,p}{1-\lambda\,\alpha} \\
 0 & -\frac{\alpha}{1-\lambda\,\alpha} & 0 \\
 \frac{\lambda\,p}{1- \lambda\,\alpha} & 0 & \frac{\lambda}{1- \lambda\,\alpha} \end{pmatrix} },\qquad
 \hat{\beta}^{ab} = {\footnotesize\begin{pmatrix}
 0 & \frac{p}{1-\lambda\,\alpha} & 0 \\
 -\frac{p}{1-\lambda\,\alpha} & 0 & -\frac{1}{1-\lambda\,\alpha} \\
 0 & \frac{1}{1-\lambda\,\alpha} & 0 \end{pmatrix} },
\end{align}
where we have chosen $\zeta_{\sfi}{}^{\sfj}=0$\,. 
This is precisely the same as the $\OO(3,3)$ transformation of Eq.~\eqref{eq:Eab-81} (without setting $\lambda=1$). 
Then by taking the limit $\lambda\to 0$\,, we can reproduce Eq.~\eqref{eq:Echeck-57}. 
Of course, if we use the formula \eqref{eq:reduced-E}, it is not necessary to introduce $\lambda$\,. 
We can reproduce Eq.~\eqref{eq:Echeck-57} for arbitrary values of $\lambda$\,. 

\subsubsection{$(\vect{5}|\vect{2.ii})$}

We here consider
\begin{align}
 C_A{}^B= {\footnotesize\left(\begin{array}{ccc|ccc}
 -1 & -1 & 1 & 0 & -1 & -1 \\
 0 & 0 & 0 & 1 & -1 & 0 \\
 -1 & -1 & 0 & 0 & 0 & -1 \\\hline
 0 & 0 & 0 & 1 & -1 & 1 \\
 1 & 0 & -1 & 0 & 1 & 0 \\
 0 & 0 & 0 & -1 & 0 & -1\end{array}\right)},
\end{align}
which gives
\begin{align}
 f_{12}{}^2 = -1\,,\qquad
 f_{13}{}^3 = -1\,,\qquad
 f_3{}^{12} = 1\,.
\end{align}
Using, $(x^m)=(x,\,y,\,z)$ and
\begin{align}
 g= \Exp{(y-1)\,T_3} \Exp{(x-\frac{1+y^2}{2})\,T_2} \Exp{z\,T_1},
\end{align}
we find
\begin{align}
 e_a{}^m= {\footnotesize\begin{pmatrix}
 y-x+\frac{1-y^2}{2} & 1-y & 1 \\
 1 & 0 & 0 \\
 y & 1 & 0\end{pmatrix}} ,\quad
 \pi^{ab} = {\footnotesize\begin{pmatrix}
 0 & y-1 & 0 \\
 1-y & 0 & -\frac{(1-y)^2}{2} \\
 0 & \frac{(1-y)^2}{2} & 0\end{pmatrix}} ,
\end{align}
and $\det(\ell^a_m)=\Exp{2\,z}$\,. 
The gauge generator is
\begin{align}
 \VV^A = (1,\,0,\,-1\,;\,-1,\,0,\,-1)\,,\qquad 
 \VV^M = \bigl(-x,\, -y,\, 1\,;\, 0,\, -1,\, -y\bigr)\,,
\end{align}
and we can choose the gauge-fixing condition as $z=0$\,. 
Here, again the gauge group $F$ is neither a subgroup of $G$ nor $\tilde{G}$\,. 
The reduced background fields are
\begin{align}
\begin{split}
 \rmd\check{s}^2 =\frac{-(\alpha^2-p^2)\,\rmd x^2-2\,p\,\rmd x\,\rmd y+\rmd y^2}{\alpha\,y^2}\,, \qquad
 \check{B}_2 = \frac{\rmd x \wedge \rmd y}{y^2}\,,
\end{split}
\end{align}
and they are uplifted to a ten-dimensional supergravity solution,
\begin{align}
 \rmd s^2 = \frac{-(\rmd x+ \hat{w}\,\rmd y)^2 + \rmd y^2}{y^2}
 + \rmd s^2_{\text{S}^2\times \text{T}^6}\,,\quad
 B_2 = \frac{\rmd x \wedge \rmd y}{y^2} + B_{\text{S}^2}\,,\quad
 \Exp{-2d}=\Exp{-2\,\hat{d}}. 
\end{align}

\subsubsection{$(\vect{2.ii}|\vect{5})$}

Finally, we consider
\begin{align}
 C_A{}^B= {\footnotesize\left(\begin{array}{ccc|ccc}
 0 & 0 & 0 & 1 & -1 & 1 \\
 1 & 0 & -1 & 0 & 1 & 0 \\
 0 & 0 & 0 & -1 & 0 & -1 \\\hline
 -1 & -1 & 1 & 0 & -1 & -1 \\
 0 & 0 & 0 & 1 & -1 & 0 \\
 -1 & -1 & 0 & 0 & 0 & -1\end{array}\right)}.
\end{align}
The structure constants become
\begin{align}
 f_{12}{}^3 = 1\,,\qquad
 f_2{}^{12} = -1\,,\qquad
 f_3{}^{13} = -1\,.
\end{align}
Using the coordinates $(x^m)=(x,\,y,\,z)$ and the parameterization
\begin{align}
 g= \Exp{(x + 1)\,T_3} \Exp{y\,T_1} \Exp{z\,T_2},
\end{align}
we find
\begin{align}
 e_a{}^m= {\footnotesize\begin{pmatrix}
 0 & 1 & 0 \\
 -y & 0 & 1 \\
 1 & 0 & 0\end{pmatrix}} ,\qquad
 \pi^{ab} = {\footnotesize\begin{pmatrix}
 0 & -z & -1-x-y\,z \\
 z & 0 & -\frac{z^2}{2} \\
 1+x+y\,z & \frac{z^2}{2} & 0\end{pmatrix}} ,\qquad
 \det(\ell^a_m)=1\,.
\end{align}
The gauge generators are
\begin{align}
 \VV^A = (-1,\, 0,\, -1\,;\,1,\, 0,\, -1)\,,\qquad 
 \VV^M = \bigl(x - \tfrac{y\,z^2}{2},\, x + y\,z,\, \tfrac{z\,(2 + z)}{2}\,;\,-1,\, 1,\, -y\bigr)\,.
\end{align}
Again, the coordinates $x$ and $y$ are not gauge invariant, but we can choose the gauge $z=1$\,. 
Then the reduced background fields can be found as
\begin{align}
\begin{split}
 \rmd\check{s}^2 &= \frac{3\,\alpha\,(-4\,\rmd x^2 + \rmd y^2)}{\alpha^2\,(2\,x+2\,y-3)^2-[p\,(2\,x+2\,y-3)+3]^2}\,,
\\
 \check{B}_2 &= \frac{[6\,p + [6 - 4\,(x + y)]\,(\alpha^2-p^2)]\,\rmd x\wedge\rmd y}{\alpha^2\,(2\,x+2\,y-3)^2-[p\,(2\,x+2\,y-3)+3]^2}\,.
\end{split}
\end{align}
This can embedded into a ten-dimensional background
\begin{align}
\begin{split}
 \rmd s^2 &= \frac{3\,(-4\,\rmd x^2+\rmd y^2)}{9\,\hat{w}^2 -9 -6\,\hat{w}\, (2\,x+2\,y-3) + (2\,x+2\,y-3)^2} + \rmd s^2_{\text{S}^2\times \text{T}^6}\,,
\\
 B_2 &= \frac{[6+6\,\hat{w}-4\,(x-y)]\,\rmd x \wedge \rmd y}{9\,\hat{w}^2 -9 -6\,\hat{w}\, (2\,x+2\,y-3) + (2\,x+2\,y-3)^2} + B_{\text{S}^2}\,,\qquad
 \Exp{-2d}=\Exp{-2\,\hat{d}},
\end{split}
\end{align}
that satisfies the generalized supergravity equations of motion with the Killing vector field
\begin{align}
 I^m = v_{1}^m = (-z,\, 1,\, 0)\quad \overset{z=1}{\to}\quad I = \partial_y - \partial_x \,.
\end{align}

\subsection{Poisson--Lie $T$-plurality of AdS$_2$ with Ramond--Ramond fields}
\label{sec:PL-AdS2-RR}

Here we study an example with the R--R fields. 
For simplicity, we use the same geometries as those used in the previous subsection and focus only on the R--R fields. 

\subsubsection{$(\vect{8}|\vect{1})$}

For the Manin triple $(\vect{8}|\vect{1})$, we use a slightly different parameterization from Eq.~\eqref{eq:81-param},
\begin{align}
 g=\Exp{y\,T_3}\Exp{\arcsinh\rho\,T_2}\Exp{t\,T_1},
\end{align}
and the gauge-fixing condition is $y=0$\,. 
The matrix $\hat{E}^{ab}$ is the same as Eq.~\eqref{eq:Eab-81}, but here
\begin{align}
 \alpha = 1\,,\qquad p= 0\,.
\end{align}
Then the reduced background fields are those of the $\text{AdS}_2$ space
\begin{align}
 \rmd\check{s}^2 = -(1+\rho^2)\,\rmd t^2 + (1+\rho^2)^{-1}\,\rmd\rho^2 \,,\qquad 
 \check{B}_2 = 0 \,.
\end{align}
Using $\det(\ell_m^a)=1$\,, we can embed this into an $\text{AdS}_2\times \text{S}^2\times \text{T}^6$ solution of type IIA supergravity
\begin{align}
\begin{split}
 \rmd s^2 &= -(1+\rho^2)\,\rmd t^2 + (1+\rho^2)^{-1}\,\rmd\rho^2 + \rmd s^2_{\text{S}^2\times \text{T}^6}\,,\qquad
 B_2=0\,,\qquad \Exp{-2\,d} = 1\,,
\\
 G &= F = \rmd t\wedge \rmd \rho\wedge(1+\nu_4) -\omega_2\wedge (\nu_2 +\nu_6) \,,
\end{split}
\label{eq:8-1-RR}
\end{align}
where we have defined
\begin{align}
\begin{split}
 &\rmd s^2_{\text{S}^2\times \text{T}^6} \equiv (1 - r^2)^{-1}\,\rmd r^2 + (1 - r^2)\,\rmd \theta^2 + (\rmd z^1)^2 + \cdots + (\rmd z^6)^2\,,
\\
 &\nu_2\equiv \rmd z^1\wedge\rmd z^2 +\rmd z^3\wedge\rmd z^4+\rmd z^5\wedge\rmd z^6\,, 
\\
 &\nu_4 \equiv \rmd z^1\wedge\rmd z^2\wedge \rmd z^3\wedge\rmd z^4+\rmd z^1\wedge\rmd z^2\wedge \rmd z^5\wedge\rmd z^6+\rmd z^3\wedge\rmd z^4\wedge \rmd z^5\wedge\rmd z^6\,,
\\
 &\nu_6 \equiv \rmd z^1\wedge\cdots \wedge\rmd z^6\,,\qquad 
 \omega_2 \equiv \rmd r\wedge \rmd \theta\,.
\end{split}
\end{align}

In the following, we shall study the PL $T$-plurality of this solution. 
For this purpose, let us construct the generalized frame fields $\check{\CE}_A{}^M$ and $\vert\hat{\cF}\rangle$ appearing in Eq.~\eqref{eq:cF-dualizability}. 
We can construct the internal part of the generalized vielbeins $\check{\CE}_{\bcheck{A}}{}^{\check{M}}$ as
\begin{align}
 (\check{\CE}_{\bcheck{A}}{}^{\check{M}}) = \frac{1}{\sqrt{2}}{\footnotesize\begin{pmatrix}
 -\frac{1}{\sqrt{1+\rho^2}} & 0 & \sqrt{1+\rho^2} & 0 \\
 0 & \sqrt{1+\rho^2} & 0 & \frac{1}{\sqrt{1+\rho^2}} \\
 \frac{1}{\sqrt{1+\rho^2}} & 0 & \sqrt{1+\rho^2} & 0 \\
 0 & -\sqrt{1+\rho^2} & 0 & \frac{1}{\sqrt{1+\rho^2}} \end{pmatrix}},
\end{align}
but this is not an $\OO(2,2)$ element. 
We choose the matrix $T_A{}^{\bcheck{B}}$ defined in Eq.~\eqref{eq:def-T} as
\begin{align}
 \bigl(T_A{}^{\bcheck{B}}\bigr) = \frac{1}{\sqrt{2}}{\footnotesize\begin{pmatrix}
 -1 & 0 & 1 & 0 \\
 0 & 1 & 0 & -1 \\
 1 & 0 & 1 & 0 \\
 0 & 1 & 0 & 1 \end{pmatrix}},
\label{eq:T-def}
\end{align}
and then we obtain
\begin{align}
 \bigl(\check{\CE}_{A}{}^M\bigr) = {\footnotesize\begin{pmatrix}
 \frac{1}{\sqrt{1+\rho^2}} & 0 & 0 & 0 \\
 0 & \sqrt{1+\rho^2} & 0 & 0 \\
 0 & 0 & \sqrt{1+\rho^2} & 0 \\
 0 & 0 & 0 & \frac{1}{\sqrt{1+\rho^2}} \end{pmatrix}} \in \OO(2,2)\,.
\label{eq:81-CE}
\end{align}
To reproduce the R--R field strength given in Eq.~\eqref{eq:8-1-RR}, we choose $\vert\hat{\cF}\rangle$ as
\begin{align}
 \vert\hat{\cF}\rangle = \bigl[\Gamma^{01}\,(1+\Gamma^{4567}+\Gamma^{4578}+\Gamma^{6789})-\Gamma^{23}\,(\Gamma^{45}+\Gamma^{67}+\Gamma^{89}+\Gamma^{456789})\bigr]\,\ket{0}\,,
\end{align}
where $(x^0,\dotsc,x^9)=(t,\,\rho,\,r,\,\theta,\,z^1,\,\cdots,\,z^6)$. 
The R--R field strength for each Manin triple can be found by twisting this $\vert\hat{\cF}\rangle$ by using the frame fields $\check{\CE}_M{}^A$ as described in Eq.~\eqref{eq:cF-dualizability}. 

Using the generalized frame fields \eqref{eq:81-CE}, we obtain the generalized fluxes \eqref{eq:bm-f} as
\begin{align}
 \tfrac{1}{3!}\,\bm{f}_{ABC}\,\Gamma^{ABC} = \tfrac{\rho}{2\sqrt{1+\rho^2}}\,\Gamma^1\,[\Gamma^0,\,\Gamma_0]\,,\qquad
 -\tfrac{1}{2}\,\bm{f}_A\,\Gamma^A = \tfrac{\rho}{2\sqrt{1+\rho^2}}\,\Gamma^1 \,.
\end{align}
Here, one might notice that these two fluxes look similar. 
Interestingly, in all of the Manin triples studied in the following, we always find that when $-\tfrac{1}{2}\,\bm{f}_A\,\Gamma^A$ takes the form
\begin{align}
 -\tfrac{1}{2}\,\bm{f}_A\,\Gamma^A = \bm{d}_0\,\Gamma^0 + \bm{d}_1\,\Gamma^1 + \bm{d}^0\,\Gamma_0 + \bm{d}^1\,\Gamma_1\,,
\end{align}
the other flux takes the form
\begin{align}
 \tfrac{1}{3!}\,\bm{f}_{ABC}\,\Gamma^{ABC} = \bm{d}_0\,\Gamma^0\,[\Gamma^1,\,\Gamma_1] + \bm{d}_1\,\Gamma^1\,[\Gamma^0,\,\Gamma_0] + \bm{d}^0\,\Gamma_0\,[\Gamma_1,\,\Gamma^1] + \bm{d}^1\,\Gamma_1\,[\Gamma_0,\,\Gamma^0]\,.
\end{align}
Then we find
\begin{align}
 \sla{D} = \sla{\partial} + (\bm{d}_0\,\Gamma^0 + \bm{d}_1\,\Gamma^1)\,(\Gamma^0\,\Gamma_0+\,\Gamma^1\,\Gamma_1) + (\bm{d}^0\,\Gamma_0+\bm{d}^1\,\Gamma_1)\,(\Gamma_0\,\Gamma^0+\Gamma_1\,\Gamma^1) \,.
\label{eq:sla-D}
\end{align}
Since $\vert\hat{\cF}\rangle$ is constant, $\sla{\partial}$ vanishes, and we also find the gamma matrices annihilate $\vert\hat{\cF}\rangle$ for any values of $\{\bm{d}_0,\, \bm{d}_1,\, \bm{d}^0,\,\bm{d}^1\}$. 
Thanks to this property, we can easily check the Bianchi identity. 
There will be some nice structure behind this property, but unfortunately, we have not been able to find it in this paper. 

\subsubsection{$(\vect{1}|\vect{8})$}

Here and hereafter, we use the same parameterization as the previous subsection.
The fields in the NS--NS sector are the same as Eq.~\eqref{eq:DG-18} with $\alpha=1$ and $p=0$\,,
\begin{align}
 \rmd s^2 = -\rmd \rho^2 + \frac{2\,z\, \rmd \rho\,\rmd z}{\rho}+\frac{(1 - z^2)\,\rmd z^2}{\rho^2} + \rmd s^2_{\text{S}^2\times \text{T}^6}\,, \qquad
 B_2 = 0\,,\qquad
 \Exp{-2d} = \rho\,,
\end{align}
where $\rho$ corresponds to $r$ of Eq.~\eqref{eq:DG-18}. 
Again we use the same matrix $T_A{}^{\bcheck{B}}$ as Eq.~\eqref{eq:T-def} and construct the generalized frame fields $\check{\CE}_{A}{}^M$. 
The inverse matrix can be decomposed as
\begin{align}
 (\check{\CE}_{M}{}^A)
 = {\footnotesize\begin{pmatrix}
 1 & 0 & 0 & 0 \\
 -\frac{z}{\rho} & -\frac{1}{\rho} & 0 & 0 \\
 0 & 0 & 1 & -z \\
 0 & 0 & 0 & -\rho\end{pmatrix} \begin{pmatrix}
 0 & 0 & 1 & 0 \\
 0 & 1 & 0 & 0 \\
 1 & 0 & 0 & 0 \\
 0 & 0 & 0 & 1\end{pmatrix}}.
\end{align}
This is a combination of the factorized timelike $T$-duality and a $\GL(2)$ transformation. 
Under the first $T$-duality, we find
\begin{align}
\begin{split}
 &\vert\hat{\cF}\rangle = \bigl[\Gamma^{01}\,(1+\Gamma^{4567}+\Gamma^{4578}+\Gamma^{6789})-\Gamma^{23}\,(\Gamma^{45}+\Gamma^{67}+\Gamma^{89}+\Gamma^{456789})\bigr]\,\ket{0}
\\
 &\to \vert\hat{\cF}\rangle = \bigl[-\Gamma^{1}\,(1+\Gamma^{4567}+\Gamma^{4578}+\Gamma^{6789})-\Gamma^{23}\,(\Gamma^{45}+\Gamma^{67}+\Gamma^{89}+\Gamma^{456789})\,\Gamma^0\bigr]\,\ket{0}\,.
\end{split}
\end{align}
The second transformation maps
\begin{align}
 \Gamma^{0}\to \Gamma^{0}-\tfrac{z}{\rho}\,\Gamma^{1}\,,\qquad
 \Gamma^{1}\to -\tfrac{1}{\rho}\,\Gamma^{1}\,, \qquad
 \ket{0} \to \sqrt{\rho}\,\ket{0}\,,
\end{align}
which also changes the orientation $\epsilon_{\rho zr\theta1\cdots 6}=-1$\,. 
We then find
\begin{align}
\begin{split}
 \ket{\cF} &= \rho^{-\frac{1}{2}}\,\bigl[\Gamma^{1}\,(1+\Gamma^{4567}+\Gamma^{4578}+\Gamma^{6789})
\\
 &\qquad\quad\ -\Gamma^{23}\,(\Gamma^{45}+\Gamma^{67}+\Gamma^{89}+\Gamma^{456789})\,(\rho\,\Gamma^{0}-z\,\Gamma^{1})\bigr]\,\ket{0}\,.
\end{split}
\end{align}
In terms of the polyform, we find $F=\Exp{-d}\cF$ as
\begin{align}
 F = \rmd z\wedge (1+\nu_4)
 -(\rho\,\rmd \rho -z\,\rmd z)\wedge \omega_2 \wedge (\nu_2+\nu_6) =G \,.
\end{align}
Due to the timelike $T$-duality, this solves the equations of motion of type IIB$^*$ supergravity. 
Moreover, due to the timelike $T$-duality, the sign of the self-duality relation is different from the standard sign (e.g.~$G_5=-*G_5$). 
The generalized fluxes are computed as
\begin{align}
 -\tfrac{1}{2}\,\bm{f}_A\,\Gamma^A = \tfrac{1}{2\rho}\,\bigl(\Gamma_0 - z\,\Gamma^1\bigr) \,,
\end{align}
and the Bianchi identity can be checked as we have explained below Eq.~\eqref{eq:sla-D}. 

\subsubsection{$(\vect{8}|\vect{5.iii})$}

The NS--NS fields are
\begin{align}
 \rmd s^2 = \frac{-\rmd t^2 + \rmd z^2}{z^2 - \eta\,(t^2-z^2)^2}+ \rmd s^2_{\text{S}^2\times \text{T}^6}\,, \qquad
 B_2 = \frac{\eta\,(t^2-z^2)\, \rmd t\wedge\rmd z}{z\,[z^2 - \eta\,(t^2-z^2)^2]}\,,\qquad
 \Exp{-2d}= z^{-2}\,,
\end{align}
and the Killing vector field is
\begin{align}
 I = \eta\,\bigl[(t^2+z^2)\,\partial_t + 2\,t\,z\,\partial_z \bigr]\,.
\end{align}
As we have already mentioned, this Manin triple is a homogeneous Yang--Baxter deformation of $(\vect{8}|\vect{1})$, and correspondingly, we find that the inverse of the generalized frame fields is given by a rescaling and the TsT transformation
\begin{align}
 (\check{\CE}_{\check{M}}{}^{\bcheck{A}})
 = {\footnotesize\begin{pmatrix}
 0 & 0 & 1 & 0 \\
 0 & 1 & 0 & 0 \\
 1 & 0 & 0 & 0 \\
 0 & 0 & 0 & 1 \end{pmatrix}\begin{pmatrix}
 1 & \eta\,z\,(t^2-z^2) & 0 & 0 \\
 0 & 1 & 0 & 0 \\
 0 & 0 & 1 & 0 \\
 0 & 0 & -\eta\,z\,(t^2-z^2) & 1 
\end{pmatrix} \begin{pmatrix}
 0 & 0 & 1 & 0 \\
 0 & 1 & 0 & 0 \\
 1 & 0 & 0 & 0 \\
 0 & 0 & 0 & 1 \end{pmatrix} \begin{pmatrix}
 \frac{1}{z} & 0 & 0 & 0 \\
 0 & \frac{1}{z} & 0 & 0 \\
 0 & 0 & z & 0 \\
 0 & 0 & 0 & z \end{pmatrix}}.
\end{align}
The first rescaling makes $\Gamma^{01}\to \frac{1}{z^2}\,\Gamma^{01}$ and $\ket{0}\to z\,\ket{0}$ and the TsT transformation makes a shift $\Gamma^{01}\to \Gamma^{01}-\eta\,z\,(t^2-z^2)$. 
Multiplying the factor $\Exp{-d}$\,, we obtain
\begin{align}
 F = \frac{\rmd t\wedge \rmd z-\eta\,z\,(t^2-z^2)}{z^2} \wedge (1+\nu_4) -\omega_2 \wedge (\nu_2+\nu_6)\,.
\end{align}
Then we can check that this is a solution of the generalized massive type IIA supergravity. 
It is noted that the 0-form field strength (or the Romans mass parameter) $G_0=-\frac{\eta\,(t^2-z^2)}{z}$ is not a constant. 
This is possible because the Bianchi identity is modified as
\begin{align}
 \rmd G_0 - \iota_I B_2 \wedge G_0 - \iota_I G_2 = 0 \,.
\end{align}
We can also compute the generalized fluxes as
\begin{align}
 -\tfrac{1}{2}\,\bm{f}_A\,\Gamma^A = -\tfrac{1}{2}\,\Gamma^1 -\eta\,(z\,\Gamma_0+t\,\Gamma_1)\,.
\end{align}

\subsubsection{$(\vect{5.iii}|\vect{8})$}

The NS--NS fields are found as
\begin{align}
\begin{split}
 \rmd s^2 &= \tfrac{-(1-\eta^2)\,\rmd x^2 - \eta\,(1 - 2\,x +2\,\eta^2\,x)\,\rmd x\,\rmd y - \eta^2\,[\frac{1}{4} - x + (1-\eta^2)\,x^2]\,\rmd y^2}{\eta^2\,(\Exp{2\,\eta\,y} - \Exp{\eta\,y} + x^2)} + \rmd s^2_{\text{S}^2\times \text{T}^6}\,,
\\
 B_2 &= \frac{\rmd x \wedge \rmd y}{\Exp{2\,\eta\,y}-\Exp{\eta\,y} + x^2 - 2 x +1} - \frac{\rmd x \wedge \rmd y}{2\,(\Exp{2\,\eta\,y}-\Exp{\eta\,y}+x^2)} \,,\qquad
 \Exp{-2d}= \tfrac{\Exp{-2\,\eta\,y}}{2\,\eta^2} \,. 
\end{split}
\end{align}
We can decompose the inverse of the generalized frame fields as
\begin{align}
 (\check{\CE}_{\check{M}}{}^{\bcheck{A}})
 &= {\footnotesize\begin{pmatrix}
 1 & -\eta & 0 & 0 \\
 \frac{\eta\,(1 -2\, x)}{2} & \eta^2\,x & 0 & 0 \\
 0 & 0 & 2\,x & -\frac{1-2\,x}{\eta} \\
 0 & 0 & \frac{2}{\eta} & \frac{2}{\eta^2} \end{pmatrix}\begin{pmatrix}
 0 & 0 & 1 & 0 \\
 0 & 1 & 0 & 0 \\
 1 & 0 & 0 & 0 \\
 0 & 0 & 0 & 1 \end{pmatrix}}{\large\begin{pmatrix} A & 0 \\ 0 & A^{-\rmT} \end{pmatrix}},
\\
 A&\equiv {\footnotesize\begin{pmatrix} 
 -\frac{\eta}{2}\,(1-\Exp{\eta\, y}+\Exp{2\,\eta\, y}+x^2) & \frac{\eta}{2}\,(1+\Exp{\eta\, y}-\Exp{2\,\eta\, y}-x^2) \\
 \frac{2\,x}{\eta\,[\Exp{2\,\eta\, y}-\Exp{\eta\, y}+(x-1)^2]} & \frac{2\,(x-1)}{\eta\,[\Exp{2 \eta\, y}-\Exp{\eta\, y}+(x-1)^2]} \end{pmatrix}},\quad \det A=1\,.
\end{align}
The first $\GL(2)$ does not change the R--R field strength, and the second transformation is the timelike $T$-duality. 
Then, the final $\GL(2)$ transformation and the dilaton factor $\Exp{-d}$ gives
\begin{align}
 F =G= \frac{\Exp{-\eta\,y} (\rmd x -\eta\,x\,\rmd y)}{\eta}\wedge (1+\nu_4)
 -\frac{\Exp{-\eta\,y} [\rmd x + \frac{\eta\,(1-2\,x)}{2}\,\rmd y]}{\eta^2}\wedge\omega_2\wedge(\nu_2+\nu_6)\,.
\end{align}
This solves the equations of motion of type IIB$^*$ supergravity. 
The generalized fluxes become
\begin{align}
\begin{split}
 -\tfrac{1}{2}\,\bm{f}_A\,\Gamma^A &= \tfrac{1+\Exp{2\eta\,y}-x+x^2}{2}\,\Gamma^0 -\tfrac{-1+\Exp{2\eta\,y}-x+x^2}{2}\,\Gamma^1
\\
 &\quad -\tfrac{\eta\,(-1+\Exp{2\eta\,y}+x^2)}{2}\,\Gamma_0 - \tfrac{\eta\,(1+\Exp{2\eta\,y}+x^2)}{2}\,\Gamma_1\,.
\end{split}
\end{align}

\subsubsection{$(\vect{6_0}|\vect{5.i})$ and $(\vect{7_0}|\vect{5.i})$}
\label{sec:65-75-RR}

As we found in the previous subsection, $(\vect{6_0}|\vect{5.i})$ and $(\vect{7_0}|\vect{5.i})$ give the same result. 
The NS--NS fields and the Killing vector field $I^m$ become
\begin{align}
 \rmd s^2 = \frac{\rmd x^2-\rmd y^2}{1-y^2} + \rmd s^2_{\text{S}^2\times \text{T}^6}\,, \qquad
 B_2 = - \frac{y\,\rmd x \wedge \rmd y}{1-y^2} \,,\quad 
 \Exp{-2d}=1 \,,\qquad I = \partial_x\,.
\end{align}
We can decompose the inverse of the generalized frame fields as
\begin{align}
 (\check{\CE}_{\check{M}}{}^{\bcheck{A}})
 = {\footnotesize\begin{pmatrix}
 0 & 1 & 0 & 0 \\
 -1 & 0 & 0 & 0 \\
 0 & 0 & 0 & 1 \\
 0 & 0 & -1 & 0 \end{pmatrix} \begin{pmatrix}
 0 & 0 & 1 & 0 \\
 0 & 1 & 0 & 0 \\
 1 & 0 & 0 & 0 \\
 0 & 0 & 0 & 1 \end{pmatrix} \begin{pmatrix}
 1 & -y & 0 & 0 \\
 0 & 1 & 0 & 0 \\
 0 & 0 & 1 & 0 \\
 0 & 0 & y & 1 \end{pmatrix} \begin{pmatrix}
 1 & 0 & 0 & 0 \\
 0 & 0 & 0 & 1 \\
 0 & 0 & 1 & 0 \\
 0 & 1 & 0 & 0 \end{pmatrix}}.
\end{align}
A short computation gives
\begin{align}
 F=1+\nu_4 + (y+\rmd x\wedge \rmd y)\wedge \omega_2 \wedge (\nu_2+\nu_6)\,.
\end{align}
Since the timelike $T$-duality is involved, this satisfies the equations of motion of the generalized massive type IIA$^*$ supergravity. 
The generalized fluxes are
\begin{align}
 -\tfrac{1}{2}\,\bm{f}_A\,\Gamma^A = -\tfrac{1}{2}\,\Gamma^1\,.
\end{align}

\subsubsection{$(\vect{5.i}|\vect{6_0})$}

The NS--NS fields become
\begin{align}
 \rmd s^2 = - \rmd x^2 + \frac{\rmd y^2}{\sin^2 x} + \rmd s^2_{\text{S}^2\times \text{T}^6}\,, \qquad
 B_2 = \frac{\rmd x \wedge \rmd y}{\cos y}\,,\qquad
 \Exp{-2d}= \sin x\,. 
\end{align}
We can decompose the inverse of the generalized frame fields as
\begin{align}
 (\check{\CE}_{\check{M}}{}^{\bcheck{A}})
 = {\footnotesize\begin{pmatrix}
 -1 & 0 & 0 & 0 \\
 0 & -\frac{1}{\sin x} & 0 & 0 \\
 0 & 0 & -1 & 0 \\
 0 & 0 & 0 & -\sin x \end{pmatrix} \begin{pmatrix}
 1 & 0 & 0 & 0 \\
 0 & 0 & 0 & 1 \\
 0 & 0 & 1 & 0 \\
 0 & 1 & 0 & 0 \end{pmatrix} \begin{pmatrix}
 1 & \tan x & 0 & 0 \\
 0 & 1 & 0 & 0 \\
 0 & 0 & 1 & 0 \\
 0 & 0 & -\tan x & 1 \end{pmatrix}}.
\end{align}
The first $\GL(2)$ does not transform $\ket{\hat{\cF}}$ and the second one is the spatial $T$-duality. 
The last $\GL(2)$ make a rescaling, and we find
\begin{align}
 F=G= -\sin x\,\rmd x\wedge (1+\nu_4) + \rmd y\wedge\omega_2\wedge(\nu_2+\nu_6)\,.
\end{align}
Since this does not involve the timelike $T$-duality, this solves the equations of motion of the standard type IIB supergravity. 
The generalized fluxes are
\begin{align}
 -\tfrac{1}{2}\,\bm{f}_A\,\Gamma^A = -\tfrac{1}{2\tan x}\,\Gamma^0\,.
\end{align}

\subsubsection{$(\vect{5.i}|\vect{7_0})$}

The NS--NS fields are the intriguing asymptotically $\text{AdS}_2\times \text{S}^2\times \text{T}^6$ space mentioned in the previous subsection,
\begin{align}
 \rmd s^2 = -\frac{\rmd x^2}{\cosh^2y} + \rmd y^2 + \rmd s^2_{\text{S}^2\times \text{T}^6}\,, \qquad
 B_2 = 0\,,\qquad
 \Exp{-2d}= \cosh y\,.
\end{align}
We decompose the inverse of the generalized frame fields as
\begin{align}
 (\check{\CE}_{\check{M}}{}^{\bcheck{A}})
 = {\footnotesize\begin{pmatrix}
 0 & 0 & 1 & 0 \\
 0 & 1 & 0 & 0 \\
 1 & 0 & 0 & 0 \\
 0 & 0 & 0 & 1 \end{pmatrix} \begin{pmatrix}
 \cosh y & 0 & 0 & 0 \\
 0 & -1 & 0 & 0 \\
 0 & 0 & \frac{1}{\cosh y} & 0 \\
 0 & 0 & 0 & -1 \end{pmatrix}},
\end{align}
and the first transformation gives a rescaling and a change in the orientation $\epsilon_{xyr\theta 1\cdots 6}=-1$\,. 
The second transformation is the timelike $T$-duality, and we obtain
\begin{align}
 F=G= \cosh y\,\rmd y\wedge (1+\nu_4) -\rmd x\wedge \omega_2\wedge(\nu_2+\nu_6)\,.
\end{align}
This is a solution of type IIB$^*$ supergravity.
The generalized fluxes can be computed as
\begin{align}
 -\tfrac{1}{2}\,\bm{f}_A\,\Gamma^A = -\tfrac{1}{2}\tanh y\,\Gamma^1\,.
\end{align}
We note that this asymptotically $\text{AdS}_2\times \text{S}^2\times \text{T}^6$ solution is a non-trivial dressing coset. 
The PL $T$-dual of this solution is also a non-trivial dressing coset found in section \ref{sec:65-75-RR}. 

\subsubsection{$(\vect{5}|\vect{2.ii})$}

Here we obtain the $\text{AdS}_2\times \text{S}^2\times \text{T}^6$ space,
\begin{align}
 \rmd s^2 =\frac{-\rmd x^2+\rmd y^2}{y^2} + \rmd s^2_{\text{S}^2\times \text{T}^6}\,, \qquad
 B_2 = \frac{\rmd x \wedge \rmd y}{y^2}\,,\qquad
 \Exp{-2d}=1\,. 
\end{align}
The $B$-field is exact and the $H$-flux is absent, but the dilaton $\Phi$ is not a constant, unlike the original solution. 
Moreover, unlike the original solution, this is a non-trivial dressing coset. 
The inverse of the generalized frame fields can be decomposed as
\begin{align}
 (\check{\CE}_{\check{M}}{}^{\bcheck{A}})
 = {\footnotesize\begin{pmatrix}
 0 & 0 & 1 & 0 \\
 0 & 1 & 0 & 0 \\
 1 & 0 & 0 & 0 \\
 0 & 0 & 0 & 1 \end{pmatrix} \begin{pmatrix}
 \frac{1+y^2}{2} & \frac{y^2-1}{2} & 0 & 0 \\
 -1 & -1 & 0 & 0 \\
 0 & 0 & 1 & -1 \\
 0 & 0 & \frac{y^2-1}{2} & -\frac{1+y^2}{2} \end{pmatrix}}.
\end{align}
The first $\GL(2)$ matrix has the determinant $-1$ and the orientation is changed $\epsilon_{xyr\theta 1\cdots 6}=-1$\,. 
The second transformation is the timelike $T$-duality, and we obtain
\begin{align}
 F=G= \rmd y\wedge (1+\nu_4) -\rmd x\wedge\omega_2\wedge(\nu_2+\nu_6)\,.
\end{align}
This is a solution of type IIB$^*$ supergravity. 
The generalized fluxes are obtained as
\begin{align}
 -\tfrac{1}{2}\,\bm{f}_A\,\Gamma^A = \tfrac{y}{2}\,(\Gamma^0-\Gamma^1)\,.
\end{align}

\subsubsection{$(\vect{2.ii}|\vect{5})$}

The NS--NS fields are given by
\begin{align}
 \rmd s^2 = \frac{3\,(4\,\rmd x^2 - \rmd y^2)}{4\,(3 - x - y)\,(x + y)} + \rmd s^2_{\text{S}^2\times \text{T}^6}\,,\quad
 \check{B}_2 = \frac{[2\,(x + y)-3]\,\rmd x\wedge\rmd y}{2\,(3 - x - y)\,(x + y)}\,,\quad
 \Exp{-2d}=1\,,
\end{align}
and the Killing vector field is
\begin{align}
 I = \partial_y - \partial_x \,.
\end{align}
The inverse of the generalized frame fields can be decomposed as
\begin{align}
 (\check{\CE}_{\check{M}}{}^{\bcheck{A}})
 = {\footnotesize\begin{pmatrix}
 -\frac{2}{3} & -\frac{4}{3} & 0 & 0 \\
 -\frac{2}{3} & -\frac{1}{3} & 0 & 0 \\
 0 & 0 & \frac{1}{2} & -1 \\
 0 & 0 & -2 & 1 \end{pmatrix} \begin{pmatrix}
 1 & 0 & 0 & 0 \\
 0 & 0 & 0 & 1 \\
 0 & 0 & 1 & 0 \\
 0 & 1 & 0 & 0 \end{pmatrix} \begin{pmatrix}
 1 & 0 & 0 & 0 \\
 \frac{2\,(x+y)-3}{3} & 1 & 0 & 0 \\
 0 & 0 & 1 & -\frac{2\,(x+y)-3}{3} \\
 0 & 0 & 0 & 1 \end{pmatrix} \begin{pmatrix}
 0 & 0 & 1 & 0 \\
 0 & 1 & 0 & 0 \\
 1 & 0 & 0 & 0 \\
 0 & 0 & 0 & 1 \end{pmatrix}}.
\end{align}
The final $\GL(2)$ changes the orientation and we obtain $\epsilon_{xyr\theta 1\cdots 6}=-1$\,. 
A straightforward computation gives
\begin{align}
 F = -\sqrt{\frac{3}{2}} \,(1+\nu_4) + \frac{2\,\rmd x \wedge \rmd y -2\,(x+y)+3}{\sqrt{6}}\wedge\omega_2\wedge(\nu_2+\nu_6)\,.
\end{align}
This is a solution of the generalized massive type IIA$^*$ supergravity. 
The generalized fluxes can be found as
\begin{align}
 -\tfrac{1}{2}\,\bm{f}_A\,\Gamma^A = -\tfrac{1}{2}\,\Gamma^1 \,.
\end{align}

\section{Discussion}

Using $T$-duality-covariant approaches, we have clarified the PL $T$-plurality for dressing cosets. 
Proposing a gauged sigma model, we found a general formula to reduce a $D$-dimensional background fields $E_{mn}$ to the $d$-dimensional reduced background fields $\check{E}_{mn}=\check{g}_{mn}+\check{B}_{mn}$\,.
Combining this and the formula for the dilaton $d$, we have shown that the PL $T$-plurality with or without spectator fields is a solution-generating transformation in DFT. 
In the presence of the R--R fields, we have failed to show the covariance of the Bianchi identity, but this is indeed always satisfied in our examples. 
We have obtained the transformation rule of the R--R field strength such that the other equations of motion are satisfied. 

In this paper, we considered only low-dimensional examples, but it is interesting to consider higher-dimensional examples and study the PL $T$-plurality of various backgrounds. 
For this purpose, classifications of higher-dimensional Drinfel'd doubles similar to \cite{math:0202210} will be useful. 

One of the motivations of this paper is to find out an extension of the dressing cosets. 
Recently, the exceptional Drinfel'd algebras (EDA) $\mathfrak{d}(\mathfrak{e}_{n(n)})$ of various dimensions $n\leq 8$ have been proposed in \cite{1911.06320,1911.07833,2007.08510,2009.04454}.
Using the EDA, we can study the PL-like extension of $U$-duality (see \cite{2001.09983,2003.06164,2006.12452,2011.11424,2012.13263,2103.01139,2107.00091} for recent progress). 
Unlike the case of the Drinfel'd double, the EDA $\mathfrak{d}(\mathfrak{e}_{n(n)})$ is a Leibniz algebra and its group-like extension $\cD(\text{E}_{n(n)})$ has not been clearly understood. 
Nonetheless, at least at the level of supergravity, the non-Abelian $U$-duality between two coset spaces $\cD(\text{E}_{n(n)})/\tilde{G}$ and $\cD(\text{E}_{n(n)})/\tilde{G}'$ has been expected to be a symmetry of the equations of motion. 
Here, $\tilde{G}$ and $\tilde{G}'$ are certain subgroup of $\cD(\text{E}_{n(n)})$ of co-dimension $n$ or $(n-1)$\,.
Indeed, in \cite{2012.13263}, by using various specific examples, it is checked that if $\cD(\text{E}_{n(n)})/\tilde{G}$ is a solution of the supergravity equations of motion, $\cD(\text{E}_{n(n)})/\tilde{G}'$ is also a solution. 
Then, we expect that this non-Abelian $U$-duality can be extended to the non-Abelian $U$-duality for dressing cosets $F\backslash \cD(\text{E}_{n(n)})/\tilde{G}$ where $F$ is a certain subgroup of $\cD(\text{E}_{n(n)})$ that may need to be isotropic. 

At the level of supergravity, this kind of extension may not be difficult. 
Introducing the parameter $\lambda$\,, we can construct a standard coset space in the $U$-duality setup. 
By performing the non-Abelian $U$-duality transformations similar to \cite{2012.13263}, the standard coset will be mapped to some backgrounds, which may be regarded as the generalized dressing cosets. 

Using the brane sigma model proposed in \cite{1607.04265,1712.10316,2004.09486}, it will be possible to describe the brane sigma models on $\cD(\text{E}_{n(n)})/\tilde{G}$ in a duality-covariant manner. 
The brane actions of \cite{1607.04265,1712.10316,2004.09486} play a similar role to the gauged sigma model by Hull and Reid-Edwards, and by introducing additional gauge fields associated with $F$, we may find the sigma model on the $U$-duality-extension of dressing cosets $F\backslash \cD(\text{E}_{n(n)})/\tilde{G}$.

We can also consider other duality groups, such as the duality groups in the half-maximal supergravity. 
In this case, the extended Drinfel'd algebra has been studied in \cite{2106.02041}. 
Another simple extension of the Drinfel'd double has been studied in \cite{2104.00007}, where the PL $T$-plurality has been extended to the Jacobi--Lie $T$-plurality. 
Even for these extensions, we will find some generalizations of the dressing cosets. 
It is also interesting to consider the supersymmetric extension. 
In \cite{1603.04684,2101.10328}, supersymmetric extensions of DFT have been studied. 
By following the procedure of \cite{2009.04454}, it should be possible to find the extended Drinfel'd algebras associated with these theories. 
Then we may also find the PL $T$-plurality for super-dressing cosets. 
This may contain the fermionic $T$-duality as a specific case, and may give a non-Abelian extension of the fermionic $T$-duality (see \cite{1706.10169,2101.08206,2112.12168} for related recent studies).

\subsection*{Acknowledgments}

We thank Jun-ichi Sakamoto for useful discussion on the gauged sigma model. 
This work is supported by JSPS Grant-in-Aids for Scientific Research (C) 18K13540 and (B) 18H01214. 

\appendix

\section{Gauge invariance of the action}
\label{app:left-mult}

In this appendix, we compute how the background fields transform under the left multiplication. 
In section \ref{sec:left-mult}, we have obtained the transformation rule \eqref{eq:left-gen-Lie} by taking the variation of the action \eqref{eq:geugedDSM}. 
Here, we find the same transformation rule by taking the variation of the standard action \eqref{eq:string-action}. 
For this purpose, we compute the transformation rule of $E_{mn}=r_m^a\,[(\hat{E}+\pi)^{-1}]_{ab}\,r^b_n$\,, namely, $r^a_m(x)$ and $\pi^{ab}(x)$. 

Under the decomposition $l=g\,\tilde{g}$\,, an infinitesimal left multiplication can be expressed as
\begin{align}
 l\to l' &= \bigl(1+\epsilon^a\,T_a + \epsilon_a\,T^a \bigr)\,g\,\tilde{g} 
 = \bigl(1+\epsilon^a\,T_a)\,g\,g^{-1}\,(1 + \epsilon_a\,T^a \bigr)\,g\,\tilde{g} 
\nn\\
 &=\bigl(1+\epsilon^a\,T_a)\,g\, \bigl[1 + \epsilon_a\,\bigl(-\pi^{ac}\,a_c{}^b\,T_b + (a^{-1})_b{}^a\,T^b \bigr)\bigr]\,\tilde{g} 
\nn\\
 &=\bigl(1+\epsilon^a\,T_a)\,g\, (1 - \epsilon_a\, \pi^{ac}\,a_c{}^b\,T_b)\,\bigl(1 + \epsilon_a\,(a^{-1})_b{}^a\,T^b \bigr)\,\tilde{g} 
\nn\\
 &=\bigl(1+\epsilon^a\,T_a + \pi^{ca}\,\epsilon_a\, T_c\bigr)\,g\,\bigl(1 + \epsilon_a\,(a^{-1})_b{}^a\,T^b \bigr)\,\tilde{g} \,.
\end{align}
Namely, the left multiplication corresponds to
\begin{align}
 g \to g' = \bigl[1+ \bigl(\epsilon^a + \pi^{ab}\,\epsilon_b\bigr)\, T_a\bigr]\,g \in G \,,\qquad 
 \tilde{g} \to \tilde{g}' = \bigl[1 + \epsilon_a\,(a^{-1})_b{}^a\,T^b \bigr]\,\tilde{g} \in \tilde{G}\,.
\label{eq:dressing-action}
\end{align}
Since the details of $\tilde{g}$ are not relevant to the definition of $r^a_m(x)$ and $\pi^{ab}(x)$, we only need to care about the transformation $g\to g'$\,, which is called the dressing action. 

In general, under $g\to h(x)\,g$ with $h\equiv 1+\bm{\epsilon}^a(x)\,T_a$\,, $r^a_m$ and $e_a^m$ are transformed as
\begin{align}
 \delta r^a = r^b\,\bigl(\bm{\epsilon}^c f_{cb}{}^a + D_b \bm{\epsilon}^a \bigr)\,, \qquad
 \delta e_a^m = - e_b^m \,\bigl( \bm{\epsilon}^c f_{ca}{}^b + D_a \bm{\epsilon}^b \bigr)\,,
\end{align}
where $D_a\equiv e_a^m\,\partial_m$\,. 
In addition, under the infinitesimal left multiplication, $\pi^{ab}$ transforms as
\begin{align}
 \pi^{ab} \to \pi^{ab}_h + (a_h^{-1})_c{}^a\,(a_h^{-1})_d{}^b\,\pi^{cd}\,,
\end{align}
where $(a_h)_a{}^b$ and $\pi^{ab}_h$ are defined through
\begin{align}
 h^{-1}\,T_a\,h=a_a{}^b\,T_b\,,\qquad h^{-1}\,T^a\,h= - \pi_h^{ac}\,(a_h)_c{}^b\,T_b + (a_h^{-1})_b{}^a\,T^b\,,
\end{align}
and are found to be
\begin{align}
 (a_h)_a{}^b = \delta_a^b - \bm{\epsilon}^c f_{ca}{}^b\,,\qquad \pi_h^{ab}=\bm{\epsilon}^c\,f_c{}^{ab}\,. 
\end{align}
Then we obtain
\begin{align}
 \delta \pi^{ab} = \bm{\epsilon}^c\,\bigl(f_c{}^{ab} +2\,f_{cd}{}^{[a}\,\pi^{|d|b]}\bigr)\,.
\end{align}
Substituting $\bm{\epsilon}^a = \epsilon^a + \pi^{ab}(x)\,\epsilon_b$ [recall Eq.~\eqref{eq:dressing-action}] into these transformation rules, we obtain
\begin{align}
\begin{split}
 \delta E_{mn} &= \delta r^c_m\,E_{cd}\,r^d_n
 -r_m^a\,E_{ac}\,\delta \pi^{cd}\,E_{db}\,r^b_n
 +r_m^c\,E_{cd}\,\delta r^d_n
\\
 &= r_m^a\,\bigl[\epsilon^{e}\,f_{ea}{}^c + (D_a\pi^{cf} + f_{ea}{}^c\,\pi^{ef})\,\epsilon_f\bigr] \,E_{cd}\,r^d_n
\\
 &\quad -r_m^a\,E_{ac}\, \bigl(\epsilon^{e}+\pi^{eg}\,\epsilon_g\bigr)\,\bigl(f_e{}^{cd} +2\,f_{ef}{}^{[c}\,\pi^{|f|d]}\bigr) \,E_{db}\,r^b_n
\\
 &\quad +r_m^a\,E_{ae}\,\bigl[\epsilon^{c}\,f_{cb}{}^e + (D_b\pi^{ef} + f_{cb}{}^e\,\pi^{cf})\,\epsilon_f\bigr] \,r_n^{b}\,.
\end{split}
\end{align}
Using the identity $D_a \pi^{bc} = f_a{}^{bc} + 2\,f_{ad}{}^{[b}\,\pi^{|d|c]}$ (see Eq.~\eqref{eq:identities}) and $E^{ab}=\hat{E}^{ab}+\pi^{ab}$\,, we find
\begin{align}
\begin{split}
 \delta E_{mn} &= E_{mc}\,\bigl\{
 - \epsilon^a\,\bigl(f_a{}^{cd}
 - f_{ab}{}^c\,\hat{E}^{bd}
 - f_{ab}{}^d\,\hat{E}^{cb}\bigr)
\\
 &\quad\qquad\ \,
 + \epsilon_a\,\bigl(
 f_{ef}{}^a\,\hat{E}^{ce}\,\hat{E}^{fd} 
 - f_b{}^{ac}\,\hat{E}^{bd}
 - f_b{}^{ad}\,\hat{E}^{cb} \bigr)
\\
 &\quad\qquad\ \,
 - \epsilon_g\,f_{ef}{}^g\,E^{ce}\,E^{fd} 
 + 3\,\epsilon_g\,\bigl(f_e{}^{[cd}\,\pi^{g]e}
 + f_{ef}{}^{[c}\,\pi^{d|e|}\,\pi^{g]f}\bigr)
 \bigr\}\,E_{dn} \,.
\end{split}
\end{align}
By further using $f_{d}{}^{[ab}\,\pi^{c]d} + f_{de}{}^{[a}\,\pi^{b|d|}\,\pi^{c]e} = 0$ (see Eq.~\eqref{eq:identities}), we obtain
\begin{align}
 \delta E_{mn} = -E_{ma}\,\Omega^{ab}\,E_{bn} - \epsilon_c\,r_m^a\,f_{ab}{}^c\,r^b_n \,,
\end{align}
where we have defined
\begin{align}
 \Omega^{ab} \equiv \epsilon^c\,\bigl(f_c{}^{ab} - f_{cd}{}^a\,\hat{E}^{db} - f_{cd}{}^b\,\hat{E}^{ad}\bigr)
 - \epsilon_c\,\bigl(f_{de}{}^c\,\hat{E}^{ad}\,\hat{E}^{eb} - f_d{}^{ca}\,\hat{E}^{db} - f_d{}^{cb}\,\hat{E}^{ad} \bigr)\,.
\end{align}
By using $2\,\partial_{[m} r^a_{n]}= f_{bc}{}^a\,r^b_m\,r^c_n$\,, we finally obtain
\begin{align}
 \delta E_{mn} = -E_{ma}\,\Omega^{ab}\,E_{bn} - 2\,\epsilon_a\,\partial_{[m} r^a_{n]}\,.
\label{eq:delta-E}
\end{align}
The second term corresponds to a gauge transformation of the $B$-field, and it does not change the string action. 
Therefore, the action is invariant if $\Omega^{ab}=0$\,. 
One can check that this condition is the same as the invariance condition \eqref{eq:H-inv-inf}. 

To check the consistency with the transformation rule \eqref{eq:left-gen-Lie}, let us further continue the computation. 
In Eq.~\eqref{eq:E-param}, the generalized frame fields have been parameterized as
\begin{align}
 E_M{}^A(x) = \begin{pmatrix} E_m{}^a & E_{ma} \\ E^{ma} & E^m{}_a \end{pmatrix} = \begin{pmatrix} r^a_m & 0 \\ \pi^{ab}\,e_b^m & e_a^m \end{pmatrix} .
\end{align}
Let us compute its variation similar to the above computation. 
We find
\begin{align}
\begin{split}
 \delta E_a{}^m &= - \epsilon^c\,f_{ca}{}^b\,e_b^m - \epsilon_c \,\bigl(f_a{}^{bc}\,e_b^m - f_{ad}{}^c\,\pi^{db}\,e_b^m \bigr) \,, \qquad
 \delta E_{am} = 0 \,,
\\
 \delta E^{am} &= - \epsilon^c\,\bigl(f_c{}^{ab}\,e_b^m + f_{cd}{}^{a}\,\pi^{db}\,e_b^m \bigr) 
 - \epsilon_b\, f_d{}^{ab}\,\pi^{dc} \,e_c^m \,,
\\
 \delta E^a{}_m &= \epsilon^c\,f_{cb}{}^a\,r_m^b + \epsilon_c\,\bigl(f_b{}^{ac} - f_{bd}{}^c\,\pi^{da} \bigr)\,r_m^b \,.
\end{split}
\end{align}
This variation can be expressed as $\delta E_A{}^M = - \gLie_{\epsilon^C \cX_C}E_A{}^M$ by using
\begin{align}
 \cX_A \equiv \bigl(\cX_a,\,\cX^a\bigr)\,,\qquad 
 \cX_a{}^M \equiv \begin{pmatrix} e_a^m \\ 0\end{pmatrix},\qquad
 \cX^{aM} \equiv \begin{pmatrix} -\pi^{ab}\,e_b^m \\ 0\end{pmatrix}.
\end{align}
As one can easily see, these vector fields $\cX_A$ differ from $E_A$ by generators of $B$-field gauge transformations $\cY_A$:
\begin{align}
 \cX_A = E_A - \cY_A\,,\qquad 
 \cY_A \equiv \bigl(\cY_a,\,\cY^a\bigr)\,,\qquad
 \cY_a{}^M \equiv \begin{pmatrix} 0 \\ 0 \end{pmatrix},\qquad
 \cY^{aM} = \begin{pmatrix} 0 \\ r^a_m \end{pmatrix}.
\end{align}
This $B$-field gauge transformation corresponds to the second term of Eq.~\eqref{eq:delta-E}, and up to this gauge transformation, the left multiplication is generated by $\gLie_{\epsilon^C E_C}$\,, as expected. 
The difference in the $B$-field gauge transformation has arisen when converting the gauge-invariant action \eqref{eq:geugedDSM} into the standard action \eqref{eq:string-action}. 

\section{Formula in general case}
\label{app:general-case}

In this appendix, we derive the formula \eqref{eq:KS-reproduced}, relaxing the assumption that the metric $\langle \cdot ,\,\hat{\cH}(\cdot)\rangle$ restricted to the gauge algebra $\mathfrak{f}$ is non-degenerate. 
Here we denote the generators of $\mathfrak{f}$ as $\{T_I\}$ $(I=1,\dotsc,r + n)$ and decompose these into $\{T_I\}=\{T_{\cI},\,T_\sfi\}$ ($\cI=1,\dotsc,r$, $\sfi=1,\dotsc,n$). 
We suppose that the matrix $\hh_{\sfi\sfj}=\langle T_{\sfi},\,\hat{\cH}(T_{\sfj})\rangle$ is non-degenerate, but $\langle T_I,\,\hat{\cH}(T_{\cJ})\rangle=0$\,. 
We introduce $\{T_{\bcheck{A}}\}\equiv \{T_{\ubar{a}},\, T_{\bar{a}}\}$ such that $\mathfrak{f}^\perp$ is spanned by $(2d+n)$ generators $\{T_{\bcheck{A}},\,T_{I}\}$\,. 
The remaining $n$ generators are denoted as $\{T^I\}\equiv \{T^{\cI},\,T^{\sfi}\}$\,. 
Similar to the discussion given in section \ref{sec:non-deg}, we assume that $\hat{\cH}(T_{\bcheck{A}})$ can be expanded by $T_{\bcheck{A}}$ and that $\langle T_{\bcheck{A}},\,T^{I}\rangle=0$\,. 
We can express $\hat{\cH}(T_{\cI})\in\mathfrak{f}^\perp$ as a linear combination of $T_{\cI}$ because $\hat{\cH}(T_{\cI})$ satisfies
\begin{align}
\begin{split}
 \langle \hat{\cH}(T_{\cI}),\,T_{\bcheck{A}} \rangle&=\langle T_{\cI},\,\hat{\cH}(T_{\bcheck{A}}) \rangle = 0\,,
\\
 \langle \hat{\cH}(T_{\cI}),\, T^{\sfi} \rangle&= \hh^{\sfi\sfj}\,\langle \hat{\cH}(T_{\cI}),\,\hat{\cH}(T_{\sfj}) \rangle
 =\hh^{\sfi\sfj}\,\langle T_{\cI},\,T_{\sfj} \rangle = 0\,.
\end{split}
\end{align}
Then, by diagonalizing the action of $\hat{\cH}$\,,\footnote{If $a\in \text{span}\{T_{\cI}\}$ is mapped to $\hat{\cH}(a)=b\in \text{span}\{T_{\cI}\}$ ($b\neq \pm a$), we redefine $a$ and $b$ as $a\pm b$\,, which satisfy $\hat{\cH}(a\pm b)=\pm (a\pm b)$. Repeating this redefinition, we can diagonalize $\hat{\cH}$ in the subspace $\text{span}\{T_{\cI}\}$\,.} we redefine $\{T_{\cI}\}$ into $\{T_{\ubar{\bm{i}}},\,T_{\bar{\bm{i}}}\}$ ($\ubar{\bm{i}}=1,\dotsc, \bm{n}$\,, $\bar{\bm{i}}=1,\dotsc,\bar{\bm{n}}$\,, $r=\bm{n}+\bar{\bm{n}}$) that satisfy
\begin{align}
 \hat{\cH}(T_{\ubar{\bm{i}}}) = + T_{\ubar{\bm{i}}}\,,\qquad
 \hat{\cH}(T_{\bar{\bm{i}}})= - T_{\bar{\bm{i}}}\,.
\end{align}
The $\bm{n}+\bar{\bm{n}}$ generators $\{T^{\ubar{\bm{i}}},\,T^{\bar{\bm{i}}}\}$ can be chosen to satisfy
\begin{align}
\begin{split}
 &\langle T_{\ubar{\bm{i}}},\, T^{\ubar{\bm{j}}}\rangle = \delta_{\ubar{\bm{i}}}^{\ubar{\bm{j}}}\,,\quad
 \langle T_{\bar{\bm{i}}},\, T^{\bar{\bm{j}}}\rangle = -\delta_{\bar{\bm{i}}}^{\bar{\bm{j}}}\,,\quad
 \langle T_{\ubar{\bm{i}}},\, T^{\bar{\bm{j}}}\rangle = 0 = \langle T_{\bar{\bm{i}}},\, T^{\ubar{\bm{j}}}\rangle\,,
\\
 &\langle T^{\ubar{\bm{i}}},\, T^{\bm{j}}\rangle = \langle T^{\bar{\bm{i}}},\, T^{\bar{\bm{j}}}\rangle = \langle T^{\ubar{\bm{i}}},\, T^{\bar{\bm{j}}}\rangle = \langle T^{\bar{\bm{i}}},\, T^{\bm{j}}\rangle =0 \,,\quad 
 \hat{\cH}(T^{\ubar{\bm{i}}})= +T^{\ubar{\bm{i}}}\,,\quad 
 \hat{\cH}(T^{\bar{\bm{i}}})= -T^{\bar{\bm{i}}} \,.
\end{split}
\end{align}
This is the most general case considered in this paper. 
The situation discussed in section \ref{sec:non-deg} corresponds to the specific case $r=0$\,. 

Similar to Eq.~\eqref{eq:P-expand}, we can expand $\mathbb{P}$ as
\begin{align}
 \mathbb{P}(\sigma) = \VV^{\ubar{\bm{a}}}\,\alpha_{+\ubar{\bm{a}}}(\sigma) + \bar{\VV}^{\bar{\bm{a}}}\,\alpha_{-\bar{\bm{a}}}(\sigma) + \VV_{\ubar{\bm{i}}}\,\alpha_+^{\ubar{\bm{i}}}(\sigma) + \bar{\VV}_{\bar{\bm{i}}}\,\alpha_-^{\bar{\bm{i}}}(\sigma) + \VV_{\sfi}\,\alpha^{\sfi}(\sigma) \,,
\end{align}
and the self-duality relation gives
\begin{align}
 *\alpha_{+\ubar{\bm{a}}} = + \alpha_{+\ubar{\bm{a}}}\,,\qquad 
 *\alpha_{-\bar{\bm{a}}} = - \alpha_{-\bar{\bm{a}}}\,,\qquad 
 *\alpha_+^{\ubar{\bm{i}}} = + \alpha_+^{\ubar{\bm{i}}}\,,\qquad
 *\alpha_-^{\bar{\bm{i}}} = - \alpha_-^{\bar{\bm{i}}}\,,\qquad
 \alpha^{\sfi} = 0 \,.
\end{align}
We then obtain
\begin{align}
\begin{split}
 r^a - \pi^{ab}\,P_b + \mathsf{\kk}_{\sfi}^a \,\cA^{\sfi}&= \hat{E}^{ab}\,\kk^{\ubar{\bm{a}}}_b\,\alpha_{+\ubar{\bm{a}}} - \bar{\kk}^{\bar{\bm{a}}}_b\,\hat{E}^{ba}\,\alpha_{-\bar{\bm{a}}} + \mathsf{\kk}_{\ubar{\bm{i}}}^a \,\alpha_+^{\ubar{\bm{i}}} + \bar{\mathsf{\kk}}_{\bar{\bm{i}}}^a \,\alpha_-^{\bar{\bm{i}}} \,,
\\
 P_a + \tilde{\kk}_{\sfi a}\,\cA^{\sfi} &= \kk^{\ubar{\bm{a}}}_a\,\alpha_{+\ubar{\bm{a}}}+\bar{\kk}^{\bar{\bm{a}}}_a\,\alpha_{-\bar{\bm{a}}} + \tilde{\kk}_{\ubar{\bm{i}} a} \,\alpha_+^{\ubar{\bm{i}}} + \tilde{\kk}_{\bar{\bm{i}} a} \,\alpha_-^{\bar{\bm{i}}}\,,
\end{split}
\end{align}
and these can be rewritten as
\begin{align}
\begin{split}
 r^a &= E^{ab}\,\kk^{\ubar{\bm{a}}}_b\,\alpha_{+\ubar{\bm{a}}} - \bar{\kk}^{\bar{\bm{a}}}_b\,E^{ba}\,\alpha_{-\bar{\bm{a}}}- \kk_{\sfi}^a \, \cA^{\sfi} + \kk_{\ubar{\bm{i}}}^a \,\alpha_+^{\ubar{\bm{i}}} + \bar{\kk}_{\bar{\bm{i}}}^a \,\alpha_-^{\bar{\bm{i}}}\,,
\\
 P_a &= \kk^{\ubar{\bm{a}}}_a\,\alpha_{+\ubar{\bm{a}}}+\bar{\kk}^{\bar{\bm{a}}}_a\,\alpha_{-\bar{\bm{a}}} - \tilde{\kk}_{\sfi a}\, \cA^{\sfi} + \tilde{\kk}_{\ubar{\bm{i}} a} \,\alpha_+^{\ubar{\bm{i}}} + \tilde{\bar{\kk}}_{\bar{\bm{i}} a} \,\alpha_-^{\bar{\bm{i}}}\,.
\end{split}
\end{align}
Then the action \eqref{eq:r-p-action} becomes
\begin{align}
 S &= \frac{1}{4\pi\alpha'}\int_\Sigma \bigl[-2\,\bar{\kk}^{\bar{\bm{a}}}_c\,E^{cd} \,\kk^{\ubar{\bm{b}}}_d\, \alpha_{-\bar{\bm{a}}} \wedge\alpha_{+\ubar{\bm{b}}} 
 +2\,\kk_{\sfi}^a\, \bigl(\kk^{\ubar{\bm{a}}}_a\,\alpha_{+\ubar{\bm{a}}} + \bar{\kk}^{\bar{\bm{a}}}_a\,\alpha_{-\bar{\bm{a}}}\bigr) \wedge \cA^{\sfi} + \kk_{\sfi}^a\,\tilde{\kk}_{\sfj a} \,\cA^{\sfi} \wedge \cA^{\sfj} \bigr] 
\nn\\
 &\quad + \frac{1}{2\pi\alpha'}\int_\Sigma \bigl[- \bigl( \bar{\kk}^{\bar{\bm{a}}}_a\, \alpha_{-\bar{\bm{a}}} - \tilde{\kk}_{\sfj a} \,\cA^{\sfj}\bigr)\,\kk_{\ubar{\bm{i}}}^a \wedge \alpha_+^{\ubar{\bm{i}}}
 + \alpha_-^{\bar{\bm{i}}} \wedge \bar{\kk}_{\bar{\bm{i}}}^a \,\bigl( \kk^{\ubar{\bm{a}}}_a\, \alpha_{+\ubar{\bm{a}}} - \tilde{\kk}_{\sfj a} \,\cA^{\sfj}\bigr) \bigr] 
\nn\\
 &\quad + \frac{1}{2\pi\alpha'}\int_\Sigma \bar{\kk}_{\bar{\bm{i}}}^a\,\tilde{\kk}_{\ubar{\bm{j}}a}\,\alpha_-^{\bar{\bm{i}}}\wedge \alpha_+^{\ubar{\bm{j}}}\,,
\end{align}
where we have used the orthogonality, such as $\langle K^{\ubar{\bm{a}}},\,\VV_I \rangle=0=\langle \bar{K}^{\bar{\bm{a}}},\,\VV_I \rangle$ and $\langle \VV_I,\,\VV_J \rangle=0$\,. 
Using the relation
\begin{align}
\begin{split}
 r^a + * r^a &= {}+ 2\,E^{ab}\,\kk_b^{\ubar{\bm{c}}}\,\alpha_{+\ubar{\bm{c}}} - \kk_{\sfi}^a \,\bigl(\cA^{\sfi}+ *\cA^{\sfi}\bigr) +2\,\kk_{\ubar{\bm{i}}}^a\,\alpha_+^{\ubar{\bm{i}}}\,,
\\
 r^a - * r^a &= {}- 2\,\bar{\kk}_b^{\bar{\bm{c}}}\,E^{ba}\,\alpha_{-\bar{\bm{c}}} - \kk_{\sfi}^a \,\bigl(\cA^{\sfi}- *\cA^{\sfi}\bigr) +2\,\bar{\kk}_{\bar{\bm{i}}}^a\,\alpha_-^{\bar{\bm{i}}}\,,
\end{split}
\end{align}
and a certain matrix $\MM_{ab}$\,, we obtain
\begin{align}
 &\frac{1}{4\pi\alpha'}\int_\Sigma \MM_{ab}\,\bigl(r^a\wedge * r^b + r^a \wedge r^b \bigr)
 =\frac{1}{8\pi\alpha'}\int_\Sigma \MM_{ab}\,\bigl(r^a - * r^a\bigr)\wedge *\bigl(r^b + * r^b\bigr)
\nn\\
 &=-\frac{1}{2\pi\alpha'}\int_\Sigma \bar{\kk}_a^{\bar{\bm{a}}}\,(E\,\MM\,E)^{ab}\,\kk_b^{\ubar{\bm{b}}}\,\alpha_{-\bar{\bm{a}}} \wedge \alpha_{+\ubar{\bm{b}}}
\nn\\
 &\quad + \frac{1}{2\pi\alpha'}\int_\Sigma \bigl[\kk_{\sfi}^a \,(\MM\,E)_a{}^b\,\kk_b^{\ubar{\bm{a}}}\, \alpha_{+\ubar{\bm{a}}} + \bar{\kk}_a^{\bar{\bm{a}}}\, (E\,\MM)^a{}_b\,\kk_{\sfi}^b\,\alpha_{-\bar{\bm{a}}} \bigr] \wedge \cA^{\sfi}
\nn\\
 &\quad +\frac{1}{4\pi\alpha'}\int_\Sigma \kk_{\sfi}^a \,\MM_{ab}\, \kk_{\sfj}^b \, \bigl(\cA^{\sfi}\wedge \cA^{\sfj} + \cA^{\sfi}\wedge *\cA^{\sfj}\bigr) 
\nn\\
 &\quad -\frac{1}{2\pi\alpha'}\int_\Sigma \bigl(\alpha_{-\bar{\bm{a}}}\,\bar{\kk}_c^{\bar{\bm{a}}}\,E^{ca} + \kk_{\sfi}^a \, \cA^{\sfi} \bigr)\,\MM_{ab}\,\kk_{\ubar{\bm{i}}}^b \wedge \alpha_+^{\ubar{\bm{i}}}
\nn\\
 &\quad +\frac{1}{2\pi\alpha'}\int_\Sigma \alpha_-^{\bar{\bm{i}}}\wedge \bar{\kk}_{\bar{\bm{i}}}^a\,\MM_{ab}\, \bigl(\alpha_{+\ubar{\bm{a}}}\,E^{bc}\,\kk_c^{\ubar{\bm{a}}} - \kk_{\sfi}^b\, \cA^{\sfi}\bigr)
\nn\\
 &\quad +\frac{1}{2\pi\alpha'}\int_\Sigma \bar{\kk}_{\bar{\bm{i}}}^a\,\MM_{ab}\,\kk_{\ubar{\bm{j}}}^b\,\alpha_-^{\bar{\bm{i}}} \wedge \alpha_+^{\ubar{\bm{j}}}\,.
\end{align}
This reproduce the action \eqref{eq:r-p-action2} if the matrix $\MM_{ab}$ satisfies
\begin{align}
 \bar{\kk}_b^{\bar{\bm{a}}}\,E^{bc}\,\MM_{ca} = \bar{\kk}_a^{\bar{\bm{a}}} \,,\qquad
 \MM_{ac}\,E^{cb}\,\kk_b^{\ubar{\bm{a}}} = \kk_a^{\ubar{\bm{a}}}\,,\qquad
 \MM_{ab}\,\kk_I^b = \tilde{\kk}_{I a} \,,\qquad
 \bar{\kk}_I^b\,\MM_{ba} = - \tilde{\bar{\kk}}_{I a}\,,
\end{align}
where $\{\bar{\VV}_I\} \equiv \{\VV_{\sfi},\,\VV_{\bar{\bm{i}}}\}$\,, $\{\VV_I\} \equiv \{\VV_{\sfi},\,\VV_{\ubar{\bm{i}}}\}$\,, $(\bar{\VV}_I{}^A)\equiv (\mathsf{\bar{\kk}}_I^a,\,\bar{\kk}_{Ia})$, $(\VV_I{}^A)\equiv (\mathsf{\kk}_I^a,\,\kk_{Ia})$, and
\begin{align}
 \kk_I^a\equiv \mathsf{\kk}_I^a+\pi^{ab}\,\kk_{Ib}\,,\qquad 
 \bar{\kk}_I^a\equiv \mathsf{\bar{\kk}}_I^a+\pi^{ab}\,\bar{\kk}_{Ib} \,.
\end{align}

For example, if $(\kk^{\hat{\bm{a}}})\equiv (\kk^{\ubar{\bm{a}}},\,\kk_{Ia})$ and $(\bar{\kk}^{\hat{\bm{a}}})\equiv (\bar{\kk}^{\bar{\bm{a}}},\,\bar{\kk}_{Ia})$ span the linear subspace $\tilde{\mathfrak{g}}$\,, similar to Eq.~\eqref{eq:cR-1}, we can define
\begin{align}
 \cR^{ab} \equiv \mathbb{E}^{ab} - \mathsf{\kk}_I^{a}\,\kk^{bJ} \qquad
 \bigl(\mathbb{E}^{ab} \equiv E^{ab} - \bar{\kk}^{aI}\,\tilde{\kk}_{I c}\,\hat{E}^{cd}\,\tilde{\kk}_{J d}\,\kk^{bJ}\bigr)\,,
\end{align}
where $\bar{\kk}^{aI}$ and $\kk^{aJ}$ are inverse matrices of $\bar{\kk}_{Ia}$ and $\kk_{Ja}$\,, respectively. 
Then we find that the matrix $\MM_{ab}$ is given by
\begin{align}
 \MM_{ab} = \bigl[(\cR+\pi)^{-1}\bigr]_{ab}\,. 
\end{align}
The general expression \eqref{eq:MM-general} is also obtained in the same way.
Thus, the non-degeneracy of $\langle T_I,\,\hat{\cH}(T_J)\rangle$ assumed in section \ref{sec:non-deg} is not important and can be lifted. 

\section{Properties of $\vect{G}\text{Ric}_{+}$ and $\text{$\mathbb{G}$Ric}_{+}$}
\label{app:GRic}

In this appendix, we show that $\bm{G}\text{Ric}_{+}$ and $\text{$\mathbb{G}$Ric}_{+}$\,, which are defined as
\begin{align}
 \bm{G}\text{Ric}_{+}(X,\,Y)
 &\equiv \div \bigl[\bm{P}_+([\bm{P}_-(Y),\,\bm{P}_+(X)]_{\text{D}})\bigr] - \bm{P}_-(Y)\cdot \div \bm{P}_+(X)
\nn\\
 &\quad - \langle [\bm{P}_-([\bm{P}_+(\CE_{\bcheck{C}}),\,\bm{P}_-(Y)]_{\text{D}}),\,\bm{P}_+(X)]_{\text{D}},\,\bm{P}_+(\CE^{\bcheck{C}}) \rangle \,,
\label{eq:GRic-app}
\end{align}
and the same for $\text{$\mathbb{G}$Ric}_{+}$ with $\bm{P}_\pm\to \mathbb{P}_\pm$\,, have some properties similar to $\text{GRic}_{+}$\,. 

The first property is that, for any $X,\,Y\in \Gamma^{\mathfrak{f}}(\mathbb{F}^\perp)$ and $f$ satisfying $\VV_{\sfi}\cdot f=0$\,, we have
\begin{align}
 \bm{G}\text{Ric}_{+}(f X,\,Y) = f\,\bm{G}\text{Ric}_{+}(X,\,Y)\,,\qquad 
 \text{$\mathbb{G}$Ric}_{+}(f X,\,Y) = f\,\text{$\mathbb{G}$Ric}_{+}(X,\,Y)\,.
\label{eq:GRic-prop-1}
\end{align}
The first two terms of $\bm{G}\text{Ric}_{+}(f X,\,Y)$ are expanded as
\begin{align}
\begin{split}
 &\div \bigl[\bm{P}_+([\bm{P}_-(Y),\,\bm{P}_+(f X)]_{\text{D}})\bigr]
\\
 &= f\,\div \bigl[\bm{P}_+([\bm{P}_-(Y),\,\bm{P}_+(X)]_{\text{D}})\bigr]
 + \bm{P}_+([\bm{P}_-(Y),\,\bm{P}_+(X)]_{\text{D}}) \cdot f
\\
 &\quad + \bigl[\bm{P}_-(Y) \cdot f\bigr]\,\div \bigl[\bm{P}_+(X)\bigr]
 + \bm{P}_+(X) \cdot \bigl[\bm{P}_-(Y)\cdot f\bigr]\,,
\end{split}
\end{align}
and
\begin{align}
 - \bm{P}_-(Y)\cdot \div \bm{P}_+(f X)
 = - \bm{P}_-(Y)\cdot \bigl[ \bm{P}_+(X)\cdot f\bigr]
 -\bigl[\bm{P}_-(Y) \cdot f\bigr]\,\div\bigl[\bm{P}_+(X)\bigr]\,.
\end{align}
The same expansion can be done also for $\text{$\mathbb{G}$Ric}_{+}$ for $\mathbb{P}_\pm$\,. 
The last term is slightly non-trivial
\begin{align}
\begin{split}
 &- \langle [\bm{P}_-([\bm{P}_+(\CE_{\bcheck{C}}),\,\bm{P}_-(Y)]_{\text{D}}),\,\bm{P}_+(f X)]_{\text{D}},\,\bm{P}_+(\CE^{\bcheck{C}}) \rangle
\\
 &=- f\,\langle [\bm{P}_-([\bm{P}_+(\CE_{\bcheck{C}}),\,\bm{P}_-(Y)]_{\text{D}}),\,\bm{P}_+(X)]_{\text{D}},\,\bm{P}_+(\CE^{\bcheck{C}}) \rangle
\\
 &\quad - \bigl[\bm{P}_-([\bm{P}_+(\CE_{\bcheck{C}}),\,\bm{P}_-(Y)]_{\text{D}})\cdot f\bigr]\,\langle \bm{P}_+(X),\,\bm{P}_+(\CE^{\bcheck{C}}) \rangle
\\
 &=- f\,\langle [\bm{P}_-([\bm{P}_+(\CE_{\bcheck{C}}),\,\bm{P}_-(Y)]_{\text{D}}),\,\bm{P}_+(X)]_{\text{D}},\,\bm{P}_+(\CE^{\bcheck{C}}) \rangle
\\
 &\quad - \bm{P}_-([\bm{P}_+(X^{\bcheck{C}}\,\CE_{\bcheck{C}}),\,\bm{P}_-(Y)]_{\text{D}})\cdot f 
 - [\bm{P}_-(Y)\cdot X^{\bcheck{C}}\bigr]\, \bm{P}_-(\bm{P}_+(\CE_{\bcheck{C}}))\cdot f 
\\
 &\quad 
 + \langle \bm{P}_-(Y),\,\bm{P}_+(\CE_{\bcheck{C}}))\rangle\, \bm{P}_-(\partial^\bullet X^{\bcheck{C}}\bigr)\cdot f \,,
\end{split}
\end{align}
where we have expanded $X$ as $X\equiv X^{\bcheck{A}}\,\CE_{\bcheck{A}}+X^{\sfi}\,\VV_{\sfi}$ and denoted $(\partial^\bullet X^{\bcheck{C}}\bigr)^M\equiv \partial^M X^{\bcheck{C}}$.
Then, using the properties
\begin{align}
 \bm{P}_+(X)\sim \bm{P}_+(X^{\bcheck{C}}\,\CE_{\bcheck{C}})\,,\qquad 
 \bm{P}_-(\bm{P}_+(\CE_{\bcheck{C}}))\cdot f=0\,,\qquad \langle \bm{P}_-(Y),\,\bm{P}_+(\CE_{\bcheck{C}}))\rangle=0 \,,
\end{align}
we have
\begin{align}
\begin{split}
 &- \langle [\bm{P}_-([\bm{P}_+(\CE_{\bcheck{C}}),\,\bm{P}_-(Y)]_{\text{D}}),\,\bm{P}_+(f X)]_{\text{D}},\,\bm{P}_+(\CE^{\bcheck{C}}) \rangle
\\
 &=- f\,\langle [\bm{P}_-([\bm{P}_+(\CE_{\bcheck{C}}),\,\bm{P}_-(Y)]_{\text{D}}),\,\bm{P}_+(X)]_{\text{D}},\,\bm{P}_+(\CE^{\bcheck{C}}) \rangle
 - \bm{P}_-([\bm{P}_+(X),\,\bm{P}_-(Y)]_{\text{D}})\cdot f\,.
\end{split}
\end{align}
We have the same result also for $\text{$\mathbb{G}$Ric}_{+}$ and $\mathbb{P}_\pm$\,. 
Then, summing up these terms, we find the property \eqref{eq:GRic-prop-1}. 
We can similarly show
\begin{align}
 \bm{G}\text{Ric}_{+}(X,\,f\,Y) = f\,\bm{G}\text{Ric}_{+}(X,\,Y)\,,\qquad 
 \text{$\mathbb{G}$Ric}_{+}(X,\,f\,Y) = f\,\text{$\mathbb{G}$Ric}_{+}(X,\,Y)\,.
\end{align}

The second property is that $\bm{G}\text{Ric}_{+}(X,\,Y)$ and $\text{$\mathbb{G}$Ric}_{+}(X,\,Y)$ for $X,\,Y\in \Gamma^{\mathfrak{f}}(\mathbb{F}^\perp)$ are independent of the choice of the frame fields $\CE_{\bcheck{A}}$\,.
Let us consider how the second line of Eq.~\eqref{eq:GRic-app} transforms under a redefinition $\CE'_{\bcheck{A}}=\Lambda_{\bcheck{A}}{}^{\bcheck{B}}\,\CE_{\bcheck{B}}$ ($\VV_{\sfi}\cdot \Lambda_{\bcheck{A}}{}^{\bcheck{B}}=0$). 
We find
\begin{align}
\begin{split}
 &\Lambda^{\bcheck{A}\bcheck{C}}\,\langle [\bm{P}_-([\Lambda_{\bcheck{A}}{}^{\bcheck{B}}\,\bm{P}_+(\CE_{\bcheck{B}}),\,\bm{P}_-(Y)]_{\text{D}}),\,\bm{P}_+(X)]_{\text{D}},\,\bm{P}_+(\CE^{\bcheck{C}}) \rangle
\\
 &=\Lambda^{\bcheck{A}\bcheck{C}}\,\langle [\Lambda_{\bcheck{A}}{}^{\bcheck{B}}\,\bm{P}_-([\bm{P}_+(\CE_{\bcheck{B}}),\,\bm{P}_-(Y)]_{\text{D}}),\,\bm{P}_+(X)]_{\text{D}},\,\bm{P}_+(\CE^{\bcheck{C}}) \rangle
\\
 &\quad -\Lambda^{\bcheck{A}\bcheck{C}}\,\langle [(\bm{P}_-(Y)\cdot \Lambda_{\bcheck{A}}{}^{\bcheck{B}})\,\bm{P}_-\,\bm{P}_+(\CE_{\bcheck{B}}),\,\bm{P}_+(X)]_{\text{D}},\,\bm{P}_+(\CE^{\bcheck{C}}) \rangle\,,
\end{split}
\label{eq:bases-indep-2nd}
\end{align}
where we have used $\langle \bm{P}_+(\CE_{\bcheck{B}}),\,\bm{P}_-(Y)\rangle=0$\,. 
In the last line, $\bm{P}_-\,\bm{P}_+(\CE_{\bcheck{B}})$ does not vanish but becomes a section of $\mathbb{F}$, and we denote
\begin{align}
 (\bm{P}_-(Y)\cdot \Lambda_{\bcheck{A}}{}^{\bcheck{B}})\,\bm{P}_-\,\bm{P}_+(\CE_{\bcheck{B}}) \equiv \varphi_{\bcheck{A}}{}^{\sfi}\,\VV_{\sfi}\,.
\end{align}
Then, using $\langle \VV_{\sfi},\,\bm{P}_+(X)\rangle=\langle \VV_{\sfi},\,\bm{P}_+(\CE^{\bcheck{C}})\rangle=0$, we find that the last line vanishes:
\begin{align}
 \Lambda^{\bcheck{A}\bcheck{C}}\,\langle [\varphi_{\bcheck{A}}{}^{\sfi}\,\VV_{\sfi},\,\bm{P}_+(X)]_{\text{D}},\,\bm{P}_+(\CE^{\bcheck{C}}) \rangle
 =\Lambda^{\bcheck{A}\bcheck{C}}\,\varphi_{\bcheck{A}}{}^{\sfi}\,\langle [\VV_{\sfi},\,\bm{P}_+(X)]_{\text{D}},\,\bm{P}_+(\CE^{\bcheck{C}}) \rangle =0\,.
\end{align}
We also find that the second line of Eq.~\eqref{eq:bases-indep-2nd} becomes
\begin{align}
 &\Lambda^{\bcheck{A}\bcheck{C}}\,\Lambda_{\bcheck{A}}{}^{\bcheck{B}}\,\langle [\bm{P}_-([\bm{P}_+(\CE_{\bcheck{B}}),\,\bm{P}_-(Y)]_{\text{D}}),\,\bm{P}_+(X)]_{\text{D}},\,\bm{P}_+(\CE^{\bcheck{C}}) \rangle
\nn\\
 &- \Lambda^{\bcheck{A}\bcheck{C}}\,[\bm{P}_+(X)\cdot\Lambda_{\bcheck{A}}{}^{\bcheck{B}}]\,\langle \bm{P}_-([\bm{P}_+(\CE_{\bcheck{B}}),\,\bm{P}_-(Y)]_{\text{D}}) ,\,\bm{P}_+(\CE^{\bcheck{C}}) \rangle
\nn\\
 &+ \Lambda^{\bcheck{A}\bcheck{C}}\,\langle \bm{P}_-([\bm{P}_+(\CE_{\bcheck{B}}),\,\bm{P}_-(Y)]_{\text{D}}),\,\bm{P}_+(X)\rangle\,\langle \partial\Lambda_{\bcheck{A}}{}^{\bcheck{B}},\,\bm{P}_+(\CE^{\bcheck{C}}) \rangle
\nn\\
 &= \langle [\bm{P}_-([\bm{P}_+(\CE_{\bcheck{C}}),\,\bm{P}_-(Y)]_{\text{D}}),\,\bm{P}_+(X)]_{\text{D}},\,\bm{P}_+(\CE^{\bcheck{C}}) \rangle\,. 
\end{align}
This shows that $\bm{G}\text{Ric}_{+}(X,\,Y)$ is independent of the choice of the generalized frame fields $\CE_{\bcheck{A}}$\,. 
The same discussion can be applied to $\text{$\mathbb{G}$Ric}_{+}(X,\,Y)$ as well.

\end{document}